\documentclass[useAMS,usenatbib]{mn2e}
\usepackage{txfonts}
\usepackage{color, supertabular, multicol}

\usepackage{graphicx}
\usepackage{amssymb}
\usepackage{threeparttable}

\newcommand{\oiii}{[O {\sc iii}]}

\mathchardef\mhyphen="2D

\title[Observations of RQ quasar feedback -- II.]{Observations of feedback from radio-Quiet quasars -- 
II. Kinematics of ionized gas nebulae}

\author[G. Liu et al.]{Guilin Liu$^1$\thanks{E-mail: liu@pha.jhu.edu}, 
Nadia L. Zakamska$^1$\thanks{E-mail: zakamska@pha.jhu.edu}, Jenny E. Greene$^2$, 
Nicole P. H. Nesvadba$^3$
\newauthor{and Xin Liu$^{4,5}$}\\
\\
$^{1}$Department of Physics \& Astronomy, Johns Hopkins University, 3400 N. Charles St., Baltimore, MD 21218, USA \\
$^{2}$Department of Astrophysical Sciences, Princeton University, Princeton, NJ 08544, USA \\
$^{3}$Institut d'Astrophysique Spatiale, CNRS, Universit\'{e} Paris-Sud, 91405 Orsay, France \\
$^{4}$Department of Physics and Astronomy, University of California, Los Angeles, CA 90095, USA \\
$^{5}$Hubble Fellow
}

\begin{document}

\date{Submitted to MNRAS: 2013 May 28}

\pagerange{\pageref{firstpage}--\pageref{lastpage}} \pubyear{2013}

\maketitle

\label{firstpage}

\begin{abstract}
  The prevalence and energetics of quasar feedback is a major
  unresolved problem in galaxy formation theory. In this paper, we
  present Gemini Integral Field Unit observations of
  ionized gas around eleven luminous, obscured, radio-quiet quasars at
  $z\sim0.5$ out to $\sim 15$ kpc from the quasar; specifically, we 
  measure the kinematics and morphology of [O {\sc iii}]$\lambda$5007\AA\ 
  emission. The round morphologies of the nebulae and the large line-of-sight 
  velocity widths (with velocities containing 80\% of the emission
  as high as 10$^3$ km s$^{-1}$) combined with relatively small
  velocity difference across them (from 90 to 520 km s$^{-1}$) point
  toward wide-angle quasi-spherical outflows. We use the observed
  velocity widths to estimate a median outflow velocity of 760 km s$^{-1}$, 
  similar to or above the escape velocities from the host galaxies. 
  The line-of-sight velocity dispersion declines slightly toward outer 
  parts of the nebulae (by 3\% per kpc on average). The majority of nebulae 
  show blueshifted excesses in their line profiles across most of their extents, 
  signifying gas outflows. For the median outflow velocity, we find $\dot{E}_{\rm kin}$ 
  between $4\times 10^{44}$ and $3\times 10^{45}$ erg s$^{-1}$ and $\dot{M}$ 
  between $2\times 10^3$ and $2\times 10^4$ $M_{\odot}$ yr$^{-1}$. 
  These values are large enough for the observed quasar winds to have a 
  significant impact on their host galaxies. The median rate of converting 
  bolometric luminosity to kinetic energy of ionized gas clouds is $\sim$2\%. 
  We report four new candidates for ``super-bubbles'' -- outflows that
  may have broken out of the denser regions of the host galaxy.
\end{abstract}


\section{Introduction}
\label{sec:intro}

One of the most fascinating astronomical discoveries of the last
several decades is the gradual realization that almost every massive
galaxy, including our own Milky Way, contains a super-massive black
hole in its center \citep{mago98}. Several lines of evidence suggest
that there is a fundamental connection between the black holes
residing in galaxy centers and formation and evolution of their host
galaxies. One such observation is the tight correlation between 
black hole masses and the velocity dispersions and masses of their
host bulges
\citep{gebh00,ferr00,trem02,marc03,hari04,gult09,mcco11}. Another is
the close similarity of the black hole accretion history and the star
formation history over the life-time of the universe \citep{boyl98}.

In addition to these observations, modern galaxy formation theory
strongly suggests that black hole activity has a controlling effect on
shaping the global properties of the host galaxies \citep{tabo93,
silk98, spri05}. This is especially true for the most massive
galaxies, whose numbers decline much more rapidly with luminosity than
the predictions of large-scale dark matter simulations would
suggest. One possibility is that the energy output of the black hole
in its most active (``quasar'') phase may be somehow coupled to the
gas from which the stars form. If the quasar launches a wind that
entrains and removes gas from the galaxy or reheats the gas, then it
can shut off star formation in its host \citep{thou95,crot06}. Thus,
quasars could be instrumental in limiting the maximal mass of
galaxies.

In recent years, this type of feedback from accreting black holes has
become a key element in modeling galaxy evolution 
\citep[e.g.,][]{hopk06, crot06, choi12}. Feedback can in principle explain 
galaxy vs. black hole correlations and the lack of overly massive blue 
galaxies in the local universe. As significant as these achievements are, 
it has been challenging to find direct observational evidence of black hole
vs. galaxy self-regulation and to obtain measurements of feedback
energetics. Direct and indirect evidence for powerful quasar-driven
winds started emerging, both for radio-quiet 
\citep{arav08,moe09,dunn10,alex10} and radio-loud objects
\citep{nesv06,nesv08} at low and high redshifts \citep{maio12,borg13}.

In the last several years, we have undertaken an observational
campaign to map out the kinematics of the ionized gas around luminous
obscured quasars \citep{zaka03, reye08} using Magellan, Gemini and
other facilities in search of signatures of quasar-driven winds
\citep{gree09, gree11, gree12, liu13a}. 
In our observations, we are focusing on the most powerful quasars
likely associated with the most massive galaxies, where feedback
effects are expected to be strongest, and we use the observational
advantages provided by circumnuclear obscuration to maximize
sensitivity to faint extended emission associated with quasar
feedback.

In December 2010, we obtained Gemini-North Multi-Object Spectrograph
(GMOS-N) Integral Field Unit (IFU) observations of a sample of
obscured luminous quasars at $z\sim0.5$. In the first paper describing
our results \citep[][hereafter Paper I]{liu13a}, we present the
analysis of the extents and morphologies of the narrow emission line
regions of these quasars. We spatially resolve emission line nebulae
in every case and find that the \oiii\ line emission from gas
photo-ionized by the hidden quasar is detected out to $14\pm4$ kpc
from the center of the galaxy. Ionized gas nebulae around radio-quiet
obscured quasars display regular smooth morphologies, in marked
contrast to nebulae around radio-loud quasars of similar line
luminosities which tend to be significantly more elongated and/or
lumpy. Surprisingly, no pronounced biconical structures, expected in a
simple quasar illumination model, are detected.

In this paper, we analyze the kinematics of the ionized gas nebulae
around these quasars. In Section \ref{sec:data} we describe
observations and modeling of line kinematics. In Section
\ref{sec:science}, we present kinematic measurements of the ionized
gas emission, in Section \ref{sec:models} we discuss kinematic models
and structure of quasar winds, in Section \ref{sec:bubbles} we present 
super-bubble candidates, in Section \ref{sec:energy} we derive
the kinetic energy of observed winds, and we summarize in Section
\ref{sec:summary}. As in Paper I, we adopt a $h$=0.71, $\Omega_m$=0.27, 
$\Omega_{\Lambda}$=0.73 cosmology throughout this paper; objects are 
identified as SDSS Jhhmmss.ss+ddmmss.s in Table~\ref{tab1} and \ref{tab2} 
and are shortened to SDSS Jhhmm+ddmm elsewhere; and the rest-frame 
wavelengths of the emission lines are given in air.

\begin{table*}
\caption{Kinematic measurements of quasar nebulae.}
\setlength{\tabcolsep}{1.4mm}
\label{tab1}
\begin{small}
\begin{center}
\begin{tabular}{llccrrrrrrrrrr}
\hline
\noalign{\smallskip}
\multicolumn{1}{c}{Object name} &
\multicolumn{1}{c}{Radio} &
\multicolumn{1}{c}{$z$} &
\multicolumn{1}{c}{$L_{\rm [O~{\scriptscriptstyle III}]}$} &
\multicolumn{1}{c}{$R_{5\sigma}$} &
\multicolumn{1}{c}{$\Delta v_{\rm max}$} &
\multicolumn{1}{c}{$\langle W_{80} \rangle$} &
\multicolumn{1}{c}{$W_{\rm 80,max}$} &
\multicolumn{1}{c}{$\nabla W_{80}$} &
\multicolumn{1}{c}{$\nabla W_{80}^{\dagger}$} &
\multicolumn{1}{c}{$\langle v_{02} \rangle$} &
\multicolumn{1}{c}{$v_{\rm 02,max}$} &
\multicolumn{1}{c}{$\langle A\rangle$} &
\multicolumn{1}{c}{$\langle K\rangle$} \\
\multicolumn{1}{c}{(1)} &
\multicolumn{1}{c}{(2)} &
\multicolumn{1}{c}{(3)} &
\multicolumn{1}{c}{(4)} &
\multicolumn{1}{c}{(5)} &
\multicolumn{1}{c}{(6)} &
\multicolumn{1}{c}{(7)} &
\multicolumn{1}{c}{(8)} &
\multicolumn{1}{c}{(9)} &
\multicolumn{1}{c}{(10)} &
\multicolumn{1}{c}{(11)} &
\multicolumn{1}{c}{(12)} &
\multicolumn{1}{c}{(13)} &
\multicolumn{1}{c}{(14)} \\
\hline

SDSS J014932.53$-$004803.7 & RQ & 0.567 & 42.87 &  9.1 & 114 & 1167 & 1406 &  $-$2.8 &     1.1 & $-$1191 & $-$1765 & $-$0.14 & 1.37 \\
SDSS J021047.01$-$100152.9 & RQ & 0.540 & 43.48 & 17.7 & 407 &  667 &  786 &  $-$1.9 &  $-$0.6 &  $-$560 &  $-$814 &    0.12 & 1.46 \\
SDSS J031909.61$-$001916.7 & RQ & 0.635 & 42.74 &  7.6 & 348 & 1845 & 2142 & $-$10.5 & $-$16.6 & $-$1474 & $-$1844 & $-$0.05 & 1.31 \\
SDSS J031950.54$-$005850.6 & RQ & 0.626 & 42.96 & 11.5 & 161 &  780 &  958 &  $-$5.1 &  $-$4.4 &  $-$934 & $-$1198 & $-$0.23 & 1.68 \\
SDSS J032144.11$+$001638.2 & RQ & 0.643 & 43.10 & 18.3 & 522 &  974 & 1092 &  $-$4.3 &  $-$6.9 &  $-$946 & $-$1102 & $-$0.18 & 2.01 \\
SDSS J075944.64$+$133945.8 & RQ & 0.649 & 43.38 & 14.1 & 122 & 1230 & 1275 &  $-$4.3 &  $-$1.3 & $-$1250 & $-$1393 & $-$0.26 & 1.58 \\
SDSS J084130.78$+$204220.5 & RQ & 0.641 & 43.31 & 11.9 & 104 &  723 &  750 &  $-$2.8 &  $-$2.4 &  $-$675 &  $-$822 & $-$0.04 & 1.40 \\
SDSS J084234.94$+$362503.1 & RQ & 0.561 & 43.56 & 15.1 & 162 &  489 &  525 &  $-$3.9 &  $-$5.2 &  $-$522 &  $-$652 &    0.00 & 1.65 \\
SDSS J085829.59$+$441734.7 & RQ & 0.454 & 43.30 & 11.7 &  89 &  876 &  920 &  $-$5.9 &  $-$4.3 &  $-$939 &  $-$970 & $-$0.24 & 1.94 \\
SDSS J103927.19$+$451215.4 & RQ & 0.579 & 43.29 & 12.2 & 126 & 1105 & 1197 &  $-$6.3 &     0.4 & $-$1046 & $-$1287 & $-$0.04 & 1.44 \\
SDSS J104014.43$+$474554.8 & RQ & 0.486 & 43.52 & 14.5 & 166 & 1315 & 1659 &  $-$4.2 &  $-$3.1 & $-$1821 & $-$2390 & $-$0.38 & 2.10 \\
3C67/J022412.30$+$275011.5 & RL & 0.311 & 42.83 & 12.2 & 388 &  688 & 1328 &    ---~ &    ---~ &  $-$681 & $-$1391 & $-$0.17 & 1.09 \\
SDSS J080754.50$+$494627.6 & RL & 0.575 & 43.27 & 19.4 & 920 &  714 &  843 &    ---~ &    ---~ &  $-$516 &  $-$938 &    0.09 & 1.19 \\
SDSS J110140.54$+$400422.9 & RI & 0.457 & 43.55 & 18.9 & 238 &  753 & 1020 &    ---~ &    ---~ &  $-$686 & $-$1160 &    0.13 & 1.34 \\

\hline
\end{tabular}
\tablenotes{{\bf Notes.} -- 
{\bf (1)} Object name.
{\bf (2)} Radio loudness (RQ: radio quiet; RL: radio loud; RI: radio intermediate).
{\bf (3)} Redshift, from \citet{zaka03} and \citet{reye08} for SDSS objects and \citet{erac04} for 3C67.
{\bf (4)} Total luminosity of the [O {\sc iii}]$\lambda$5007\AA\ line (logarithmic scale, in erg s$^{-1}$), from \citet{liu13a}.
{\bf (5)} Semi-major axis (in kpc) of the best-fit ellipse which encloses pixels with $S/N\geqslant5$ in the
   [O {\sc iii}]$\lambda$5007\AA\ line map, from \citet{liu13a}.
{\bf (6)} Maximum difference in the median velocity map (cf. Figure \ref{fig:VWAK}), in km s$^{-1}$. For each object, the 5\% tails on either side of the velocity distribution are excluded for determination of $\Delta v_{\rm max}$ to minimize the effect of the noise. 
{\bf (7, 11, 13, 14)} $W_{80}$ (km s$^{-1}$), $v_{02}$ (km s$^{-1}$), $A$ and $K$ values of the integrated [O {\sc iii}]$\lambda$5007\AA\ line, measured from the SDSS fiber spectrum (see Section \ref{sec:parameters} for the definition of these parameters).
{\bf (8, 12)} Maximum $W_{80}$ and most negative $v_{02}$ values in their respective spatially-resolved maps ((km s$^{-1}$)).
   Like for $\Delta v_{\rm max}$, the 5\% tails on either side of their respective distributions are excluded.
{\bf (9)} Observed percentage change of $W_{80}$ per unit distance from the brightness center, in units of \% kpc$^{-1}$.
   It is defined as $\nabla W_{80}\equiv\delta W_{80}/R_5$, where $R_5$ is the maximum radius for the region where the
   peak of the [O {\sc iii}]$\lambda$5007\AA\ line is detected with $S/N\geqslant 5$ (Figure \ref{fig:sigr}), and 
   $\delta W_{80}=100\times(W_{80,R_{5}}-W_{\rm 80,R=0})/W_{\rm 80,R=0}$.
{\bf (10)} Percentage change of $W_{80}$ per unit distance from the brightness center, in units of \% kpc$^{-1}$, but
   calculated with a uniform $S/N$ of 15 at the peak of the [O {\sc iii}]$\lambda$5007\AA\ line (Figure~\ref{fig:sigr},
   see Section \ref{sec:dispersion}). It is defined as $\nabla W_{80}^{\dagger}\equiv\delta W_{80}^{\dagger}/R_{15}$, where 
   $R_{15}$ is the maximum radius for $S/N\geqslant15$ in the observed map, and 
   $\delta W_{80}^{\dagger}=100\times(W_{80,R_{15}}-W_{80,R=0})/W_{80,R=0}$.}
\end{center}
\end{small}
\end{table*}

\section{Data and line profile fits}
\label{sec:data}

\subsection{Sample and observations}

The sample presented in this paper consists of eleven radio-quiet
obscured quasars at $z\sim 0.5$ selected to be among the most
\oiii$\lambda$5007-luminous objects in the catalog by \citet{reye08}. These sources
were originally identified based on their optical spectroscopic
properties. Their permitted emission lines have widths similar to
those of forbidden lines, their integrated \oiii$\lambda$5007/H$\beta$
ratios tend to be high ($\sim$10), they routinely show high-ionization
lines such as [NeV]$\lambda\lambda$3346, 3426 and they do not 
show the characteristic blue
continuum of unobscured quasars \citep{zaka03, reye08}. These
properties are classical signatures of obscured active galactic nuclei
\citep{anto93} and lead us to conclude that the emission-line region
is illuminated by a quasar-like spectrum rich in ultra-violet and
X-ray photons, but the continuum-emitting and broad-line regions of
these objects are not directly seen. Multi-wavelength observations of
the objects in this sample confirms their nature as intrinsically
luminous quasars ($L_{\rm bol}\ga 10^{46}$ erg s$^{-1}$ at the \oiii\
luminosities probed in this paper, \citealt{liu09}) with large amounts
of circumnuclear obscuration along the line of sight \citep{zaka04,
  zaka05, zaka06, zaka08, jia12}. Such obscured objects may constitute
half or more of the entire quasar population at all redshifts and
luminosities \citep{reye08, lawr10}.

Radio-quiet candidates (those without powerful jets) are selected
based on their radio luminosities and positions in the \oiii-radio
luminosity diagram \citep{xu99, zaka04, lal10}. We further supplement
our sample with one radio-intermediate and two radio-loud objects
that we use as a comparison sample in combination with other
radio-loud sources from the literature, both at low \citep{fu09} and
at high redshifts \citep{nesv08}.

We use GMOS on Gemini-North in the 2-slit IFU mode with a field of
view of 5\arcsec$\times$7\arcsec\ and with typical on-source exposure
times of 60 minutes to obtain spatially resolved spectroscopic
observations with r.m.s. surface brightness sensitivity 1.1--2.2
$\times 10^{-17}$ erg s$^{-1}$ cm$^{-2}$ arcsec$^{-2}$. The average
seeing of 0.4\arcsec--0.7\arcsec\ corresponds to a linear resolution
of $\sim$3 kpc at the redshifts of our sample. Data reductions are
performed using the IRAF-based standard GMOS pipeline. Spectro-photometric
calibrations are performed off of Sloan Digital Sky Survey (SDSS) data
and are likely good to 5\% or better \citep{adel08}. Further details
of sample selection and data reductions are given in Paper I.

The employed grating R400-G5305 has a spectral resolution of
$R$=1918. More precisely, we fit a Gaussian profile to the 
unresolved sky lines and find their full width at half maximum to 
be FWHM=137$\pm$12 km s$^{-1}$.
As the width of our observed \oiii\ line is always well above the
instrumental resolution, the velocity structure of our
objects is well resolved. Because line profiles are strongly
non-Gaussian and in most cases are much broader than the spectral
resolution, we do not correct the observed profiles for instrumental
effects and report all values as measured. For a typical line profile
in our sample, the velocity range containing 80\% of line power is
$W_{80}\gtrsim500$ km s$^{-1}$, and instrumental broadening results in
a 4\% increase in the $W_{80}$ measurement.

\subsection{Multi-Component Gaussian Fitting}
\label{sec:fitting}

At each position in the field of view, the velocity structure of the \oiii$\lambda$5007\AA\ emission line
is generally complicated, presumably because of the existence of multiple moving gas 
components whose 3-dimensional velocities are then projected onto the line of sight. To measure 
the centroid velocity and velocity dispersion of these components for our physical analysis, we 
fit a combination of multiple Gaussians to the \oiii\ line profile on a fixed wavelength range of 
the spectrum in each spatial element (spaxel) using the IDL package {\sc mpfit} developed by \citet{mark09}. 
The purpose of these fits is merely to realistically represent the velocity profiles, so that the effect 
of noise is minimized when we perform non-parametric measurements (cf. Section \ref{sec:parameters}).
Our tests of using Voigt profiles shows that their heavier wings result in fits statistically 
inferior to the Gaussian ones. In particular, the line width measures are systematically 20\% overly high, 
although the qualitative conclusions are consistent with Gaussian fits \citep{zaka13}.

\begin{figure*}%[h!]
\begin{flushleft}
\hspace*{3mm}
    \includegraphics[scale=0.35,trim=1cm 0mm 15mm 0mm]{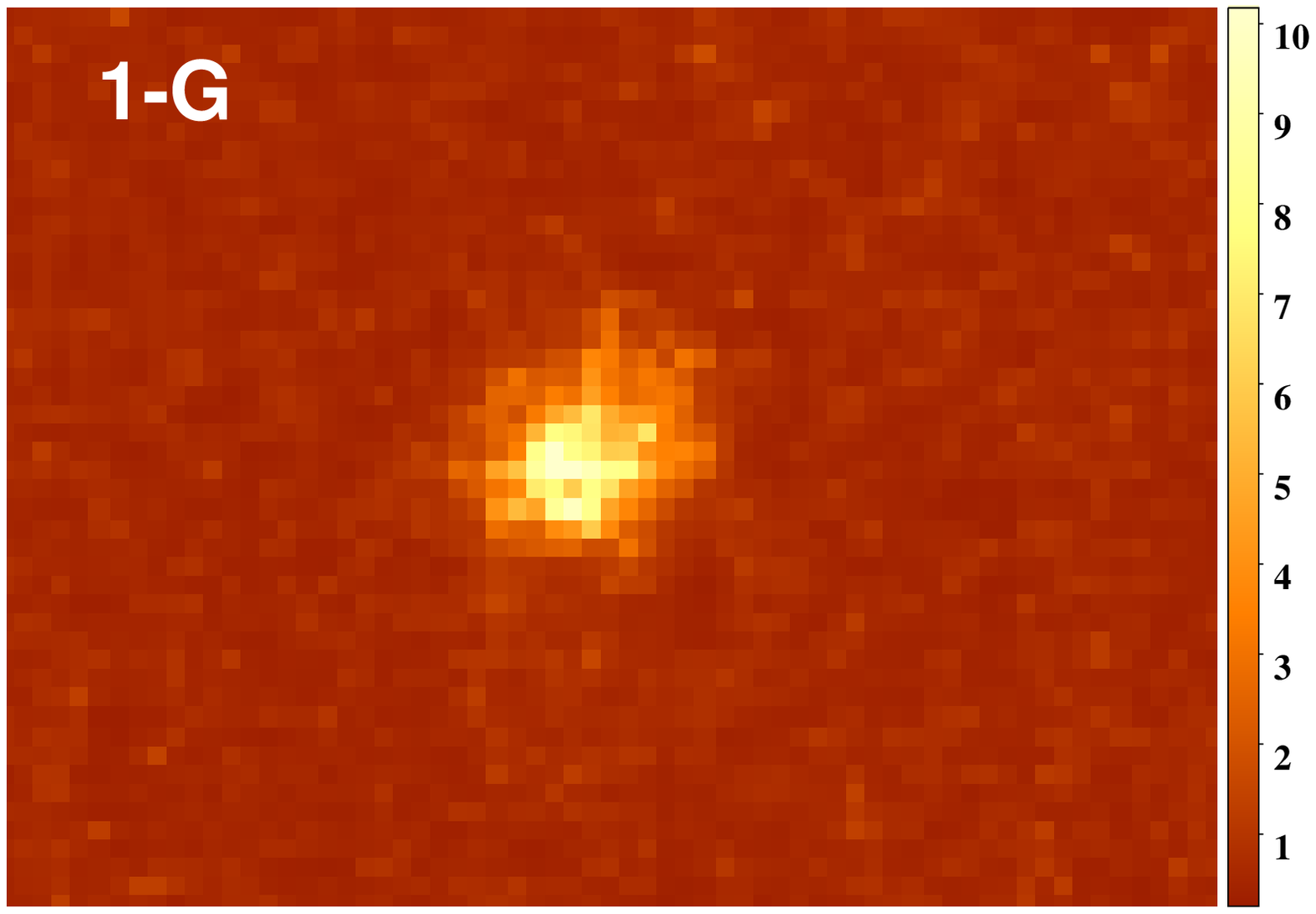}%
    \includegraphics[scale=0.35,trim=1cm 0mm 15mm 0mm]{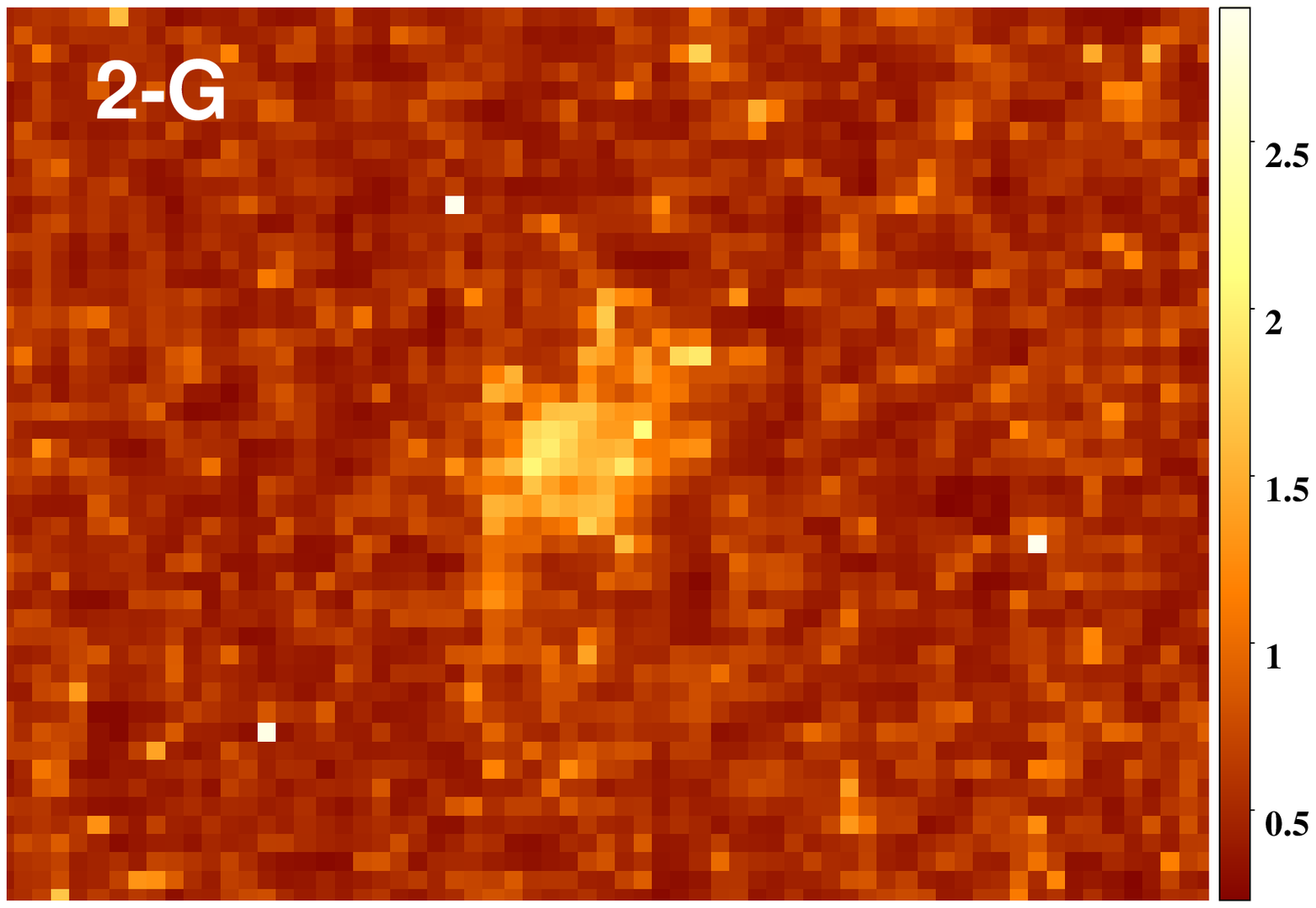}%
    \includegraphics[scale=0.35,trim=1cm 0mm 15mm 0mm]{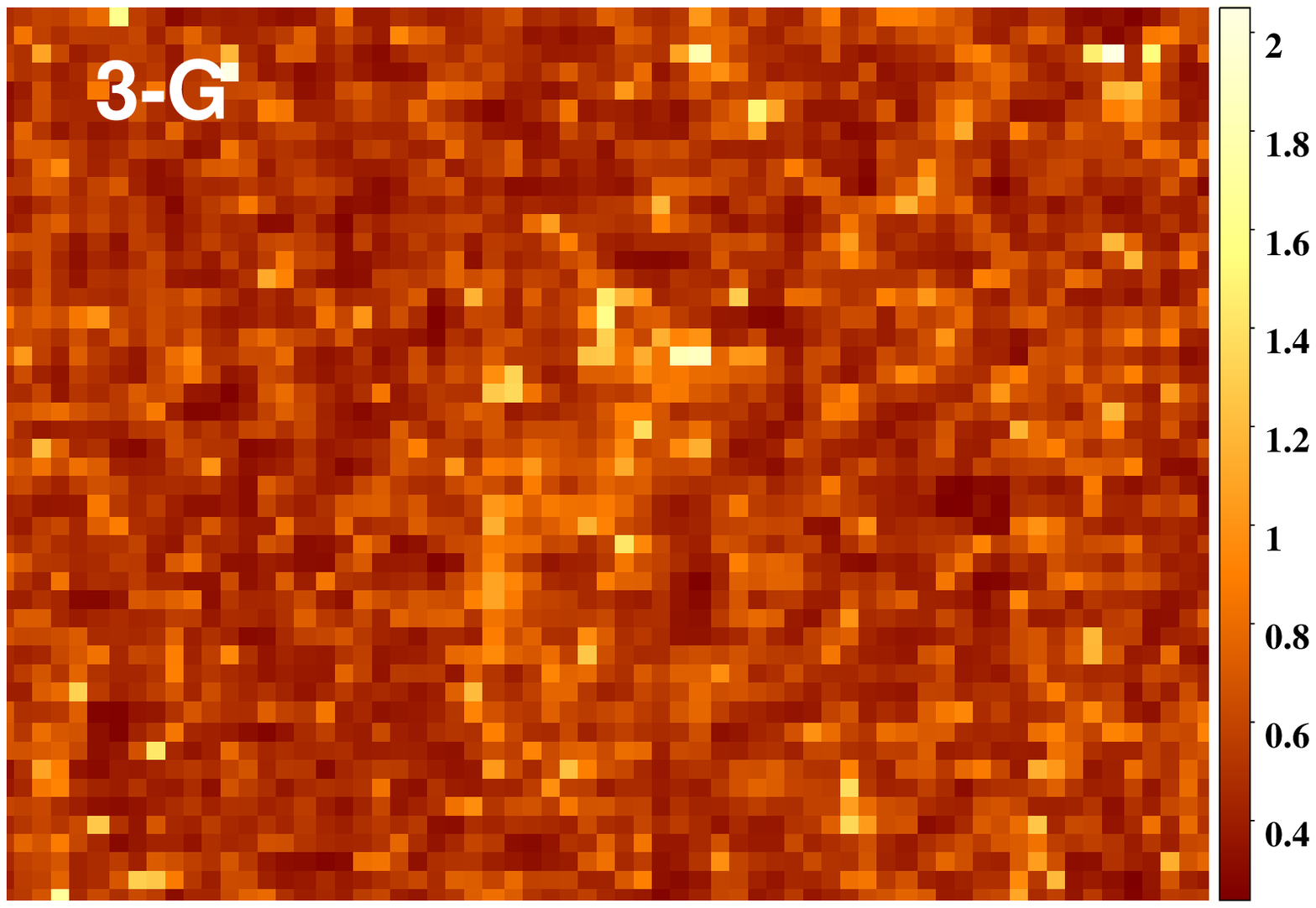}\\  	
\hspace*{3mm}
    \includegraphics[scale=0.35,trim=1cm 0mm 15mm 0mm]{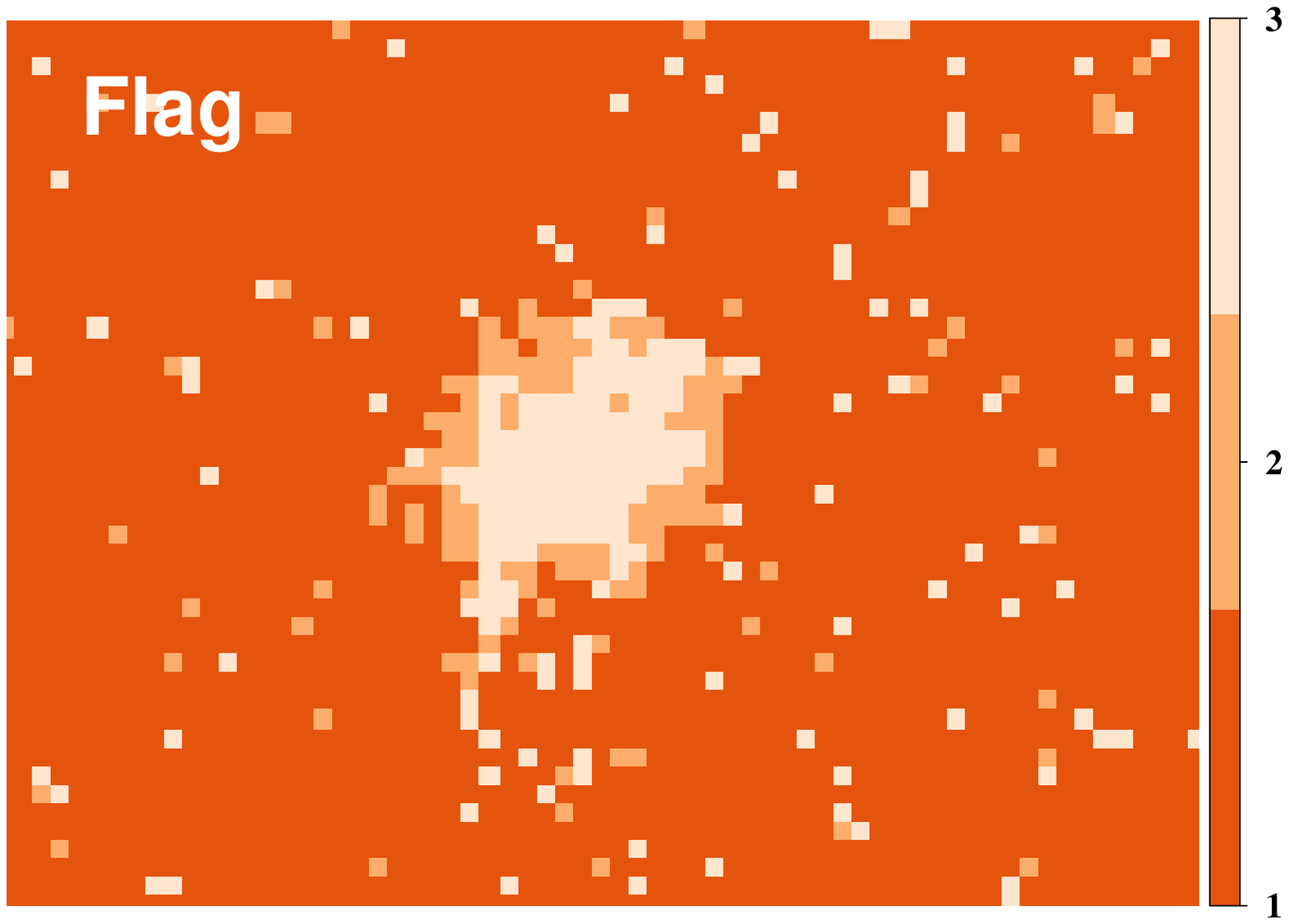}%
    \includegraphics[scale=0.35,trim=1cm 0mm 15mm 0mm]{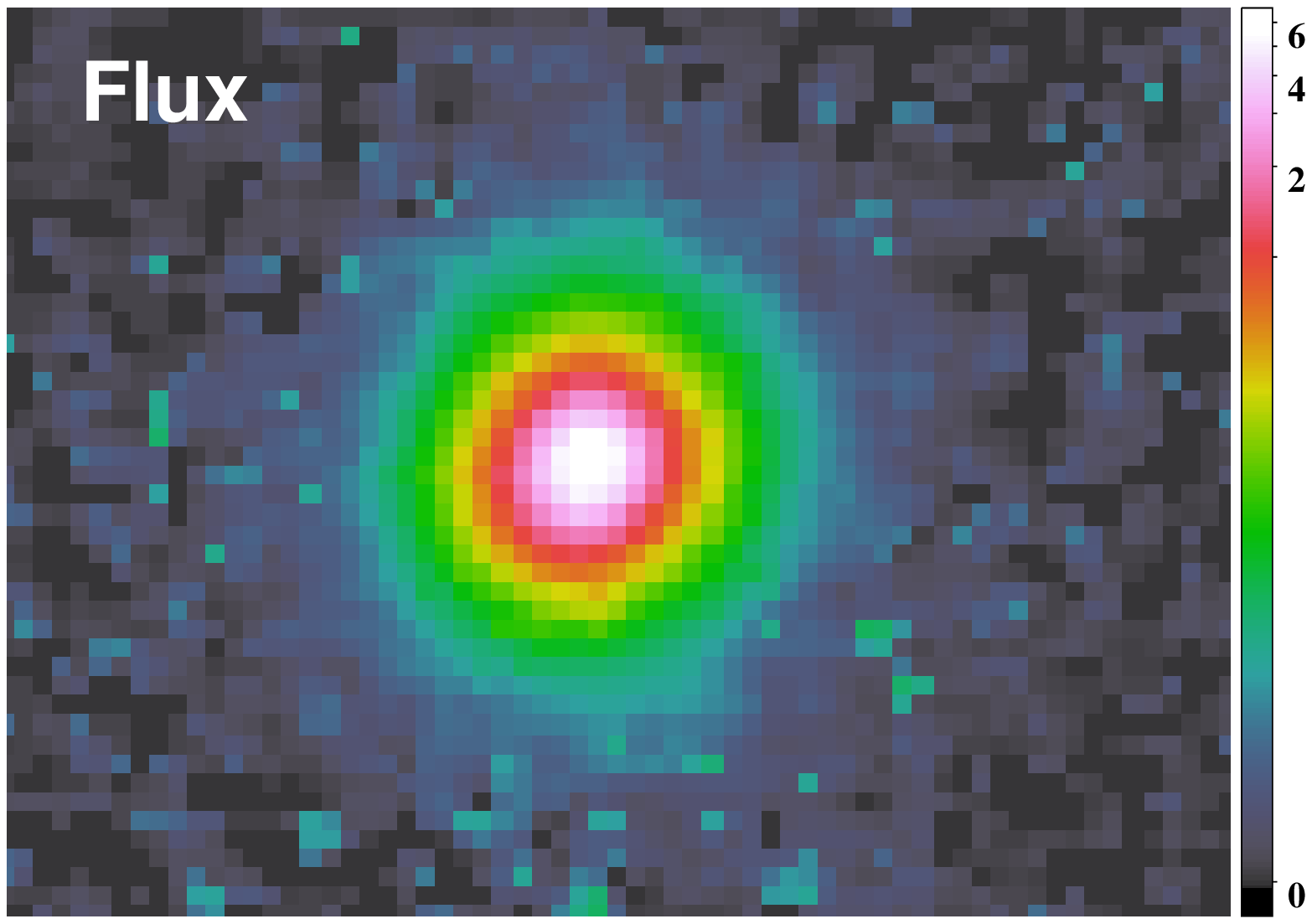}%
    \includegraphics[scale=0.35,trim=1cm 0mm 15mm 0mm]{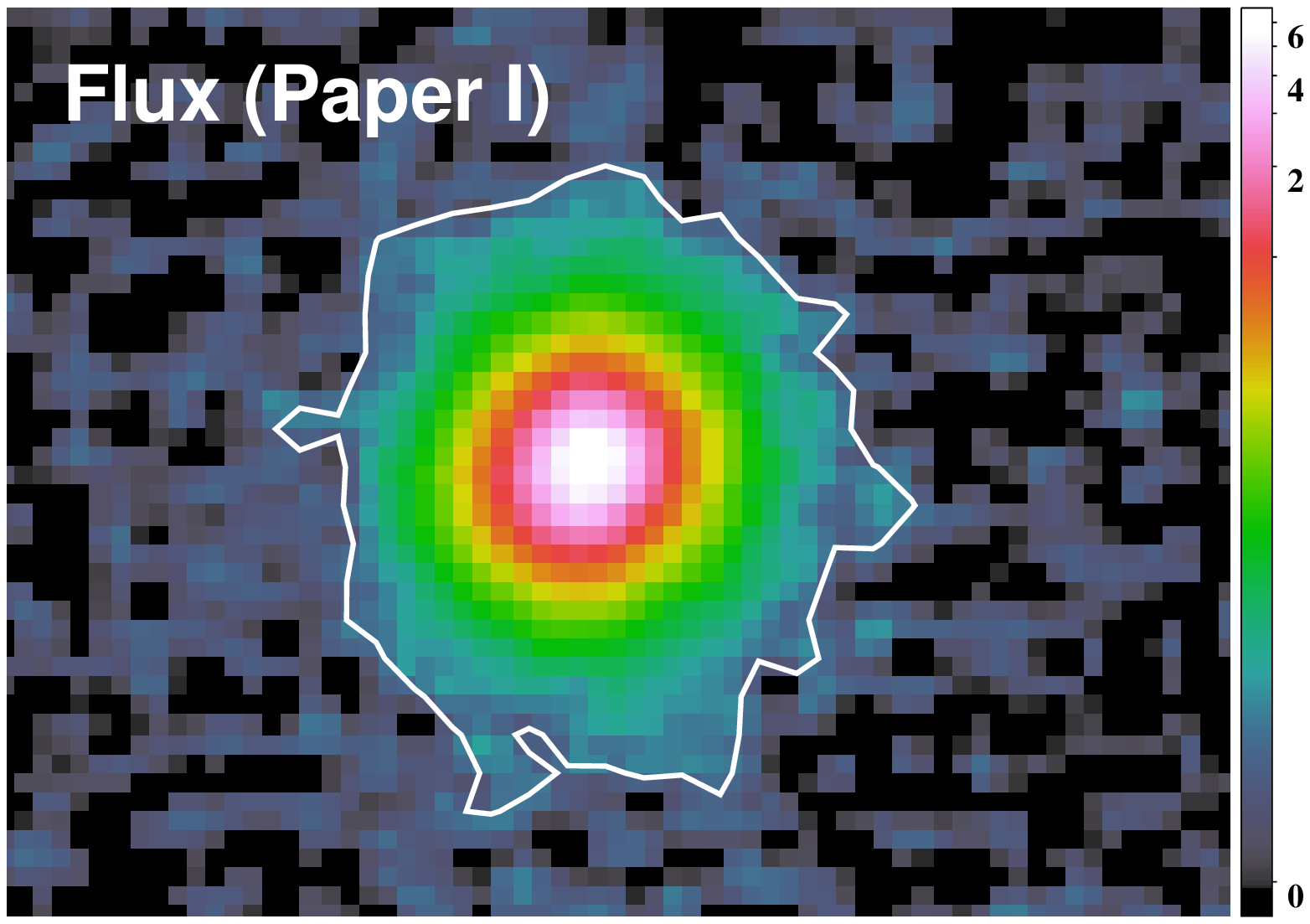}%
\end{flushleft}
\caption{SDSS J0841+2042. Upper row: the maps of reduced $\chi^2$ values for fits with 1, 2 and 3 Gaussian components. Lower row: the flag map showing the number of Gaussian components used for fits, the intensity map of the [O {\sc iii}]$\lambda$5007\AA\ recovered by our multi-Gaussian fits (logarithmic scale, in units of 10$^{-14}$ erg s$^{-1}$ cm$^{-2}$ arcsec$^{-2}$), and the [O {\sc iii}] map from Paper I \citep{liu13a} created by collapsing the continuum-subtracted IFU datacube with the $S/N=5$ threshold used in Paper I overlaid (note that $S/N$ refers to the integrated line intensity in Paper I, but to the peak of the line in Figures \ref{fig:VWAK} and \ref{fig:sigr}).}
\label{fig:gaussian}
\end{figure*}

Despite the complexity of velocity profiles, we find that for every
source in our sample, the \oiii\ line in the spectra can always be fitted by 
a combination of no more than 3 Gaussians, resulting in minimized 
reduced $\chi^2$ values that satisfy $\chi^2/\nu<2$ (where $\nu$ is 
the number of degrees of freedom) except for sporadic problematic spaxels. 
As an example, we show the reduced $\chi^2$ maps of one of our objects 
for single, double and triple Gaussian fits in Figure~\ref{fig:gaussian}. 
The number of degrees of freedom $\nu$ is the same in all the spaxels of 
each panel because the wavelength is evenly sampled. Fits with three 
Gaussians are sufficient to achieve a uniform reduced $\chi^2$ map without 
any correlated residuals. 

If a spectrum can be reasonably fitted using only one or two Gaussian components, 
we prefer the smallest possible number to avoid over-fitting. Our quantitative comparison of the
fitting quality is based on $p$-values. The $p$-value for $x_0$, denoted $p(x_0,\nu)$, is the 
probability that a random variable $x$ drawn from the $\chi^2$ distribution satisfies 
$x\leqslant x_0$, and is therefore the probability that the discrepancy between the data and 
the best-fit model is purely accidental.

To determine whether $M$ or $N$ Gaussians (for $M>N$) should be used,
we calculate the $p$-values ($p_{M}$ and $p_{N}$) that the minimized
$\chi^2$-value and $\nu$ correspond to, respectively. We then choose a
threshold of 0.01, which is a widely used fiducial criterion of
significance level for hypothesis testing in statistics. In the case
of $p_{M}\geqslant 0.01$, the data are relatively easy to fit. Thus,
we neglect the 0.01 difference and adopt the smaller $N$ as long as
$p_{M}-p_N<0.01$. If $p_{M}<0.01$, the data are more difficult to fit,
therefore we have stronger inclination to adopting $M$ components, and
adopt $N$ only if $p_M<p_N$.

\begin{figure*}%[h!]
\centering
    \includegraphics[scale=0.5]{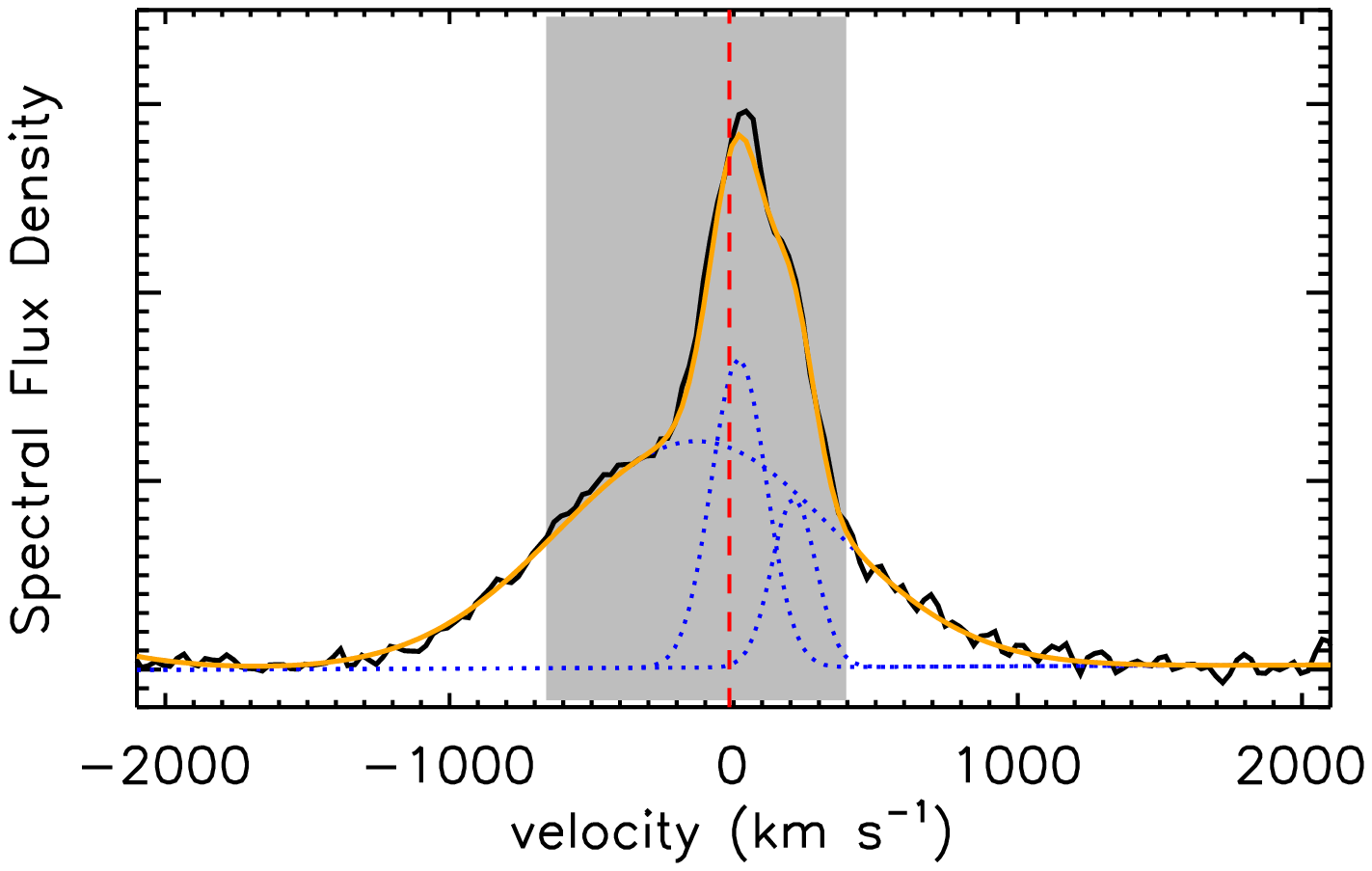}%
    \includegraphics[scale=0.5,trim=1cm 0mm 0mm 0mm]{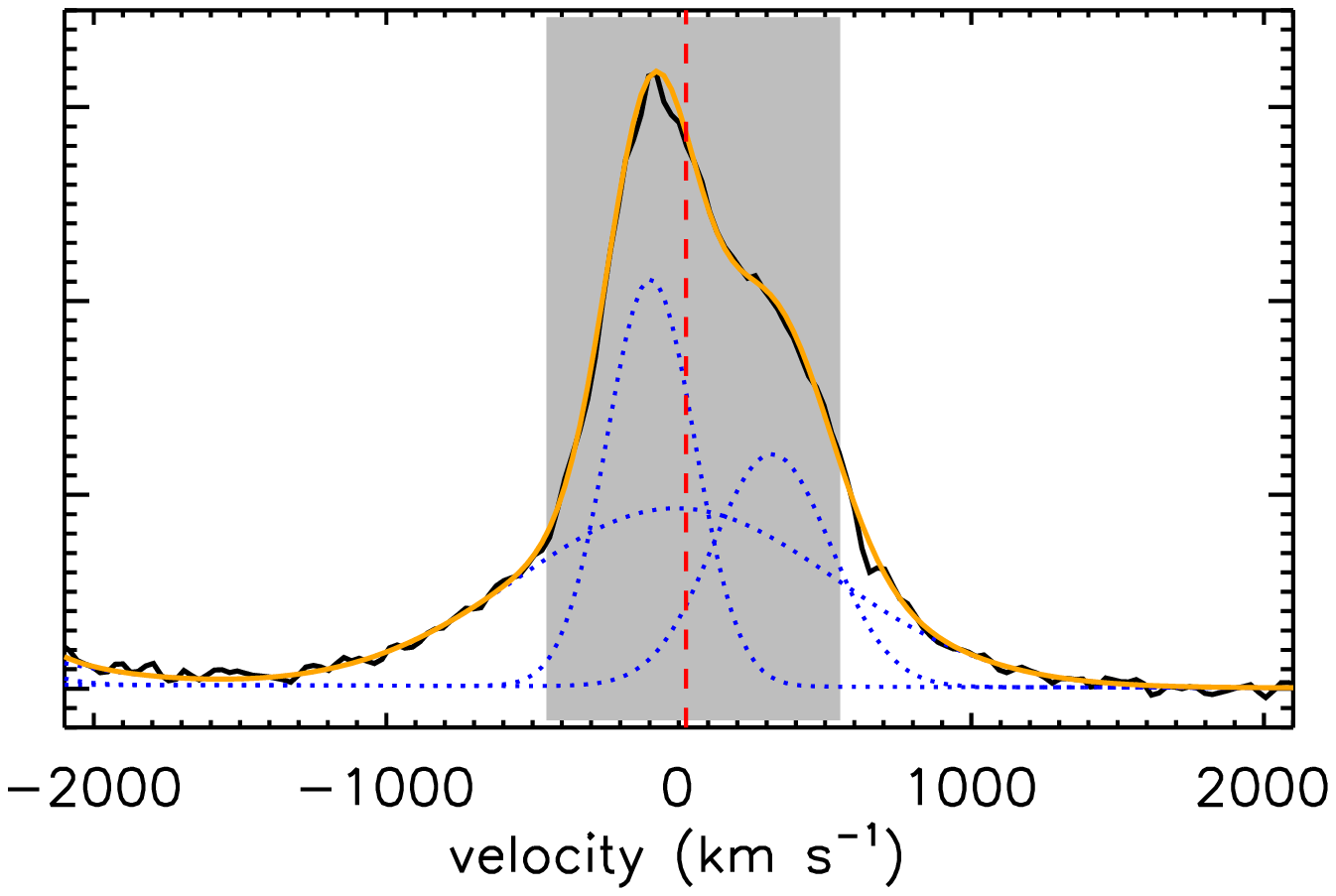}
\caption{Example Gemini IFU spectra in the vicinity of the \oiii$\lambda$5007\AA\ line in two individual spaxels, fitted by 3 Gaussian components. These spectra are from SDSS J0321+0016 and SDSS J1039+4512, respectively. The fitted line is in orange, while the blue dotted lines are the Gaussian components. The median velocity is denoted by a red dashed line, and the velocity range used for calculating $W_{80}$ is shown by a gray box.}
\label{fig:spectra}
\end{figure*}

In general, $p_M>p_N$ holds because an increased number of parameters improves 
the fits, but $p_M<p_N$ may occur when the line has multiple features due to 
the signal and / or the noise, in which case the $M$- and $N$-Gaussian fits 
may trace different (real or false) features and lead to various results. 
The result of this procedure for SDSS J0841+2042 is shown in Figure \ref{fig:gaussian}, 
where the pixel value denotes the number of Gaussians we adopt at each position. 
As revealed by this figure, the bright central area generally requires three 
Gaussians, while the outer faint regions often prefer fewer components. 
In Figure \ref{fig:spectra}, we show example spectra that are fitted by 3 
Gaussian components.

\subsection{Non-Parametric measurements}
\label{sec:parameters}

Conventionally, non-parametric measurements of emission line profiles are carried out either by measuring velocities at various fixed fractions of the peak intensity (\citealt{heck81}; full width at half maximum is the standard example), or by measuring velocities at which some fraction of the line flux is accumulated \citep{whit85}. We mainly adopt the latter parametrization in this paper because its integral nature makes it relatively insensitive to the quality of the data, as pointed out by \citet{veil91}. After each profile is fit with multiple Gaussian components, we use the fits to calculate the cumulative flux as a function of velocity:
\begin{equation}
\Phi(v)\equiv\int_{-\infty}^v \! F_v(v^{\prime}) \, \mathrm{d}v^{\prime}.
\end{equation}
With this definition, the total line flux is given by $\Phi(\infty)$. For each spectrum, we use this definition to calculate the following quantities (related to the first, second, third and forth moments of the line-of-sight velocity distribution within the line profile):

\begin{enumerate}

\item {\bf Velocity.} The line-of-sight velocity is represented by the median velocity $v_{\rm med}$, i.e. the velocity that bisects the total area underneath the \oiii\ emission line profile, so that $\Phi(v_{\rm med})=0.5\,\Phi(\infty)$. 

  Following \citet{rupk13}, we also report $v_{02}$ in Table
  \ref{tab1}, the velocity at 2\% of the cumulative flux, meaning that
  98\% of the gas is less blueshifted than this value. \citet{rupk13}
  use this value as a measure of the maximal outflow velocity. We
  evaluate both the brightness-weighted value $\langle v_{02} \rangle$
  measured from the integrated SDSS spectrum and the maximum 
  $v_{\rm 02,max}$ from the spatially resolved IFU data. To determine
  $v_{\rm 02,max}$ for each object, we create the $v_{02}$ map for all
  spaxels where the [O {\sc iii}]$\lambda$5007\AA\ line has
  $S/N\geqslant5$ at its peak, reject the 5\% most extreme negative
  values which may be contaminated by noise, and report the remaining
  maximum negative value.

\item {\bf Line width.} Many different measures of velocity dispersion are possible; we are seeking one that does not discard the information contained in the broad wings of the emission lines, but at the same time is not too sensitive to the low signal-to-noise emission at high velocities. We use the velocity width that encloses 80\% of the total flux  $W_{80}$, defined as the difference between the velocities at 10\% and 90\% of cumulative flux:
\begin{equation}
W_{80} \equiv v_{90}-v_{10}.
\end{equation}
This measurement is illustrated in Figure~\ref{fig:spectra}. For a purely Gaussian velocity profile, this value is determined entirely by the velocity dispersion and is close to the conventionally used full width at half maximum (FWHM): 
\begin{equation}
W_{80}=2.563\sigma=1.088\times{\rm FWHM},
\end{equation}
but for non-Gaussian profiles it is more sensitive to the weak broad bases of emission lines characteristic of our sample. For example, a profile composed of two Gaussians, one with dispersion $\sigma$ and another with $3\sigma$, centered at the same velocity and with flux ratio in two components of 2:1 would have a FWHM$=2.593\sigma$ (i.e., a value only 10\% above what would be measured just for the narrow component alone), but a $W_{80}$ value of $3.981\sigma$, significantly higher than the Gaussian value. 

\item {\bf Asymmetry.} We use an asymmetry parameter defined as 
\begin{equation}
A \equiv \frac{(v_{90}-v_{\rm med})-(v_{\rm med}-v_{10})}{W_{80}}.
\end{equation}
This parameter is introduced by \citet{whit85}, but with the opposite sign. With our definition, a profile with a heavy blueshifted wing has a negative $A$ value. A symmetric profile, such as a single Gaussian or a combination of multiple Gaussian components centered at the same velocity, has $A=0$. This parameter is related to the standard profile skewness. 

\item {\bf Shape parameter.} The shape parameter that we use is $K$ defined as
\begin{equation}
K\equiv \frac{W_{90}}{1.397\times {\rm FWHM}},
\end{equation}
where FWHM is the full width at half maximum and $W_{90}$ is the velocity width that encloses 90\% of the total flux, $W_{90}\equiv v_{95}-v_{05}$. This definition is the reciprocal of the parameter $K_2$ defined in \citet{whit85}. $K$ is related to line kurtosis: for a Gaussian profile, $K=1$, and profiles that have wings heavier than a Gaussian have $K>1$, whereas stubby profiles with no wings have $K<1$.  

\end{enumerate}

While it is possible to calculate these parameters directly from the observed velocity profiles, we perform these non-parametric measurements on the best-fit single/multi-Gaussian profiles instead. In this case the cumulative function is monotonically increasing and positive definite, which allows us to minimize the effect of noise, especially for faint broad wings of the emission lines. We further comment on the quality of non-parametric measurements in Section \ref{sec:dispersion}. The resulting maps of these parameters are shown for the whole sample in Figure \ref{fig:VWAK}.
 
\begin{figure*}%[h!]
    \includegraphics[scale=0.28,clip=clip,trim=0mm 0mm 3.5cm 0mm]{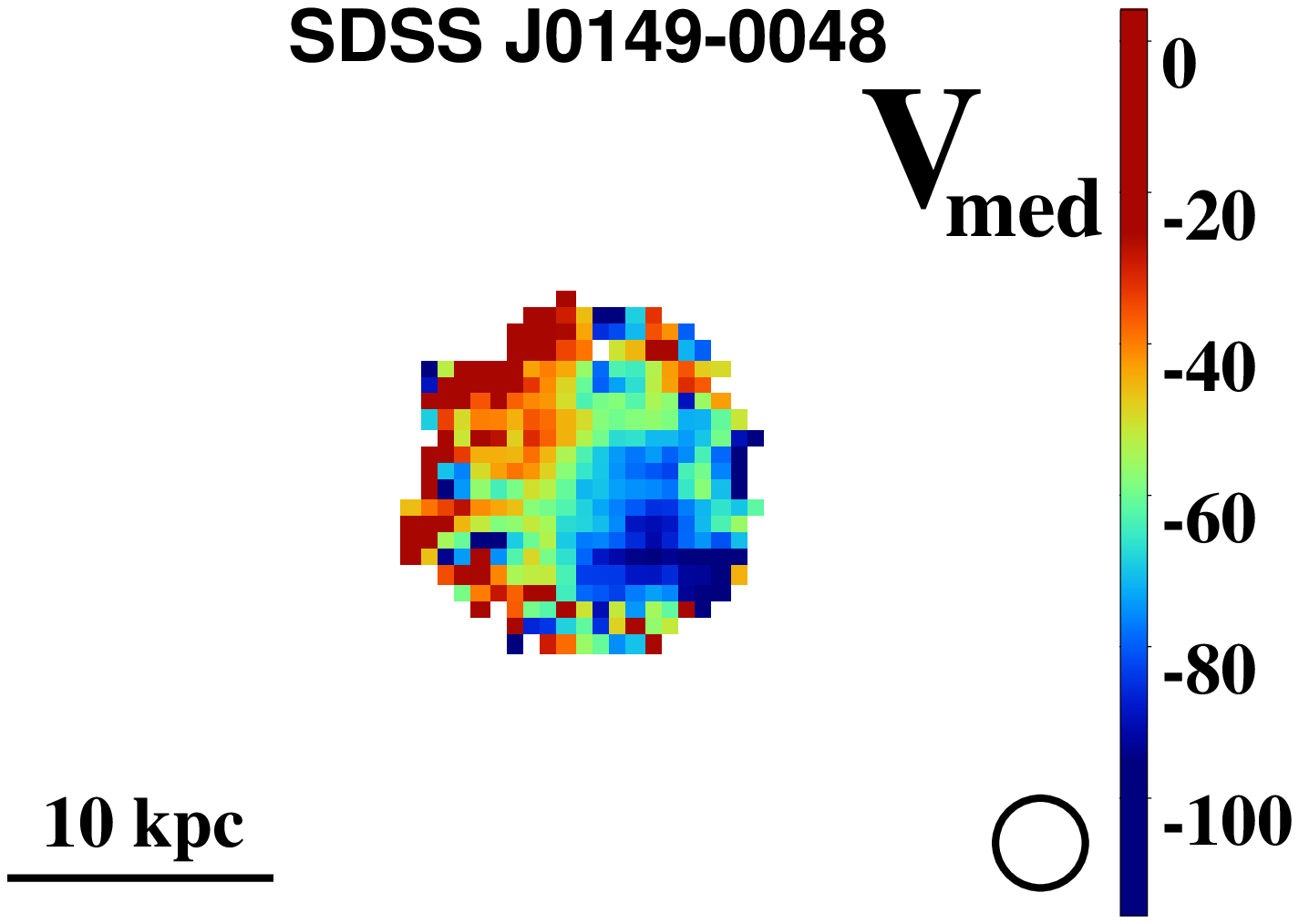}%
    \includegraphics[scale=0.28,clip=clip,trim=0mm 0mm 3.5cm 0mm]{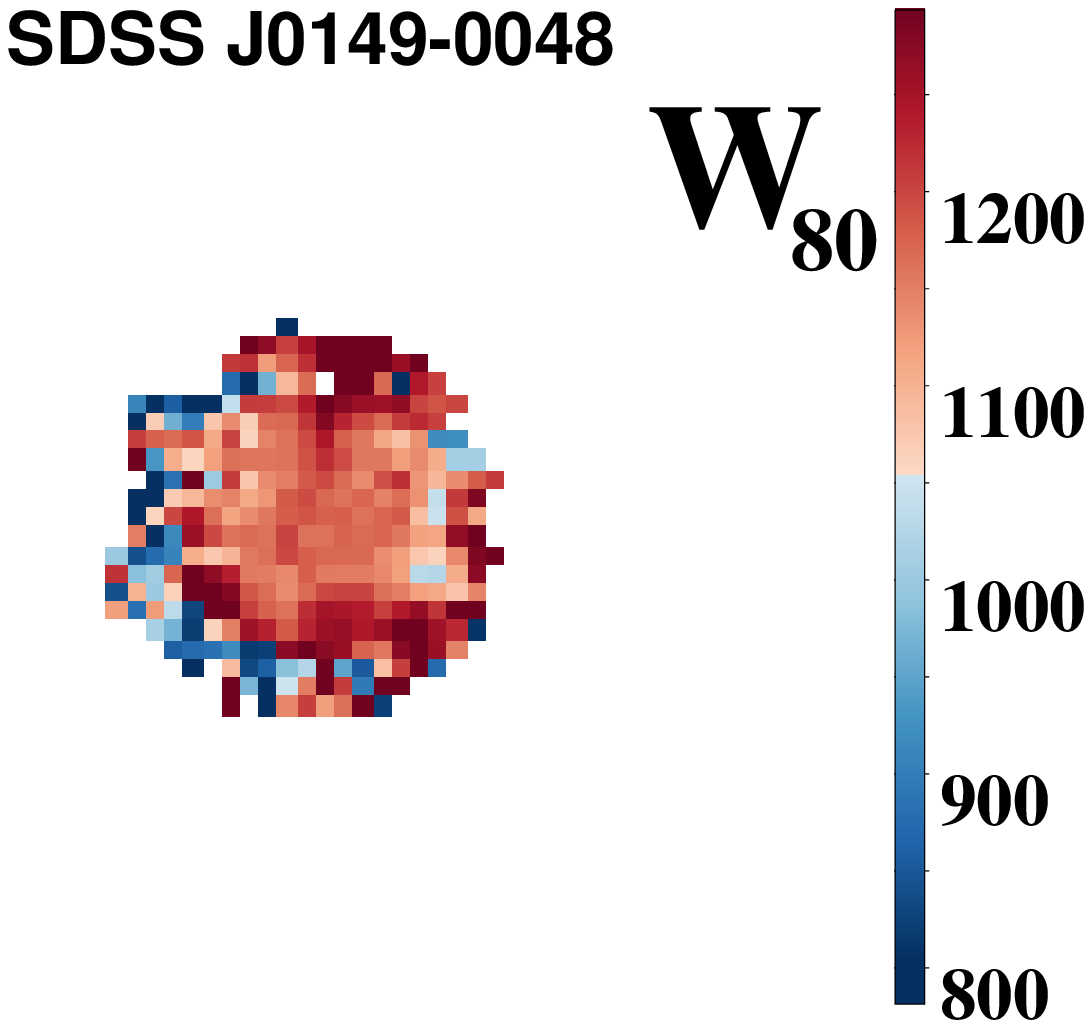}%
    \includegraphics[scale=0.28,clip=clip,trim=0mm 0mm 3.5cm 0mm]{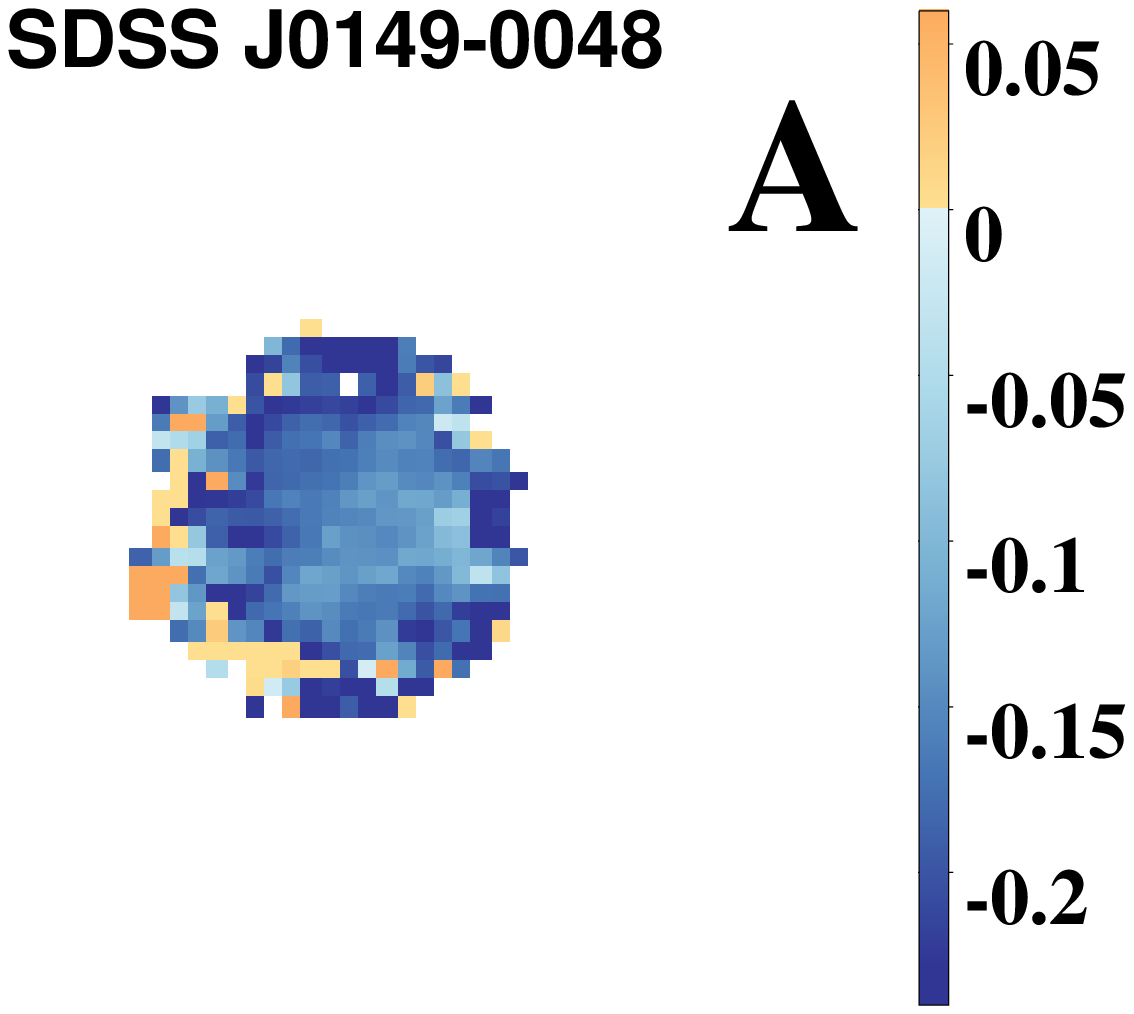}%
    \includegraphics[scale=0.28,clip=clip,trim=0mm 0mm 3.5cm 0mm]{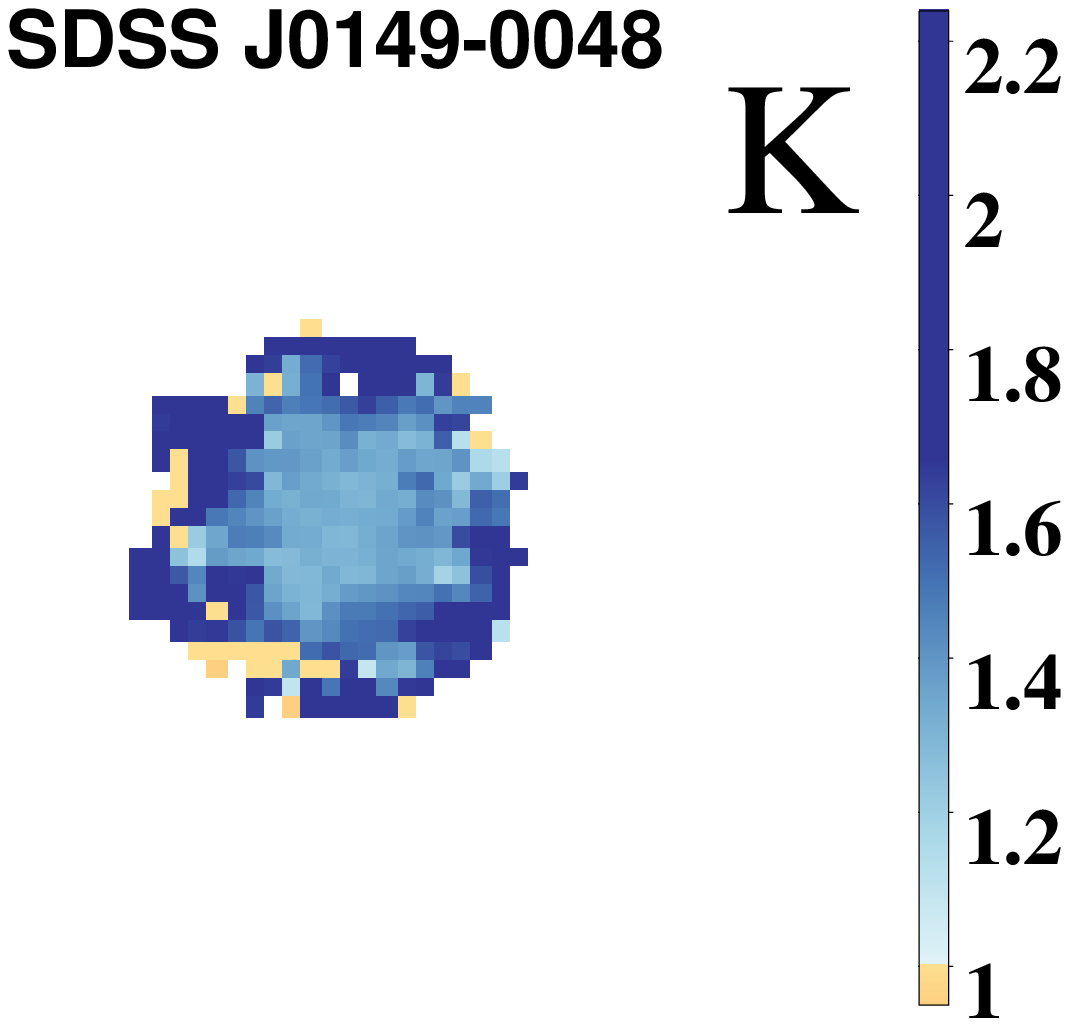}\\
    \includegraphics[scale=0.28,clip=clip,trim=0mm 0mm 3.5cm 0mm]{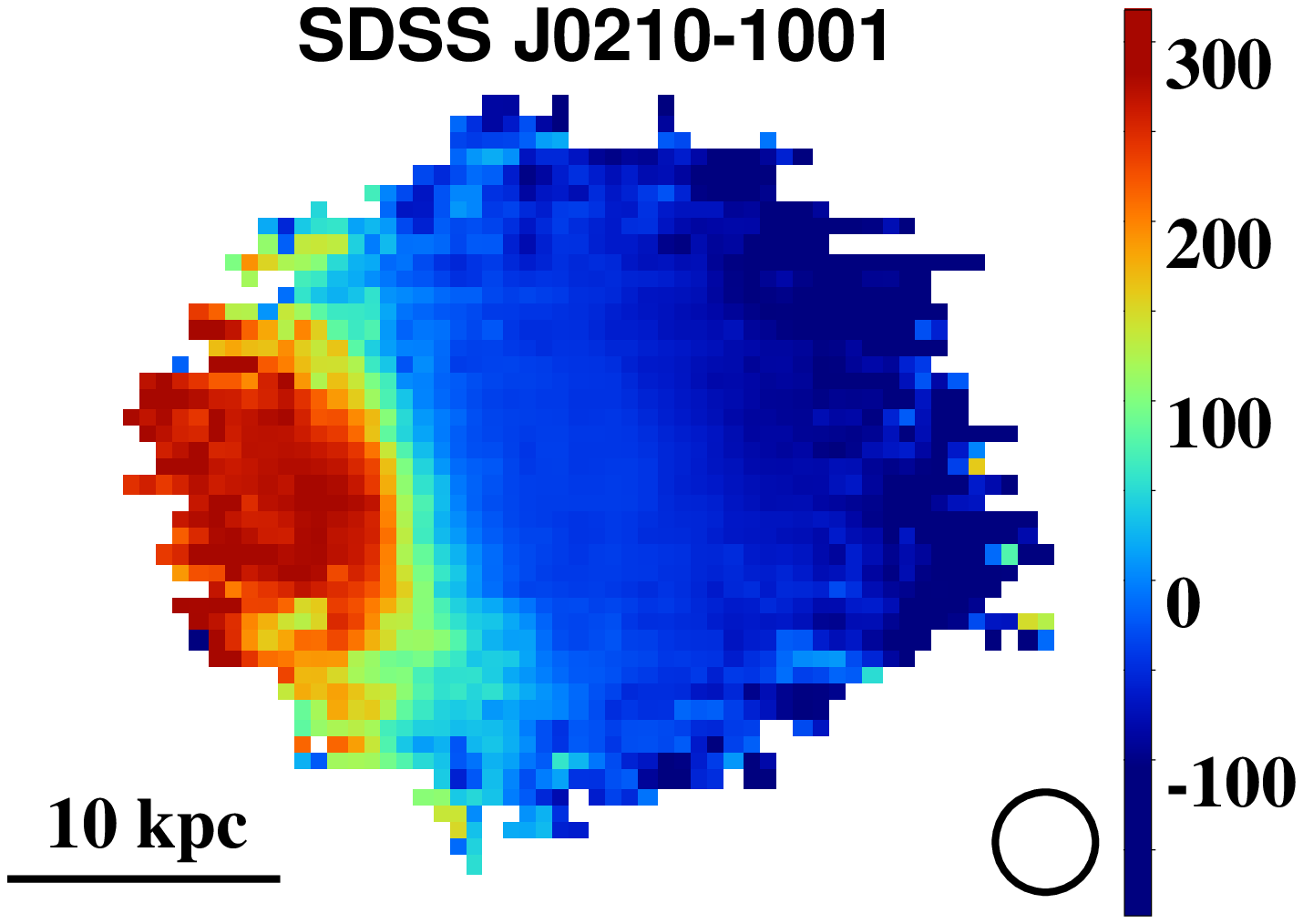}%
    \includegraphics[scale=0.28,clip=clip,trim=0mm 0mm 3.5cm 0mm]{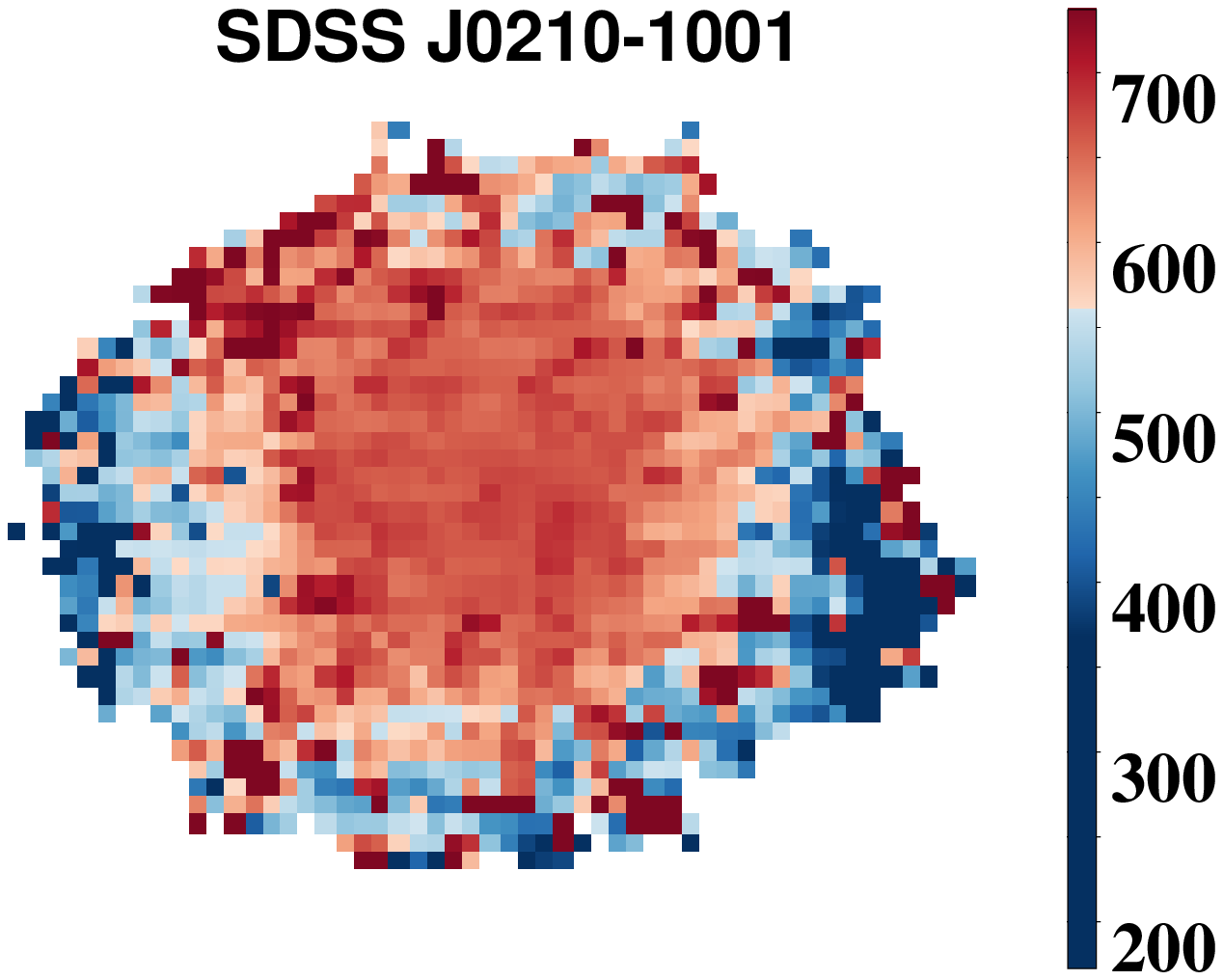}%
    \includegraphics[scale=0.28,clip=clip,trim=0mm 0mm 3.5cm 0mm]{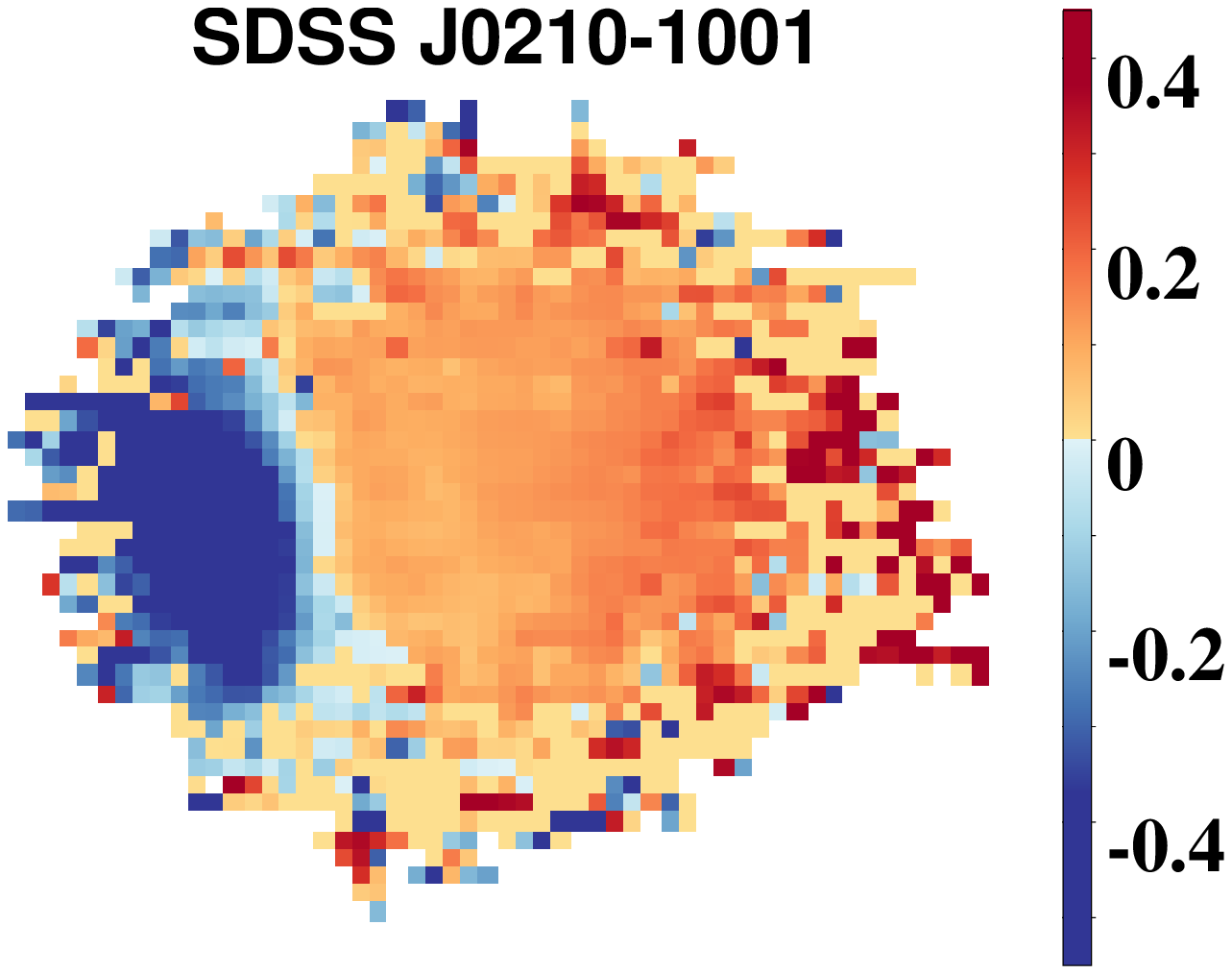}%
    \includegraphics[scale=0.28,clip=clip,trim=0mm 0mm 3.5cm 0mm]{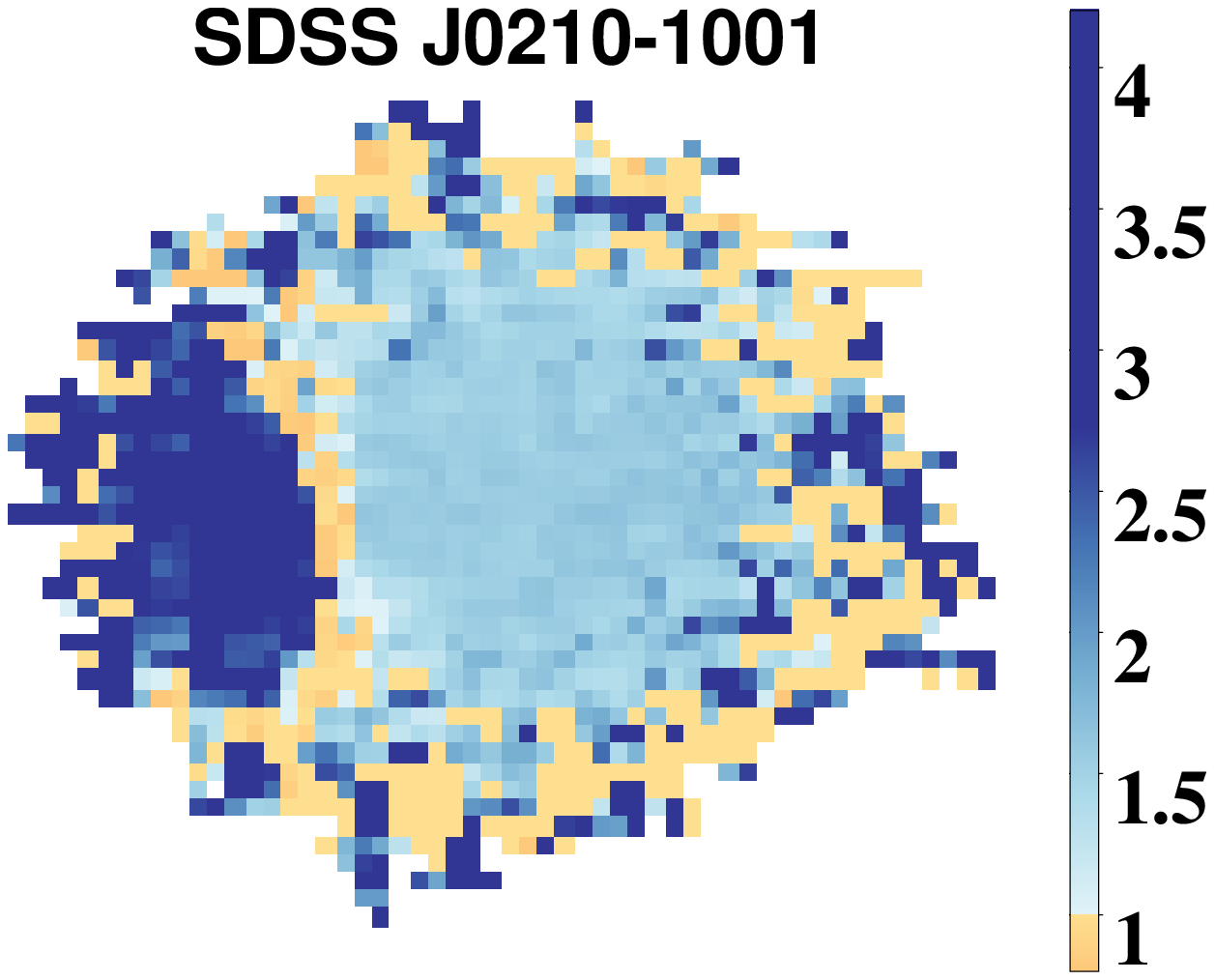}\\
    \includegraphics[scale=0.28,clip=clip,trim=0mm 0mm 3.5cm 0mm]{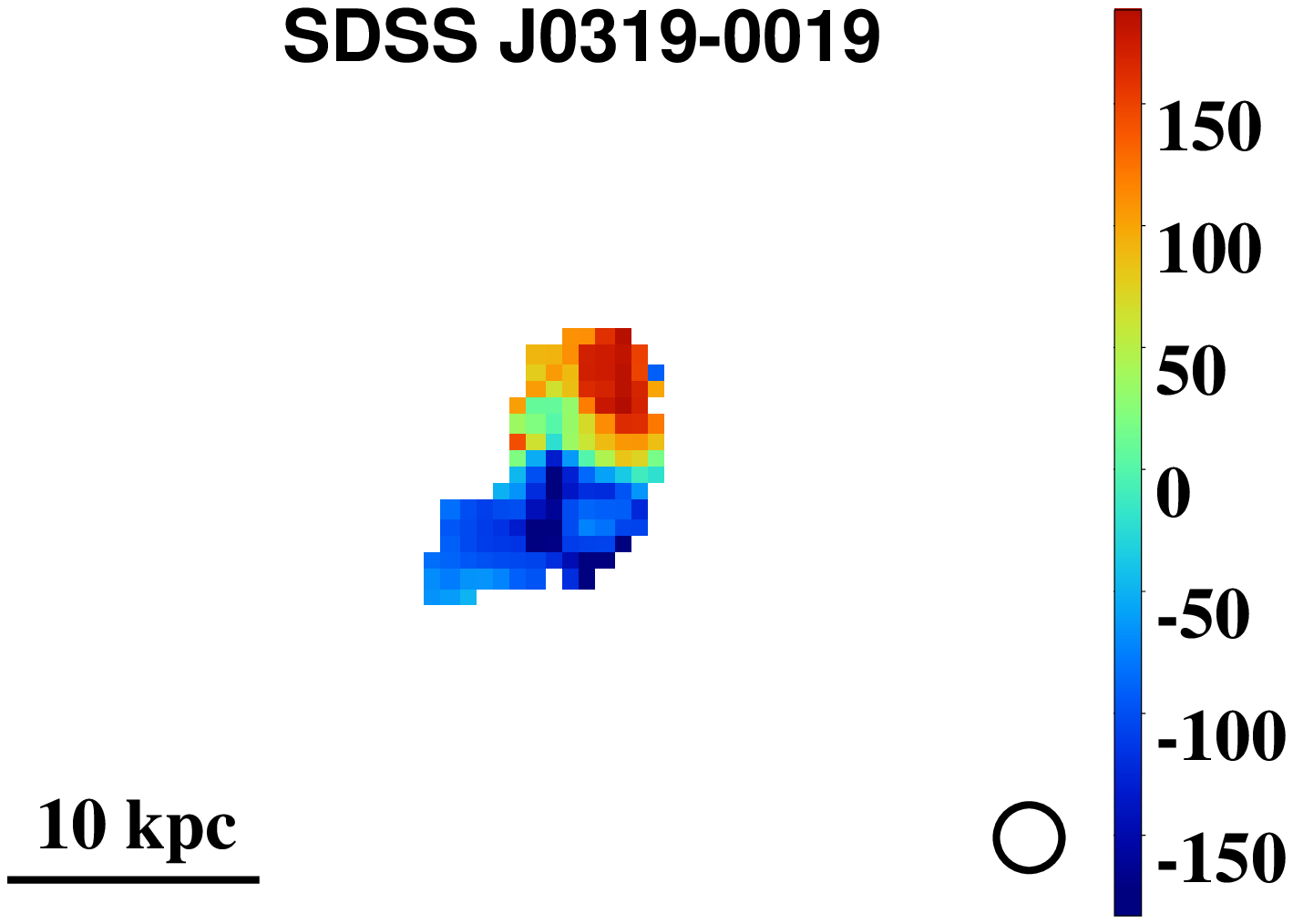}%
    \includegraphics[scale=0.28,clip=clip,trim=0mm 0mm 3.5cm 0mm]{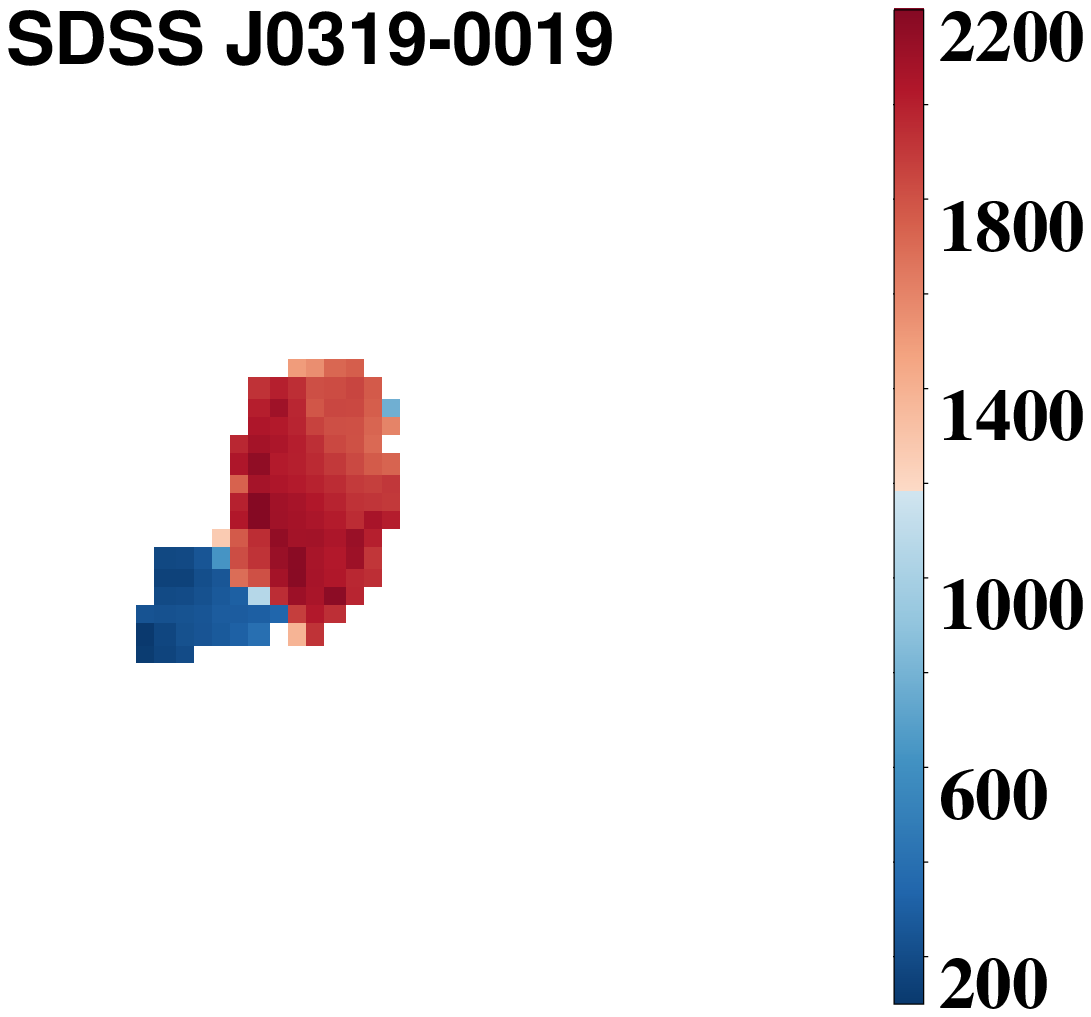}%
    \includegraphics[scale=0.28,clip=clip,trim=0mm 0mm 3.5cm 0mm]{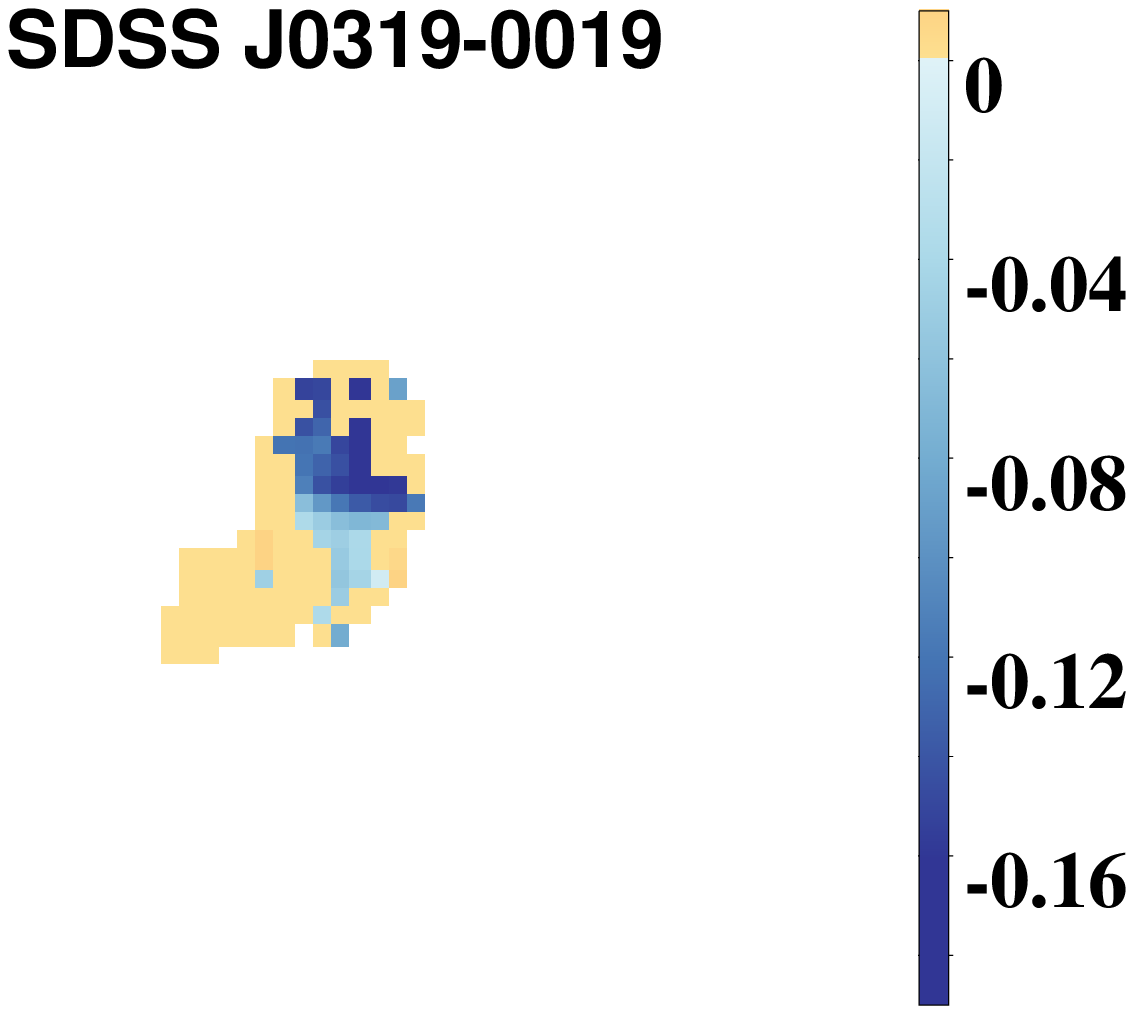}%
    \includegraphics[scale=0.28,clip=clip,trim=0mm 0mm 3.5cm 0mm]{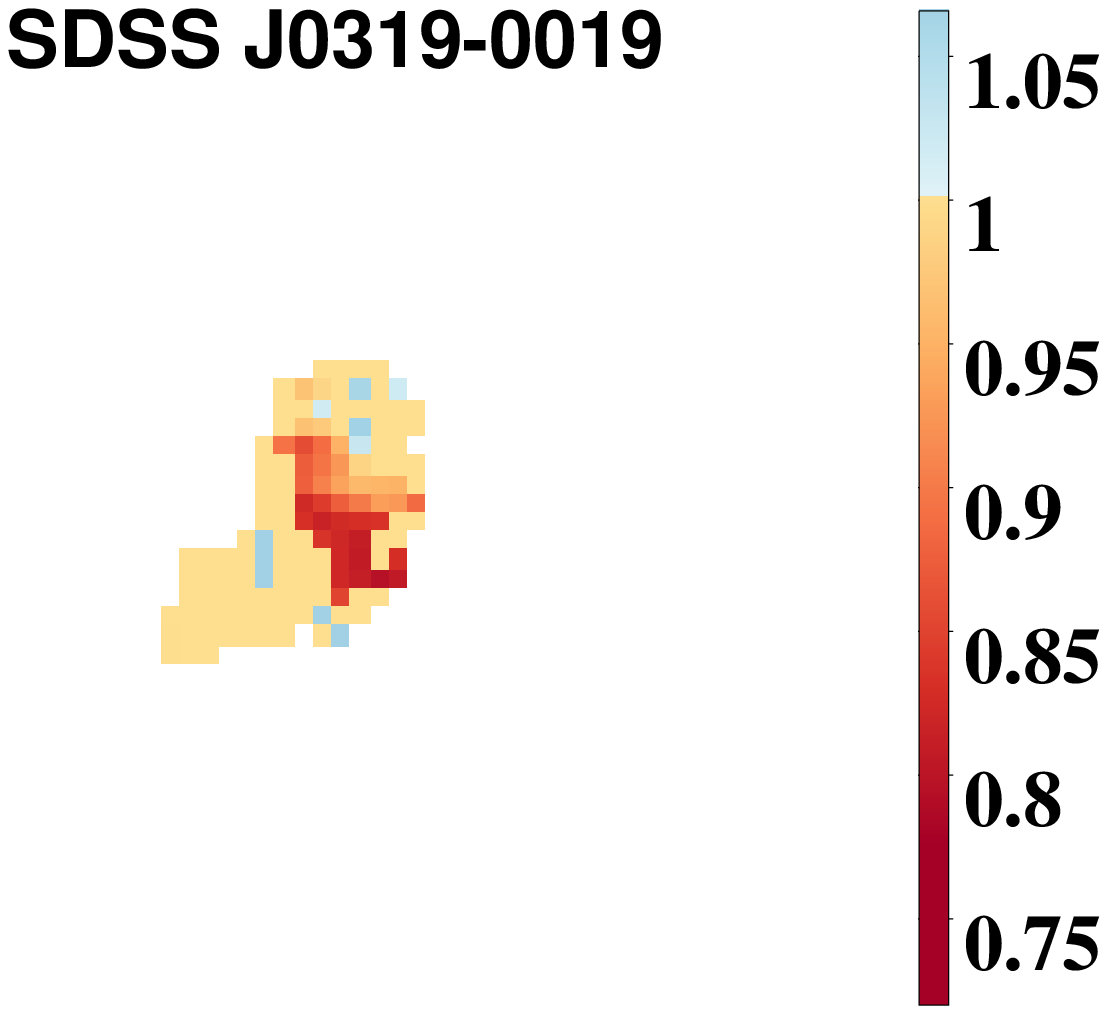}\\
    \includegraphics[scale=0.28,clip=clip,trim=0mm 0mm 3.5cm 0mm]{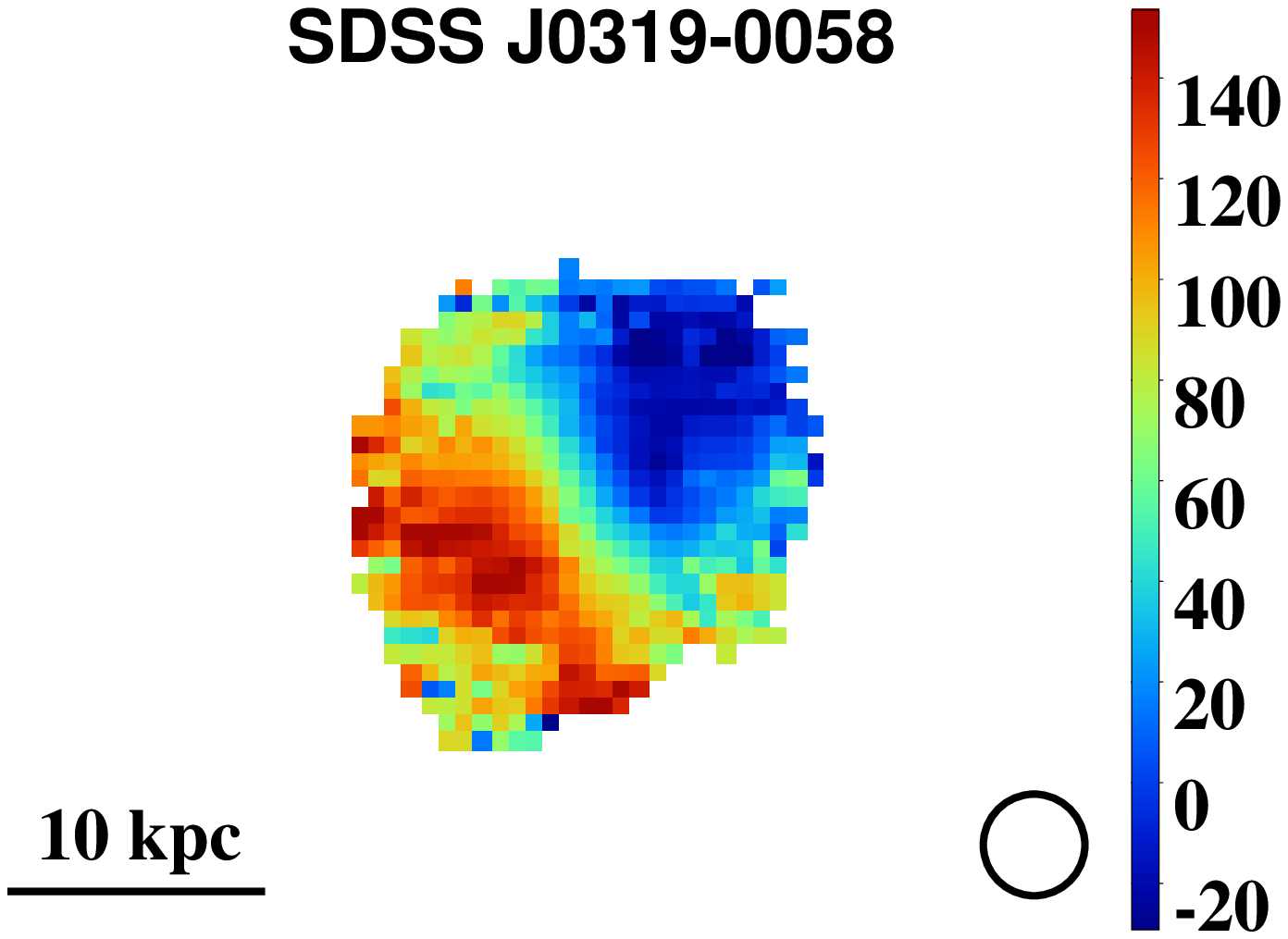}%
    \includegraphics[scale=0.28,clip=clip,trim=0mm 0mm 3.5cm 0mm]{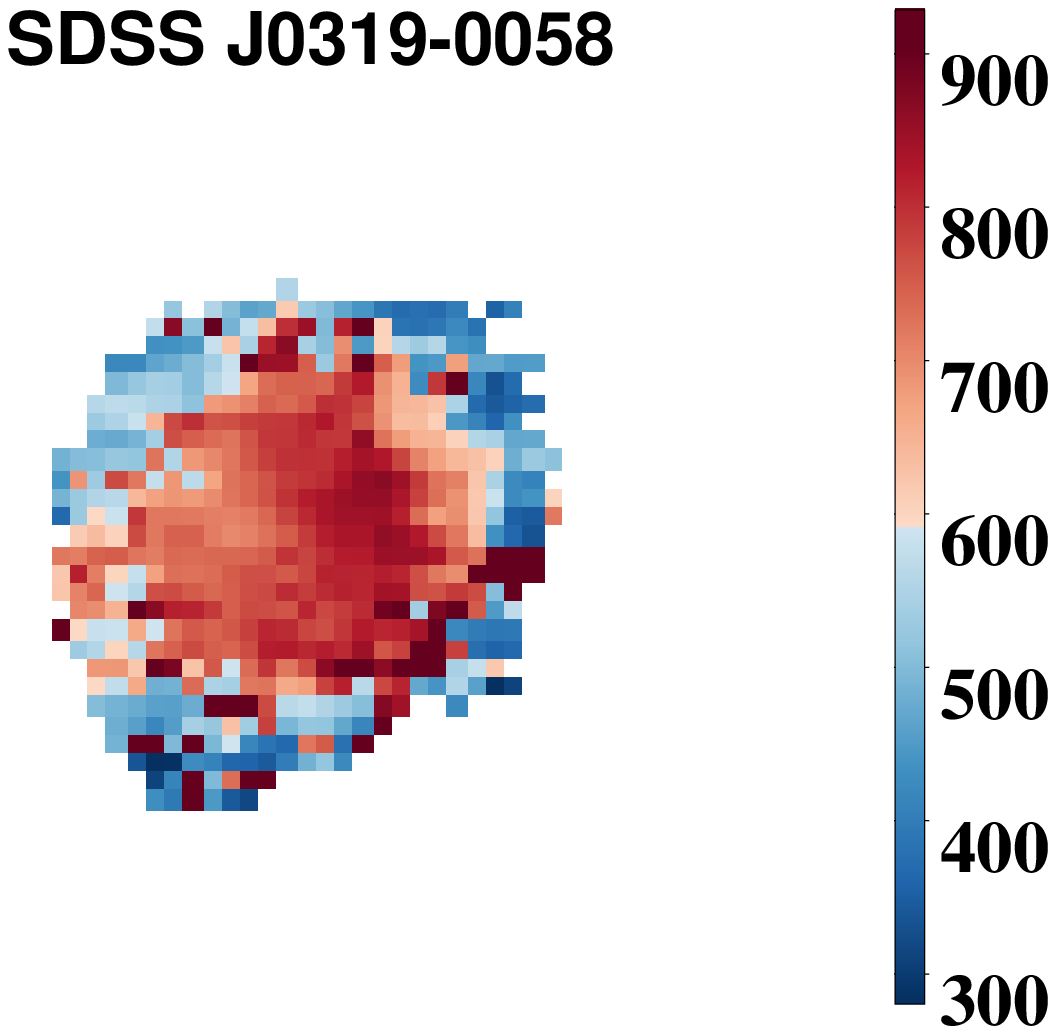}%
    \includegraphics[scale=0.28,clip=clip,trim=0mm 0mm 3.5cm 0mm]{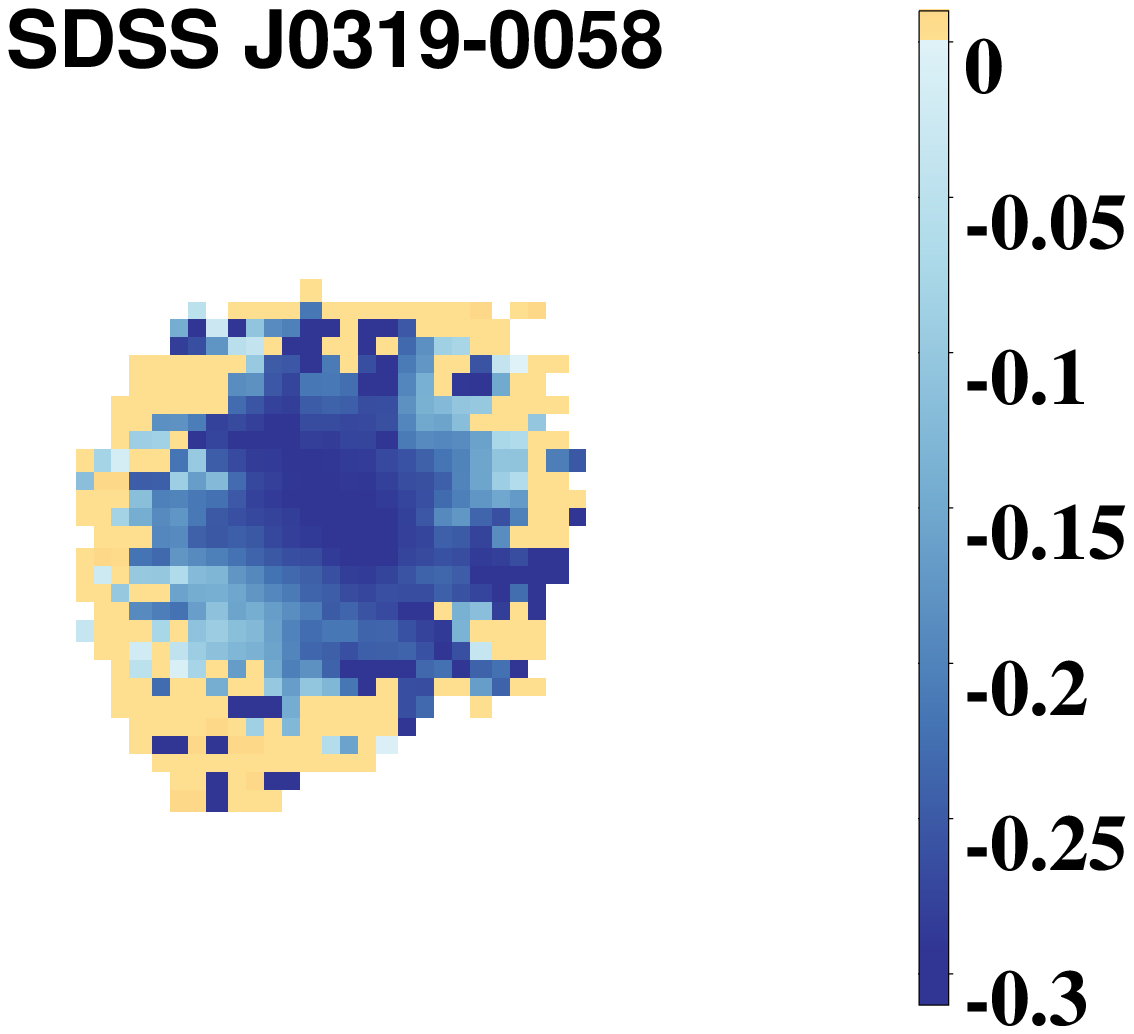}%
    \includegraphics[scale=0.28,clip=clip,trim=0mm 0mm 3.5cm 0mm]{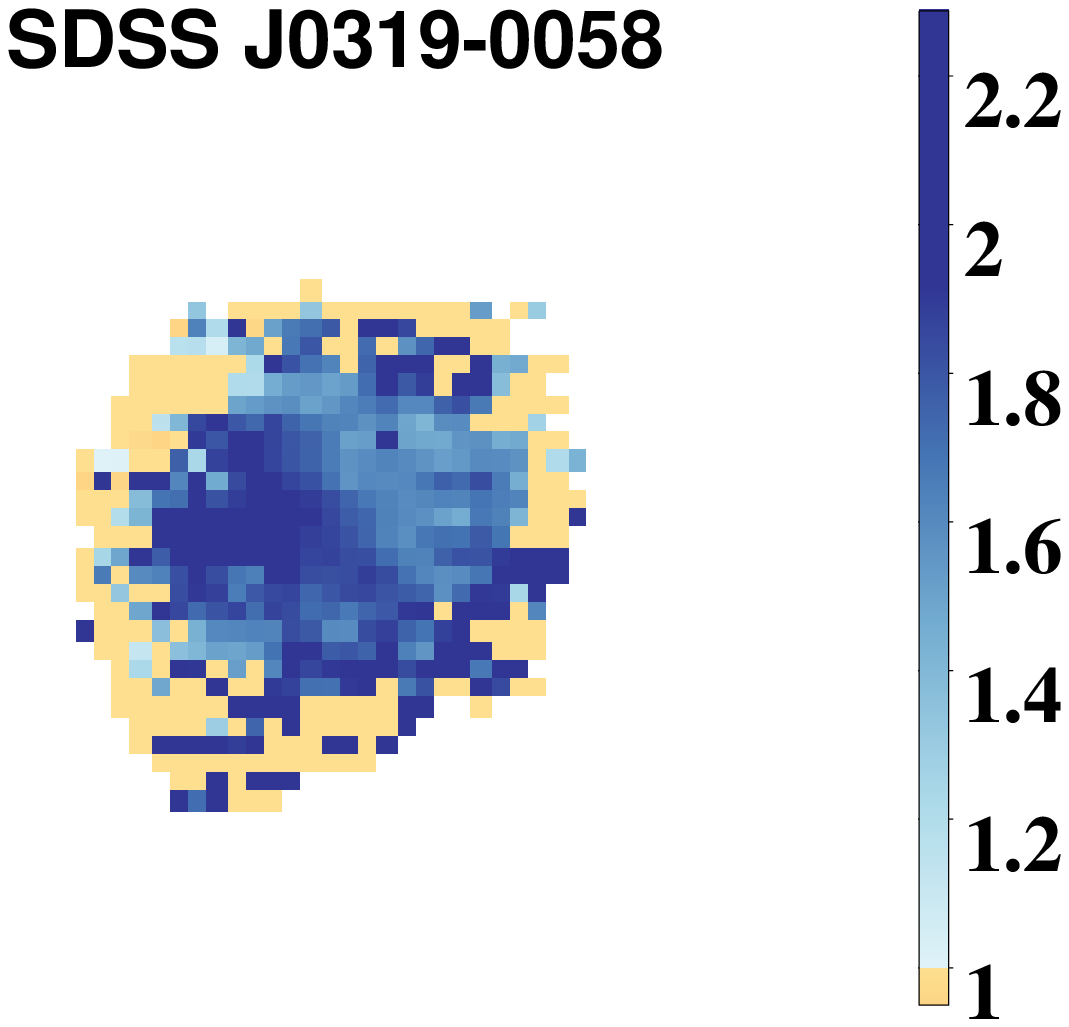}\\
    \includegraphics[scale=0.28,clip=clip,trim=0mm 0mm 3.5cm 0mm]{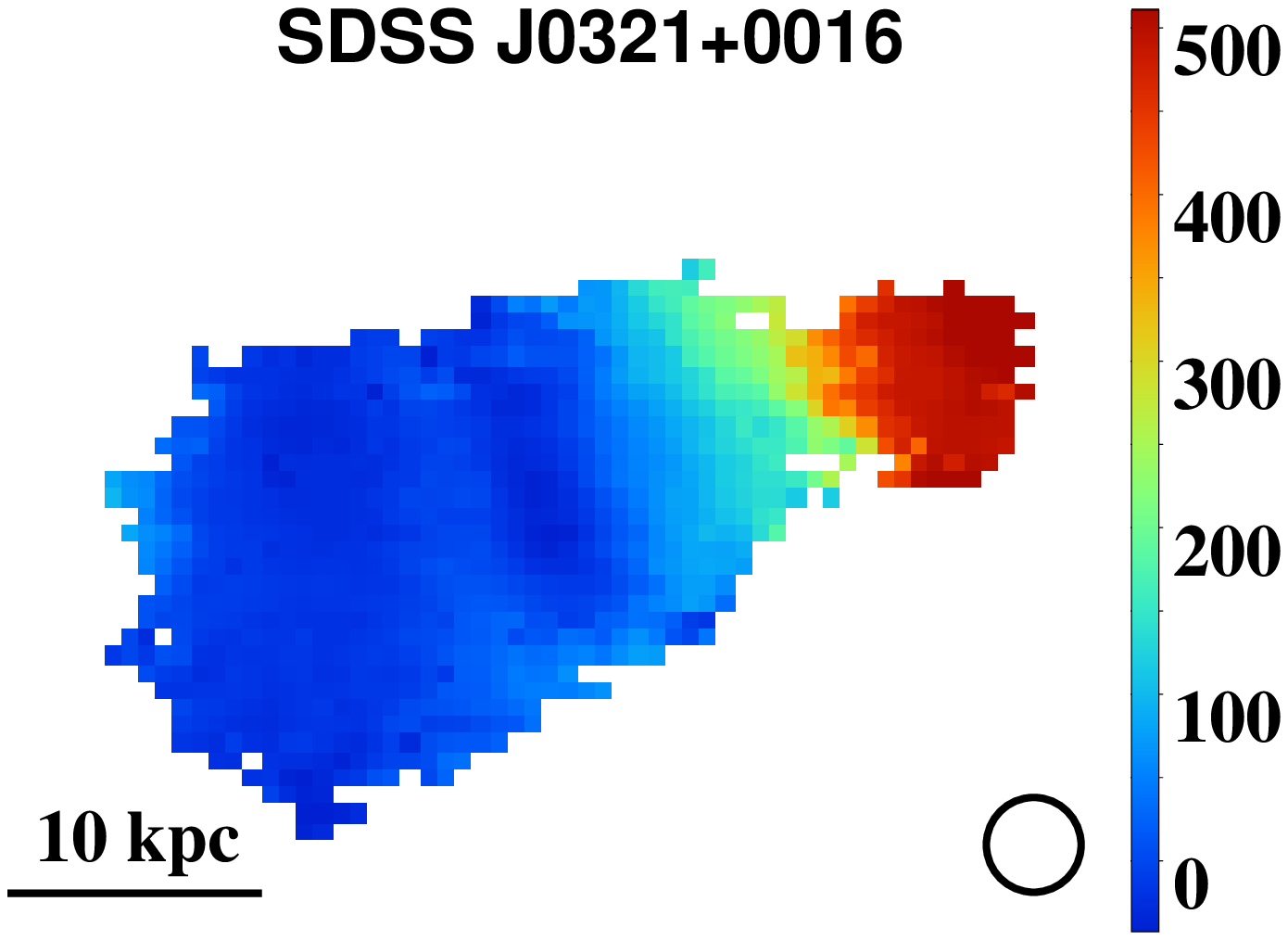}%
    \includegraphics[scale=0.28,clip=clip,trim=0mm 0mm 3.5cm 0mm]{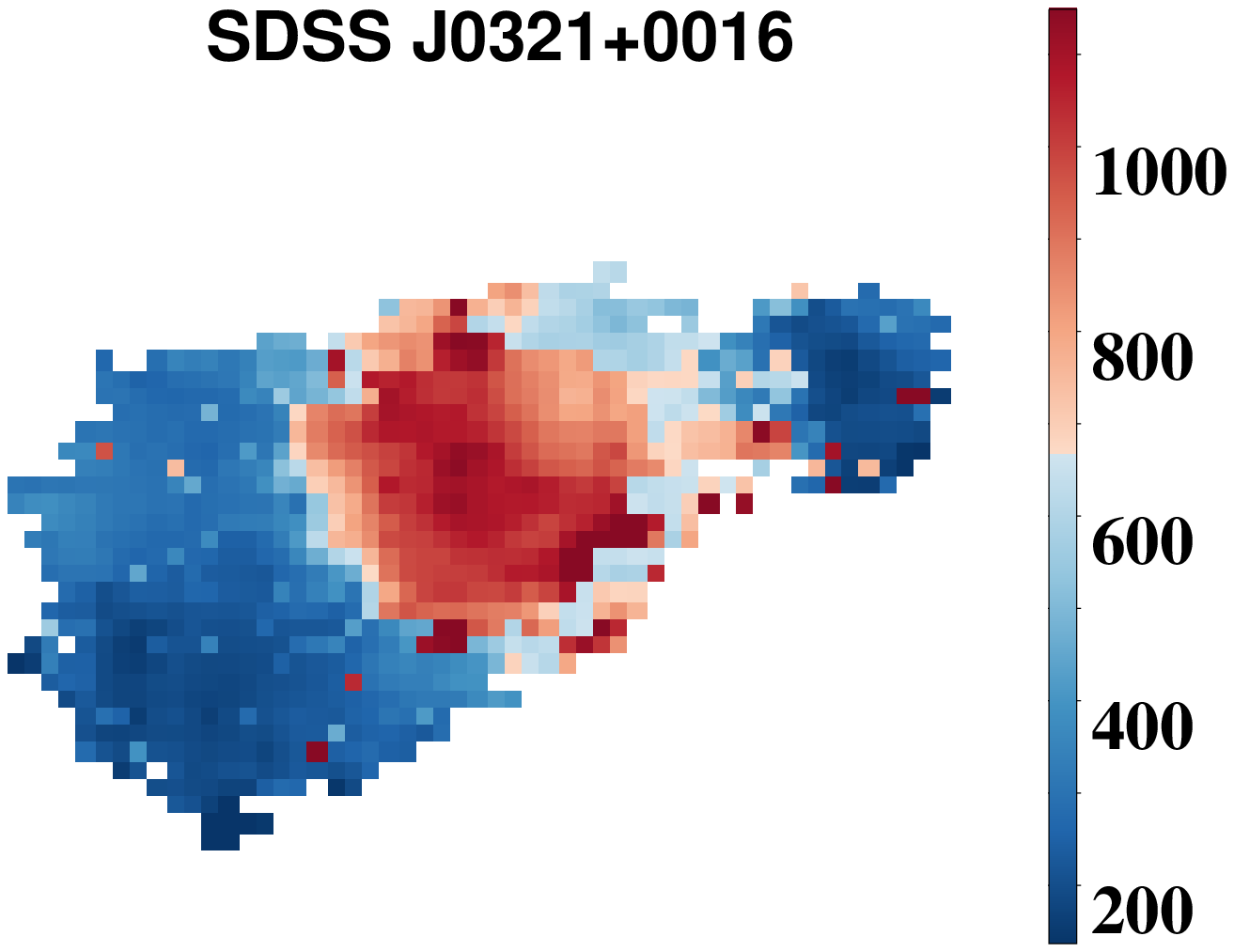}%
    \includegraphics[scale=0.28,clip=clip,trim=0mm 0mm 3.5cm 0mm]{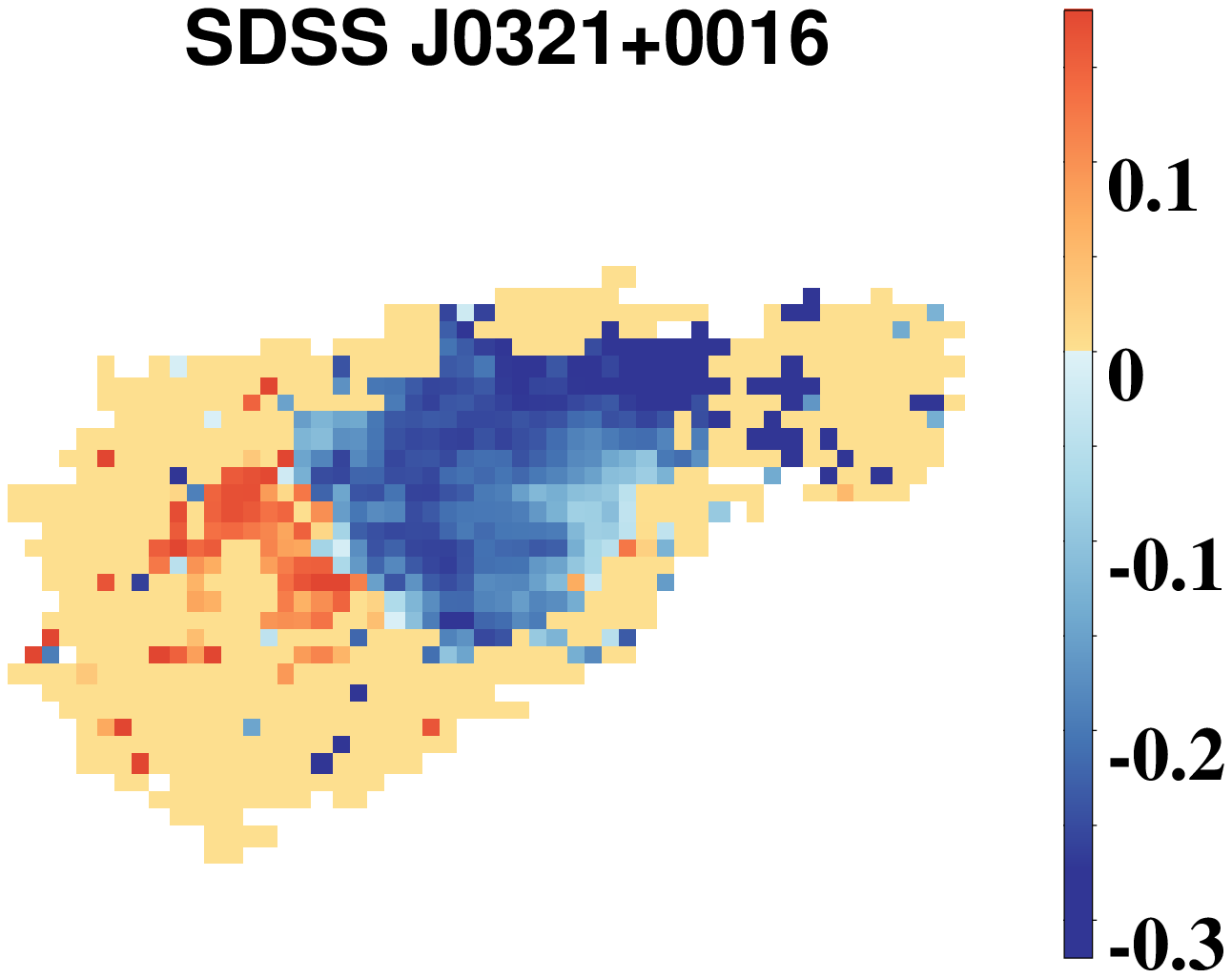}%
    \includegraphics[scale=0.28,clip=clip,trim=0mm 0mm 3.5cm 0mm]{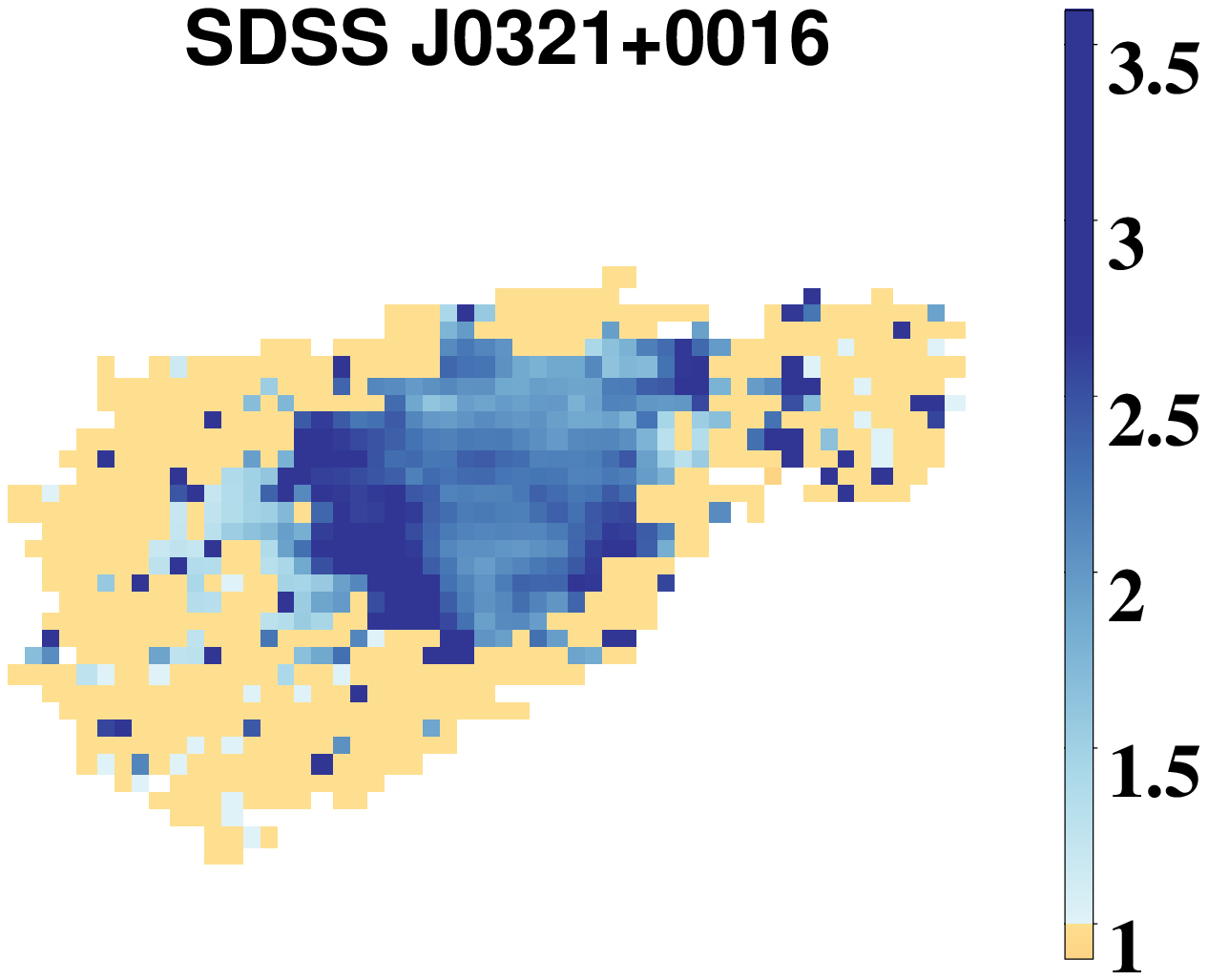}\\
    \includegraphics[scale=0.28,clip=clip,trim=0mm 0mm 3.5cm 0mm]{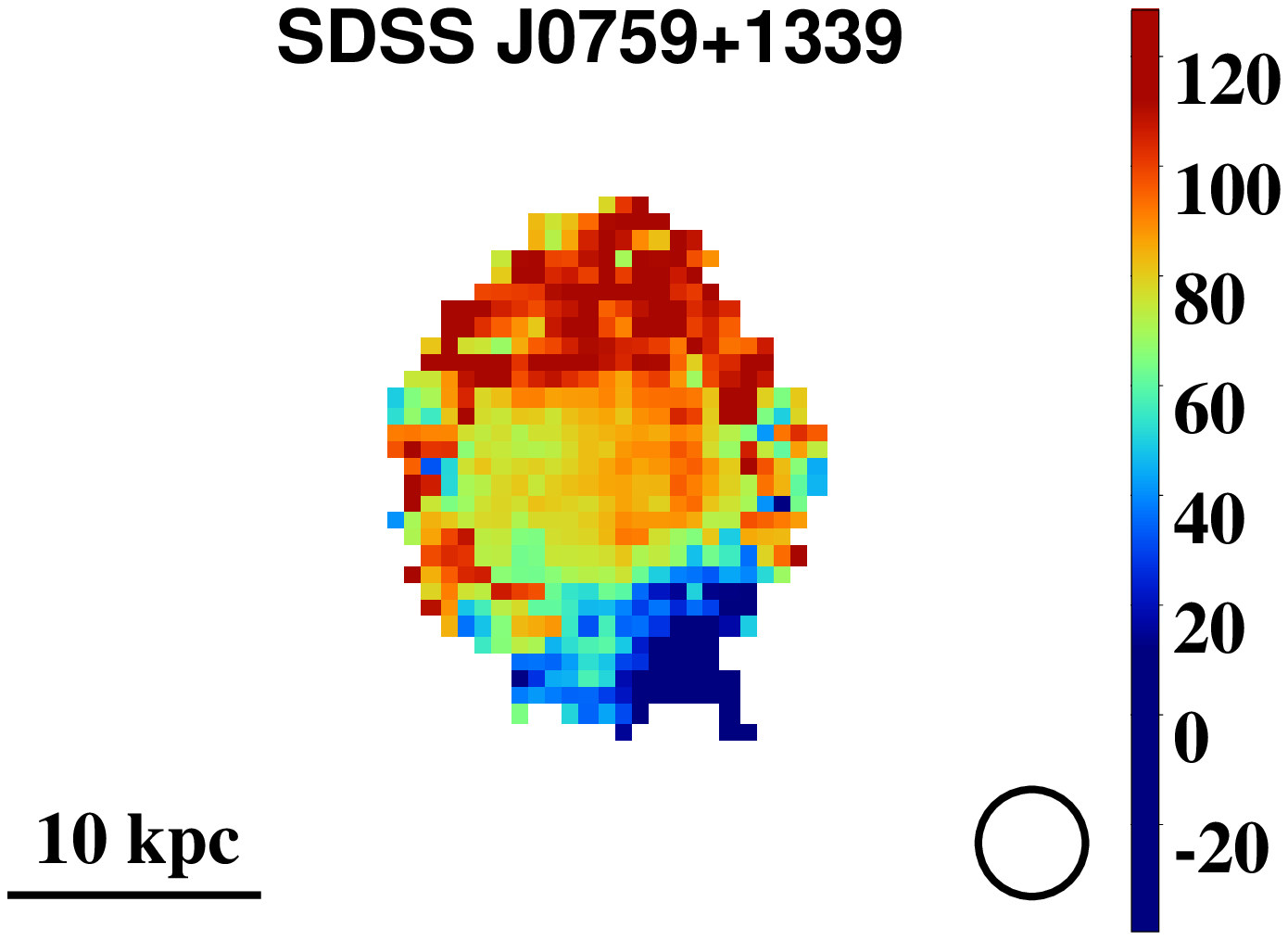}%
    \includegraphics[scale=0.28,clip=clip,trim=0mm 0mm 3.5cm 0mm]{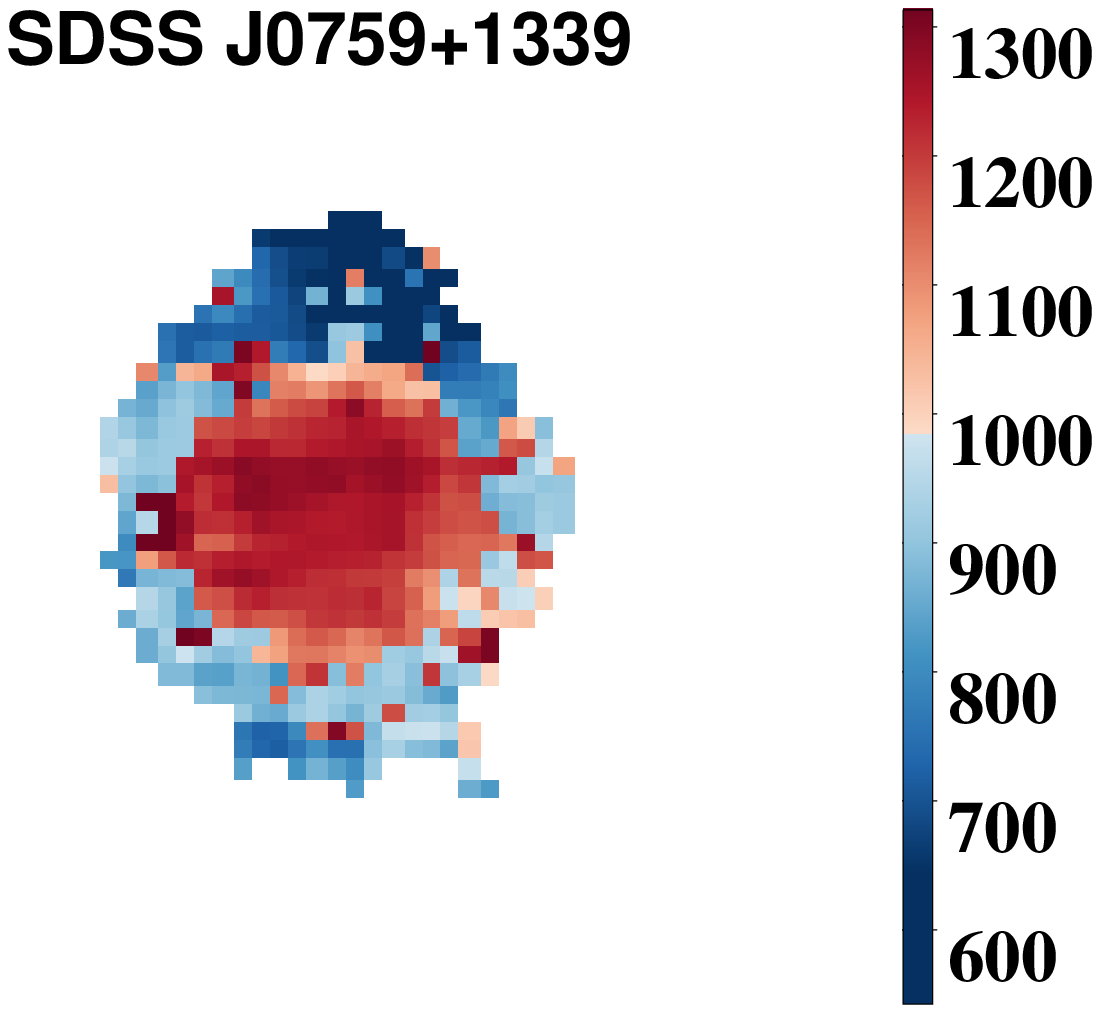}%
    \includegraphics[scale=0.28,clip=clip,trim=0mm 0mm 3.5cm 0mm]{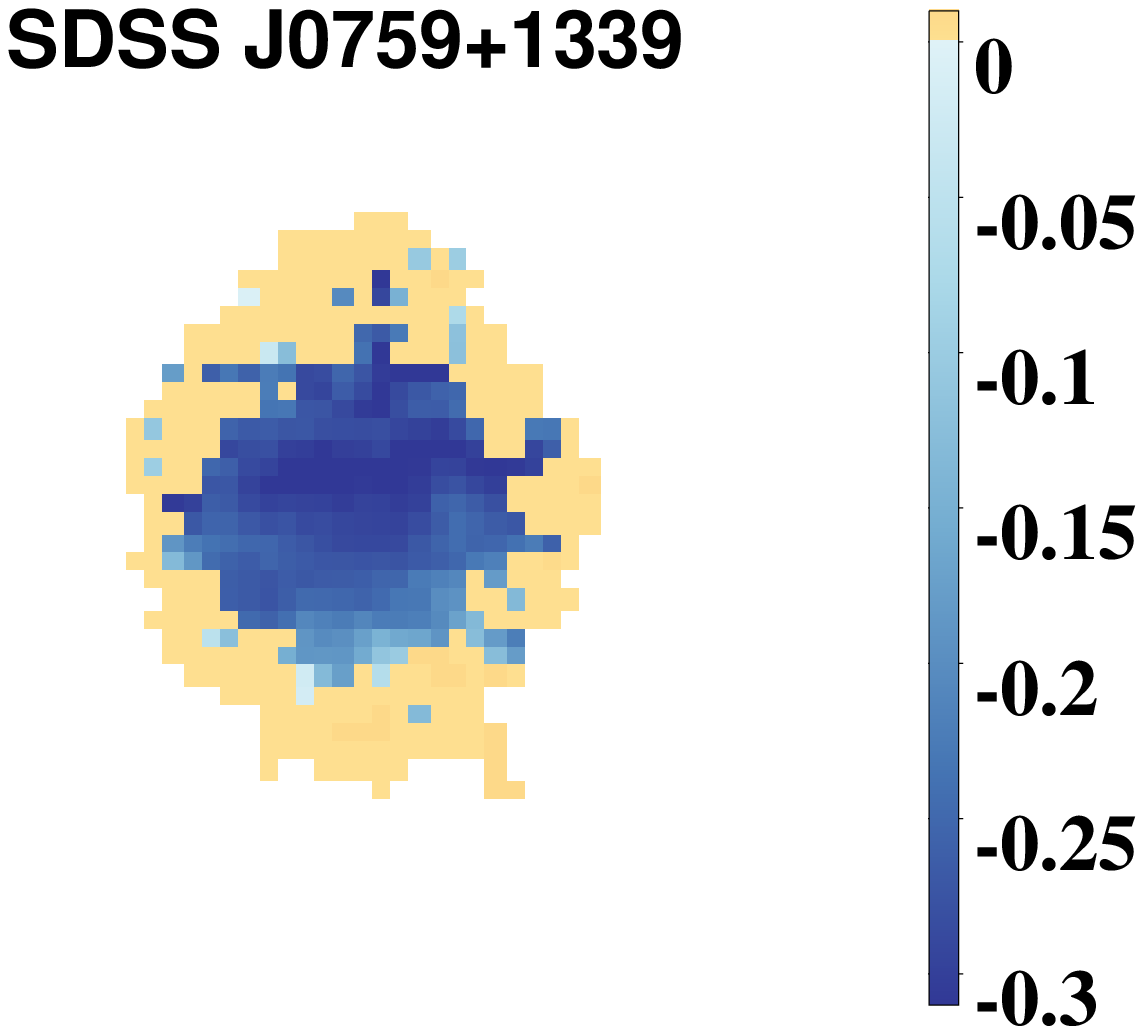}%
    \includegraphics[scale=0.28,clip=clip,trim=0mm 0mm 3.5cm 0mm]{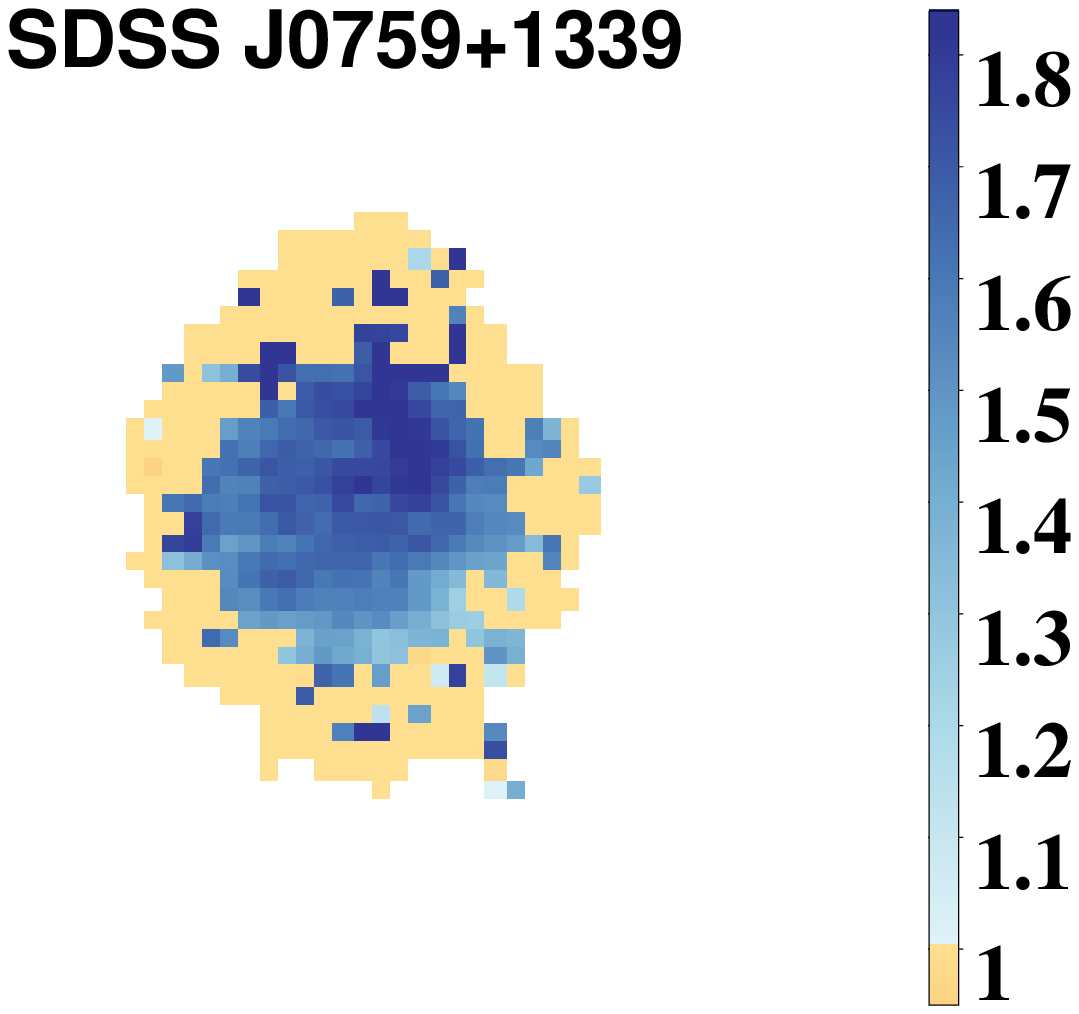}\\
    \includegraphics[scale=0.28,clip=clip,trim=0mm 0mm 3.5cm 0mm]{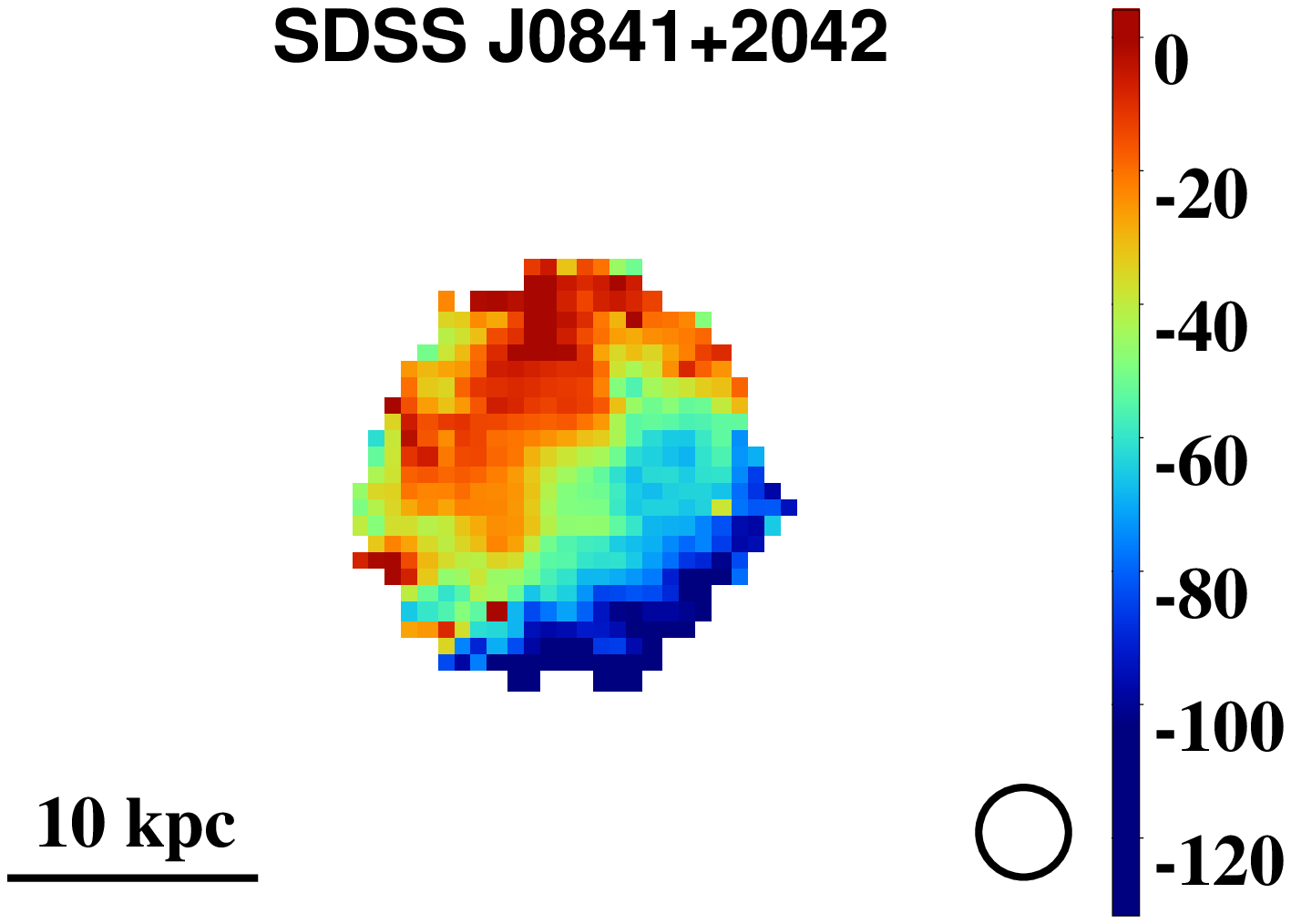}%
    \includegraphics[scale=0.28,clip=clip,trim=0mm 0mm 3.5cm 0mm]{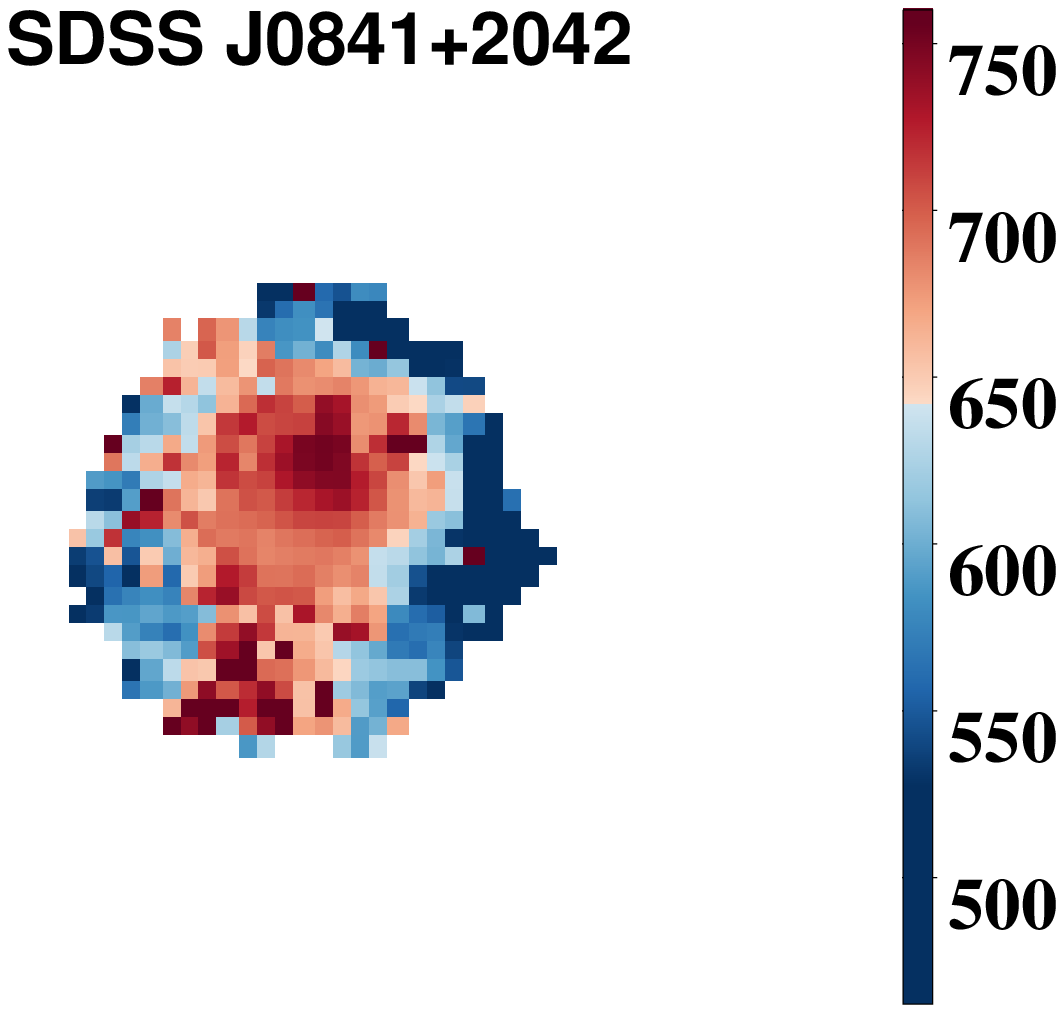}%
    \includegraphics[scale=0.28,clip=clip,trim=0mm 0mm 3.5cm 0mm]{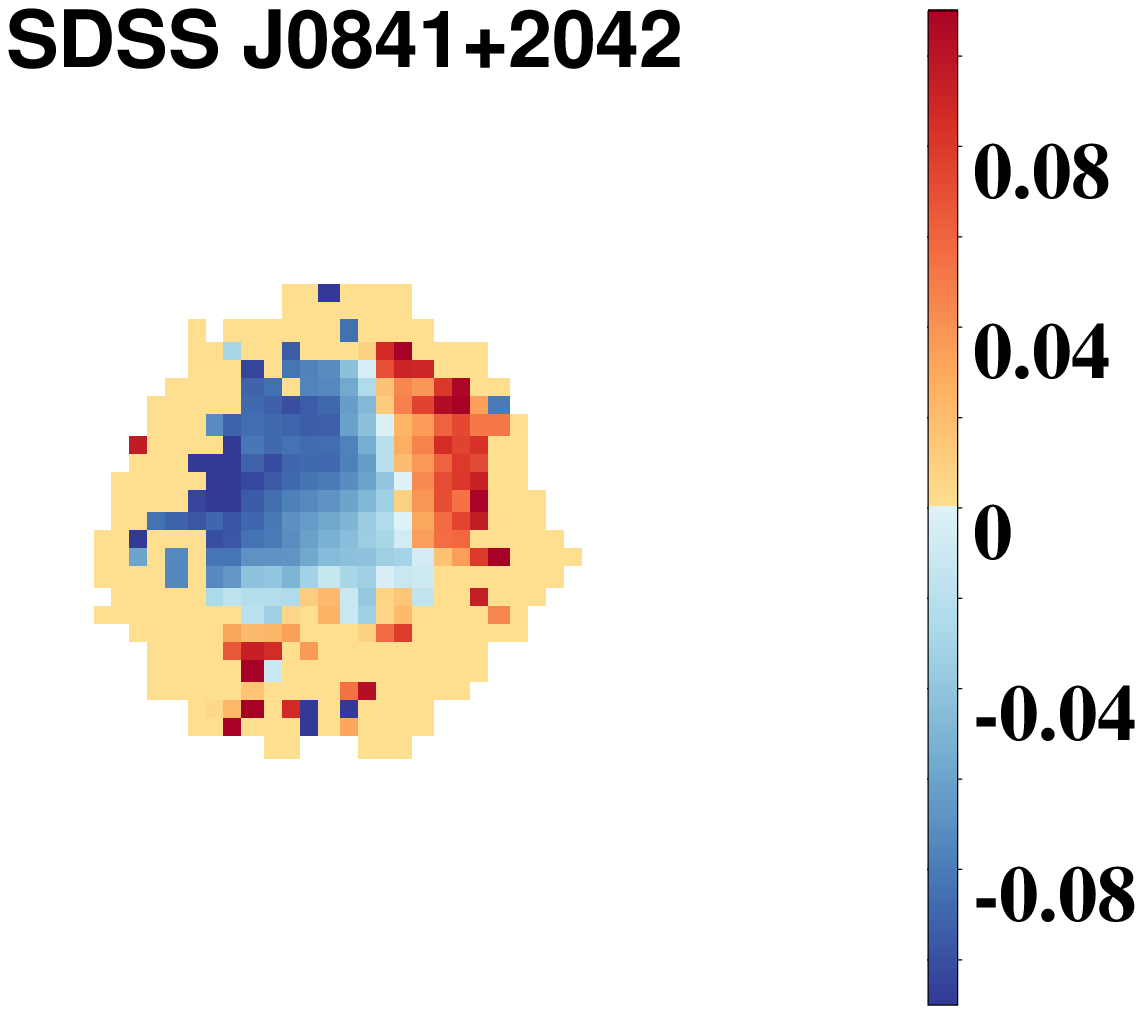}%
    \includegraphics[scale=0.28,clip=clip,trim=0mm 0mm 3.5cm 0mm]{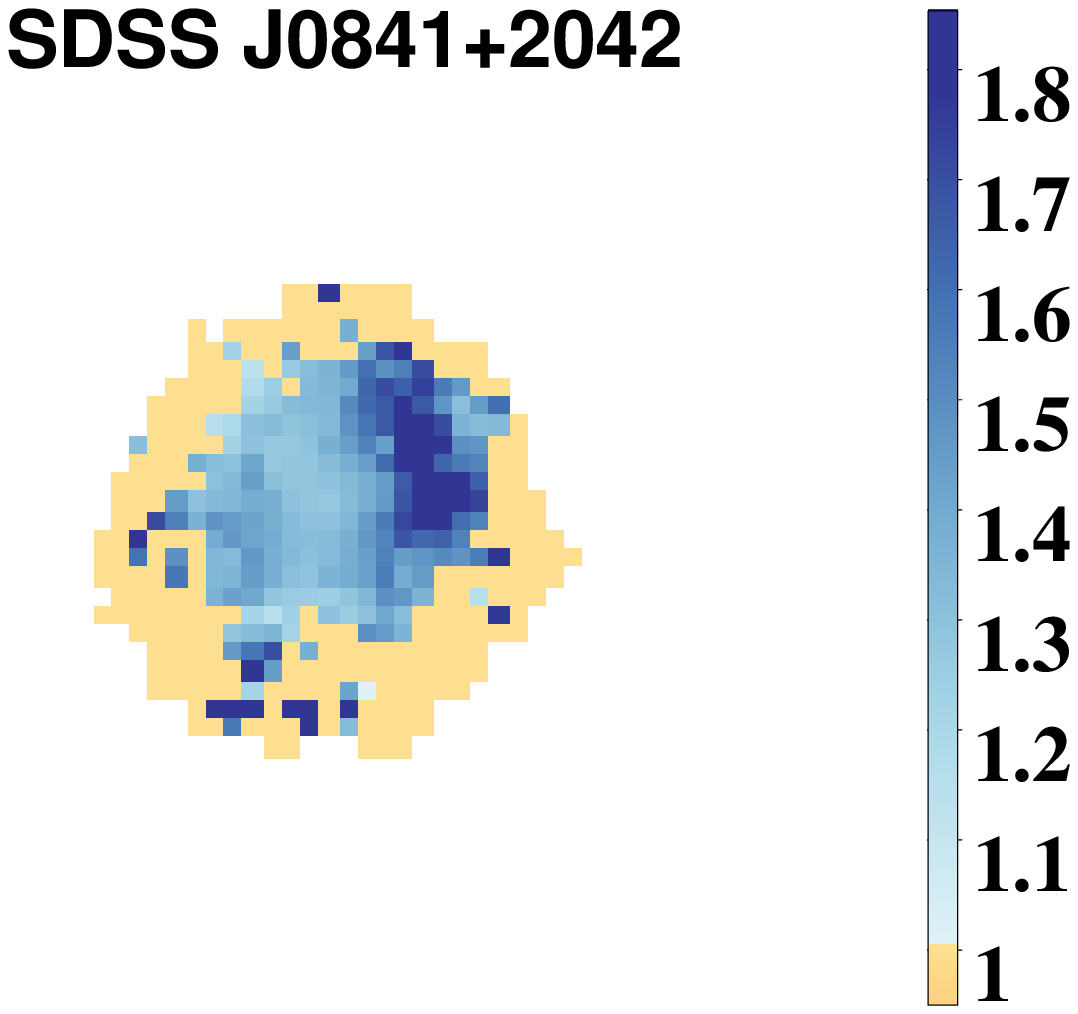}\\
\caption{Non-parametric measurement of our eleven radio-quiet and three radio-loud or radio-intermediate quasars in our sample. The four columns from left to right are: median velocity (km s$^{-1}$), line width ($W_{80}$, km s$^{-1}$), asymmetry ($A$) and shape parameter ($K$) maps. Only spaxels where the peak of the [O {\sc iii}]$\lambda$5007\AA\ line is detect with $S/N>5$ are plotted. In the $A$ maps, negative (left-tailed), zero (symmetric) and positive (right-tailed) values are color-coded blue, yellow and red, respectively; while in the $K$ maps, profiles with wings heavier than ($K>1$), as strong as ($K=1$) and weaker than ($K<1$) a Gaussian function are color-coded blue, yellow and red, respectively. The seeing at the observing site is depicted by the open circle on each median velocity map.}
\label{fig:VWAK}
\end{figure*}
    
\begin{figure*}%[h!]    
    \includegraphics[scale=0.28,clip=clip,trim=0mm 0mm 3.5cm 0mm]{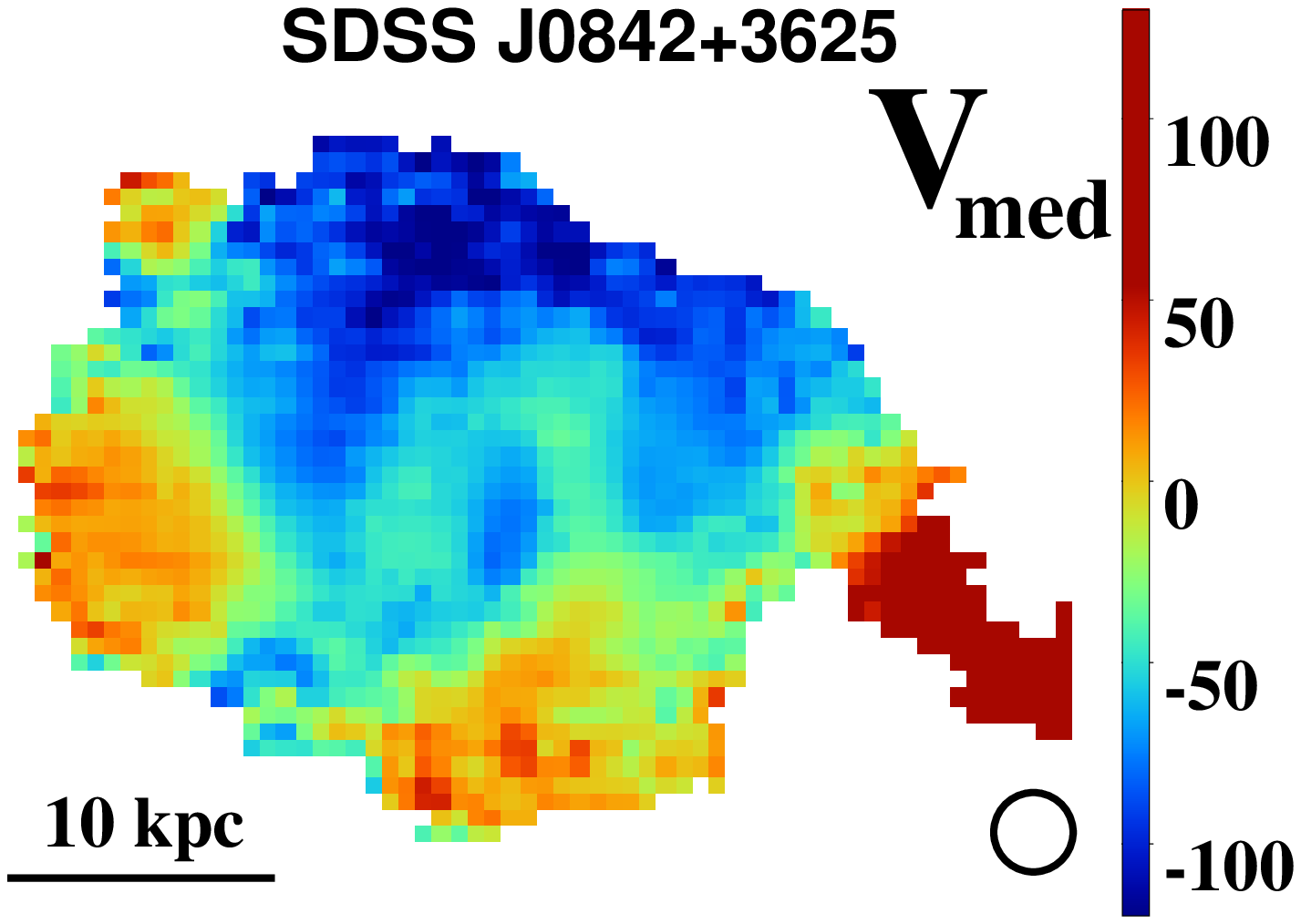}%
    \includegraphics[scale=0.28,clip=clip,trim=0mm 0mm 3.5cm 0mm]{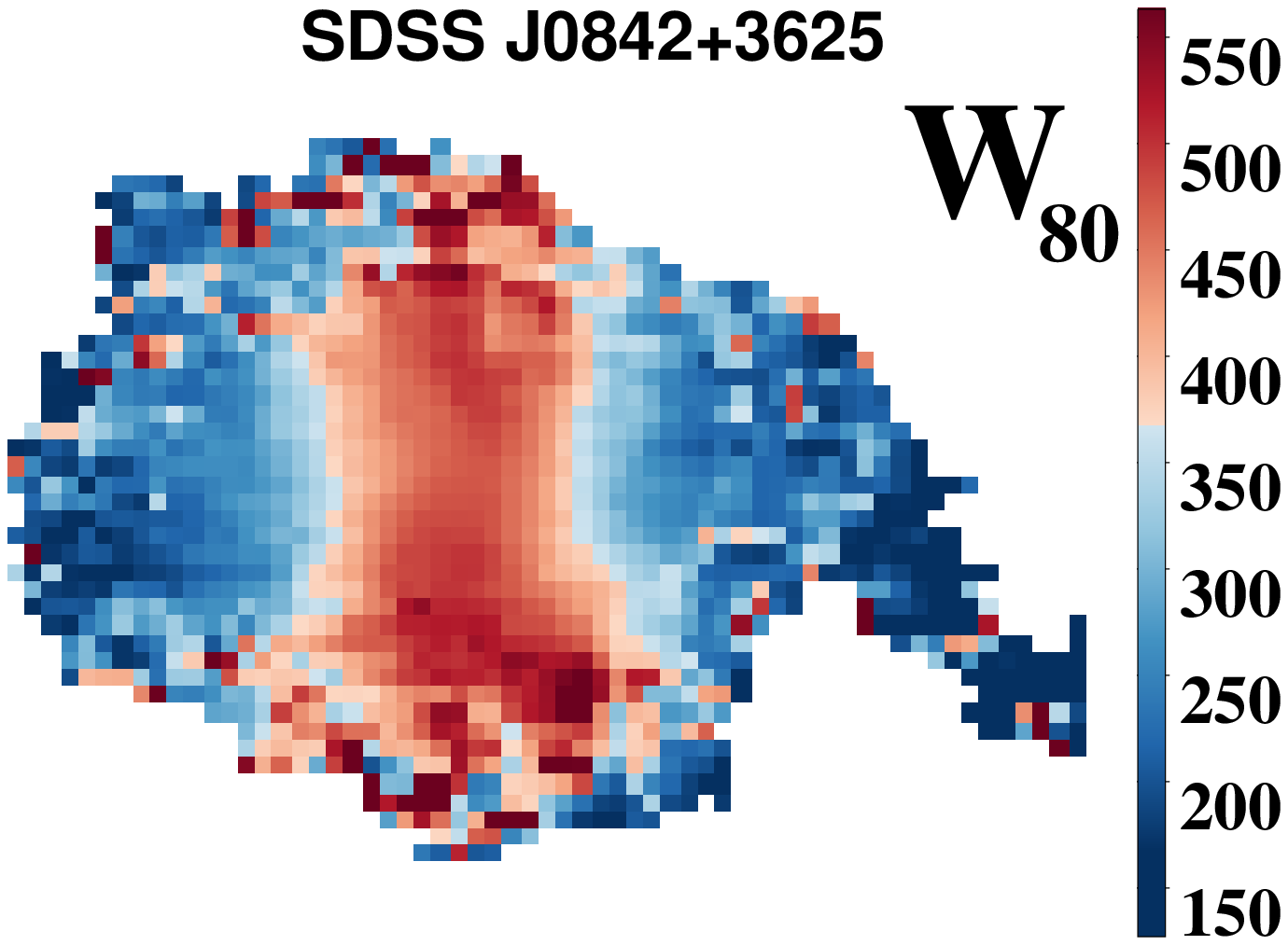}%
    \includegraphics[scale=0.28,clip=clip,trim=0mm 0mm 3.5cm 0mm]{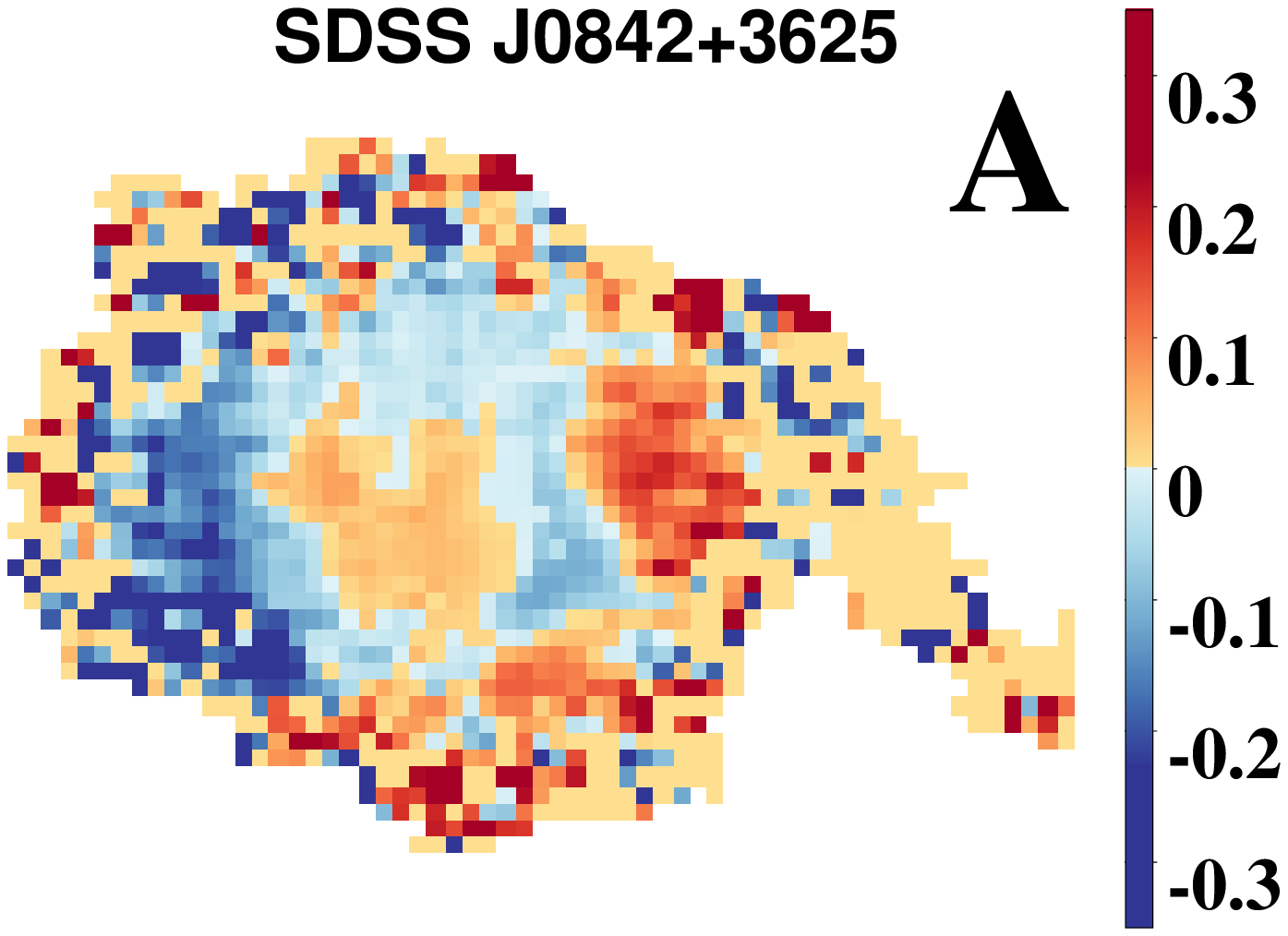}%
    \includegraphics[scale=0.28,clip=clip,trim=0mm 0mm 3.5cm 0mm]{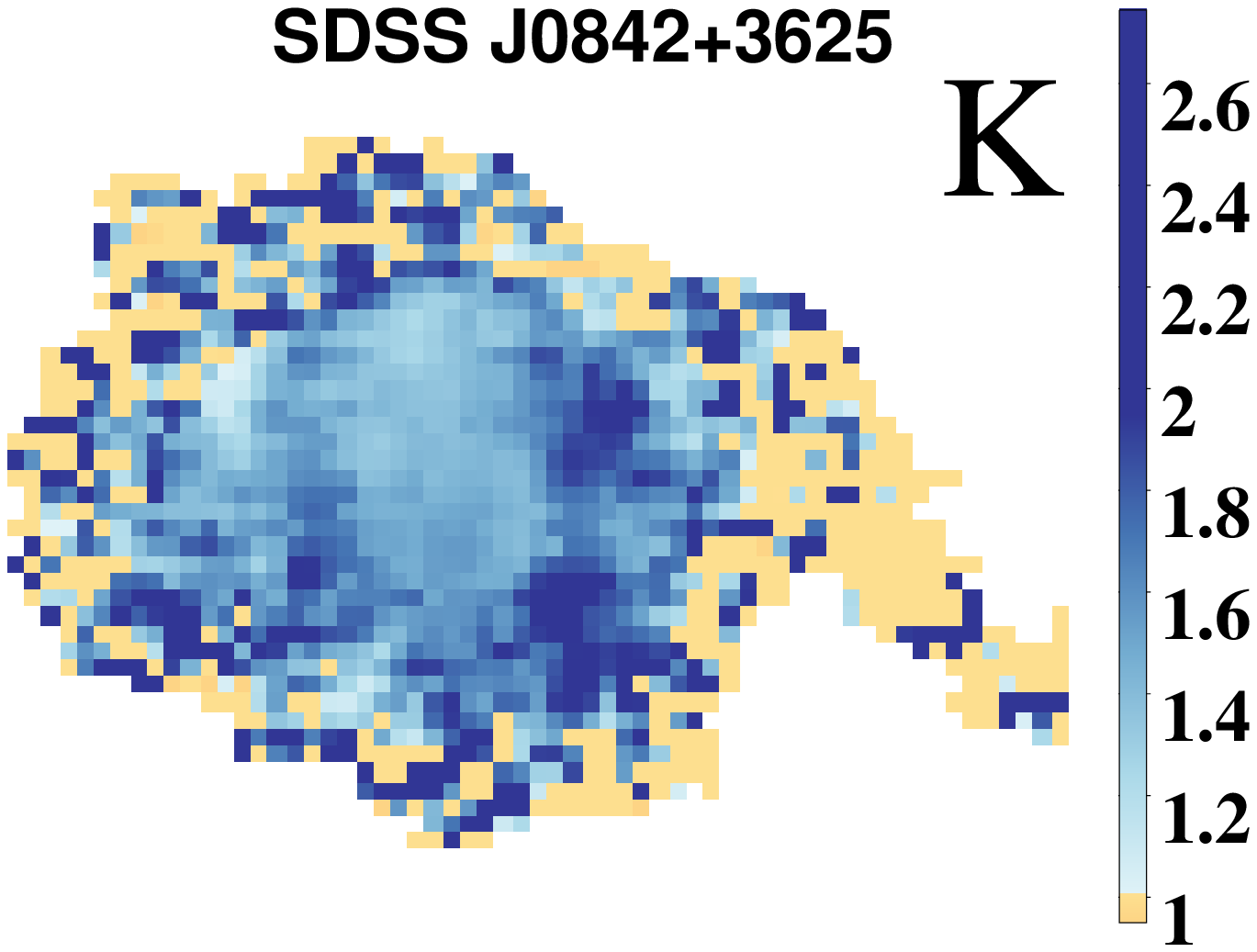}\\
    \includegraphics[scale=0.28,clip=clip,trim=0mm 0mm 3.5cm 0mm]{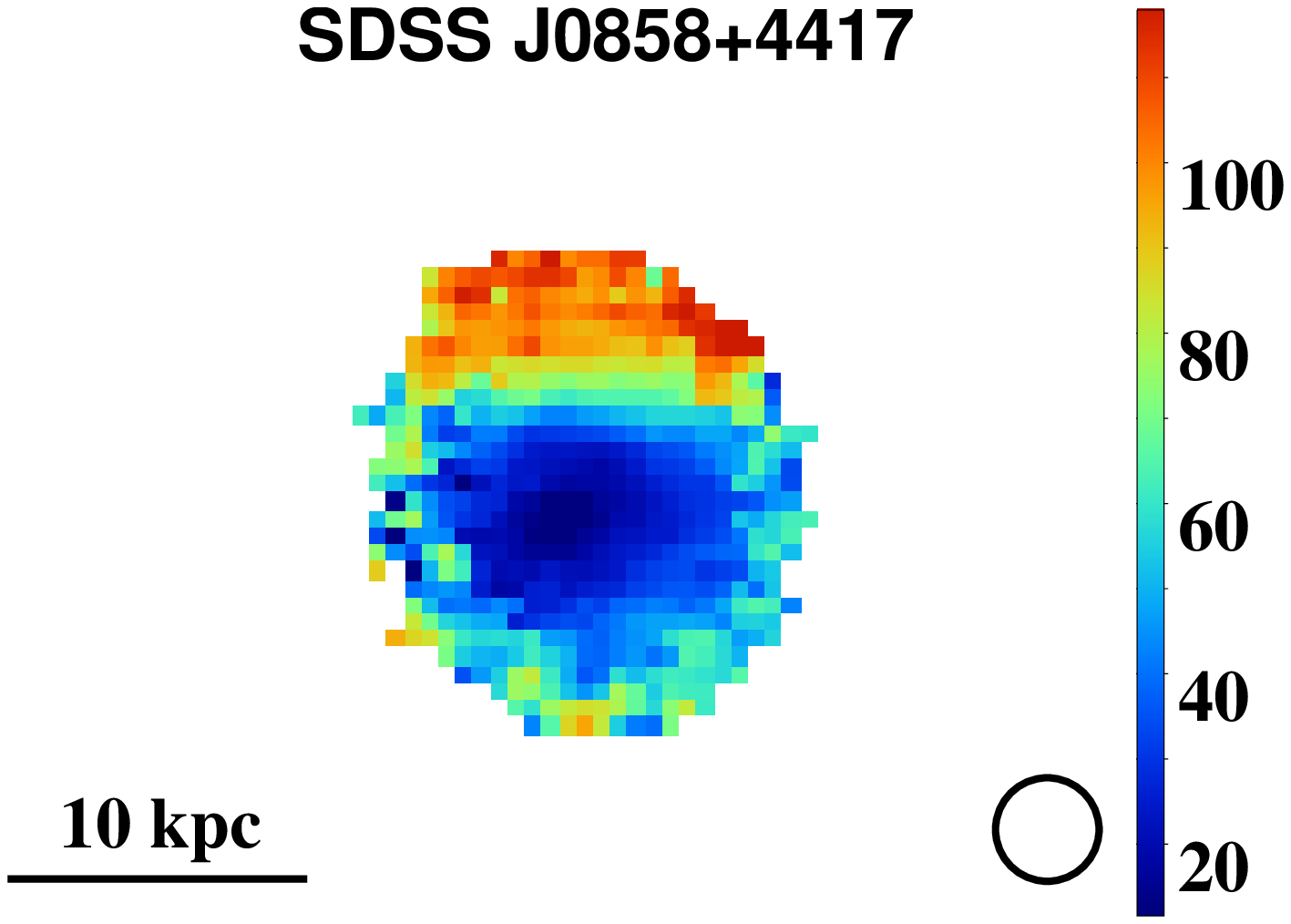}%
    \includegraphics[scale=0.28,clip=clip,trim=0mm 0mm 3.5cm 0mm]{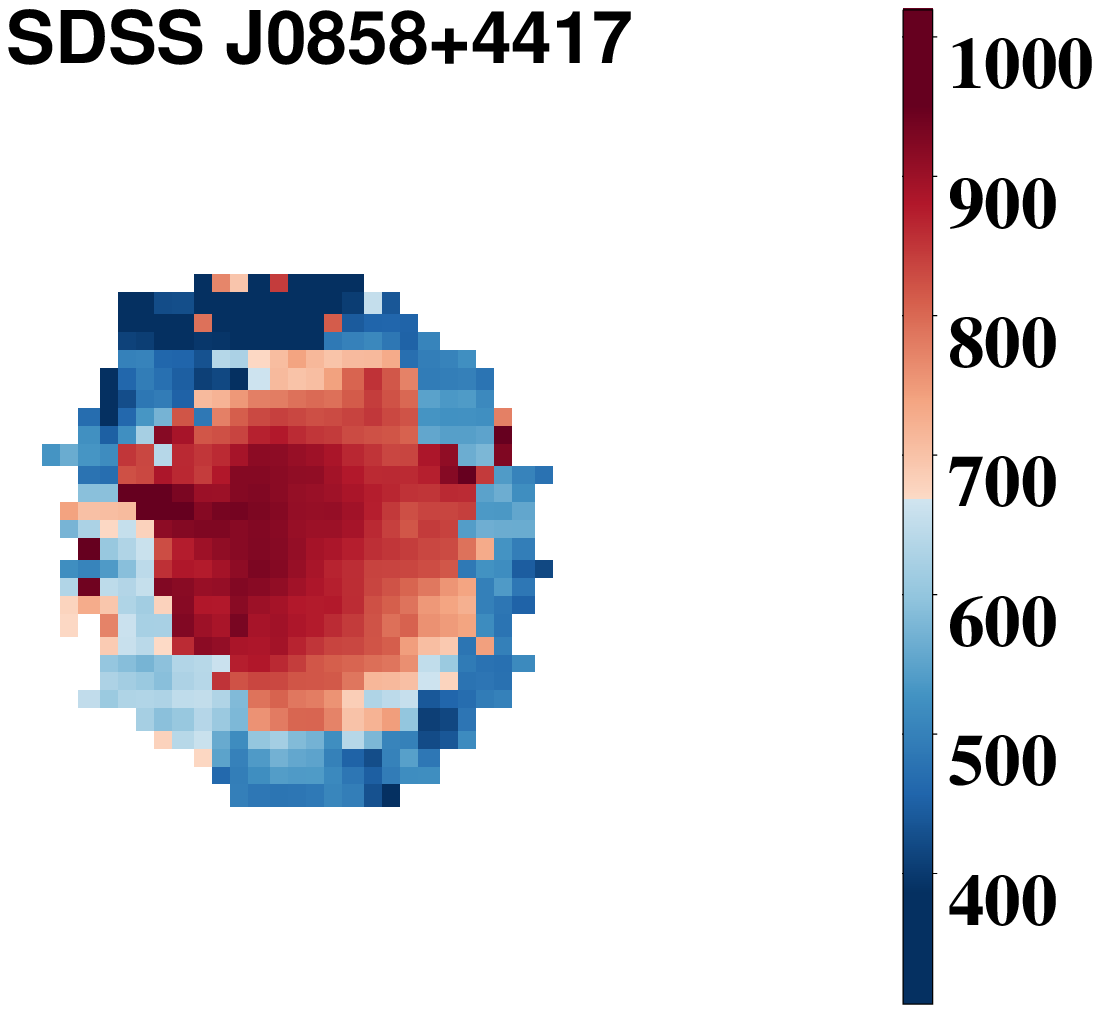}%
    \includegraphics[scale=0.28,clip=clip,trim=0mm 0mm 3.5cm 0mm]{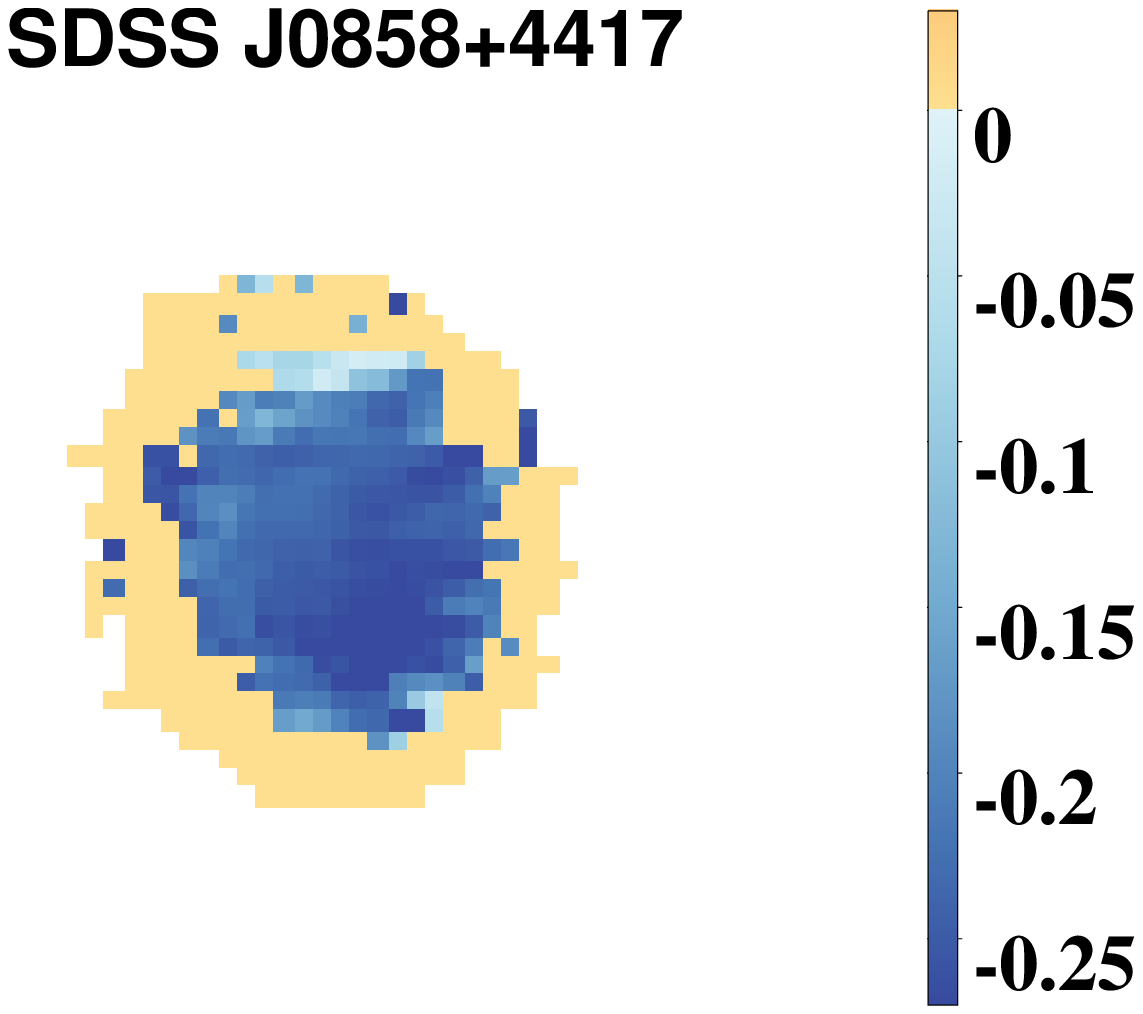}%
    \includegraphics[scale=0.28,clip=clip,trim=0mm 0mm 3.5cm 0mm]{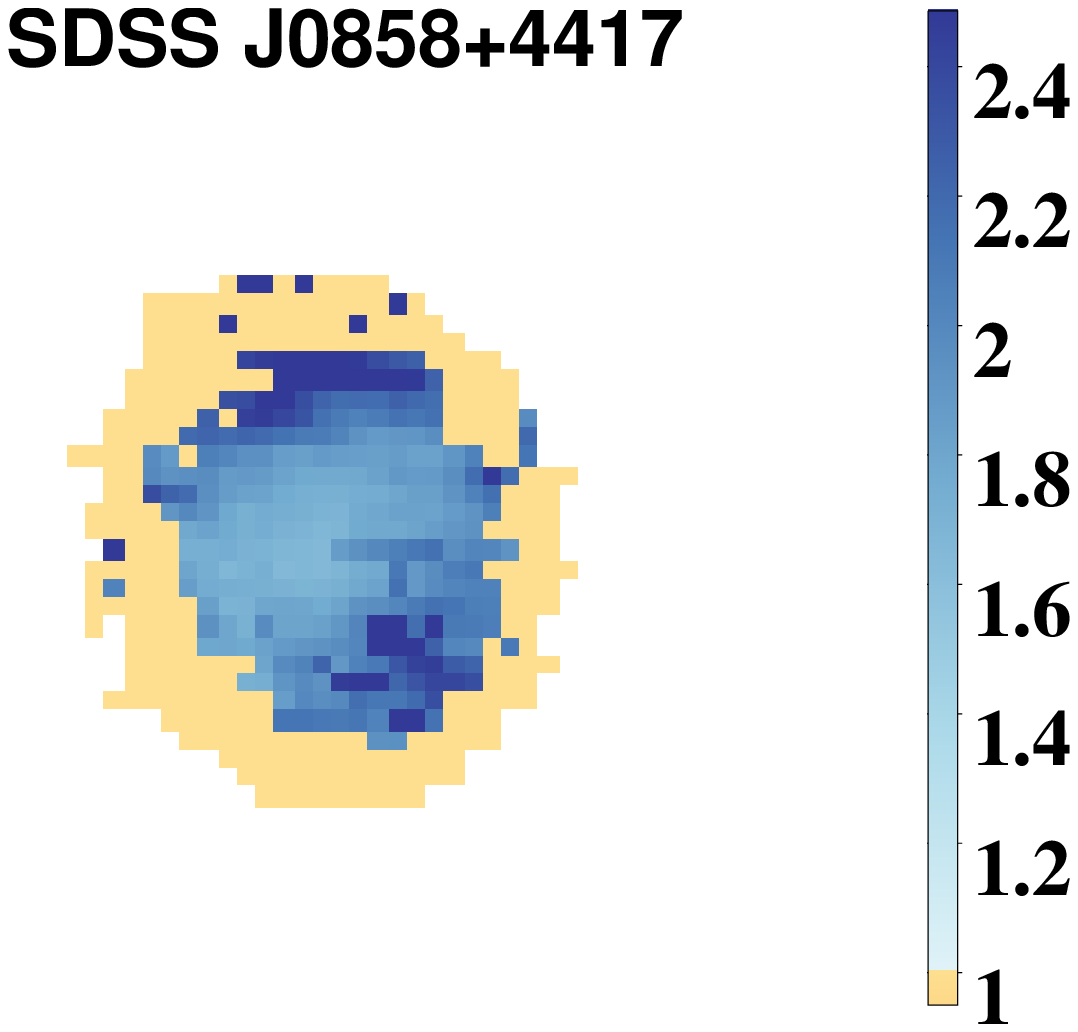}\\
    \includegraphics[scale=0.28,clip=clip,trim=0mm 0mm 3.5cm 0mm]{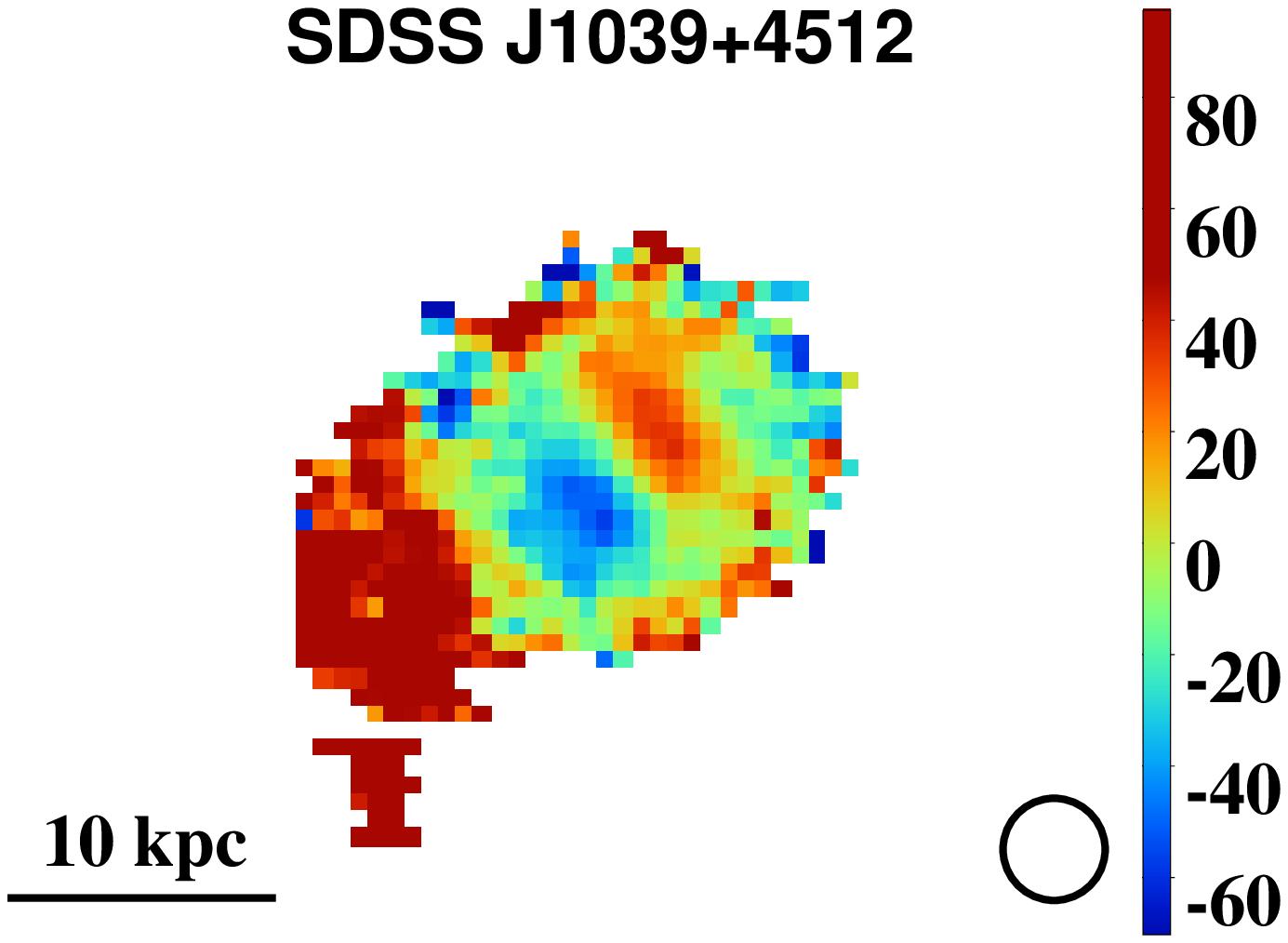}%
    \includegraphics[scale=0.28,clip=clip,trim=0mm 0mm 3.5cm 0mm]{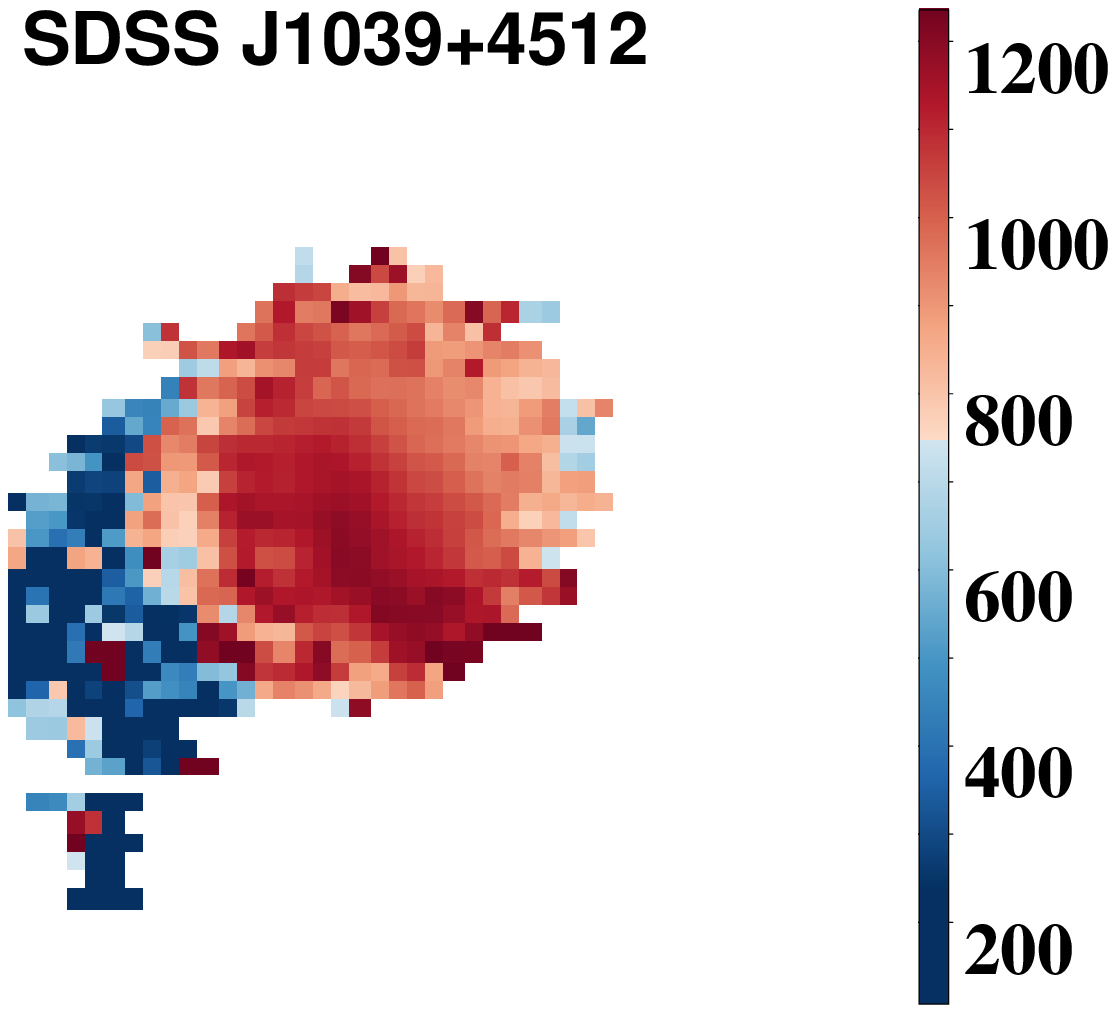}%
    \includegraphics[scale=0.28,clip=clip,trim=0mm 0mm 3.5cm 0mm]{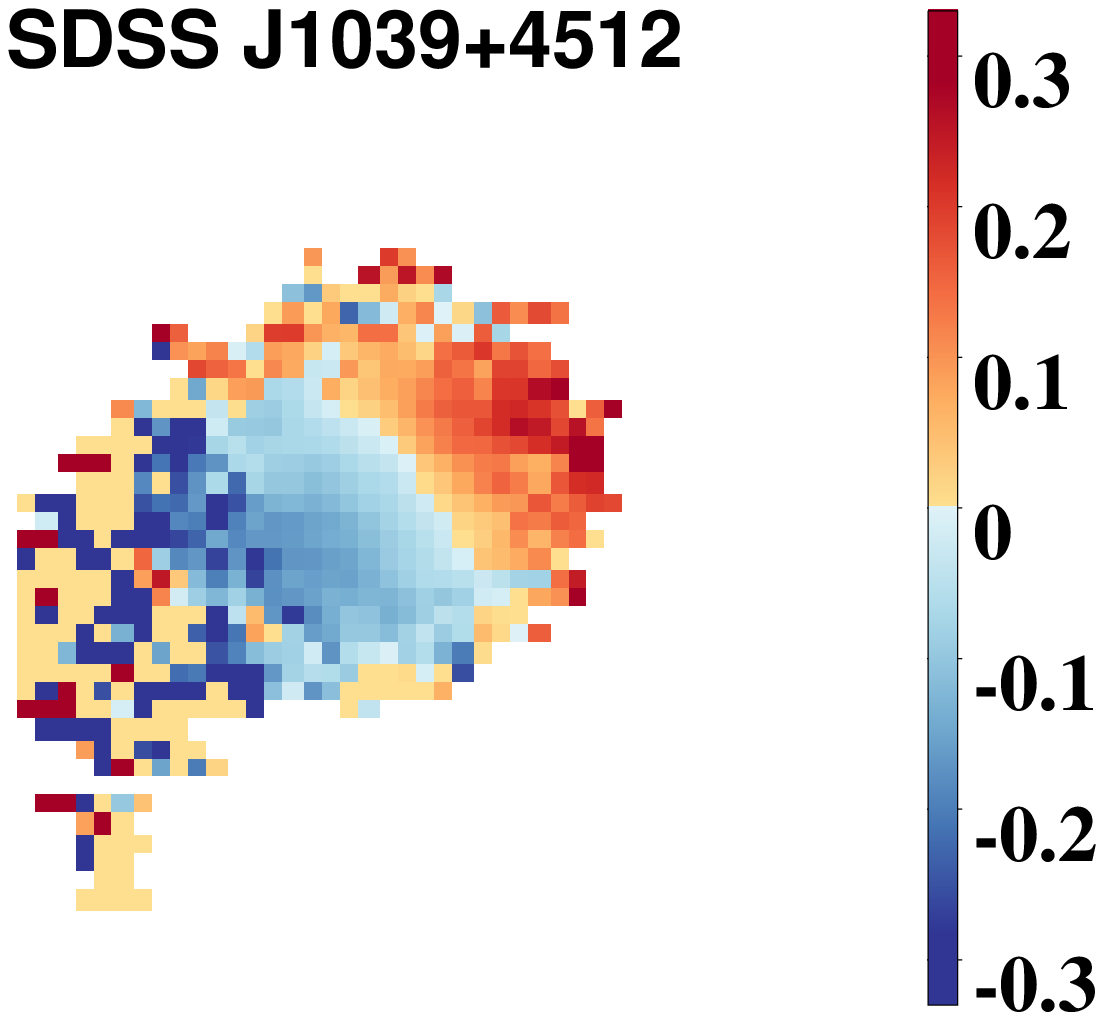}%
    \includegraphics[scale=0.28,clip=clip,trim=0mm 0mm 3.5cm 0mm]{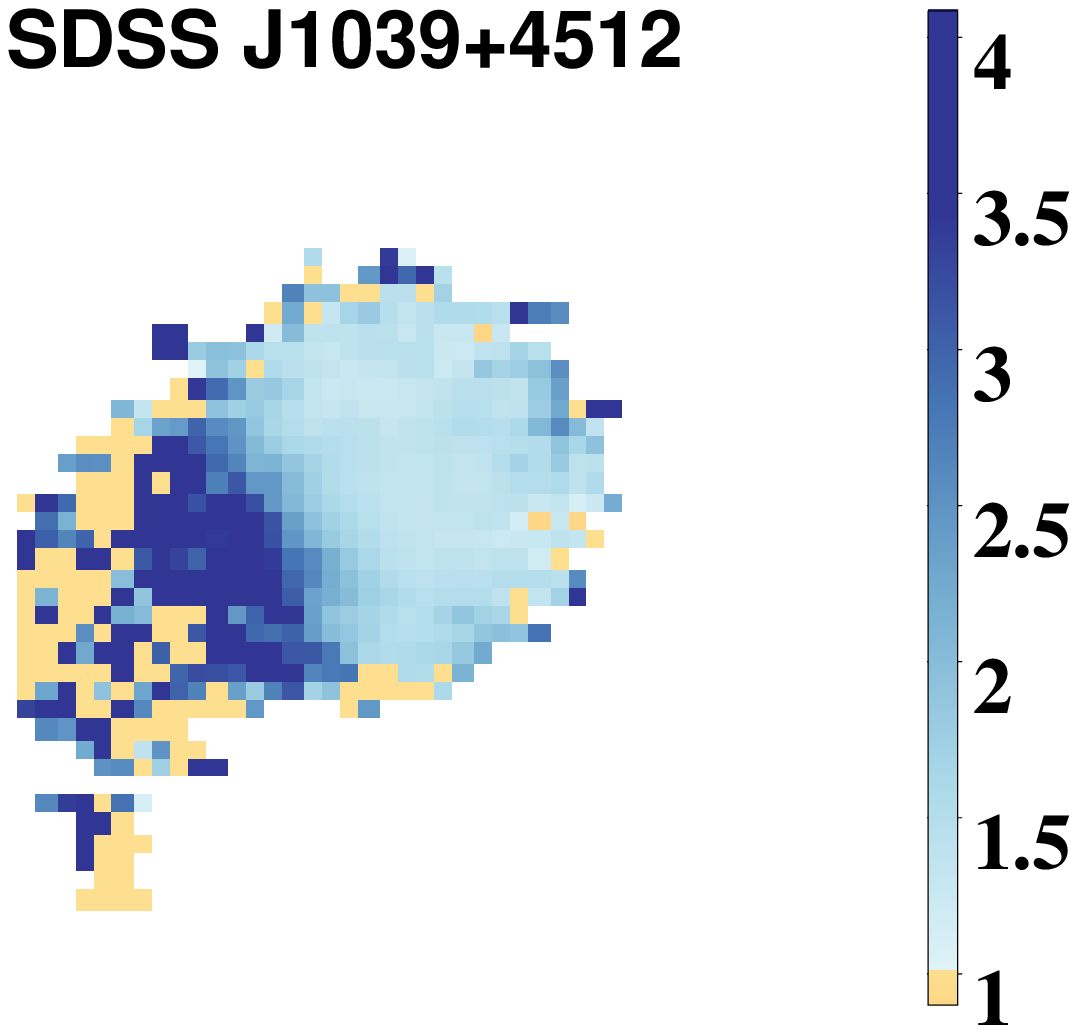}\\
    \includegraphics[scale=0.28,clip=clip,trim=0mm 0mm 3.5cm 0mm]{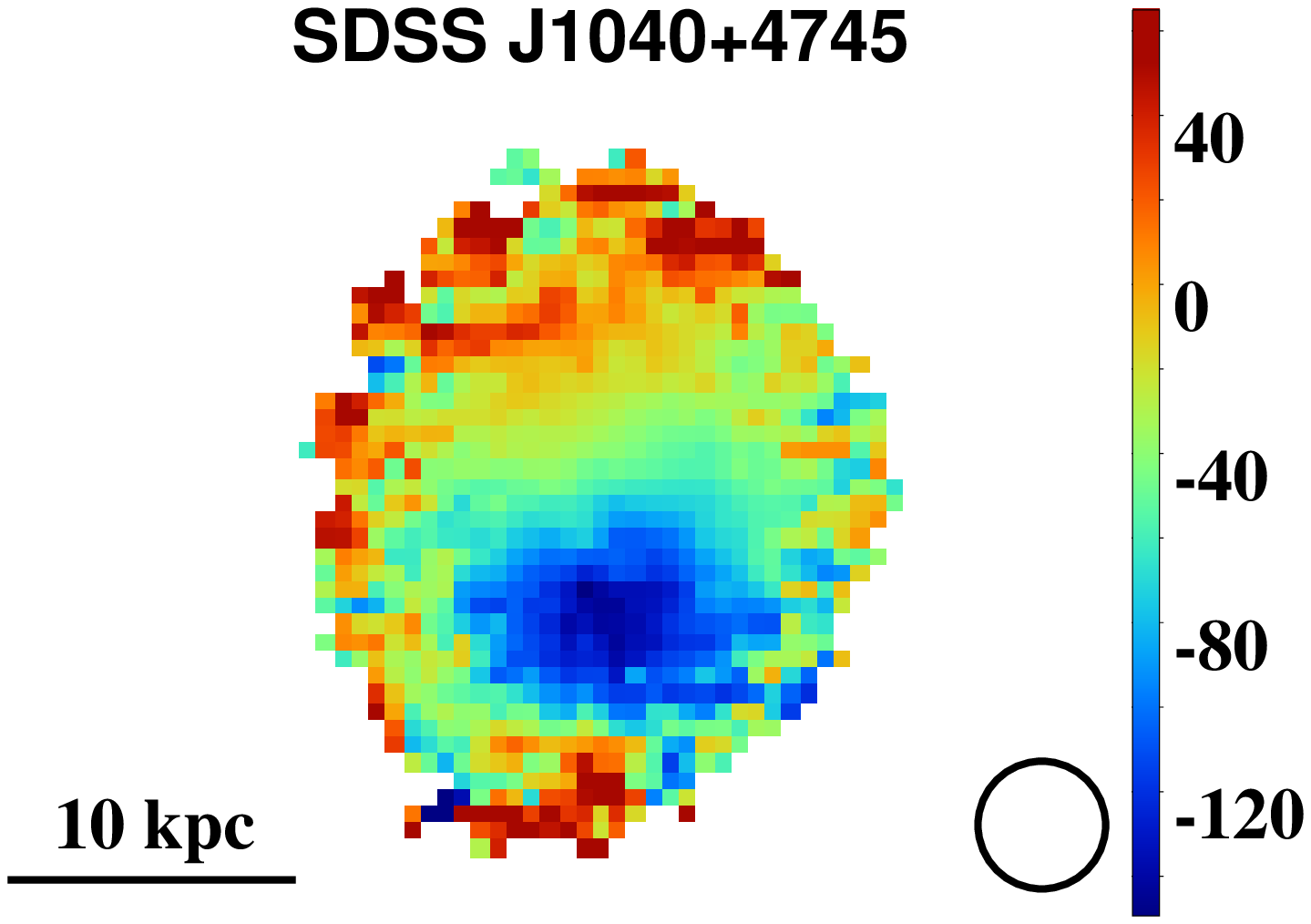}%
    \includegraphics[scale=0.28,clip=clip,trim=0mm 0mm 3.5cm 0mm]{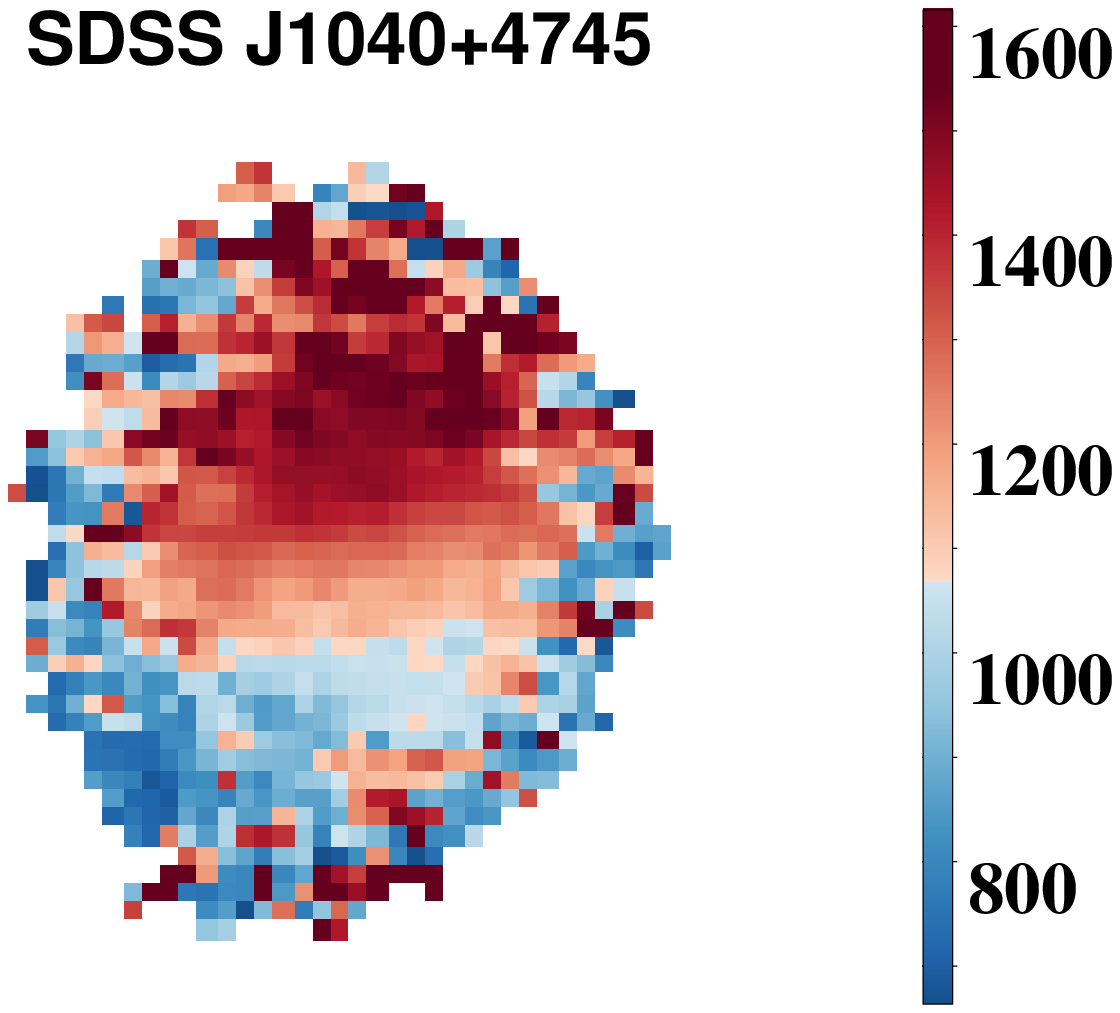}%
    \includegraphics[scale=0.28,clip=clip,trim=0mm 0mm 3.5cm 0mm]{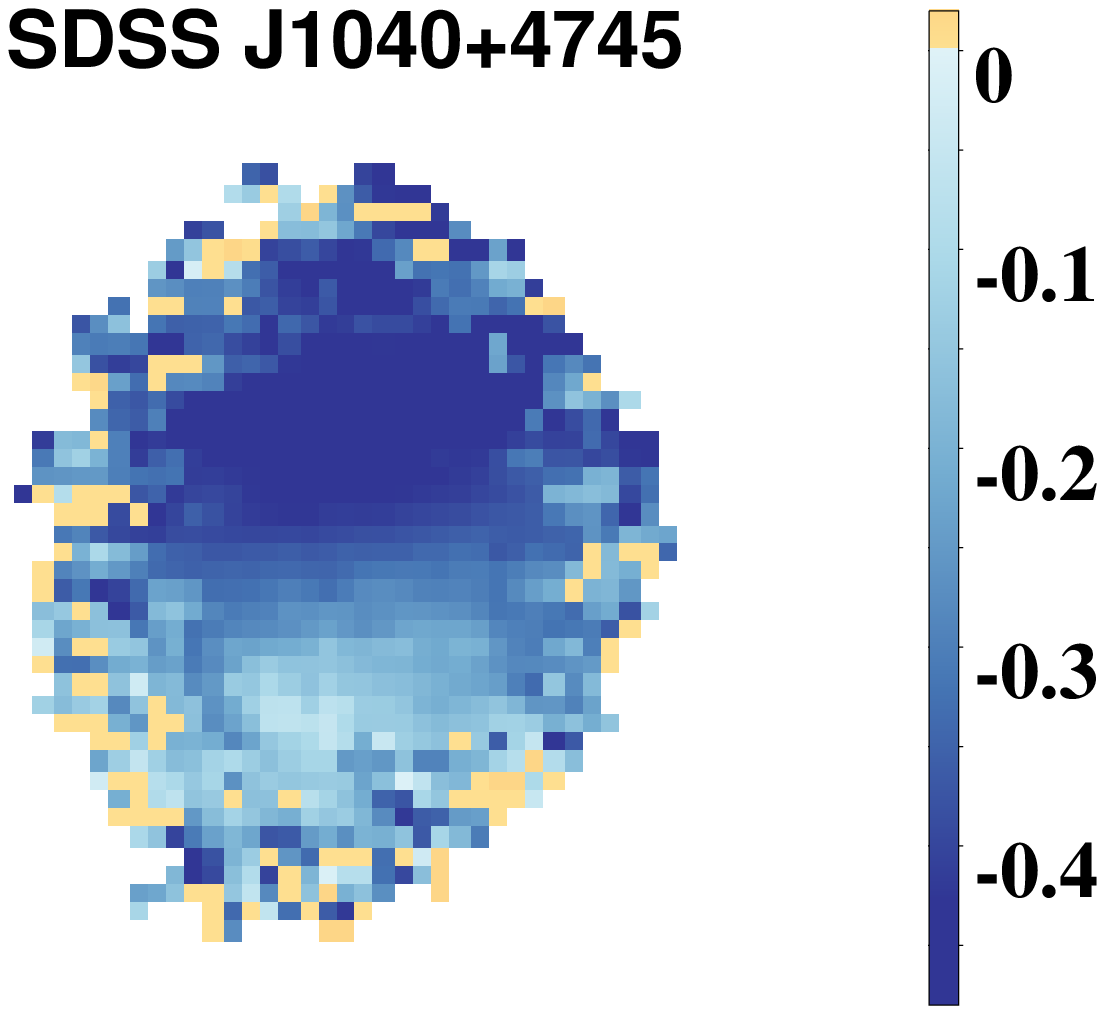}%
    \includegraphics[scale=0.28,clip=clip,trim=0mm 0mm 3.5cm 0mm]{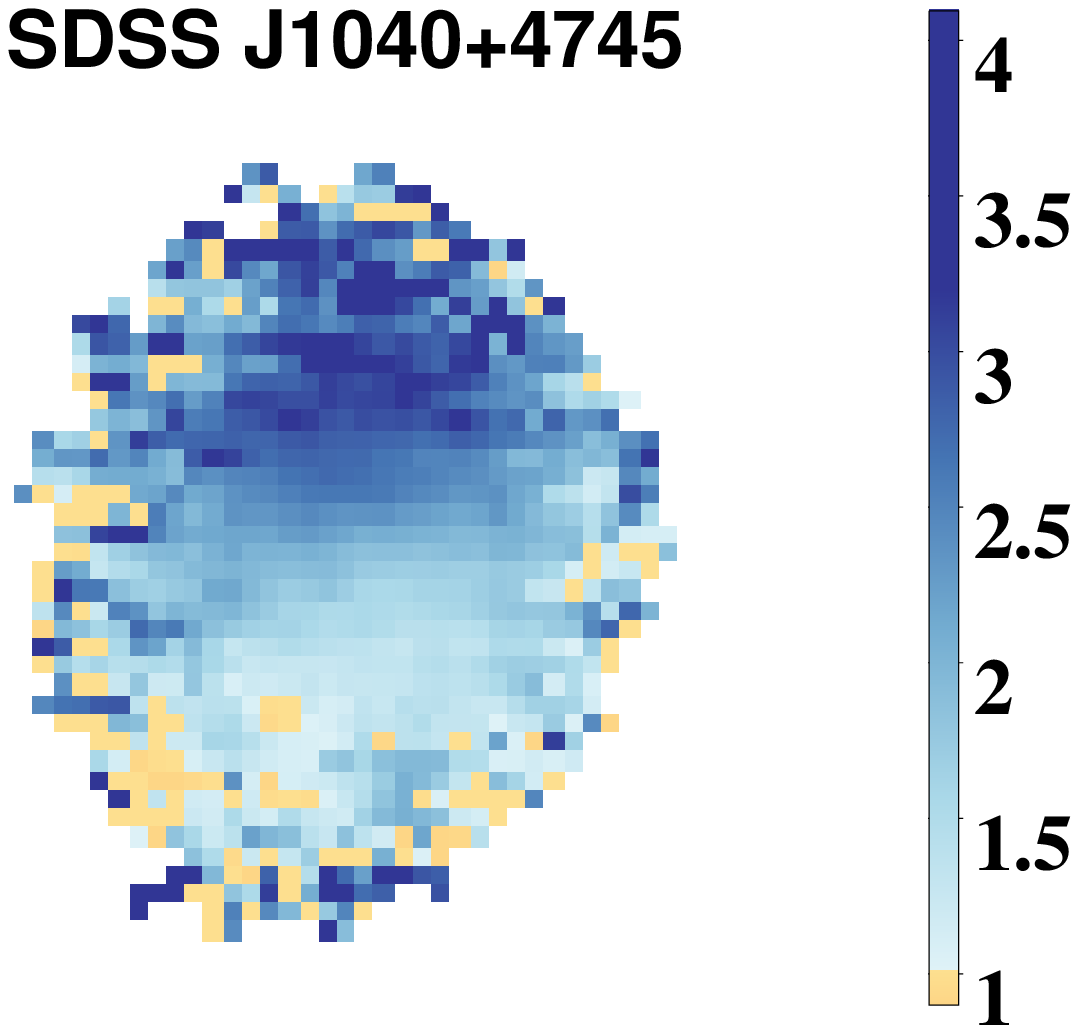}\\
    \includegraphics[scale=0.28,clip=clip,trim=0mm 0mm 3.5cm 0mm]{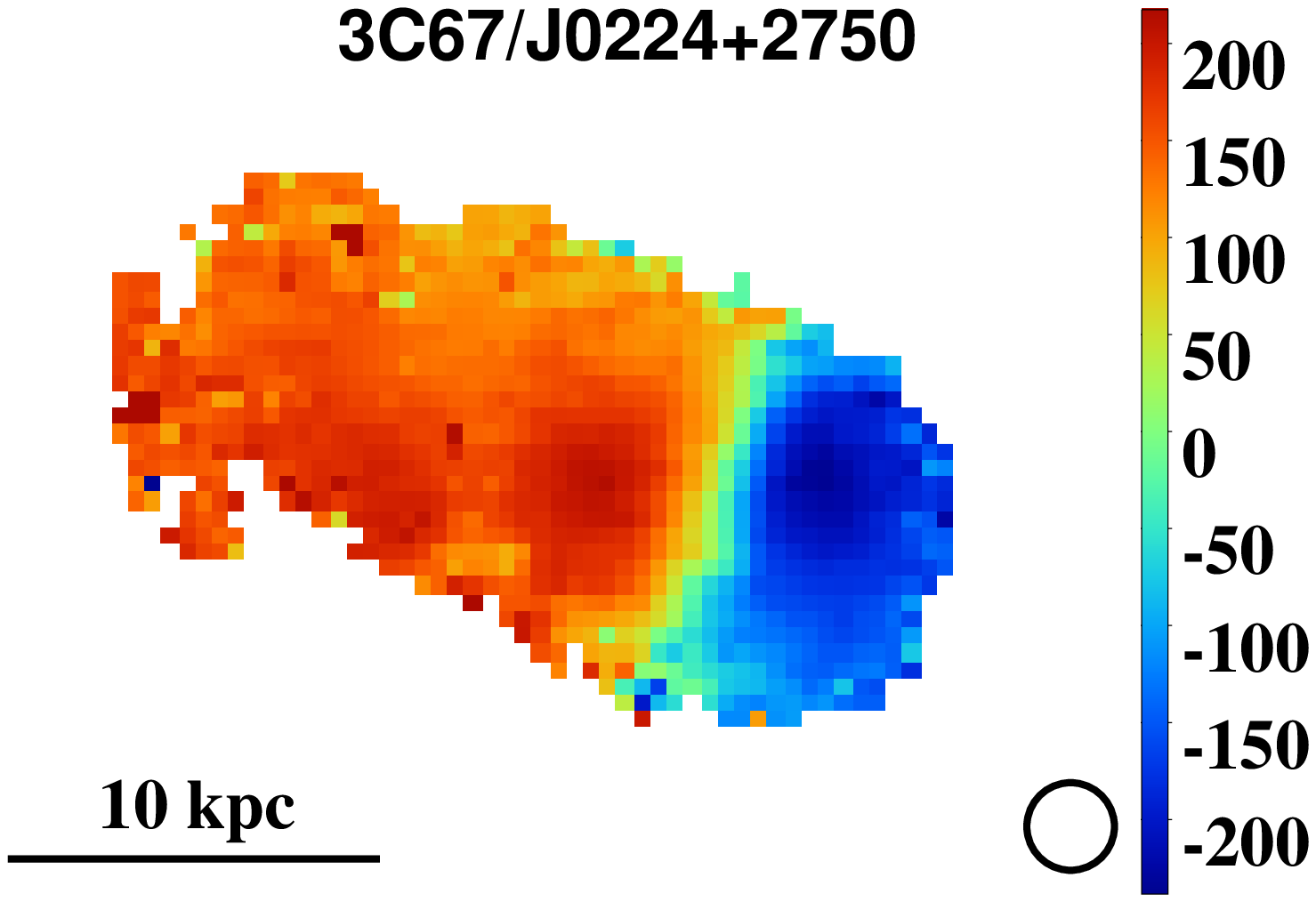}%
    \includegraphics[scale=0.28,clip=clip,trim=0mm 0mm 3.5cm 0mm]{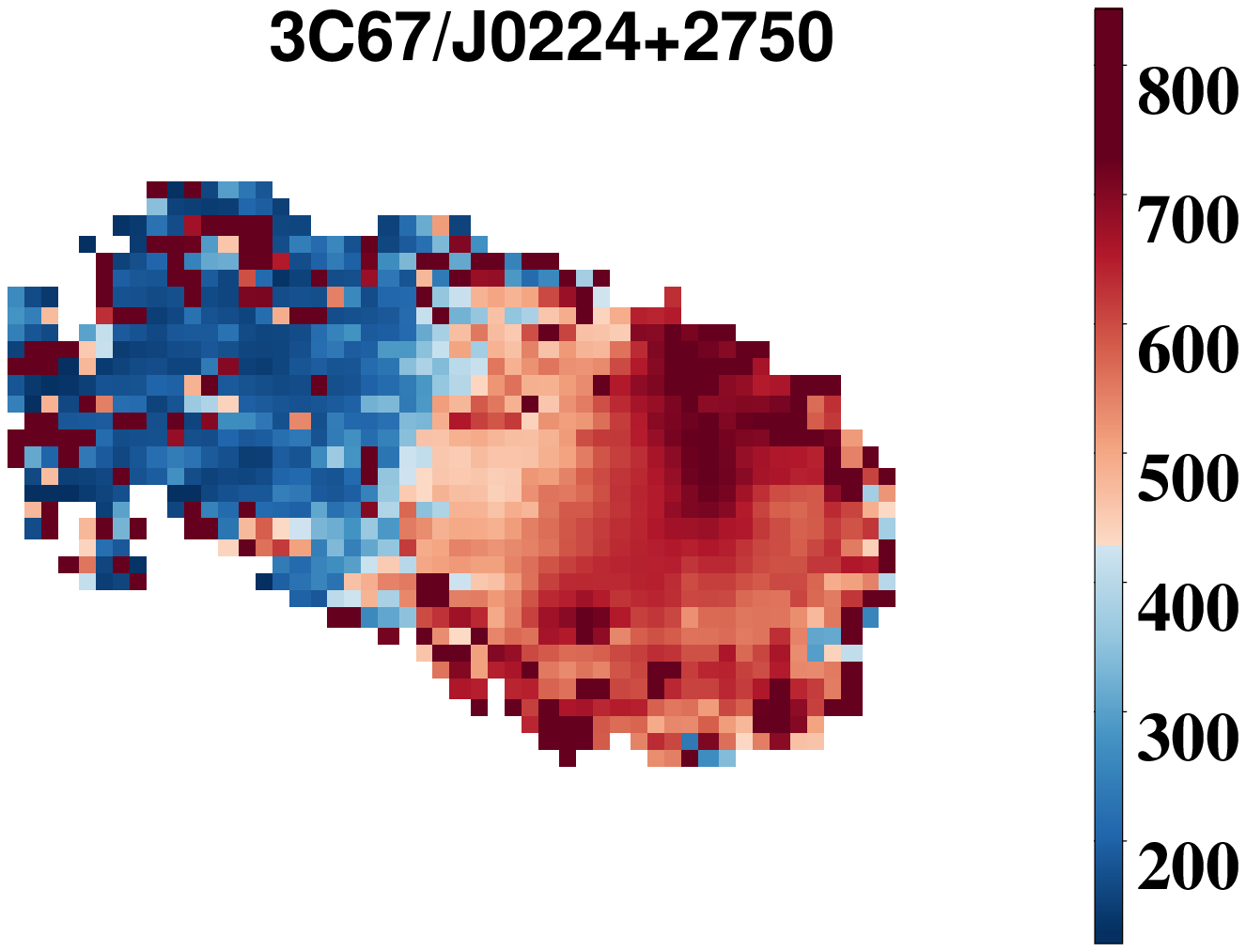}%
    \includegraphics[scale=0.28,clip=clip,trim=0mm 0mm 3.5cm 0mm]{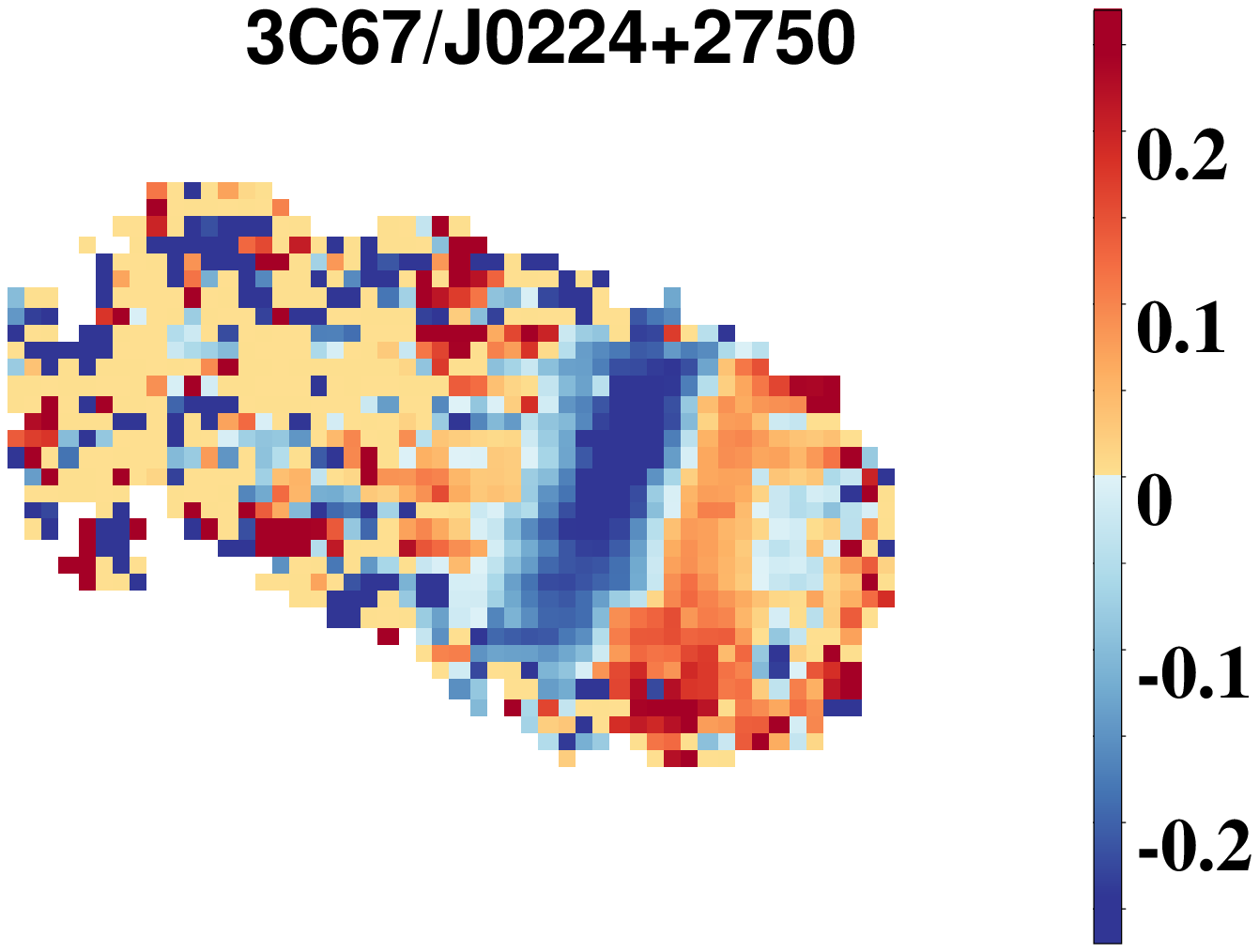}%
    \includegraphics[scale=0.28,clip=clip,trim=0mm 0mm 3.5cm 0mm]{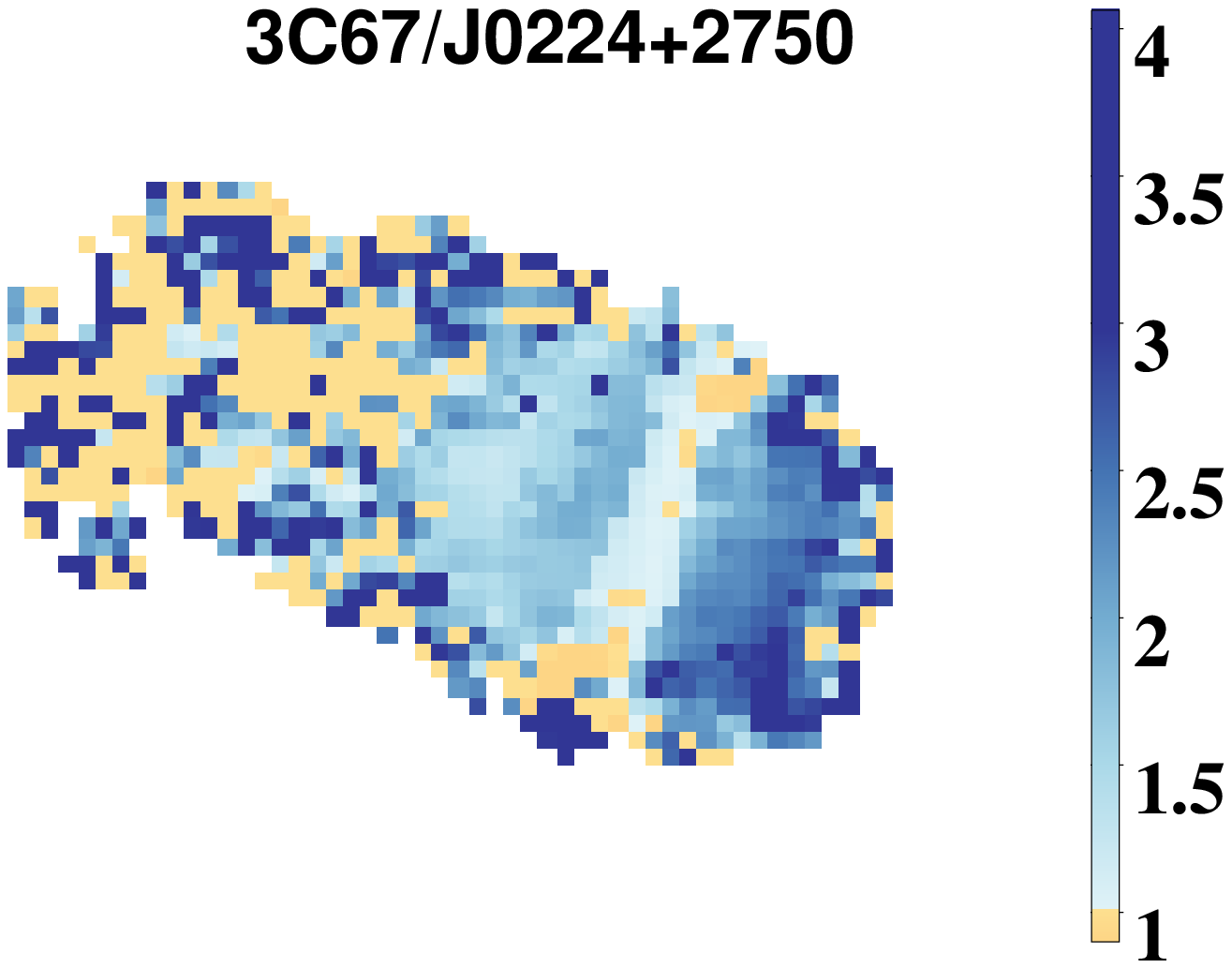}\\
    \includegraphics[scale=0.28,clip=clip,trim=0mm 0mm 3.5cm 0mm]{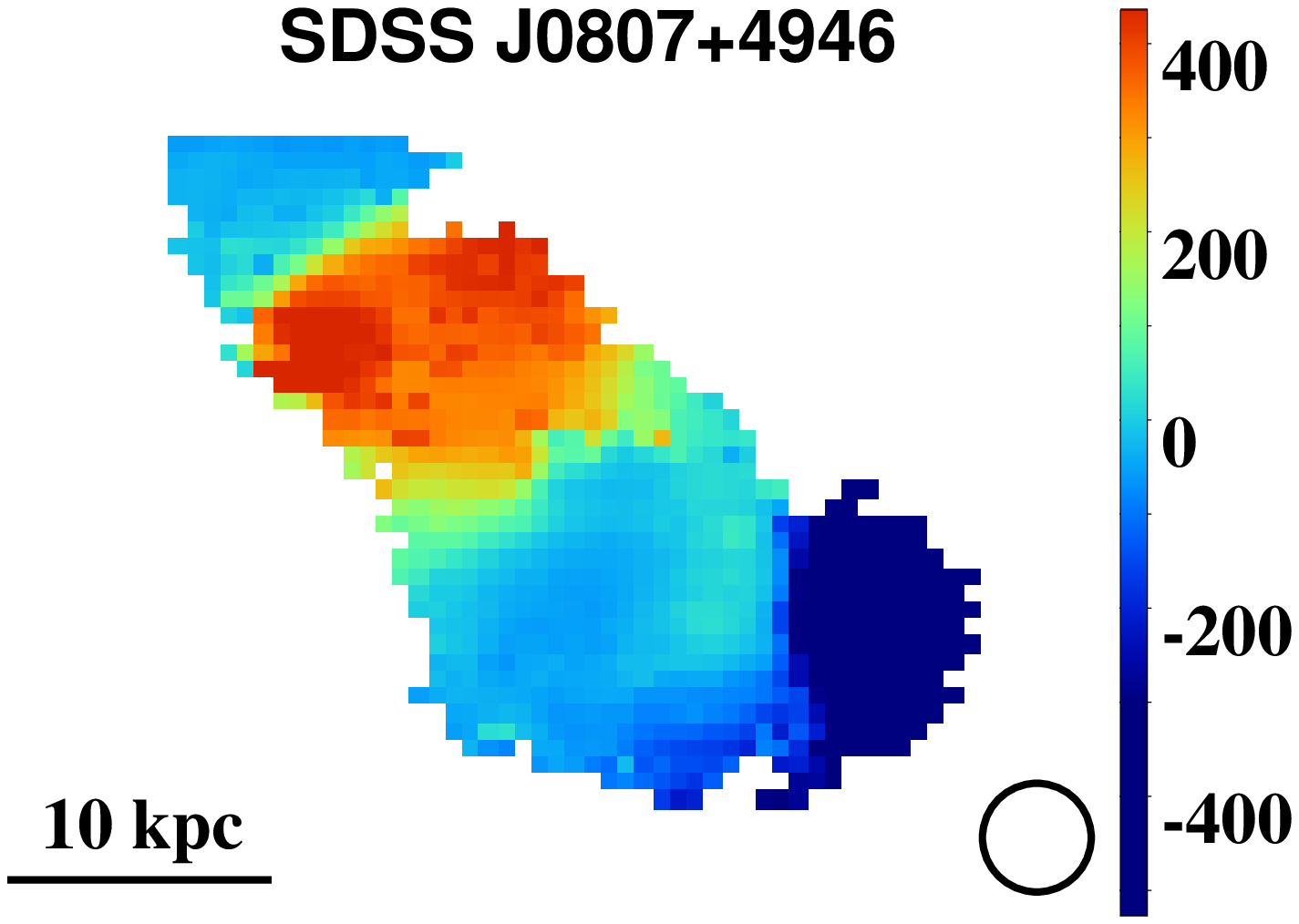}%
    \includegraphics[scale=0.28,clip=clip,trim=0mm 0mm 3.5cm 0mm]{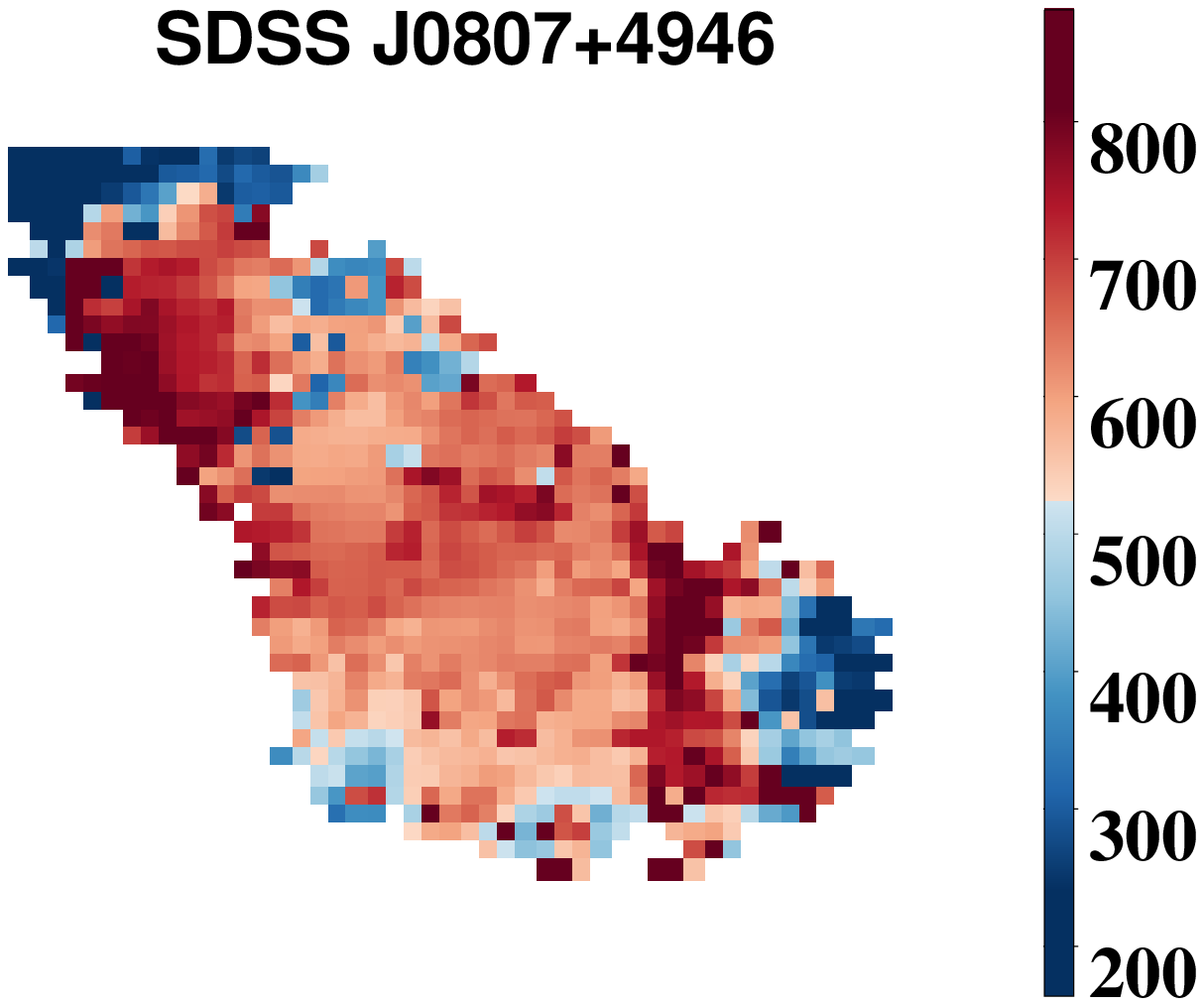}%
    \includegraphics[scale=0.28,clip=clip,trim=0mm 0mm 3.5cm 0mm]{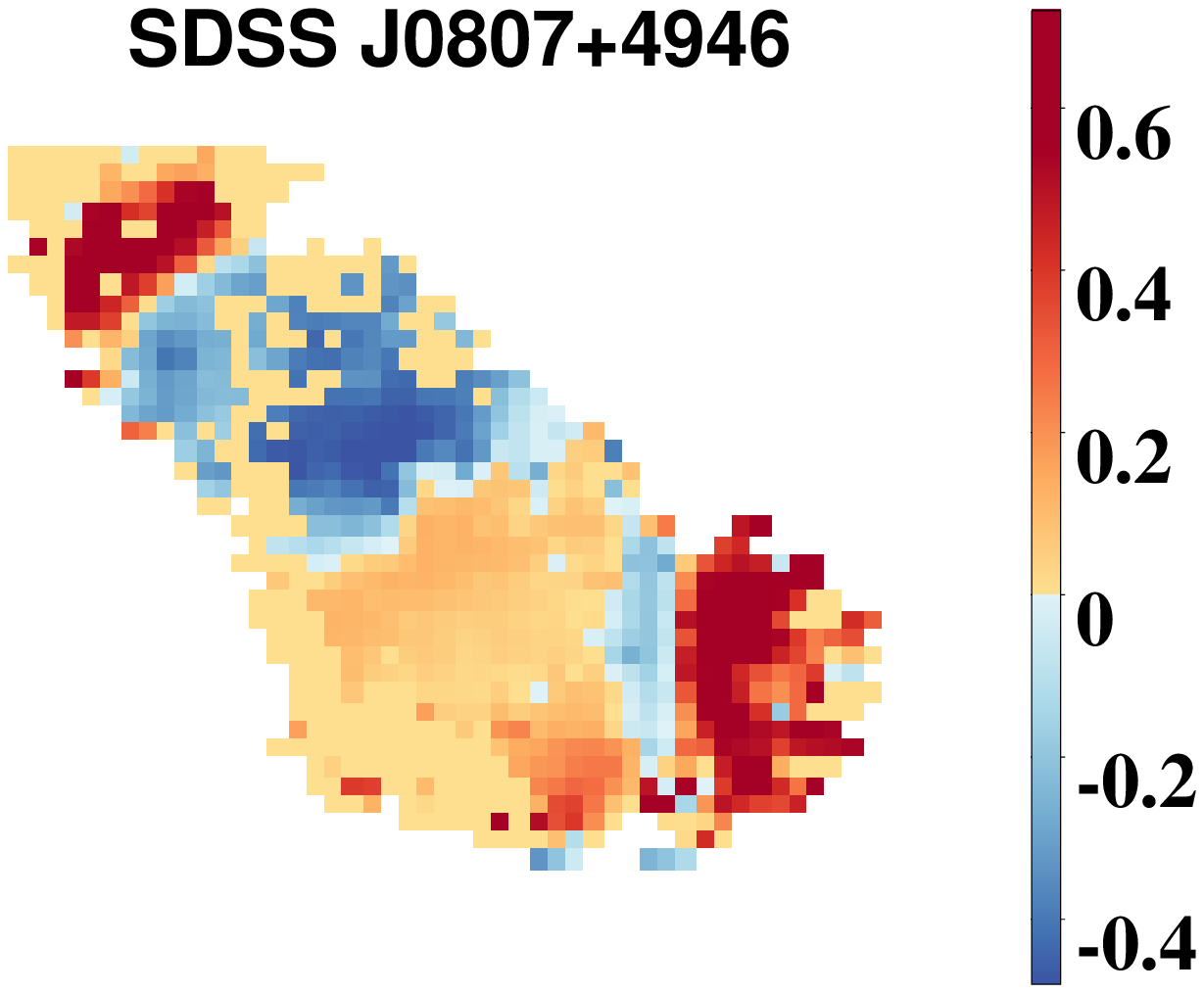}%
    \includegraphics[scale=0.28,clip=clip,trim=0mm 0mm 3.5cm 0mm]{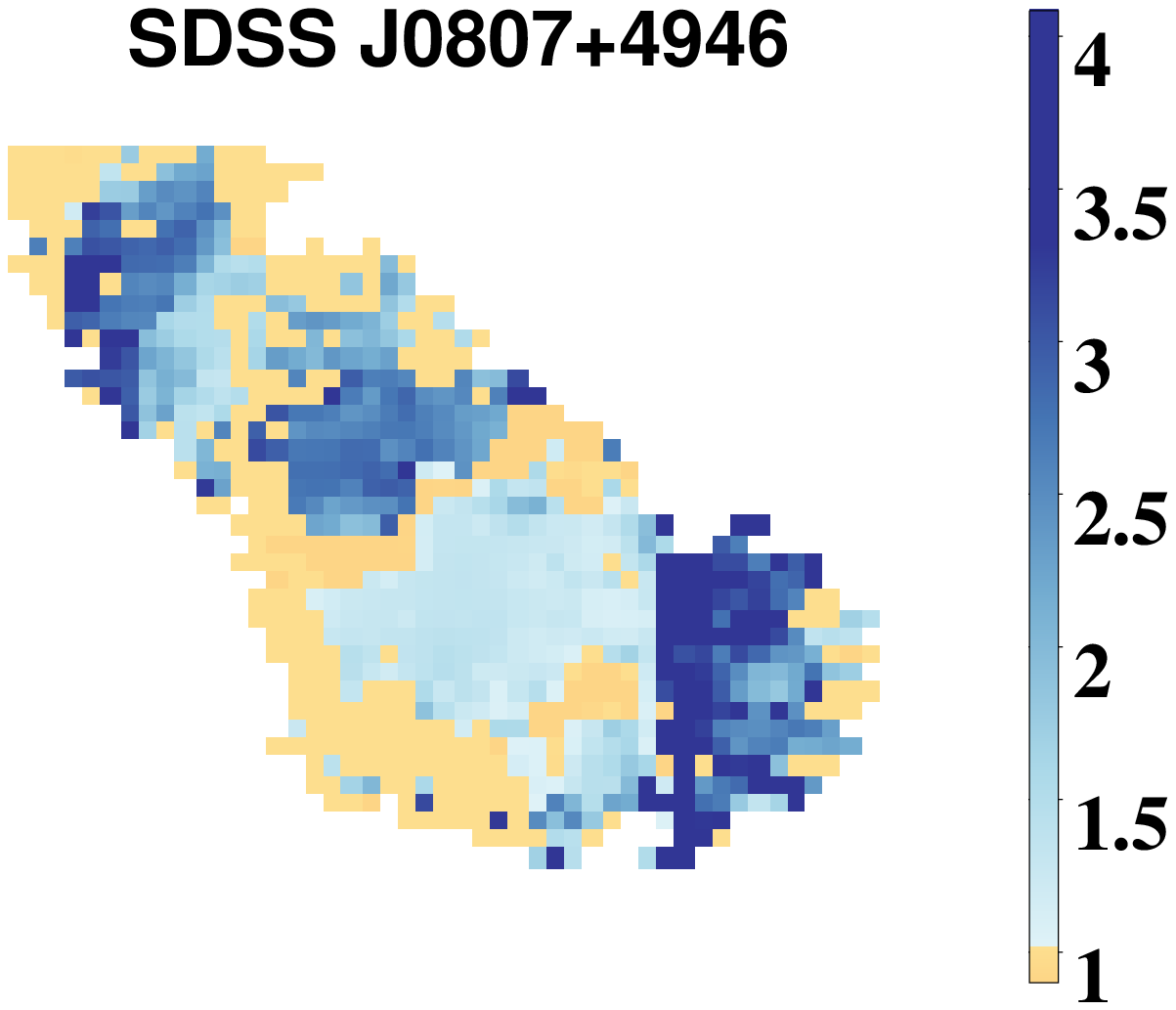}\\
    \includegraphics[scale=0.28,clip=clip,trim=0mm 0mm 3.5cm 0mm]{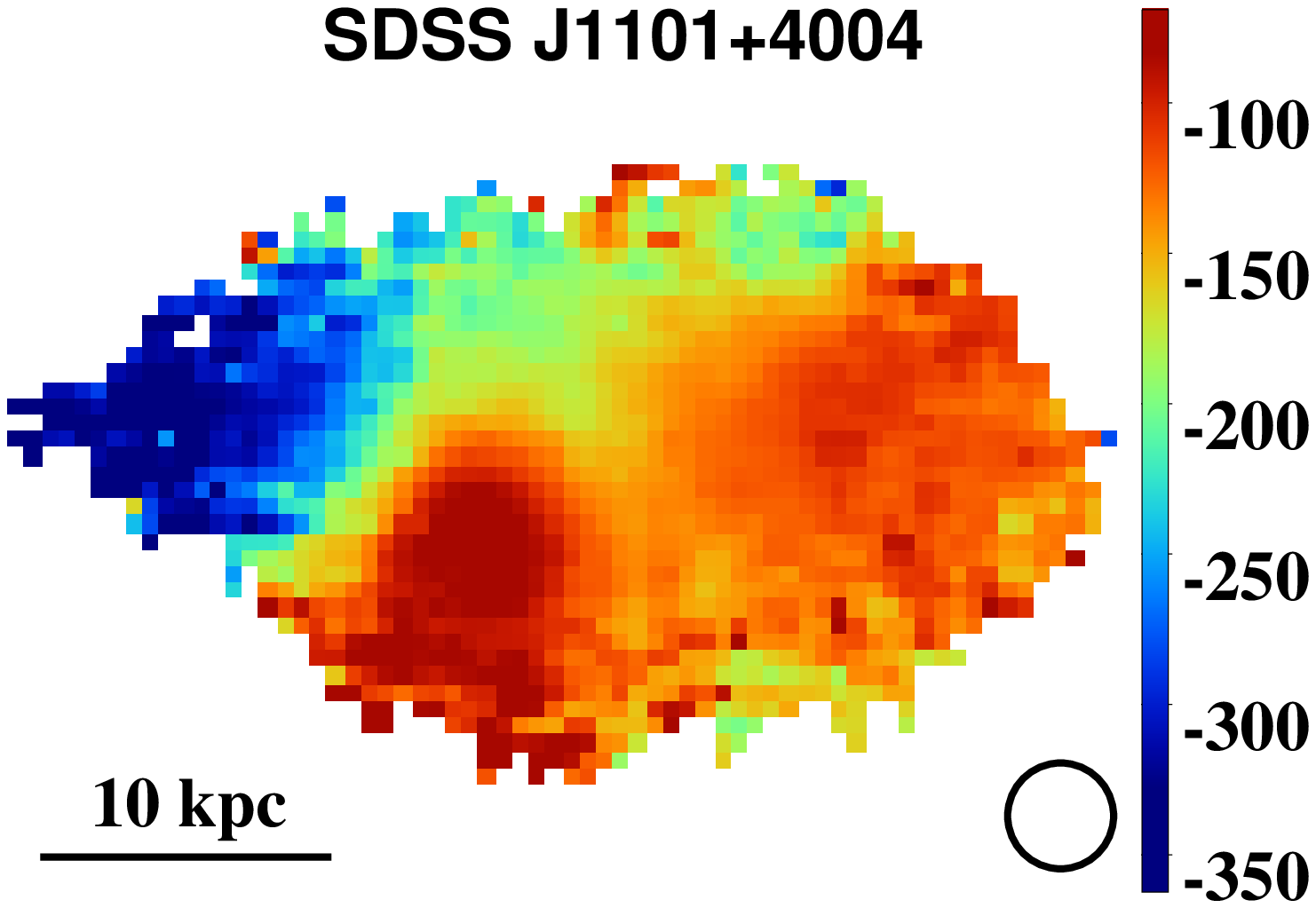}%
    \includegraphics[scale=0.28,clip=clip,trim=0mm 0mm 3.5cm 0mm]{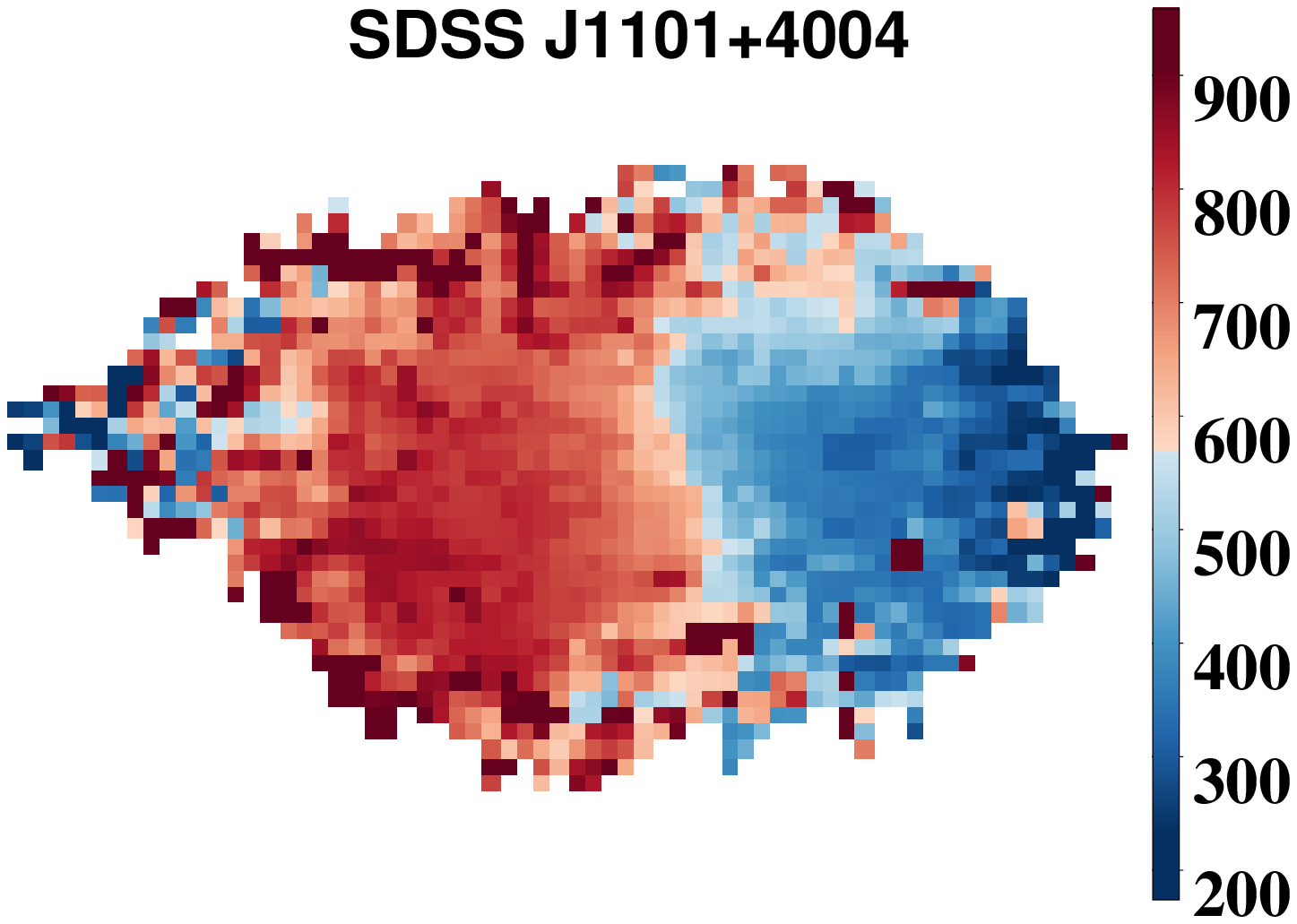}%
    \includegraphics[scale=0.28,clip=clip,trim=0mm 0mm 3.5cm 0mm]{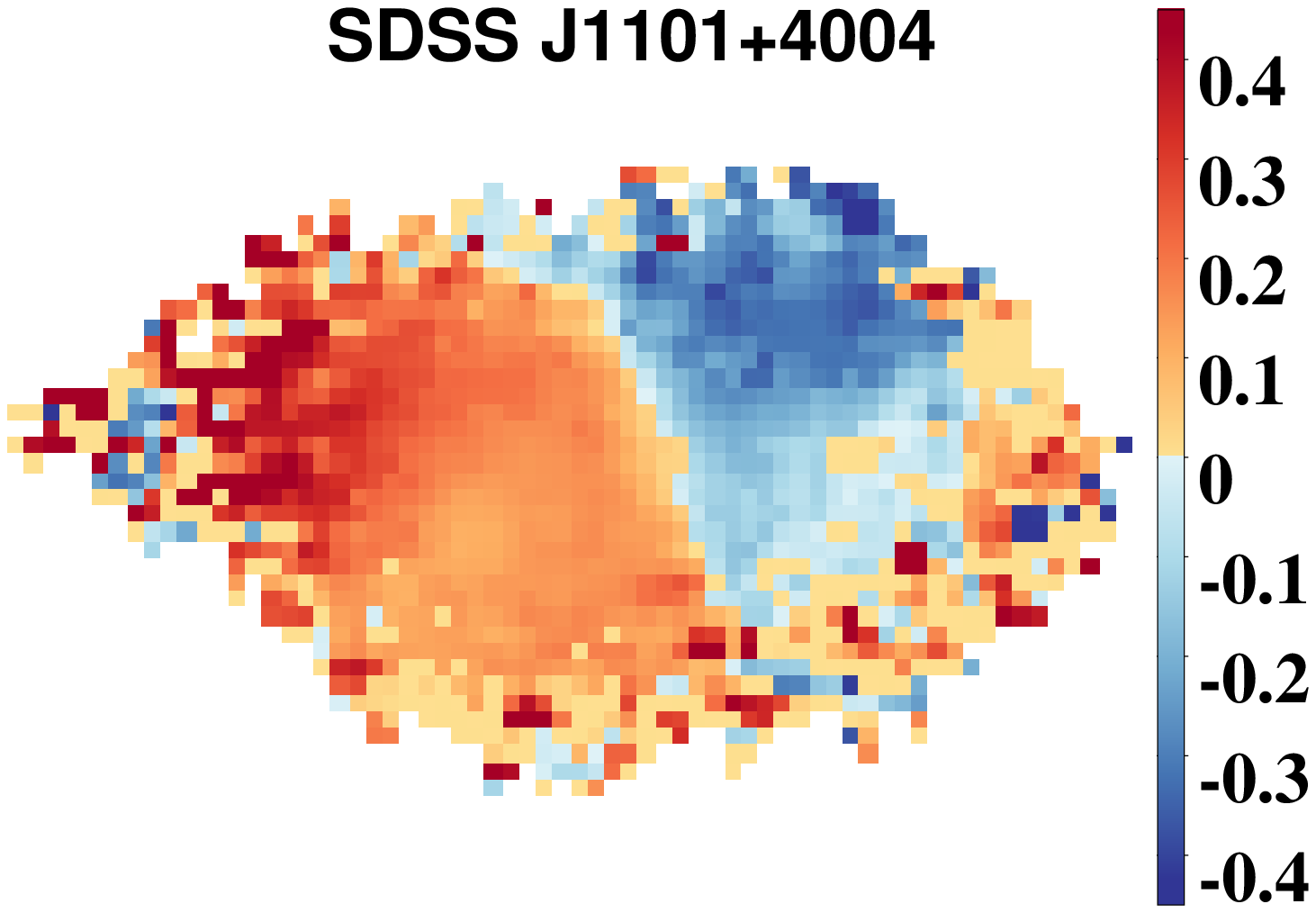}%
    \includegraphics[scale=0.28,clip=clip,trim=0mm 0mm 3.5cm 0mm]{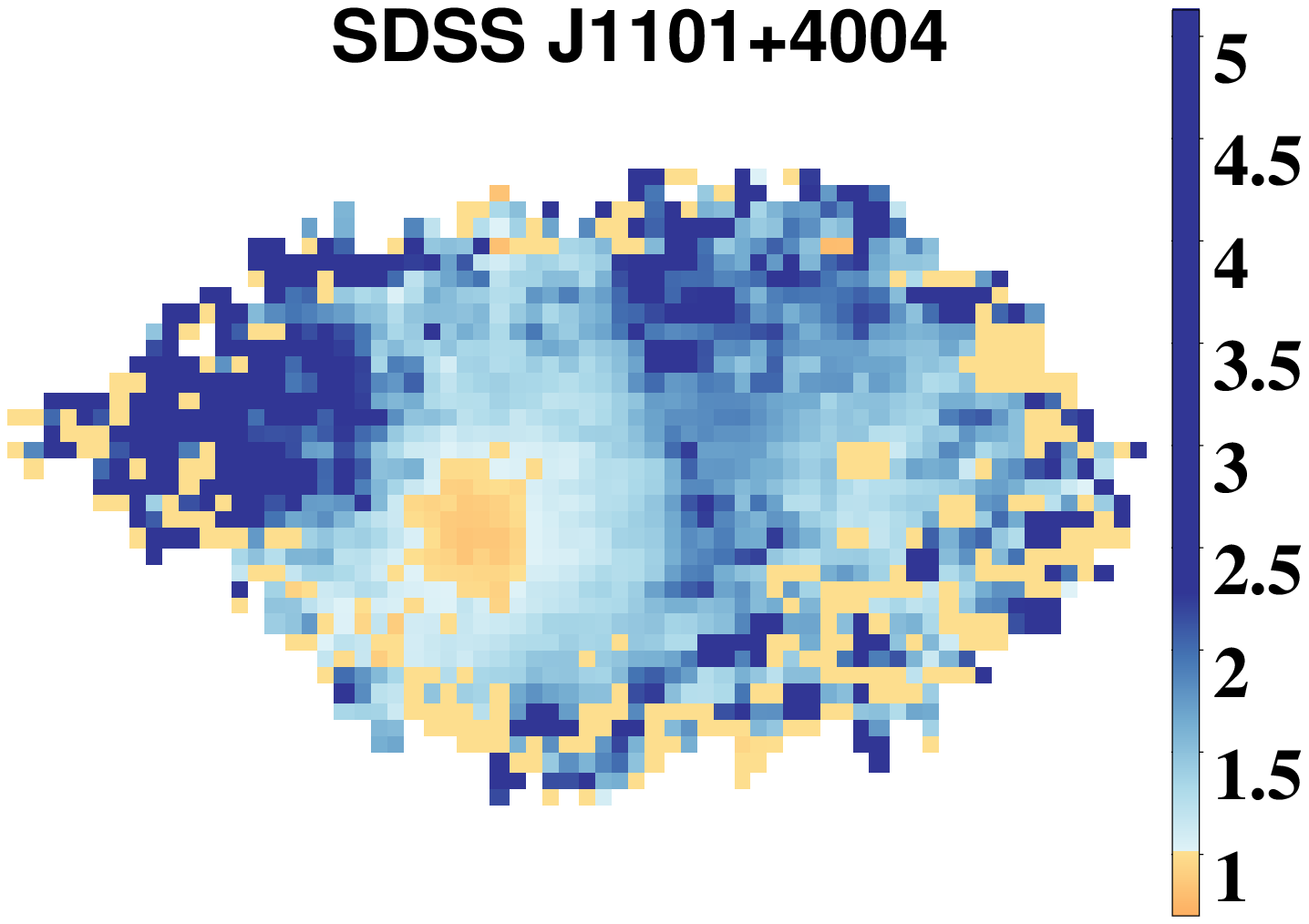}\\
\contcaption{}
\end{figure*}

While stellar absorption lines provide the cleanest measure of the systemic velocity, the stellar continua of the host galaxies are too faint to detect reliably in our data or the original SDSS spectra. In the absence of accurate host redshifts, we are forced to adopt the redshifts derived from the SDSS spectroscopic pipeline and listed in Table \ref{tab1}, which effectively trace the typical redshifts of the strong emission lines. If the gas motions relative to the host galaxy are of order a few hundred km s$^{-1}$, the velocities that we derive (particularly $v_{\rm med}$ defined above) have absolute uncertainties of this order, although the relative changes in $v_{\rm med}$ across the field of view are unaffected. The value of $v_{02}$ is in principle also subject to the uncertainty in the host redshift, but in practice these velocities tend to be so high that this uncertainty is likely negligible. All other parameters that we use (FWHM, $W_{80}$, $A$ and $K$) contain only differences in velocity from one part of the line profile to the next and are not dependent on a careful determination of the redshifts of the host galaxies.
     
Despite these caveats, we find that the standard SDSS redshifts determined 
from the ensemble of the emission lines \citep{abaz09} are in fact very
close to the host galaxy redshifts whenever those can be accurately
determined from the stellar absorption lines \citep{zaka13}. In three
of the objects in this sample, we are able to tease out weak
absorption features in the spectra and find that the absorption line
redshifts are within 30 km s$^{-1}$ of the standard SDSS ones. In
Section \ref{sec:models} we find that the outflow velocities of the gas
may reach many hundreds of km s$^{-1}$, but the outflows proceed in
quasi-spherical fashion, and therefore it may not be surprising that
the velocity centroids of the emission lines remain very close to the
host galaxy redshifts.

\section{Kinematic maps of nebulae around luminous quasars}
\label{sec:science}

\subsection{Kinematic signatures of quasar winds}

What observational signatures of quasar-driven winds are we looking for and are we seeing them in our sources? Numerical simulations of galaxy formation show that properties of massive galaxies are most successfully reproduced when gas is physically removed from the host galaxy by the quasar-driven wind \citep{spri05, hopk06, nova11}. Therefore, we should look for signs that velocities of the gas are high, preferably higher than the escape velocity from the galaxy. However, quasar-driven outflows seen in these simulations proceed in a quasi-spherical fashion, and unfortunately, spherically symmetric outflows produce zero net line-of-sight velocity and symmetric line profiles and thus do not display any obvious kinematic signatures. Even if the outflow is not spherically symmetric, but proceeds close to the plane of the sky (as we have reason to expect in type 2 quasars which may be illuminating the gas preferentially in these directions, as opposed to toward the observer), then the net line-of-sight velocity is strongly affected by projection effects and can be close to zero. 

As single-fiber and long-slit spectra demonstrate, the velocity dispersion of the ionized gas in obscured quasars tends to be very high, is uncorrelated with the stellar velocity dispersion of the host galaxy and is unrelated to its rotation \citep{gree09, gree11, vill11b}. These observations suggest that the gas is not in equilibrium with the potential of the host galaxy and may be dynamically disturbed by the quasar. But it is hard to unambiguously determine the three-dimensional geometry of gas motions just from the velocity dispersion measurements along one slit or within one fiber. Additionally, there are well-known observational degeneracies between gas outflows and inflows and rotational signatures which complicate the picture even further. 

Not all extended narrow-line emission is due to the gas that has been removed from the host galaxy. Sometimes we just happen to see tidal debris or nearby small companion galaxies that are illuminated by the quasar \citep{liu09, vill10}, and such features do not constitute proof of quasar feedback. One good radio-quiet example is SDSS J0123+0044, where a kinematically cold tail with an extent of $>$100 kpc is observed in a recent merger product that shows other tidal signatures \citep{zaka06, vill10}. In this object, the origin of the extended ionized gas as tidal debris and the direction of quasar illumination can be determined with high confidence using detailed spectroscopic, imaging (HST) and polarimetric observations. Such data are not available for most objects in our sample, but we find that it is often possible to discriminate between feedback and other possible origins of ionized gas at large distances based on seeing-limited morphology and kinematics data. 

In particular, we previously pointed out \citep{liu13a} that the nebulae in our sample are much smoother and rounder than those around radio-loud quasars in the sample of \citet{fu09}, many of which were interpreted using a model of illuminated or shocked tidal debris similar to the interpretation of SDSS J0123+0044. This pronounced morphological difference between the narrow-line regions of the radio-quiet and the radio-loud objects suggests that the origin of gas is different in these two samples. In the sections below we describe the results of our kinematic analysis of the IFU data and discuss the features that are most naturally explained by high-velocity outflows.

\subsection{Velocity fields and projection effects}

The velocity fields of the \oiii\ nebulae surrounding our radio-quiet
quasars are remarkably well organized (Figure \ref{fig:VWAK}):
typically one semi-circular part of the nebula is on average
redshifted while the other one is blueshifted. These signatures can be
interpreted as those of an outflow whose axis is inclined at some
non-zero and non-right angle $i$ to the line of sight (inclination
angles are defined relative to the plane of the sky, so that
$i=0^{\circ}$ for a face-on galaxy or a vector in the plane of the
sky). In this case, the blueshifted side is due to the gas moving away
from the quasar and somewhat toward the observer, whereas the
redshifted gas is on the far side of the quasar and moving yet further
away. However, alternative explanations for the blueshift / redshift
patterns also need to be considered. For example, an inflow with 
$i'=-i$ would produce the same maps of line-of-sight velocities on the
sky. A rotating galaxy also has one redshifted and one blueshifted
side.

In Table \ref{tab1} we report the maximum difference in velocity
between the redshifted and the blueshifted regions $\Delta v_{\rm
  max}$. To this end, we exclude the spaxels with the 5\% highest and
the 5\% lowest $v_{\rm med}$ which could be affected by noise, and
calculate the difference between the remaining highest and lowest
$v_{\rm med}$ values. The maximal projected velocity difference ranges between
90 and 520 km s$^{-1}$ among the eleven radio-quiet quasars in our
sample. If the projected velocity gradients are due to inflow/outflow, 
then the clear spatial separation of the redshifted and
blueshifted sides argues that the axis of this motion is not too close
to the line of sight. In this case, the observed radial velocities are only
a small fraction of the actual physical velocities of the flow, 
reduced to the observed values by the projection effects. We
further discuss outflow models and the resulting velocity
differences in Section \ref{sec:models}.

Regular velocity fields can also be produced by rotating gas disks, and we explore whether this scenario can explain our kinematic data. The maximal rotation speed of massive galactic disks typically does not exceed 300 km s$^{-1}$ \citep{debl08, reye11}. But a better model for our objects may be a gas disk embedded in a massive elliptical host \citep{zaka06}, which would explain the strong kpc-scale dust lanes, young stellar populations \citep{zaka06, liu09} and on-going star formation \citep{zaka08} that we observe in the host galaxies of obscured quasars. In this case the rotation speed can be estimated using an isothermal potential, with stellar velocity dispersion $\sigma_*\la 300$ km s$^{-1}$ \citep{gree09, liu09} and resulting maximal rotation speed $v_{\rm rot}=\sqrt{2} \sigma_*=424$ km s$^{-1}$. 

A face-on disk does not yield any differences in radial velocity, but an edge-on disk will appear elongated on the sky. Our objects have an ellipticity $\epsilon\leqslant0.23$ except for SDSS J0321+0016 ($\epsilon=0.42$, Section \ref{sec:bubbles}). The inclination angle $i$ of a round disk galaxy is related to the observed ellipticity via
\begin{equation}
\sin i=\sqrt{\frac{1-(1-\epsilon)^2}{1-u^2}},
\end{equation}
where $u$ is the thickness-to-diameter ratio of the galactic disk, and
is generally less than 0.1 \citep{miha81}. We consider a configuration
that produces maximal $\Delta v_{\rm max}$ while maintaining the
ellipticity just below the observed maximum. We take $\epsilon=0.23$,
$u=0.1$ and the maximal rotational velocity of $v_{\rm rot}=424$
km s$^{-1}$, and we find $i=40^{\circ}$ and $\Delta v_{\rm max}=545$ km
s$^{-1}$. Thus, neither $\Delta v_{\rm max}$ nor $v_{\rm med}$ measurements alone can exclude the possibility that we are observing a gas disk in a massive galaxy. 

\begin{figure}%[h!]
\centering
    \includegraphics[totalheight=3cm,angle=0,origin=c,scale=2]{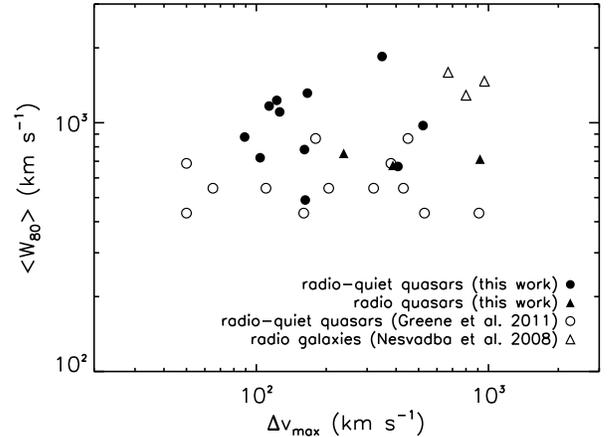}
\caption{The relationship between $\langle W_{80} \rangle$ (measured from the SDSS fiber spectrum) and $\Delta v_{\rm max}$ for our quasar sample. Each data point represents an entire quasar. Also shown are the radio-quiet obscured quasars from \citet{gree11} and the high-redshift ($z\sim2$--3) radio galaxies from \citet{nesv08}; in the latter cases, $W_{80}$ is taken to be 1.088$\times$FWHM and is thus likely underestimated.}
\label{fig:w80dv}
\end{figure}

\subsection{Velocity dispersion}
\label{sec:dispersion}

We present the spatial distribution of $W_{80}$ in Figure
\ref{fig:VWAK}. In 9 out of 11 radio-quiet objects, the distribution
of $W_{80}$ peaks at $\sim$700--1100 km s$^{-1}$ or even at $\sim$2100
km s$^{-1}$ in SDSS J0319$-$0019. Excluding the noisy sporadic
spaxels, we find the maximum $W_{80}$ value to be $\gtrsim$1000 km
s$^{-1}$ in every case, reaching $\sim$2300 km s$^{-1}$ in SDSS
J0319$-$0019.

In Figure \ref{fig:w80dv}, we show the relationship between $\Delta
v_{\rm max}$ and the linewidths $W_{80}$ measured from the SDSS fiber
spectra -- i.e., integrated over the entire nebulae. We find no
correlation between these two quantities (the Kendall rank correlation
coefficient $\tau=0.05$ with probability $p=0.82$ that no correlation
is present). As we discuss in Section \ref{sec:models}, this result is
not surprising in the outflow model, where $W_{80}$ reflects the
typical bulk velocities of the gas, whereas the observed $\Delta
v_{\rm max}$ is much more sensitive to projection and geometric
orientation effects. The three radio objects we have in our comparison
sample are situated at the low end of $W_{80}$, but this happens to be
due to low-counts statistics, and the trend for the radio-loud objects
to have lower $W_{80}$ values is not borne out either by measurements
in a larger sample at similar redshifts \citep{zaka13} or by the three
high-redshift radio galaxies from \citet{nesv08}.

The extremely high values of $W_{80}$ and their lack of correlation with $\Delta
v_{\rm max}$ are very unusual for gas-rich disk galaxies, unless they show high-velocity gaseous outflows \citep{rupk13}. The central line-of-sight velocity dispersion of the ionized gas in stellar disks does not exceed 250 km s$^{-1}$ (corresponding to $W_{80}=640$ km s$^{-1}$) and falls off rapidly away from the center to $\la 100$ km s$^{-1}$ ($W_{80}=256$ km s$^{-1}$, \citealt{vega01}). Even in the ultraluminous infrared galaxies at high redshift, which are massive mergers whose gas dynamics is expected to be most disturbed of all, the observed maximal width of \oiii\ and H$\alpha$ emission lines is $W_{80}\simeq 550$ km s$^{-1}$ \citep{harr12,alag12}. Similar argument applies if we consider a gas disk embedded in an elliptical galaxy, where we could expect a maximum of $W_{80}\simeq 580$ km s$^{-1}$. Our measured values of $W_{80}$ are clearly higher than those observed in even the most massive and most disturbed disk galaxies. 

We conclude that a rotating gas disk --- whether in the most massive
disk galaxy or in the most massive elliptical --- would have a line
profile that is much narrower than what we observe in our data. The
potential of a galaxy, even of an extremely massive one, is simply
insufficient to provide an equilibrium configuration for the high
velocities seen in our sample.  We will not consider rotating disk models 
further in this paper.

\subsection{Spatial variations in velocity dispersion}

\begin{figure*}%[h!]
\centering
    \includegraphics[origin=c,scale=0.5]{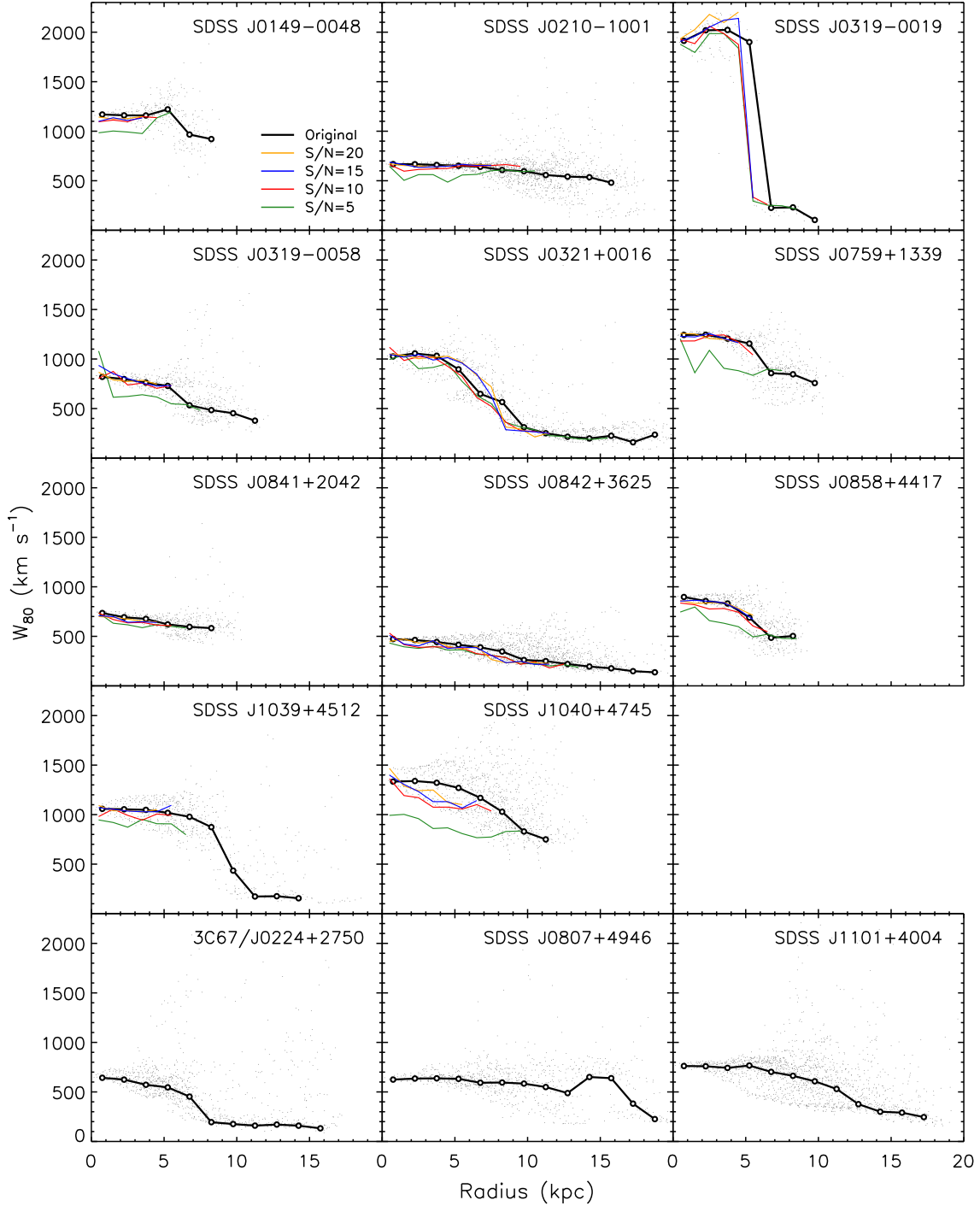}
\caption{The radial dependence of $W_{80}$ in individual objects. Shown by open circles are the modes of $W_{80}$ calculated in 1.5 kpc bins using the original data. For the 11 radio-quiet quasars, we also show the $W_{80}$ radial profiles when the peak $S/N$ of [O {\sc iii}] is downgraded to certain levels (5, 10, 15 and 20) by adding extra noise to the real data. For each $S/N$ level, only spaxels where [O {\sc iii}]$\lambda$5007\AA\ originally is detected with this significance or higher are taken into account, resulting in smaller maximal radii for higher $S/N$ levels.}
\label{fig:sigr}
\end{figure*}

In Figure \ref{fig:sigr}, we show values of $W_{80}$ in all spaxels
along with their modal values as a function of projected distance from
the center (brightness peak). The radial profiles of $W_{80}$ are almost flat
at projected distances $R \la 5$ kpc and appear to decrease at larger
radii, with the scatter increasing with $R$. The possible origin of the mild 
decline in $W_{80}$ is discussed in Section \ref{sec:dispmodels}; here we focus 
on whether this observation is reliable. 

The primary concern with the measured decline in $W_{80}$ is that it
coincides with the decrease in the surface brightness of the line
emission and thus with the decrease in the peak signal-to-noise ratio
of the data. The same two-Gaussian line profile composed of a narrow
core and a weak broad base would have a higher $W_{80}$ value in the
regions of high signal-to-noise and lower $W_{80}$ in the regions of
low signal-to-noise where the weak broad component can no longer be
identified and the entire $W_{80}$ measurement is due to the narrow
core.

To test the robustness of our detection of the $W_{80}$ decrease, we remeasure the radial profiles of $W_{80}$ at the same signal-to-noise ($S/N$) level. To this end, we degrade the observed line profiles to a specified $S/N$ value by adding the appropriate amount of Gaussian random noise to the spectral line profile in each spaxel and repeat all kinematic measurements using these profiles. The resulting radial profiles of $W_{80}$ are shown in Figure \ref{fig:sigr} at different $S/N$ levels. The profiles at higher $S/N$ are cut off at smaller $R$: only data with $S/N$ higher than 20 can be degraded down to $S/N=20$. It is clear from this simulation that noise in the data does have the expected effect on the $W_{80}$ measurement: on average, the $W_{80}$ values measured at $S/N=20$ are 20\% higher than the values measured at $S/N=5$. This is most likely due to the weak broad bases of emission lines that cannot be identified in low $S/N$ data. However, the general trend of $W_{80}$ slowly declining with $R$ is just as apparent in the constant $S/N$ profiles as in the original ones.

Finally, using individual Gaussian components, we can investigate the spatial behavior of the broad and narrow components individually. Using two-component Gaussian fits we find median $W_{80}$ values of 480 and 1240 km s$^{-1}$ for the narrow and broad components among our 11 radio-quiet objects. We find no major
differences between their spatial distribution, other than the trend
that the high-dispersion components are slightly more centrally
concentrated, in agreement with our finding that the $W_{80}$ radial
profile decreases outward. Otherwise, the spatial maps of the narrower
and the broader Gaussian components of the line profiles appear round
and featureless (the exception is SDSS~J0321+0016 discussed in
Section \ref{sec:bubbles}).

This is in contrast to the findings in low-redshift ultraluminous
galaxies with strong supernovae-driven outflows \citep{rupk13}. In
these objects, the narrow components ($W_{80}$=110--330 km s$^{-1}$
even in these massive, merger-driven systems) are preferentially
concentrated in disks which show the same orientation as the molecular
gas as well as clear rotational signatures, while the broad components
associated with outflows are preferentially oriented perpendicular to
the disk as they escape from the galaxy along the path of least
resistance. In our objects, using the maps of individual Gaussian
components we do not see either the increased ellipticity or the
differences in the spatial orientation associated with this
phenomenon. Thus, it is unlikely that the narrow components in our
data are associated with the rotation of the galaxy disk, and
furthermore the host galaxies of obscured quasars at these high
luminosities do not tend to be disk-dominated \citep{zaka06}.

Another interesting possibility is that the narrow component is due to a residual gas disk seen in numerical simulations of gas-rich mergers with quasar feedback \citep{spri05}, when the remnant galaxy has already taken its elliptical shape and when the quasar wind (broad component in this scenario) proceeds in quasi-spherical fashion. But in this case we would expect to find that the narrow component has a more compact spatial distribution than does the broad one, which is not observed. Thus it is most likely that none of the ionized gas we observe is in a disk-like component.  

\subsection{Line asymmetry and shape}

In the right-hand columns of Figure \ref{fig:VWAK} we show the maps of asymmetry parameter $A$ and the shape parameter $K$. With the sole exception of SDSS J0210$-$1001, the asymmetry parameter $A$ is uniformly negative in the bright central parts of all objects, indicating heavy blueshifted wings in the line profiles. This is the tell-tale signature of an outflow which may be proceeding in a symmetric fashion but whose redshifted part is obscured by the material in the host galaxy or near the nucleus \citep{whit85}. 

The line shape parameter is at or above the Gaussian values in the vast majority of all spaxels in all objects, indicating that line profiles with wings heavier than Gaussian values are very common in our objects. In the outer parts of the nebulae, where the peak $S/N$ of even the brightest emission line \oiii\ is just a few, typically only one Gaussian component is sufficient to fit the line profile, and therefore both the asymmetry and the shape parameter tend to be at the Gaussian values. Consistent with previous spatially integrated spectroscopic studies, the absence of prevailing stubby line profiles in our spectra with steep sides and double-horn structure suggests that the quasar nebulae are not rotating extended discs \citep{whit85}. This result will become even stronger if we fit the \oiii\ line with Voigt profiles instead, which leads to even stronger wings and thus larger $K$ values.

While the blue-shifted asymmetry of the line profiles is a strong classical outflow signature, the high velocity dispersions of the gas and the smooth morphologies of our nebulae also suggest that the gas is in an outflow. Galactic inflows, although long postulated to exist, have been very difficult to detect, but in the handful of known cases (mostly at high redshifts) the inflows occur in distinct kinematic components with small velocity dispersions \citep[e.g.,][]{bouc13}. Inflowing gas clouds illuminated by luminous quasars may be expected to produce line emission with clumpy morphology and to be split into distinct narrow kinematic components, which is not the dominant appearance of the nebulae in our sample either in the physical space or in the velocity space. Existing observations and theory suggest that inflows have a small covering factor \citep{deke09,stei10}, so we do not expect inflows to result in the ubiquitous blue wings seen in our data. Some examples of illuminated companion clumps are discussed in Section 5 (although these are not necessarily inflowing into the quasar host galaxy), and we further discuss differences between them and the kinematic and morphological features of our nebulae.

\section{Kinematic models and implications for our data}
\label{sec:models}

Since we have ruled out disk rotation and inflows as the origin of the kinematic signatures seen in our data, we proceed to models of outflows. The models are discussed in the order of increasing complexity and increasing number of observables that they are trying to reproduce. In Sections \ref{sec:sph} to \ref{sec:bicone}, we discuss IFU signatures of three simple outflow models: a spherically symmetric outflow; an outflow affected by extinction in the host galaxy; and a biconical outflow. We describe the similarities and the differences between the observed signatures and those predicted by these kinematic models in Section \ref{sec:modsum}. In Section \ref{sec:dispmodels} we discuss possible origins of the $W_{80}(R)$ profiles. 

\subsection{Spherically symmetric outflow}
\label{sec:sph}

Quasar winds propagate into an inhomogeneous interstellar medium of the host galaxy, and thus we may expect several different gas phases to be present in quasar outflows. Using our observations of optical emission lines, we are sensitive to one particular phase of this medium (at $T\sim 10^4$ K) which is likely concentrated in relatively dense ($\sim$ 100 cm$^{-3}$) clouds. We observe the combined effect of all clouds, as each individually cannot be spatially resolved by our current observations. The clouds are likely embedded in lower density, hotter gas which can potentially be observed at other wavelengths but to which our current observations are not sensitive. This picture is qualitatively similar to that suggested by \citet{heck90} for winds driven by supernova explosions and is supported by the density measurements of the line-emitting clouds and of the nebulae overall \citep{gree11}. In what follows, we assume that that the observed emission is produced by the ensemble of narrow-line-emitting clouds and we discuss their possible geometric and kinematic distributions.

\begin{figure*}%[h!]
\centering
    \includegraphics[scale=0.9,trim=5cm 10cm 4cm 0cm]{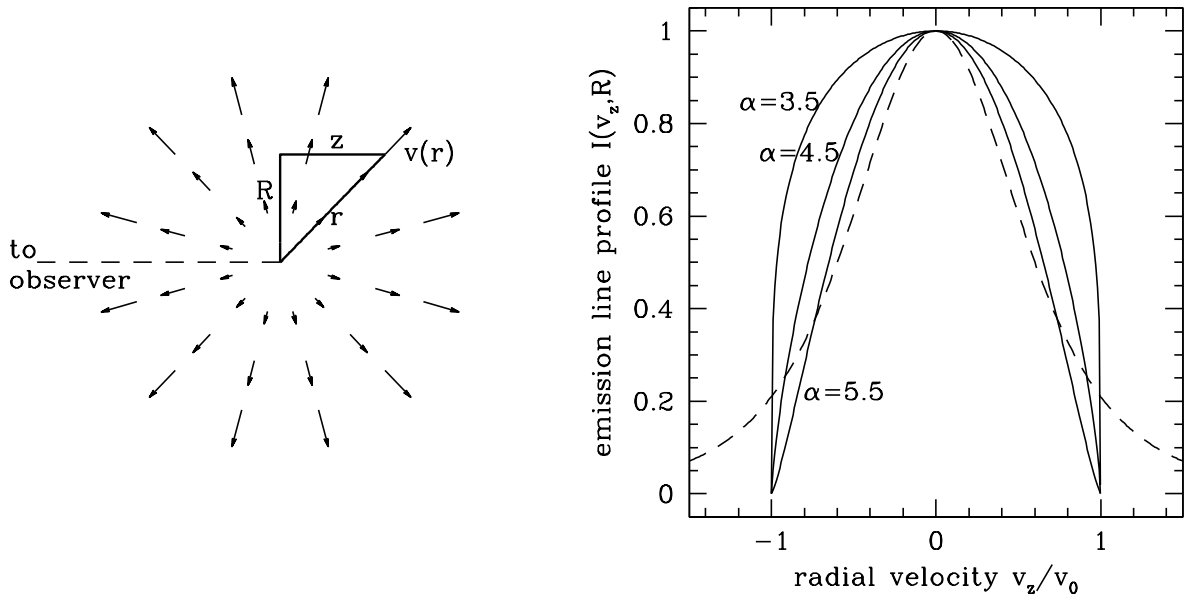}
\caption{Left: schematic of the spherical outflow models and the notation used in our calculation. Right: emission line profiles for models with $v(r)=v_0=$const (solid lines for three different values of $\alpha$ which parametrizes the luminosity density profile) and $v(r)\propto r$ (dashed; $\alpha=4.5$). For constant velocity outflow, $\alpha$ increases from 3.5 to 5.5 from the rounder to the more triangular profile, but the profile remains stubbier than a Guassian, with $K>1$.}
\label{fig:model1}
\end{figure*}

If the outflow has a three-dimensional velocity profile ${\bf v_0}({\bf r})$ and luminosity density $j({\bf r})$ as a function of the three-dimensional radius-vector from the center of the outflow ${\bf r}$, then the distribution of line profiles on the sky (our observable in the IFU data) can be calculated from the following equation:
\begin{equation}
I(v_z,{\bf R})=\int j({\bf r}) \delta ({\bf v}-{\bf v_0}({\bf r})) {\rm d}v_x {\rm d}v_y {\rm d} z. \label{eq:inten1}
\end{equation}
Here ${\bf R}$ is the two-dimensional radius-vector in the image plane $(x,y)$, while $z$ is the coordinate along the line of sight (Figure \ref{fig:model1}, left). If the outflow is spherically symmetric (so that luminosity density is only a function of spherical radius $r$), then the radial velocity profiles in the plane of the sky are
\begin{equation}
I(v_z,R)=\left.\frac{2j(r)}{|v_0'(r)(1-R^2/r^2)+v_0(r)R^2/r^3|}\right\vert_{v_z=v_0(r)\sqrt{1-R^2/r^2}}
\end{equation} 
For a constant velocity $v_0$ outflow, this further simplifies to
\begin{equation}
I(v_z,R)=\frac{2R}{v_0\left(1-v_z^2/v_0^2\right)^{3/2}}~j\left(\frac{R}{\sqrt{1-v_z^2/v_0^2}}\right). \label{eq:inten2}
\end{equation}
If we further make an assumption that the luminosity density is a power-law function of radius from the center of the outflow, $j(r)\propto r^{-\alpha}$, we find that $I(v_z,R)$ is a separable function of $v_z$ and $R$:
\begin{equation}
I(v_z,R)\propto \left(1-v_z^2/v_0^2\right)^{\frac{1}{2}(\alpha-3)} R^{1-\alpha}. \label{eq:inten3}
\end{equation}
This means that while the total intensity of the line varies across the image plane, the radial velocity profile remains exactly the same (Figure \ref{fig:model1}, right); thus, in this model $W_{80}$ is the same across the image plane. This somewhat counter-intuitive property is due to the self-similar nature of the power-law luminosity density distributions. Indeed, in a spherically symmetric outflow, the emission line profile is entirely determined by projection effects: the $v_z\approx 0$ part of the line profile is contributed by the gas that is propagating close to the plane of the sky, whereas other parts of the profile are produced by streamlines inclined at varying angles to the line of sight. For power-law luminosity density distributions, the relative contributions of these points remain the same, even though as we consider lines of sight further away from the center the total brightness declines. 

In a spherical outflow, the line profiles are symmetric everywhere and centered on $v_z=0$. Thus this model cannot describe either the line asymmetries or median velocity variations across the nebulae we observe. However, this model is useful for understanding how the range of physical velocities within the outflow relates (due to projection effects) to the observed velocity widths of the emission lines. The $W_{80}$ parameter can be calculated as a function of the physical velocity $v_0$ for the line profile (\ref{eq:inten3}) for different values of $\alpha$. From surface brightness distributions presented in Paper I, we know that $\alpha$ ranges from 4.0 to 6.7, with a mean and dispersion of $4.8\pm 0.7$ among the objects in our sample ($\alpha=\eta+1$ in the notation of Paper I, where $\eta$ is the power-law slope of the surface brightness profile). For $\alpha=4.5$, the median value of $W_{80}$ within our radio-quiet sample (974 km s$^{-1}$) corresponds to a physical velocity $v_0=760$ km s$^{-1}$; the range of $\alpha$ introduces a 15\% uncertainty in $v_0$. Therefore, if the observed $W_{80}$ values correspond to the range of projected velocities in a bulk flow, the physical velocities of the gas must be very high.

The velocity profiles of constant velocity outflows (Figure \ref{fig:model1}) are wingless, with the $K$ parameter significantly smaller than 1 (the Gaussian value). This occurs because along any line of sight the maximal range of velocities is limited to $-v_0$ to $v_0$: the projected velocities cannot exceed the physical velocity of the gas. $K$ remains $<1$ for a wide range of plausible luminosity density profiles and for the case when the gas has intrinsic isotropic velocity dispersion. On the contrary, profiles with $K>1$ are dominant in our data.

One possible solution to this discrepancy is an outflow with a velocity that is an increasing function of the distance from the center. Such velocity profiles may be established during the expansion of a wind with a steeply declining pressure profile \citep{veil94}. For example, a linearly increasing velocity profile $v=v_0(r/r_0)$ (not too far off from the velocity profile of the outflow in SDSS~J1356+1026, \citealt{gree12}) yields a line profile
\begin{equation}
I(v_z,R)\propto \left(\frac{R^2}{r_0^2}+\frac{v_z^2}{v_0^2}\right)^{-\alpha/2},
\end{equation}
which for any value of $R$ has a symmetric profile with heavy wings and a $K$ value of 1.41 (at $\alpha=4.5$), very similar to our observations (Figure \ref{fig:model1}). However, in this model the $W_{80}$ values are expected to increase outward $\propto R$, while no evidence for such increase is seen in our data. 

A more natural explanation for $K>1$ values is that at every distance from the quasar the radial velocity of narrow-line clouds have a broad (and non-Gaussian) velocity distribution, and this local velocity distribution (rather than global outflow kinematics) is the origin of the line profile shapes seen in our sample. This hypothesis can be tested with modern numerical simulations of quasar feedback in which velocity distributions of high-density regions (the analogs of the narrow-line region clouds) can be directly measured \citep{nova11}. Qualitatively, they do indeed show a wide range of velocities at every distance, and the typical distribution appears fairly independent of distance, in agreement with our observations, but a quantitative measurement of the simulated clouds' velocity distribution is necessary to determine whether they would result in line shapes consistent with observations. Even in the case of complicated velocity distributions the median $W_{80}$ still reflects the median outflow velocity in quasi-spherical outflows. 

\subsection{Spherical outflow affected by galactic extinction}
\label{sec:sphext}

A spherically symmetric outflow shows symmetric line profiles and zero velocity difference across the field of view. In practice the host galaxy of the quasar often breaks this symmetry. If the host galaxy is gas- and dust-rich or if there is some circumnuclear material, then the receding part of the outflow is preferentially extincted. The exception is the case of the obscuring disk aligned with the line of sight (inclination angle $i=90^{\circ}$), in which again the outflow appears symmetric on the sky and we return to the spherical case. We see strong observational evidence that obscuration is occurring in our objects, since the asymmetry parameter $A$ is uniformly negative (meaning that the line profiles show blueshifted asymmetry) in all our sources except one in the central brightest parts. In this section we discuss quantitative predictions of this model. 

\begin{figure*}%[h!]
\centering
    \includegraphics[scale=0.9,trim=5cm 10cm 4cm 0cm]{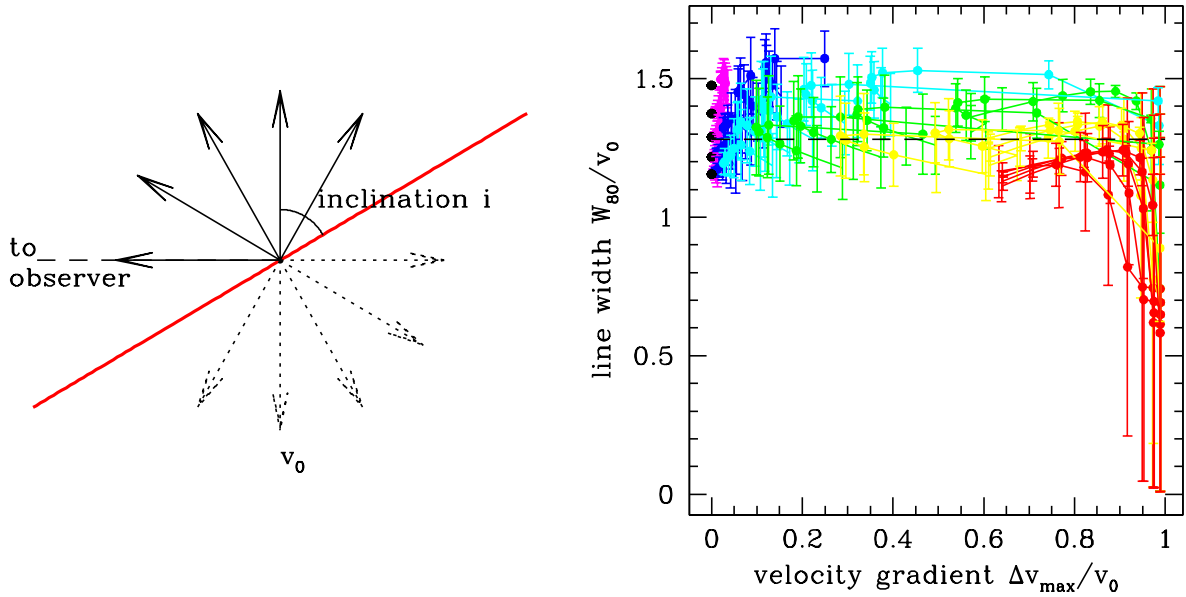}
\caption{Left: schematic of the model with a spherical outflow ($v_0=$const) obscured by a layer of dust in the host galaxy (red). Right: observed kinematic signatures of outflows affected by a dusty disk. The largest observed median velocity difference across the nebula $\Delta v_{\rm max}$ and the integrated width of the emission line $W_{80}$ are shown in units of outflow velocity $v_0$ which is assumed to be the same in all parts of the outflow. Optical depth $\tau$ increases from 0.4 to 2.4 (from magenta through blue through green to red), with $\tau=0$ models shown with black circles at $\Delta v_{\rm max}=0$. Only models with $i>45^{\circ}$ corresponding to type 2 (rather than type 1) objects are shown. Models on the right-hand side (in red) can be produced only by strong extinction on kpc scale in the host galaxy. Five curves in each color show five different values of $\alpha$ from 3.5 to 5.5. The black dashed line shows the value of $W_{80}$ in a spherically symmetric outflow with $\alpha=4.5$.}
\label{fig:model2} 
\end{figure*}

For intermediate inclination angles, we expect to see a change in median velocity across the nebula: the lower part of the outflow in Figure \ref{fig:model2}, left, is expected to be more blueshifted than the upper part. The velocity difference $\Delta v_{\rm max}$ across the nebula depends both on the optical depth of the dust disk $\tau$ and on the inclination $i$. We show velocity differences and integrated line widths in Figure \ref{fig:model2}, right, for a range of $\tau$ and $i$. In these models, we assume that the outflow has a constant velocity $v_0$ everywhere and that the optical depth $\tau$ is constant across the entire disk of the host galaxy (in practice it is likely that $\tau$ is highest in the center). 

We find that the models populate a fairly wide range of $\Delta v_{\rm max}$ from 0 to $v_0$, whereas the line widths $W_{80}$ stay within $\sim 30\%$ of the spherically symmetric case. Most models predict $\Delta v_{\rm max}/W_{80}\la 0.5$, and higher values of this ratio correspond to the most extreme case of host galaxy obscuration (Figure \ref{fig:model2}). This result is expected: the smaller the extinction, the closer the outflow is to a spherically symmetric case and the smaller is the apparent $\Delta v_{\rm max}$ value. The median observed value of $\Delta v_{\rm max}/W_{80}$ among the eleven radio-quiet objects in our IFU sample is 0.13, and all but three have $\Delta v_{\rm max}/W_{80}\le 0.2$. Comparison of the distribution of models in the $\Delta v_{\rm max}$ -- $W_{80}$ plane with the range of observed values (Figure \ref{fig:w80dv}) suggests that outflows with $v_0\approx 760$ km s$^{-1}$ and dusty disks with $\tau \la 1$ can yield the observed values of line widths and velocity differences $\Delta v_{\rm max}$. 
We note that the relationship between $W_{80}$ and $\Delta v_{\rm max}$ is purely a result of geometry, and $W_{80}$ will increase if the gas is intrinsically turbulent. 

\subsection{Bi-conical outflow with no extinction}
\label{sec:bicone}

If the outflow is bi-conical, and if the axis of the bi-cones is inclined relative to the plane of the sky, then one cone is on average blueshifted while the other one is on average redshifted. Leaving aside the geometry of the super-bubbles (discussed further in Section \ref{sec:dispmodels}), we now consider whether such a model can simultaneously explain the round morphology of the bright parts of the nebulae and the distribution of the nebulae in the $\Delta v_{\rm max}$ -- $W_{80}$ plane (Figure \ref{fig:w80dv}). These models are more commonly used when the collimation of the outflow is obvious even in projection on the plane of the sky (e.g., examples in \citealt{cren00b, rupk13}), but the combination of a wide opening angle of the cone with beam smearing can result in round morphologies. 

We consider a model of bi-conical outflow with three parameters: the outflow velocity $v_0$, the inclination angle of the bi-cones' axis $i$ ($i=0^{\circ}$ for an outflow in the plane of the sky) and half-opening angle $\theta$ ($\theta=90^{\circ}$ for the spherically symmetric limit). Unlike the hollow-cone models of \citet{cren00b} where the walls of the cone dominate the emission, we are considering filled cones where the luminosity density scales as $r^{-\alpha}$. For $\theta\ga 60^{\circ}$ and $i\la 45^{\circ}$ (i.e., wide-angle outflow relatively close to the plane of the sky), the velocity difference across the nebula is $\la 0.5 v_0$, while $W_{80}$ is not too dissimilar from the spherical case ($\simeq 1.3 v_0$) since the opening angle of the outflow is not too far from spherical. Thus, the bi-conical models can qualitatively reproduce the positions of our nebulae on the $\Delta v_{\rm max}$ -- $W_{80}$ plane as long as the opening angle of the bi-cones is wide enough that the approaching and receding bi-cones are not too widely separated in projection on the plane of the sky, thus making the nebula appear round rather than collimated. 

\subsection{Summary of basic outflow models}
\label{sec:modsum}

A spherically symmetric, constant velocity outflow with a power-law surface brightness distribution (Section \ref{sec:sph}) displays a round morphology and a constant line-of-sight velocity dispersion across the nebula. Both of these characteristics are seen in our data: the ellipticities of the nebulae around radio-quiet quasars are much smaller than those of nebulae around radio galaxies (Paper I) and the radial profiles of $W_{80}(R)$ are nearly constant. The most important result we can glean from this simple model is that it ties the observed line width to the physical velocity of the gas within the outflow since the latter is determined purely by projection of the outflow velocities onto the line of sight. The exact ratio of the two is slightly dependent on the slope of the surface brightness profile, but is typically $W_{80}\simeq 1.3 v_0$. 

A spherically symmetric outflow has zero mean velocity along the line of sight since the approaching and the receding gas contributes equally at every point on the sky. In the observed data we do see variations in radial velocity accross the nebulae, but they tend to be much smaller than the entire range of velocities as seen from the line widths. The observation of non-zero radial velocity differences across the nebulae suggests that we do need to consider non-spherical models, but the fact that $\Delta v_{\rm max}\ll W_{80}$ suggests that the deviations from the spherical symmetry are modest, that quasar outflows have large covering factors, and that the line-of-sight velocity distribution can be indeed used to estimate outflow velocities. 

The distribution of objects in the $\Delta v_{\rm max}$ -- $W_{80}$ plane can be reproduced equally well with spherical outflows whose redshifted side is obscured by a $\tau\la 1.0$ layer of dust in the host galaxy or wide bi-conical outflows with cone opening angles $2\theta \ga 120^{\rm o}$. The model with extinction predicts that the redshifted parts of the line profile should be on average slightly fainter than the blueshifted ones, whereas in the bi-conical outflow model there is no such effect. One example where this prediction appears to be borne out by our observations is SDSS J1040+4745, where the peak of the \oiii\ emission is offset by 1 kpc from the peak of the continuum emission in the direction of the radial velocity gradient, toward the blueshifted part. This object also has the highest blueshift asymmetry in the sample, suggesting a high value of extinction. In other objects we see no dependence of brightness on velocity, which would favor bi-conical models.  On the other hand, the universally negative line asymmetries seen in the central parts of the nebulae favor models with extinction. Perhaps both slight collimation and extinction are taking place. 

Where is this extinction taking place? The observed signatures that suggest partial obscuration are the differences of the radial velocity across the entire extent of the nebulae and the almost uniformly negative asymmetry parameter $A$ over the central few kpc of the nebulae (at larger distances the $S/N$ of our data may not be sufficient to determine deviations from line symmetry). Both these observables suggest that extinction is not confined to the circumnuclear region but rather operates on galaxy-wide scales. Dust embedded with a spherically symmetric outflow could produce the requisite line asymmetries, but not the patterns of radial velocities which suggest an inclined disk. As was discussed before, the host galaxies of obscured quasars tend to be ellipticals, but with unusually high presence of gas, dust and star formation \citep{zaka06, zaka08, liu09}. Perhaps obscured quasars tend to trace a particular stage in the evolution of elliptical galaxies right before they are cleared of residual gas. Therefore, despite their elliptical morphology, the host galaxies of obscured quasars may contain sufficient gas to provide the extinction required by our models. 

The spherical model, the extincted spherical model, and the bi-cone model with gas moving at the same velocity $v_0$ produce line profiles that are stubbier than a Gaussian because all are lacking high-velocity gas. The almost constant radial profiles of $W_{80}(R)$ suggest that the physical velocities of the outflow are not strongly dependent on the distance from the quasar. Thus, a range of cloud velocities at every distance from the quasar (but with similar velocity distributions at different distances) may be required to explain the line profile shape in all these models. For example, an outflow with two velocity components, one at $v_0=375$ km s$^{-1}$ and the other one at $v_0=970$ km s$^{-1}$ would reproduce the median parameters of the two-Gaussian fits to the emission lines among the radio-quiet sample, and the combined narrow core $+$ broad wing profile would have $K>1$. 

We note that discussed in this paper are purely geometrical models assuming that the bulk velocities are the dominant component. These models are highly simplified on the details of gas dynamics, neglecting gas instabilities, fragmentation process, turbulence created by the outflow as it expands into the ambient gas and other complications.

\subsection{Declining velocity dispersion profiles}
\label{sec:dispmodels}

In Section \ref{sec:dispersion} we find that the $W_{80}$ parameter is almost constant across the nebulae in most cases, perhaps declining slightly toward the outer parts. We thus confirm the previous results based on long-slit observations which indicated flat $\sigma_{\rm gas}(R)$ profiles \citep{gree11}. For objects whose morphology is close to round and whose velocity variations across the nebulae are significantly smaller than the line widths, the most natural explanation for the flat $W_{80}$ profiles is a spherical or quasi-spherical outflow with essentially constant velocity. We accept this as the ``0th order model'' for our kinematic data and now discuss the possible origins of the mild decline of $W_{80}$ with distance from the center. 
%Most of the discussion is focused on the central parts of the nebulae where the median decline is 3\% per kpc, and toward the end of this subsection we discuss the rapidly decreasing $W_{80}$ of the super-bubble candidates. 

If emission-line clouds are all thrown out from the central region of the galaxy at the same time, but with different radial velocities, then at some later time $t$ the objects with velocities $v$ end up at distances $r=vt$. Thus in the snapshot of this outflow taken at time $t$ the velocities are linearly increasing as a function of distance, even though no outflow acceleration is taking place. In Section \ref{sec:sph} we demonstrate that such outflow has a $W_{80}$ profile that is linearly increasing with the projected distance $R$, and thus this model (analogous to the Hubble flow) is ruled out by the data. We thus conclude that the observed outflow did not originate as a quasi-instantaneous explosion but rather was established over an extended period comparable to its life time $\tau=14$ kpc / 760 km s$^{-1} = 1.8 \times 10^7$ years. 

There are at least four different mechanisms capable of establishing an apparently declining $W_{80}(R)$. One is that of episodic explosions. The first explosion of quasar feedback sends out a shock wave through the interstellar medium of the galaxy and clears out some of it \citep{nova11}. The terminal velocity of this flow is determined by the amount of energy injected and the amount of resistance from the medium that needs to be cleared away. But in any subsequent episodes the resistance is smaller and smaller, and thus one might conclude that the terminal velocity becomes higher and higher. This is an attractive possibility, best explored using numerical simulations. A quantitative understanding of this process will shed light on the amount of interstellar medium remaining in the galaxy at the time of our observations and will help us understand the stages of quasar feedback.

The second possibility is that once the clouds are accelerated somewhere close to the quasar, they proceed ballistically and thus they slow down as they climb out of the potential well of the host galaxy $\Phi(r)$. If the potential is that of an isothermal sphere with a core, characterized by the circular velocity at infinity $v_{\rm circ}$ and core radius $r_0$, then the radial velocity of ballistic clouds is
\begin{equation}
v_r^2(r)=v_r^2(r_0)+\frac{4}{3}v_{\rm circ}^2\left(\frac{r_0}{r}-1\right)-2v_{\rm circ}^2\ln\left(\frac{r}{r_0}\right).
\end{equation}
Taking $v_r(r_0)=760$ km s$^{-1}$, $r_0=1$ kpc, and $v_{\rm circ}=300$ km s$^{-1}$, typical for massive galaxies, we find that at 10 kpc the clouds slow down to $v_r(10~{\rm kpc})=230$ km s$^{-1}$. This would produce a decline in $W_{80}$ that is significantly stronger than observed. So either the potential wells of the galaxies in question are not as steep as this calculation suggests, or (as is more likely) the clouds are continually pushed by the low-density unseen volume-filling component of the wind.

The third possibility is that in the central parts of the nebulae the clouds are moving with large turbulent motions, but once they are thrown out of the galaxies their motion becomes purely radial. If the clouds in the center have an isotropic velocity dispersion and they conserve their radial component $v_r$ on the way out, then the observed line-of-sight velocity dispersions are related via $\sigma_{\parallel, {\rm bulk}}=\sigma_{\parallel, {\rm turbulent}}/\sqrt{\alpha}$, where $\alpha\simeq 4.5$ is the index of the emissivity profile ($j(r)\propto r^{-\alpha}$). This simple calculation corresponds to the limiting case of an outflow that transitions from isotropically turbulent to purely radial and results in a more rapid decline in $W_{80}$ than what is observed (median among our objects is 30\% over 10 kpc). Milder changes in the degree of radial anisotropy would produce milder changes in $W_{80}$. Thus the observed $W_{80}(R)$ profiles may be due to the velocity dispersion of the clouds becoming more radially anisotropic at larger distances from the quasar. These scenarios and the origin of the more isotropic (turbulent) motions of clouds in the central parts are best probed by the numerical simulations. 

The fourth possibility is that in the central parts the wind expands in all directions from the quasar, whereas at larger distances the opening angle of the outflow is decreased, perhaps because there are low-density regions along which the wind prefers to propagate. This may occur for example in a galaxy with a flattened gaseous atmosphere or a thick gaseous disk. The wind starts expanding in all directions, but when its size reaches the typical scale-height of the galactic disk, the directions perpendicular to the disk become much less obstructed by the interstellar medium, and the wind breaks out in these directions \citep{fauc12} producing the ``bubbles'' discussed in Section \ref{sec:bubbles}. Numerical simulations show that this phenomenon is expected both for jet-driven winds \citep{suth07} and for supernovae-driven winds \citep{veil05} breaking out of a thin galaxy disk. Because the opening angle of the outflow is reduced once it becomes confined into the preferred directions along the path of least resistance, the range of projected velocities is reduced as well. Thus we would expect to see a smaller range of velocities in the confined parts than in the isotropically expanding parts. 

Quantitatively, if the outflow has a constant velocity $v_0$, the axis of the break-out cone is in the plane of the sky and the half-angle of the cone is $\theta$, then the line-of-sight velocity dispersion observed within the cone is
\begin{equation}
\sigma^2_{\parallel}(R, \theta)=v^2_0\frac{\int_0^{R \tan\theta}j(r)\frac{z^2}{r^2}{\rm d}z}{\int_0^{R \tan\theta}j(r){\rm d}z}.\label{eq:angle}
\end{equation}
Assuming a power-law luminosity density $j(r)\propto r^{-\alpha}$ ($\alpha\approx 4.5$), we can use this equation to calculate how much the observed line width changes when we switch from a spherically symmetric ($\theta=90^{\circ}$) outflow to a narrower conical one. For example, to reduce the observed $W_{80}$ from $\sim 1000$ km s$^{-1}$ to $\sim 300$ km s$^{-1}$ would require a collimation within $2\theta\approx 30^{\circ}$. Thus, a rapid decline of $W_{80}$ accompanied by an apparent narrowing of the outflow is likely due to the fact that the outflow becomes more directional or more collimated within these structures. We present possible examples of this model in Section \ref{sec:bubbles}. 

\section{Extended morphology: illuminated companions vs. super-bubbles}
\label{sec:bubbles}

The dominant morphology of the \oiii\ nebulae in our radio-quiet sample is smooth and round (Paper I), in contrast to the clumpy and irregular morphology of the three radio-intermediate and radio-loud objects that we observed as a comparison sample and of low-redshift radio quasars studied by \citet{fu09}. One exception is SDSS~J0210$-$1001, in which we detect a surface brightness peak in [O {\sc iii}] about 1.5\arcsec\ (10 kpc) away from the central source which was also seen in our previous long-slit observations of this object \citep{liu09}. The \oiii\ emission line in this region shows two peaks, one due to the main nebula and one due to the presumed companion, with the line-of-sight velocity relative to the main nebula of about 300 km s$^{-1}$ and a velocity dispersion of $\sim70$ km s$^{-1}$. The fact that this feature shows a separate surface brightness peak, its compact extent (and in particular the fact that it does not extend beyond the reaches of the main nebula) and its low velocity dispersion all suggest that it is a small gas-rich companion galaxy illuminated by the quasar, although no continuum emission from this clump is detected.

\begin{figure*}%[h!]
\centering
    \includegraphics[scale=0.32,trim=0mm 0mm 3cm 0mm]{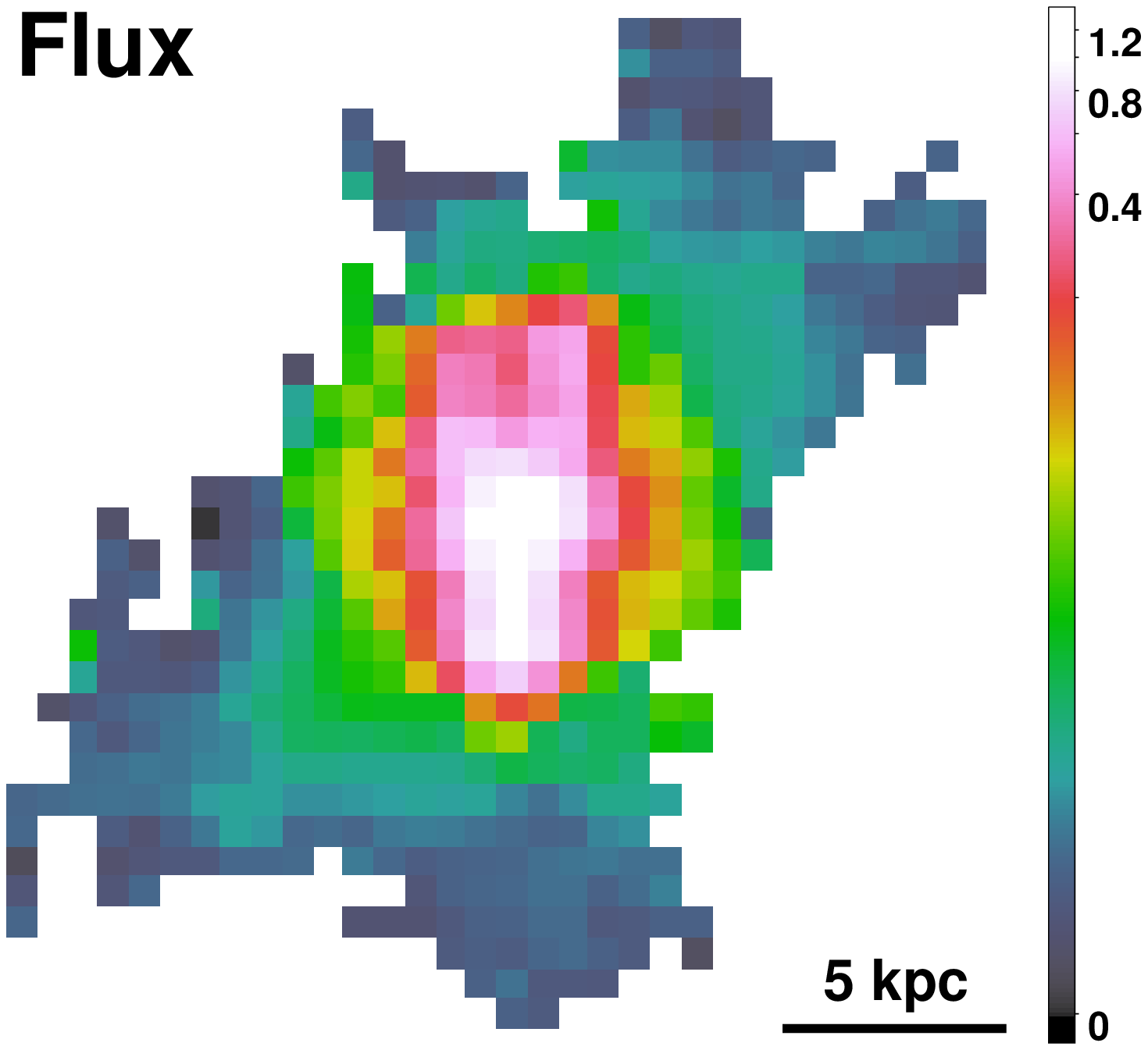}%
    \includegraphics[scale=0.32,trim=0mm 0mm 3cm 0mm]{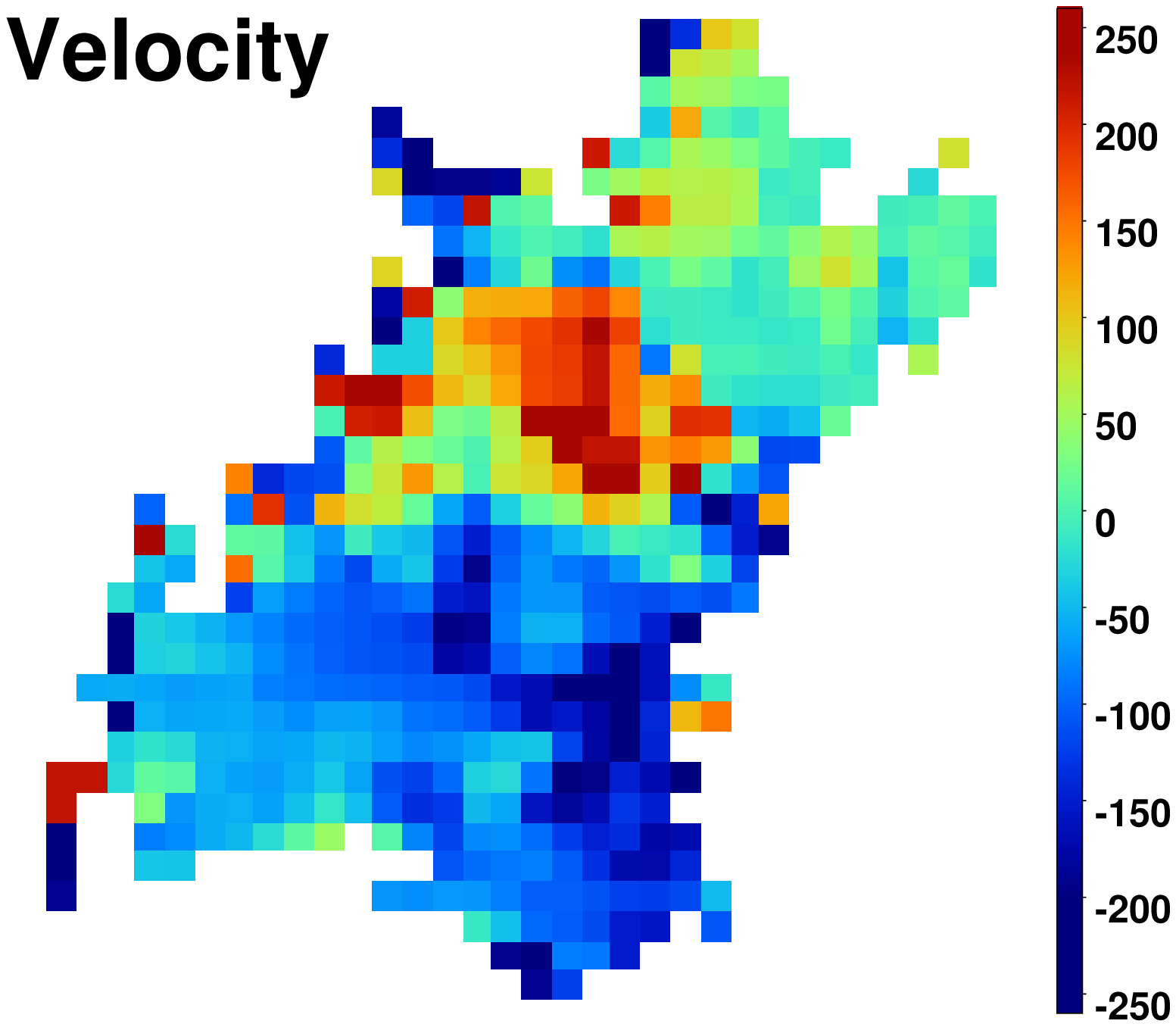}%
    \includegraphics[scale=0.32,trim=0mm 0mm 3cm 0mm]{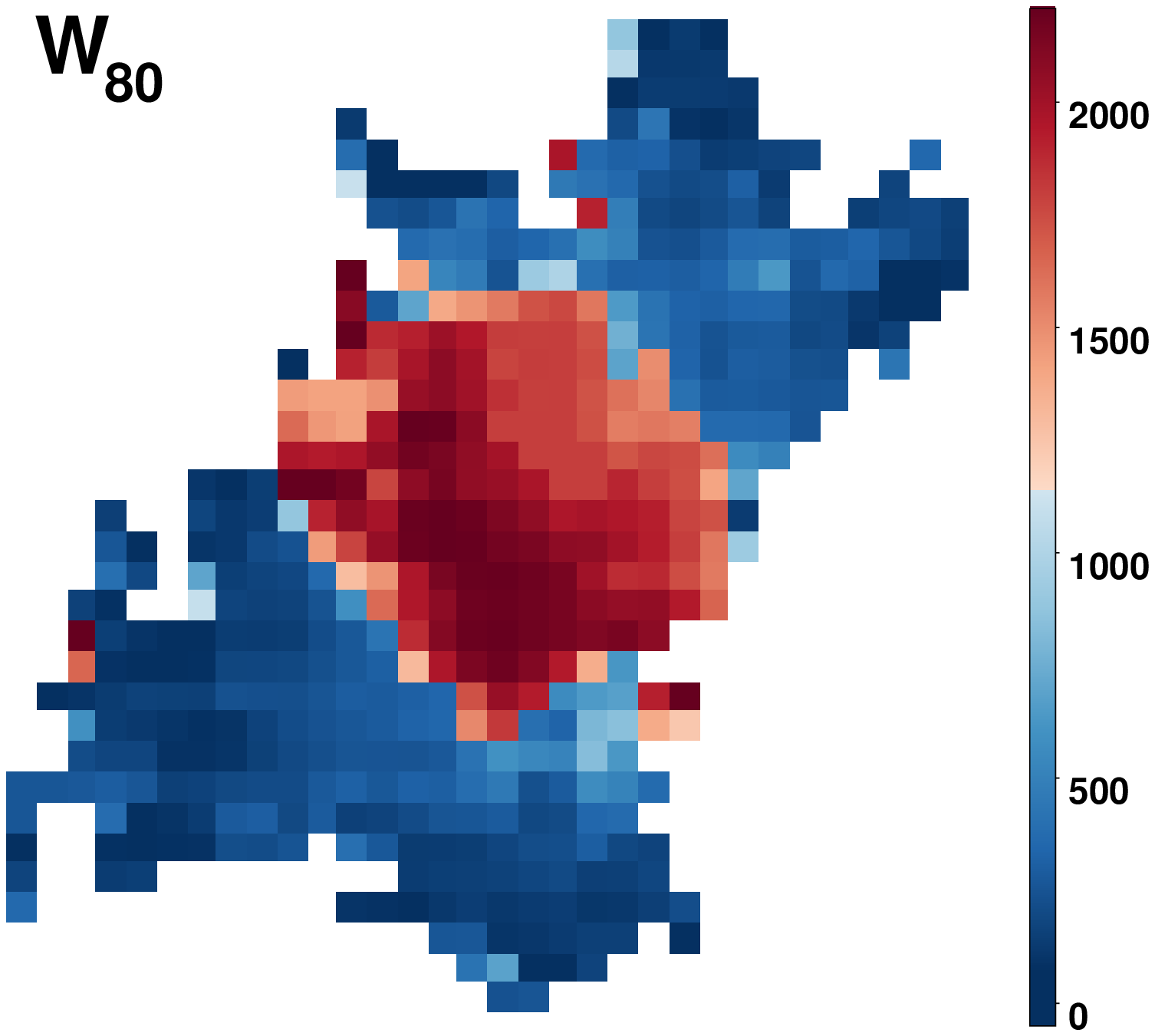}\\
    \includegraphics[scale=0.32,trim=0mm 0mm 3cm 0mm]{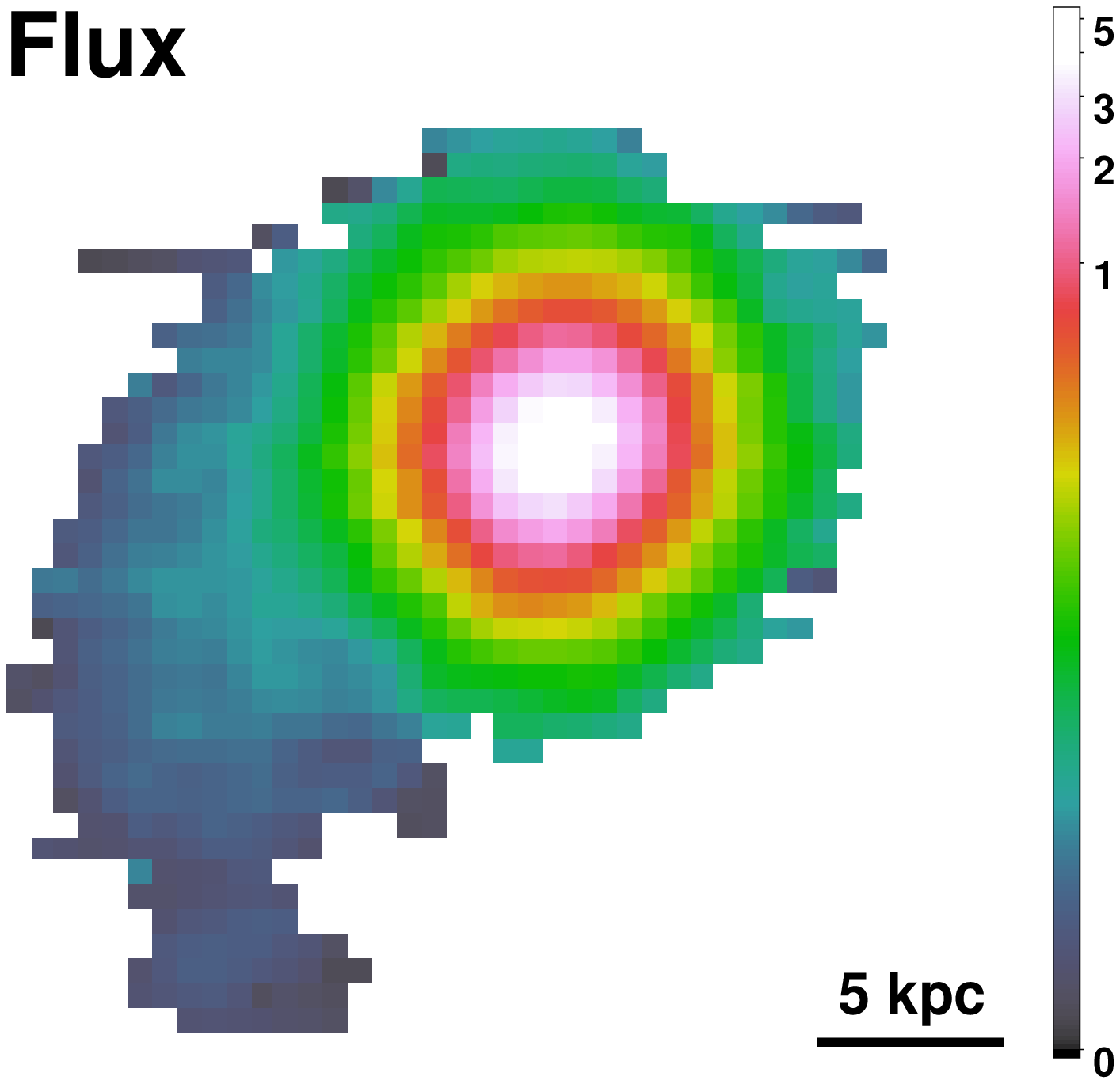}%
    \includegraphics[scale=0.32,trim=0mm 0mm 3cm 0mm]{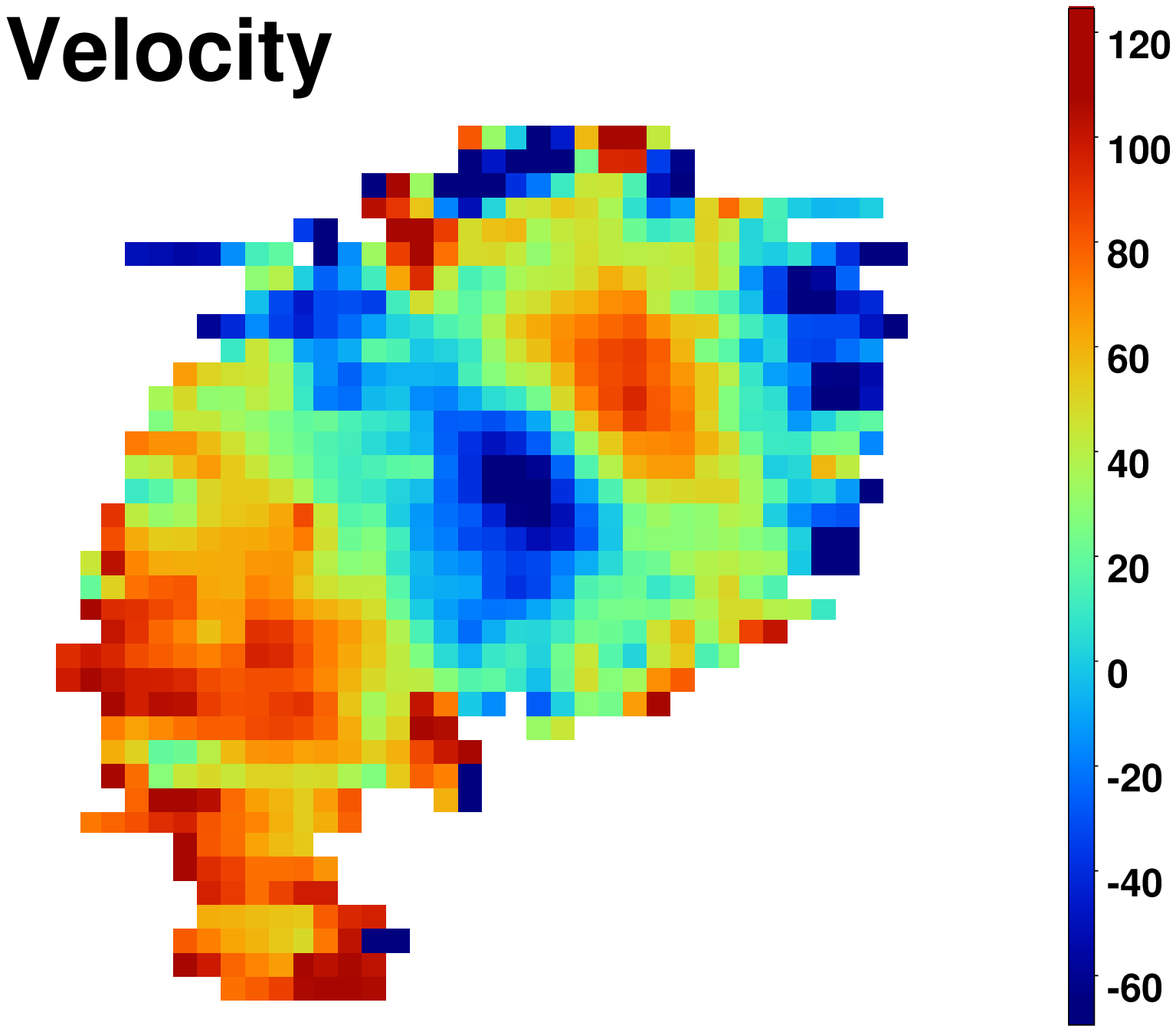}%
    \includegraphics[scale=0.32,trim=0mm 0mm 3cm 0mm]{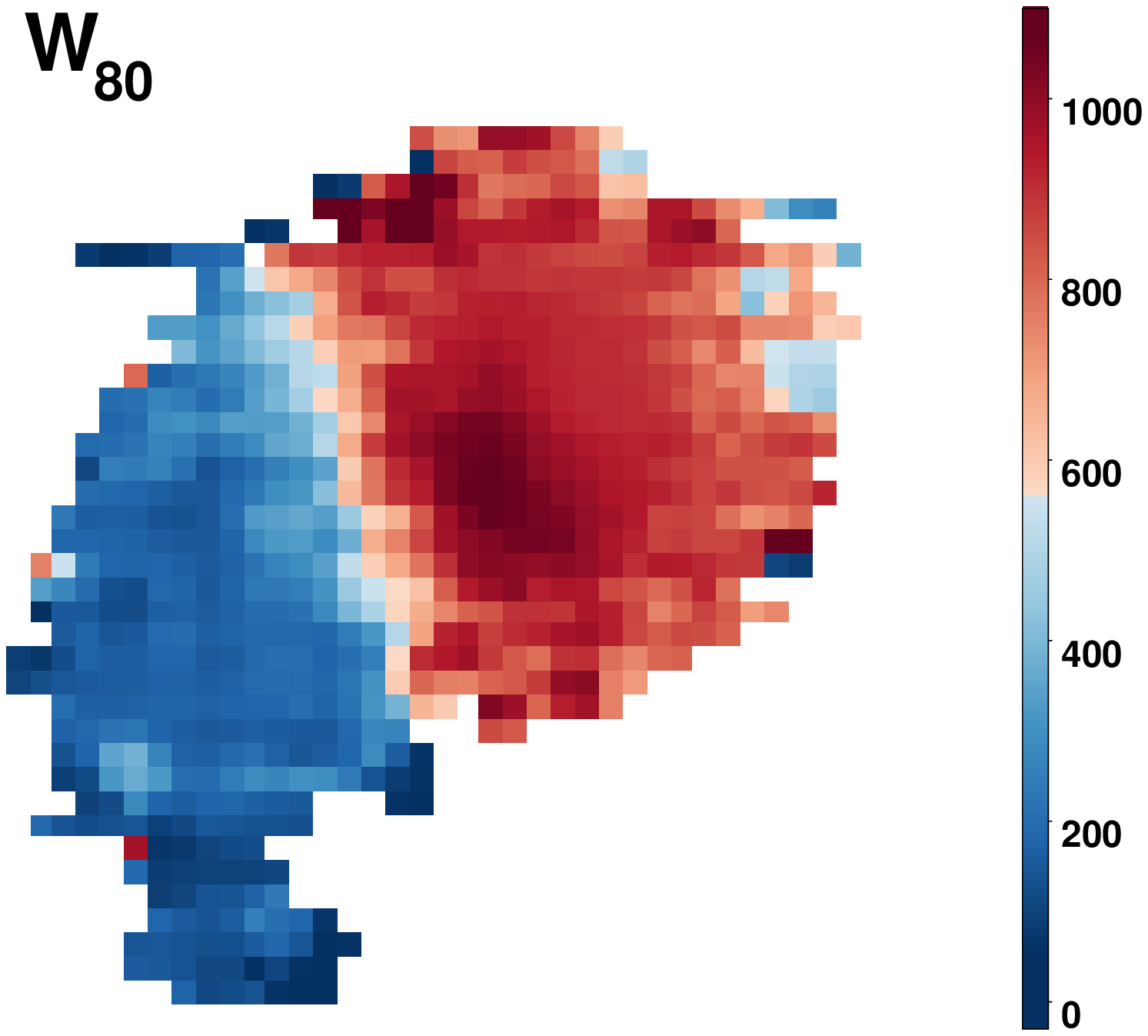}
\caption{The surface brightness (logarithmic scale, in units of $10^{-14}$ erg s$^{-1}$ cm$^{-2}$ arcsec$^{-2}$), radial velocity and $W_{80}$ maps (both in units of km s$^{-1}$) of two super-bubble candidates. The top row is for a strong candidate SDSS J0319$-$0019 recovered 
by single-Gaussian fits, measured down to $S/N=1.5$ for the [O {\sc iii}] peak, in order to map the outer regions to the largest possible galactic radii, where for most spaxels single-Gaussian fits are the best model by our standard. The bottom row is for a tentative candidate SDSS~J1039+4512 recovered by single-Gaussian fits to $S/N=2.5$.}
\label{fig:bubbles}
\end{figure*}

\begin{figure*}%[h!]
\begin{flushleft} 
\hspace*{1cm}
\vspace*{2mm}
   \includegraphics[scale=0.35,trim=0mm 0mm 3.5cm 0mm]{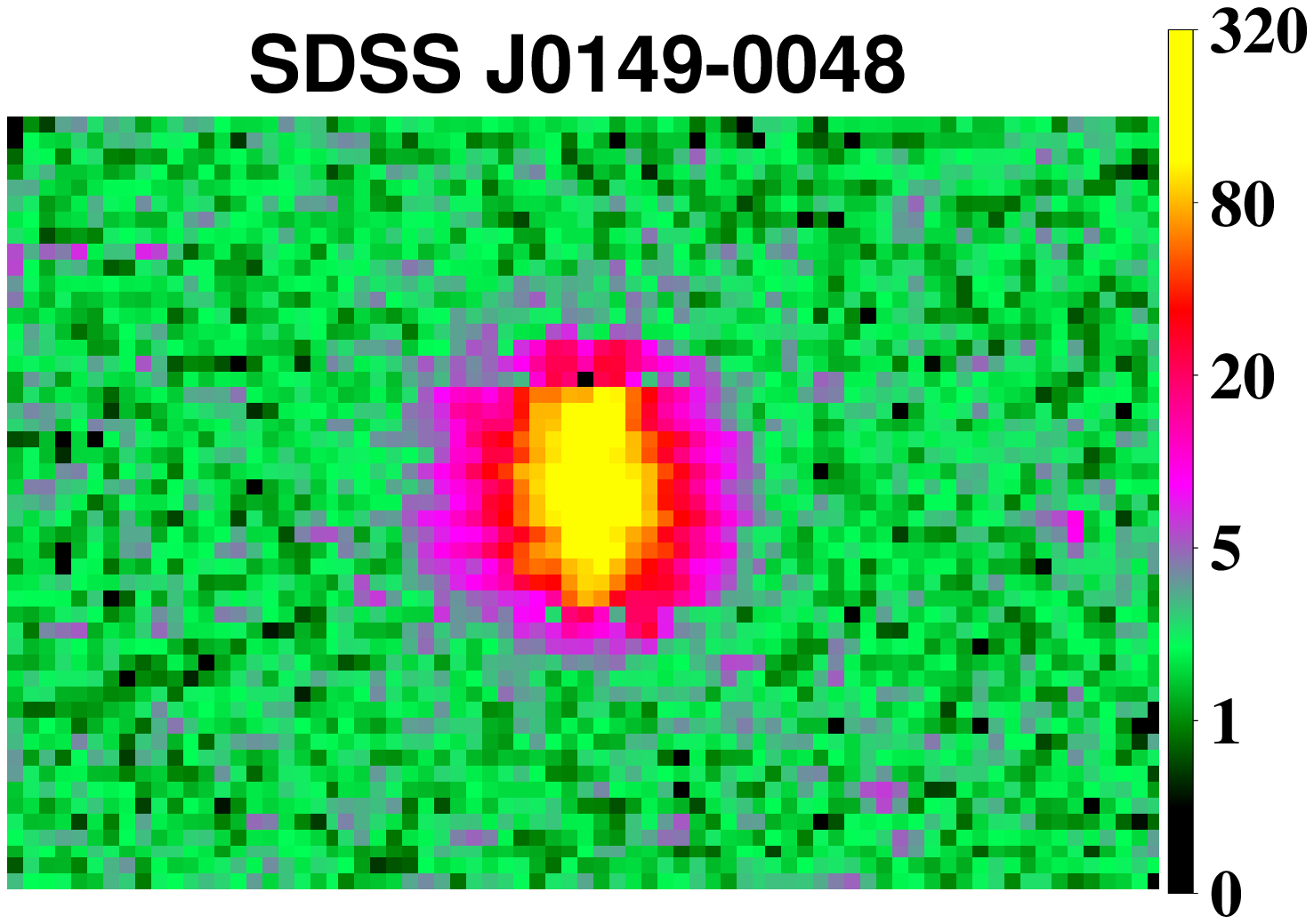}%
   \includegraphics[scale=0.35,trim=0mm 0mm 3.5cm 0mm]{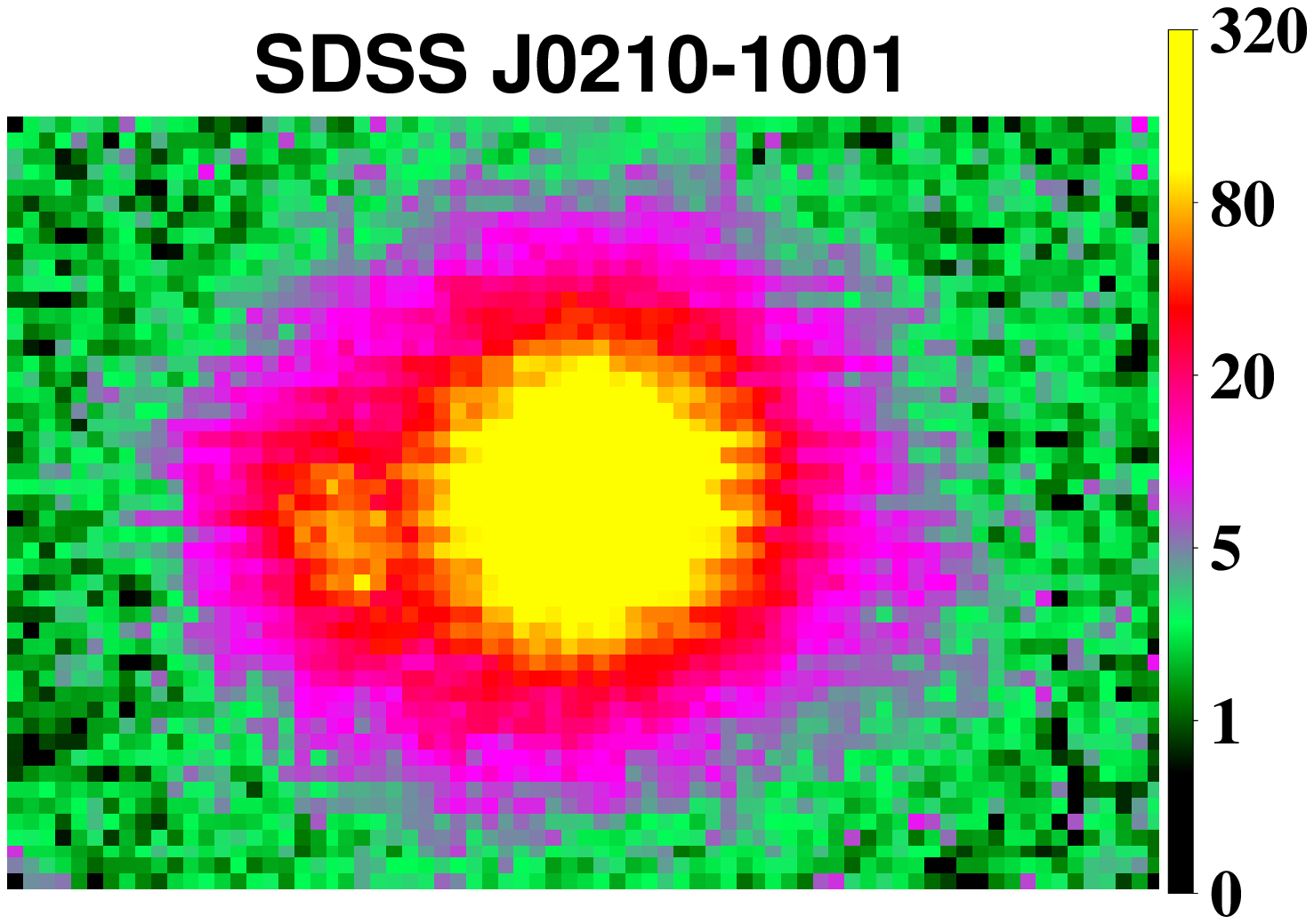}%
   \includegraphics[scale=0.35,trim=0mm 0mm 3.5cm 0mm]{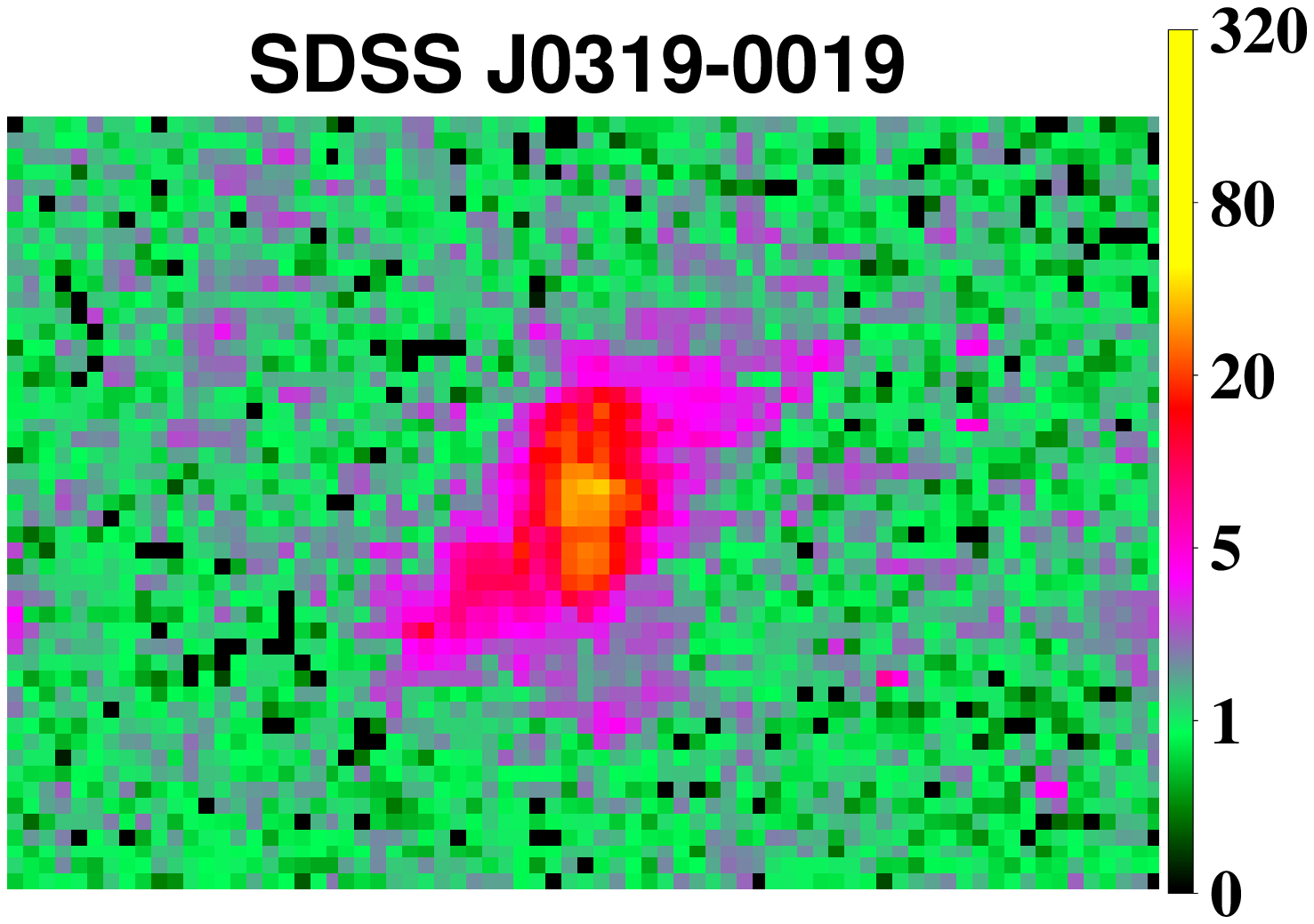}\\
\hspace*{1cm}
\vspace*{2mm}
   \includegraphics[scale=0.35,trim=0mm 0mm 3.5cm 0mm]{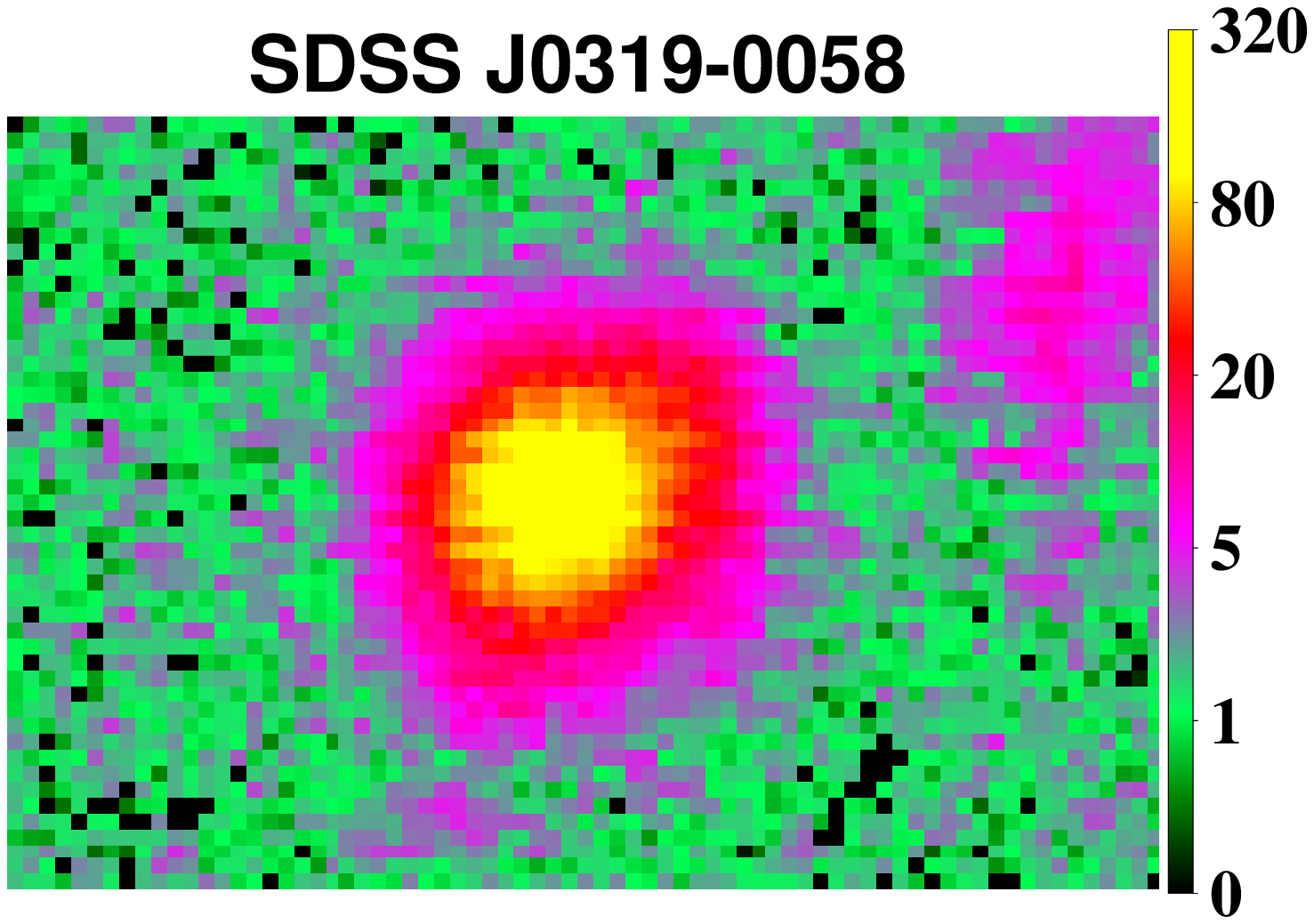}%
   \includegraphics[scale=0.35,trim=0mm 0mm 3.5cm 0mm]{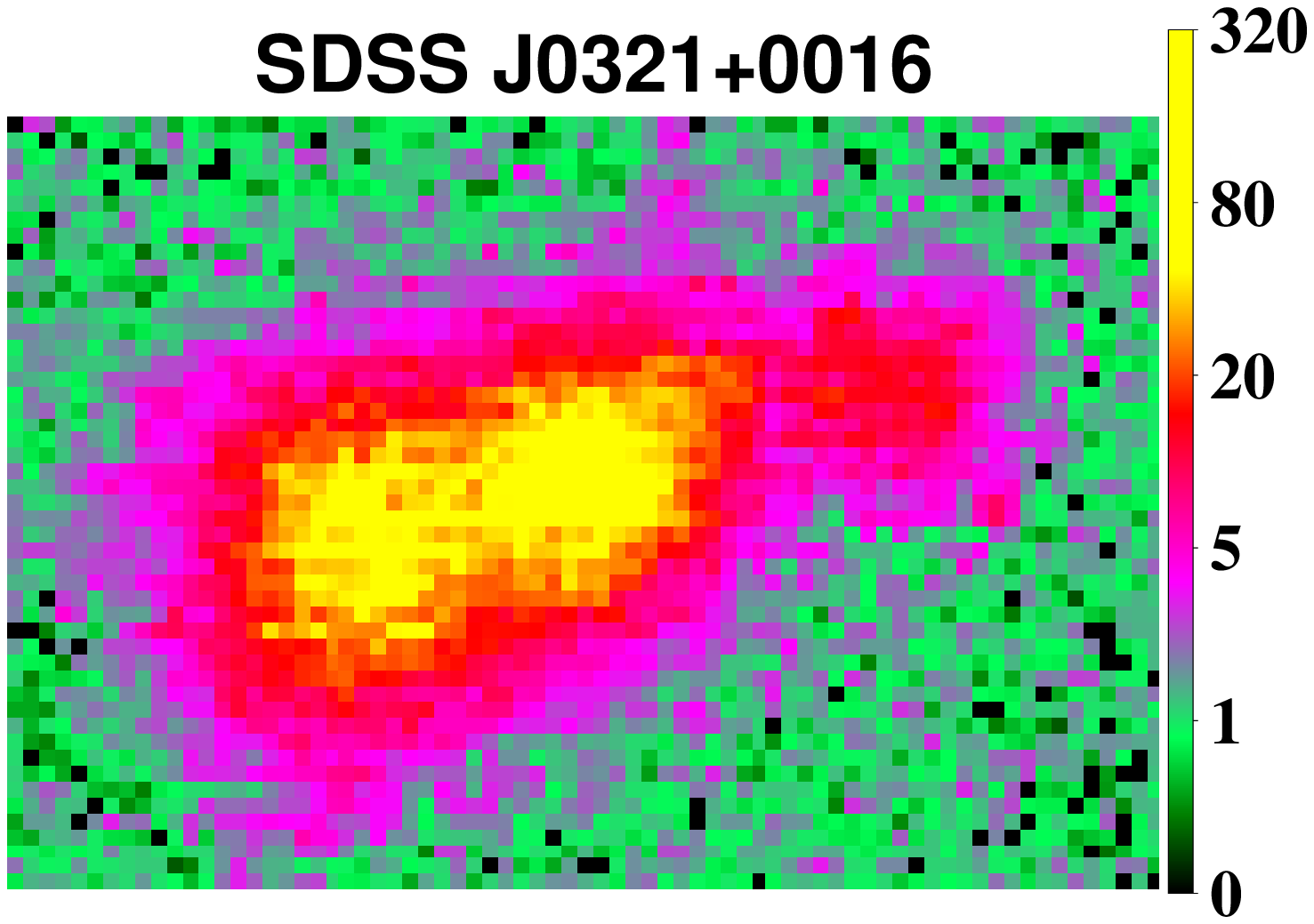}%
   \includegraphics[scale=0.35,trim=0mm 0mm 3.5cm 0mm]{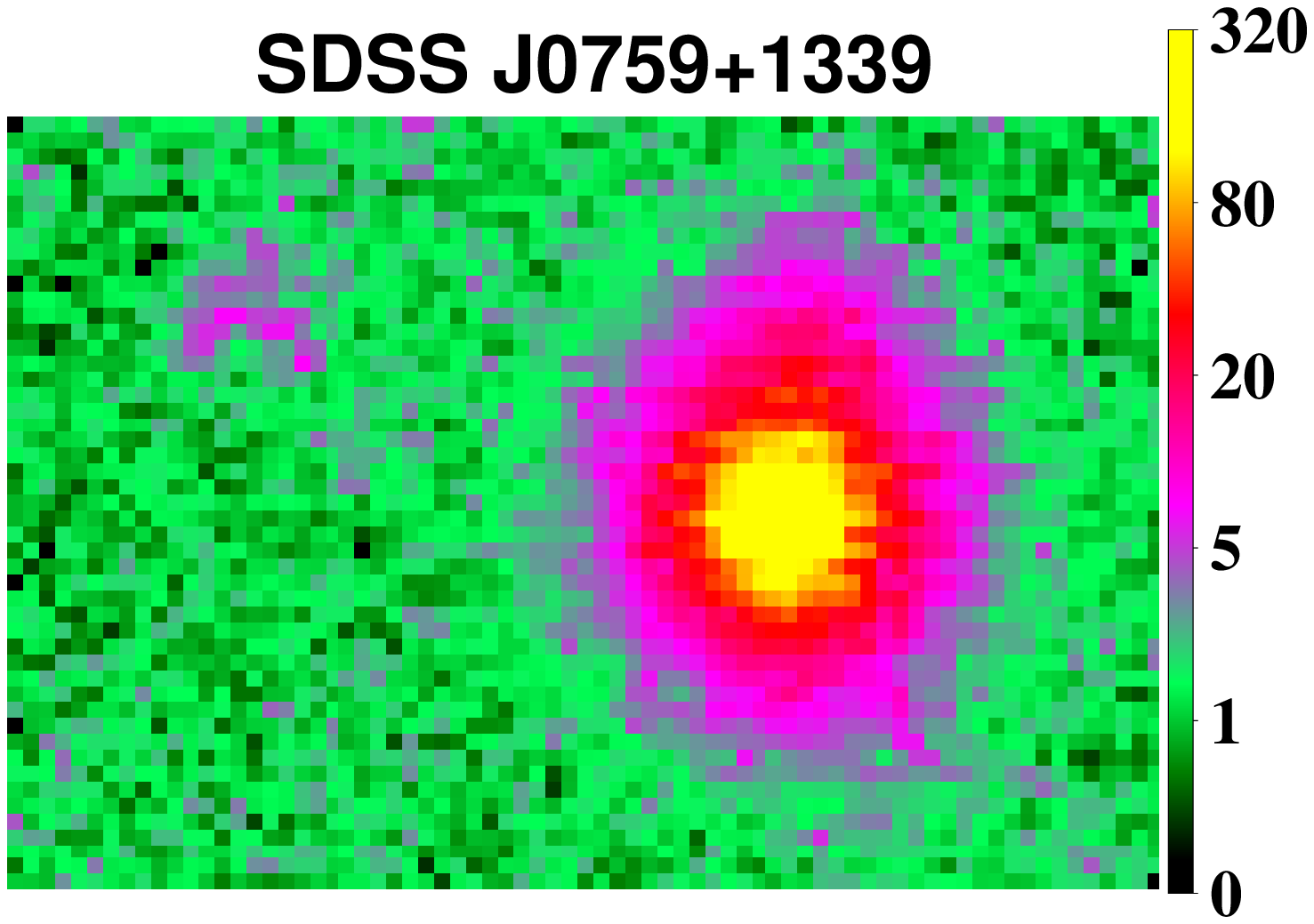}\\
\hspace*{1cm}
\vspace*{2mm}
   \includegraphics[scale=0.35,trim=0mm 0mm 3.5cm 0mm]{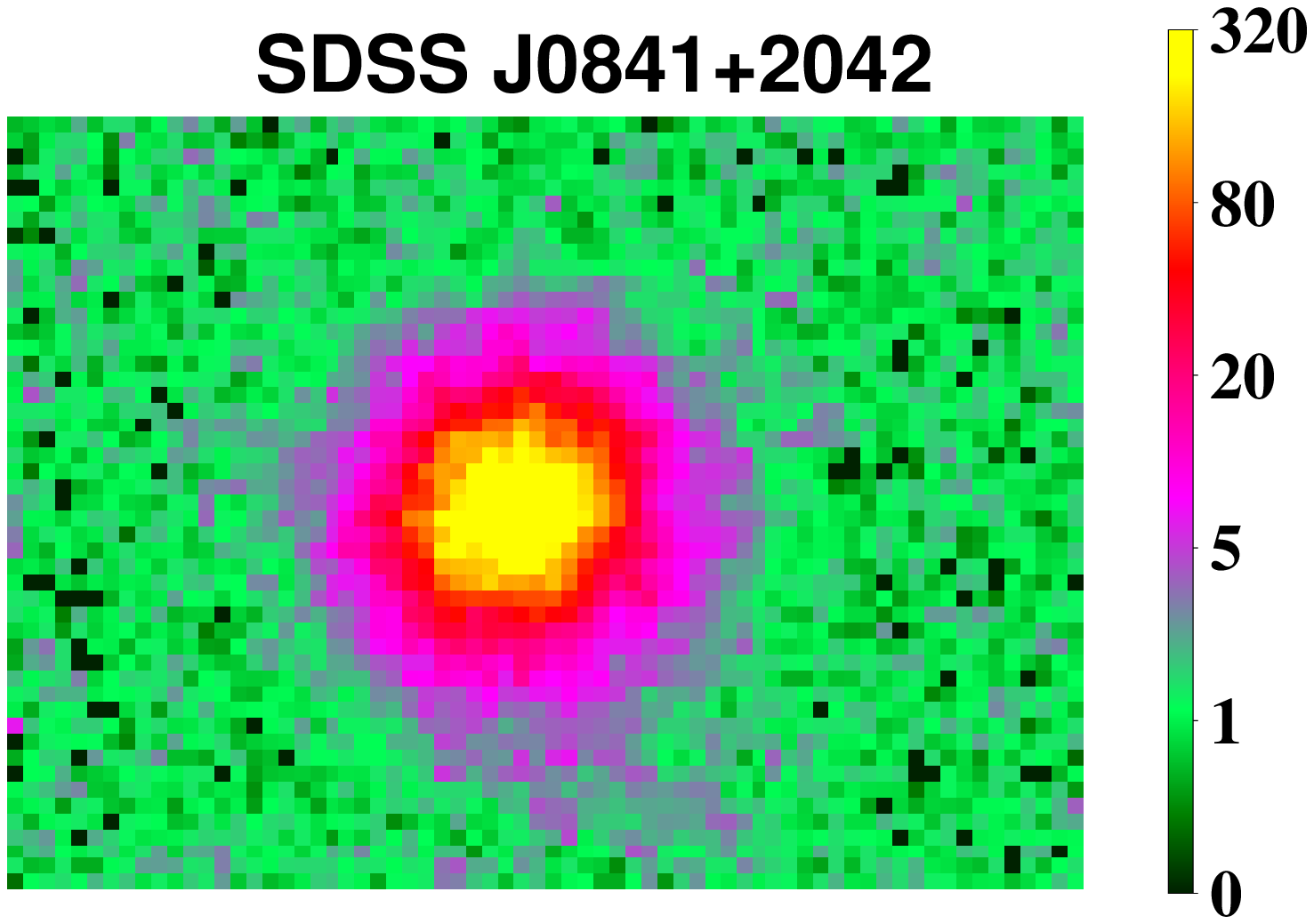}%
   \includegraphics[scale=0.35,trim=0mm 0mm 3.5cm 0mm]{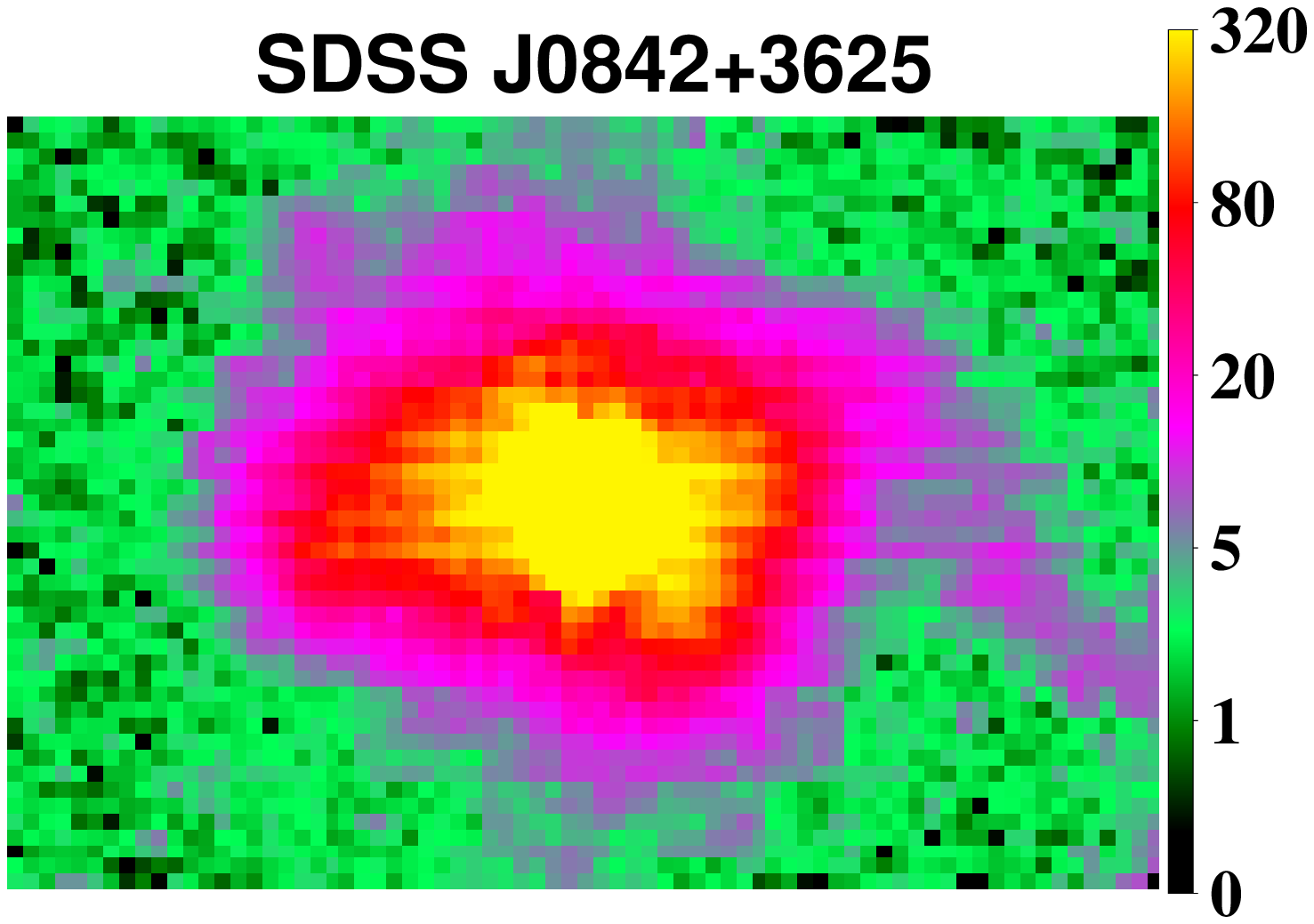}%
   \includegraphics[scale=0.35,trim=0mm 0mm 3.5cm 0mm]{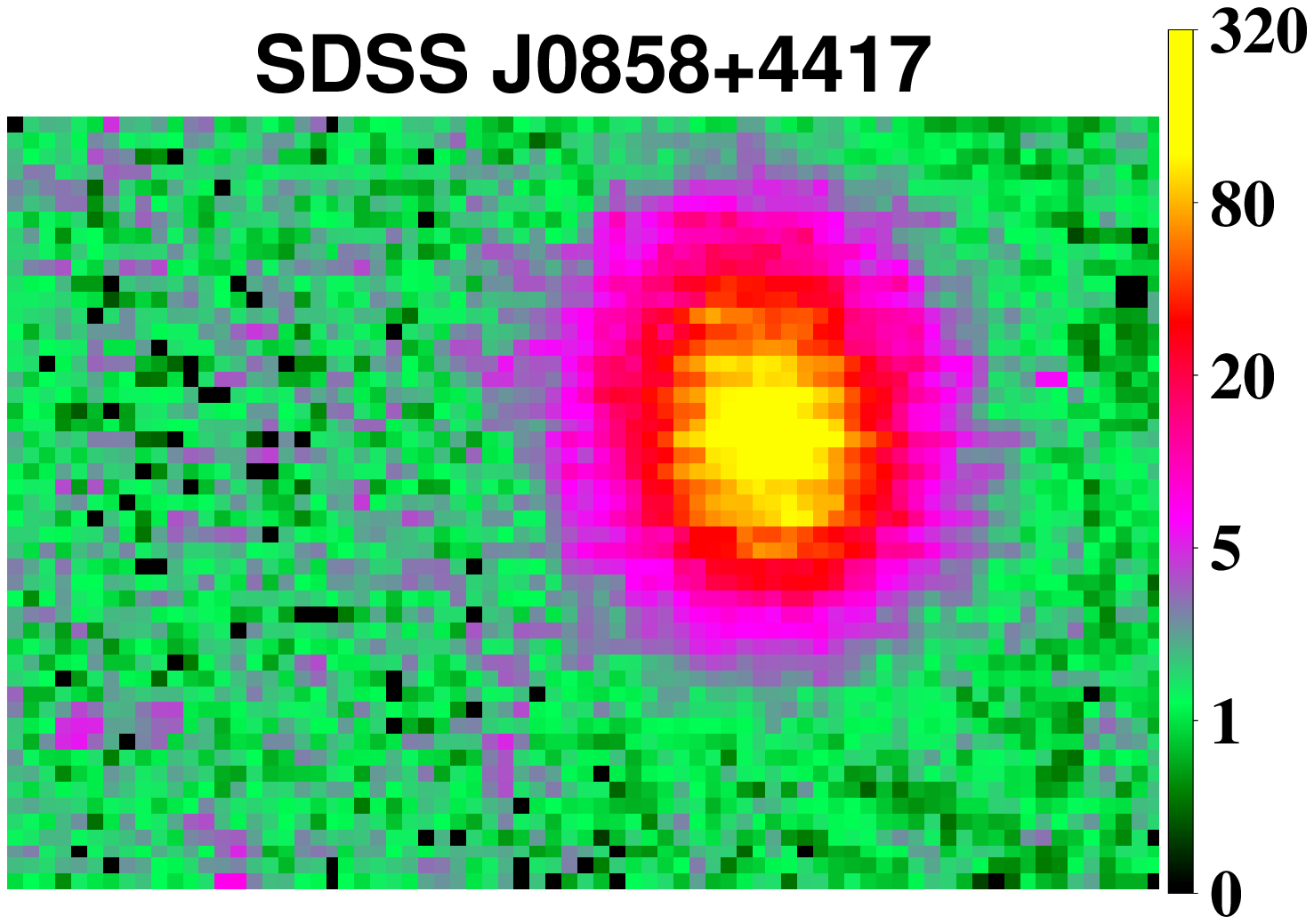}\\
\hspace*{1cm}
\vspace*{2mm}
   \includegraphics[scale=0.35,trim=0mm 0mm 3.5cm 0mm]{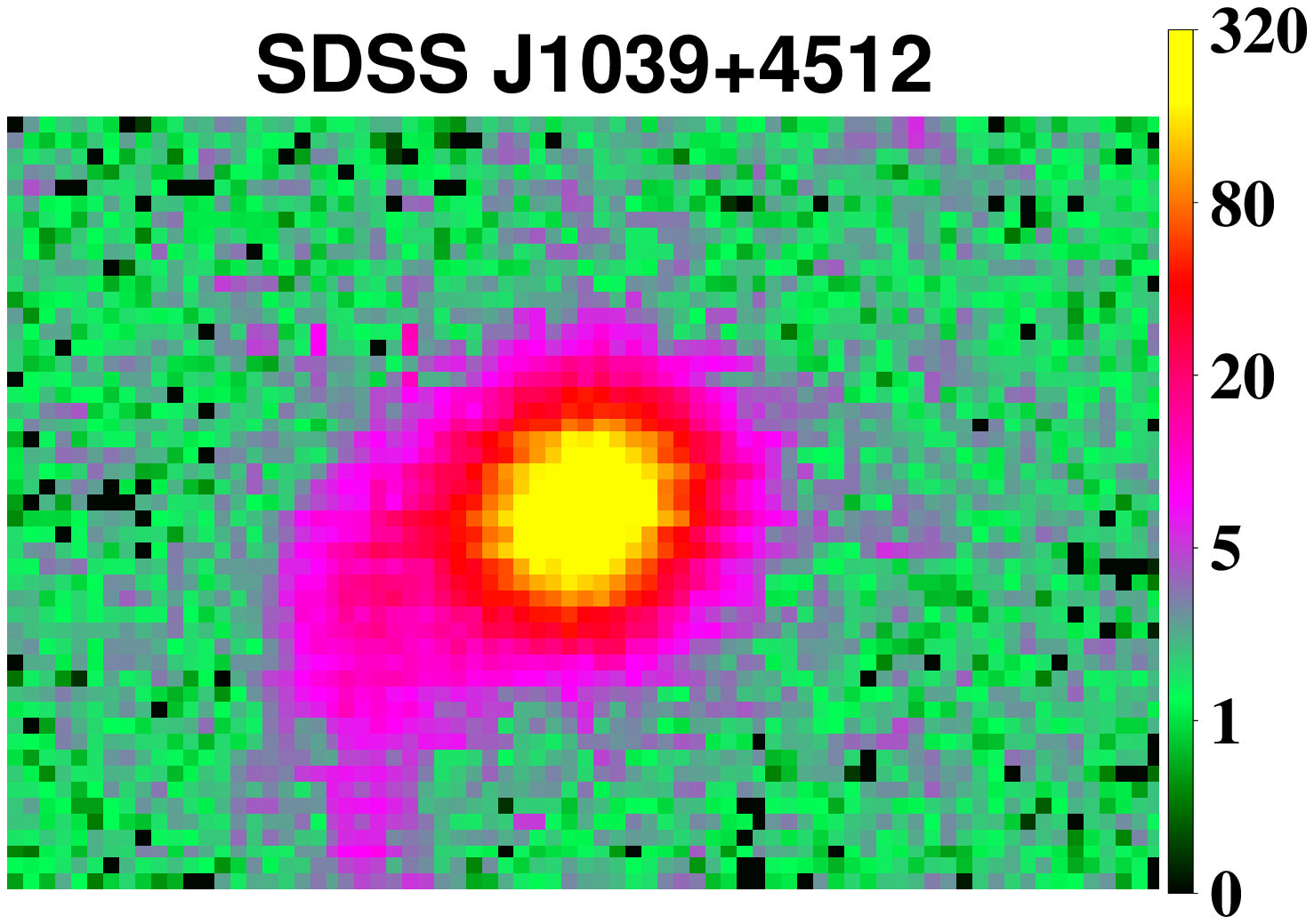}%
   \includegraphics[scale=0.35,trim=0mm 0mm 3.5cm 0mm]{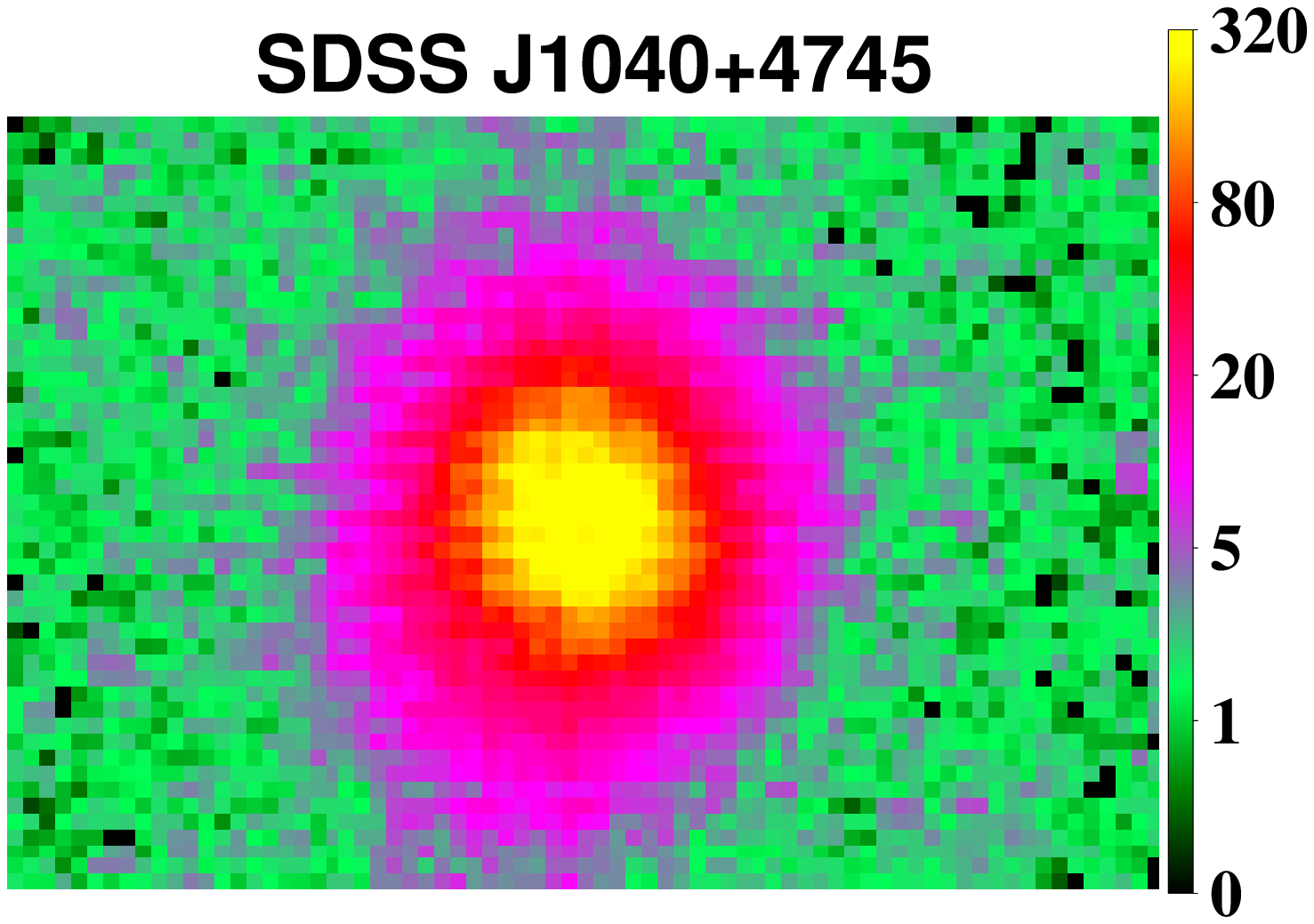}\\
\hspace*{1cm}
   \includegraphics[scale=0.35,trim=0mm 0mm 3.5cm 0mm]{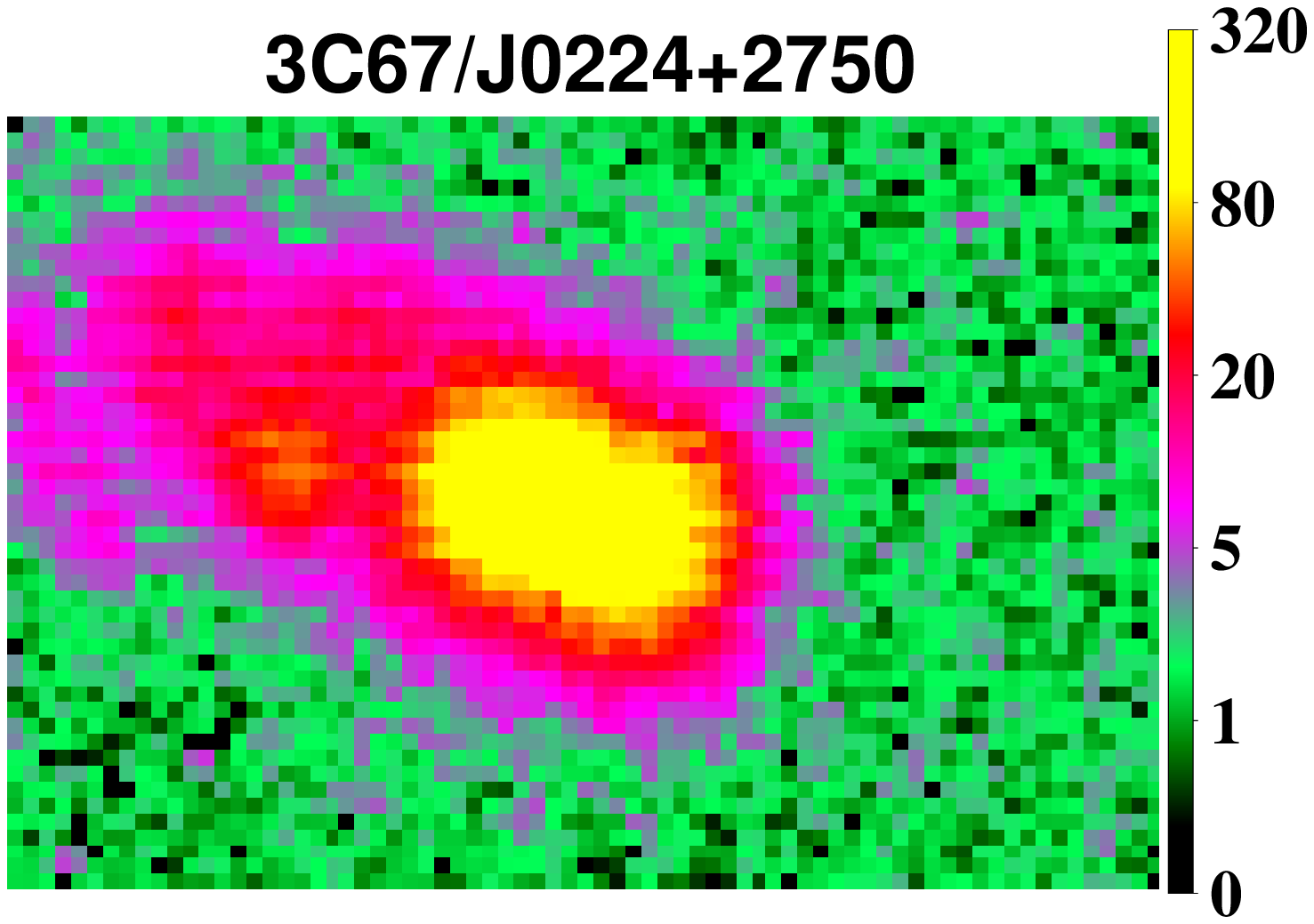}%
   \includegraphics[scale=0.35,trim=0mm 0mm 3.5cm 0mm]{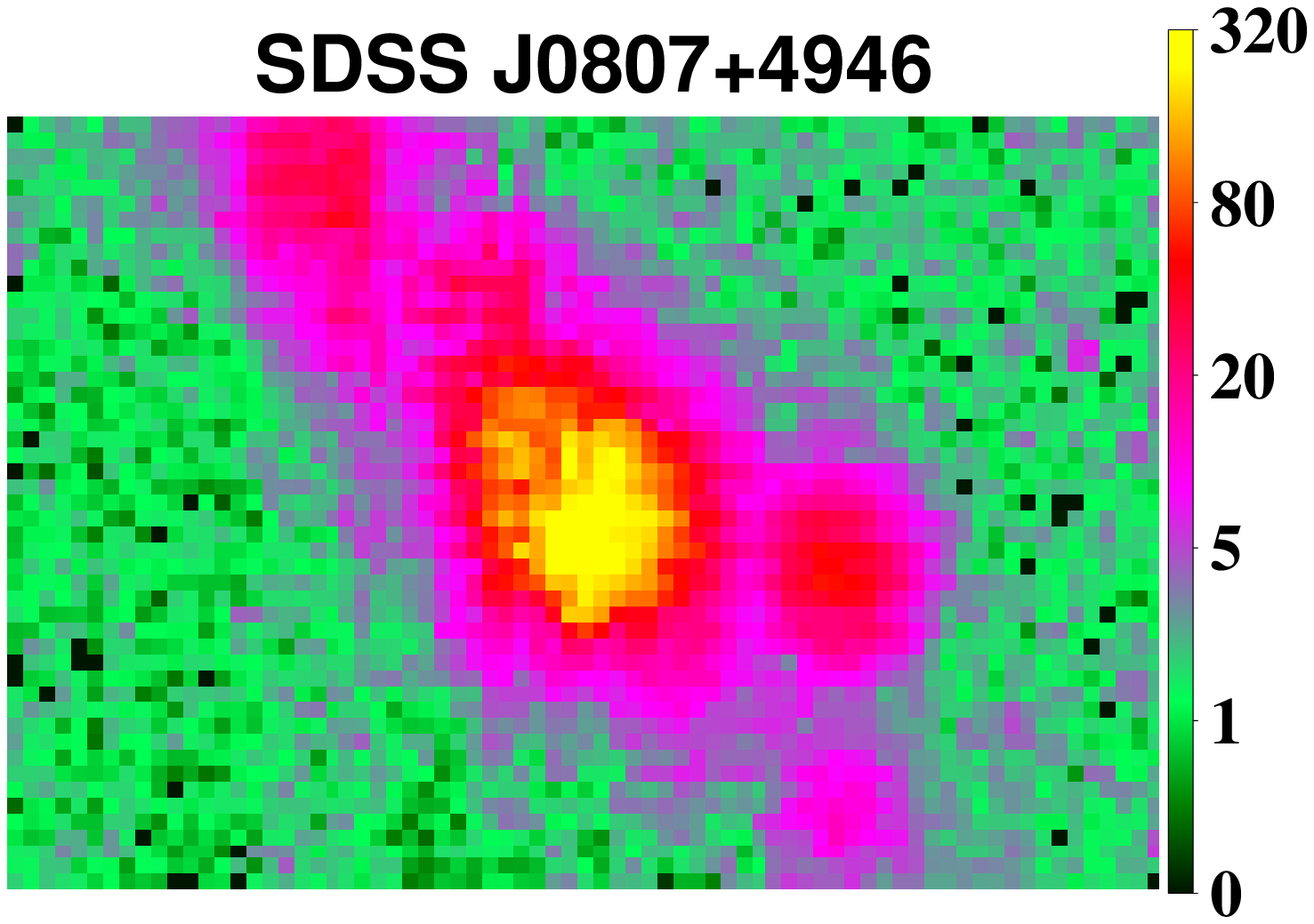}%
   \includegraphics[scale=0.35,trim=0mm 0mm 3.5cm 0mm]{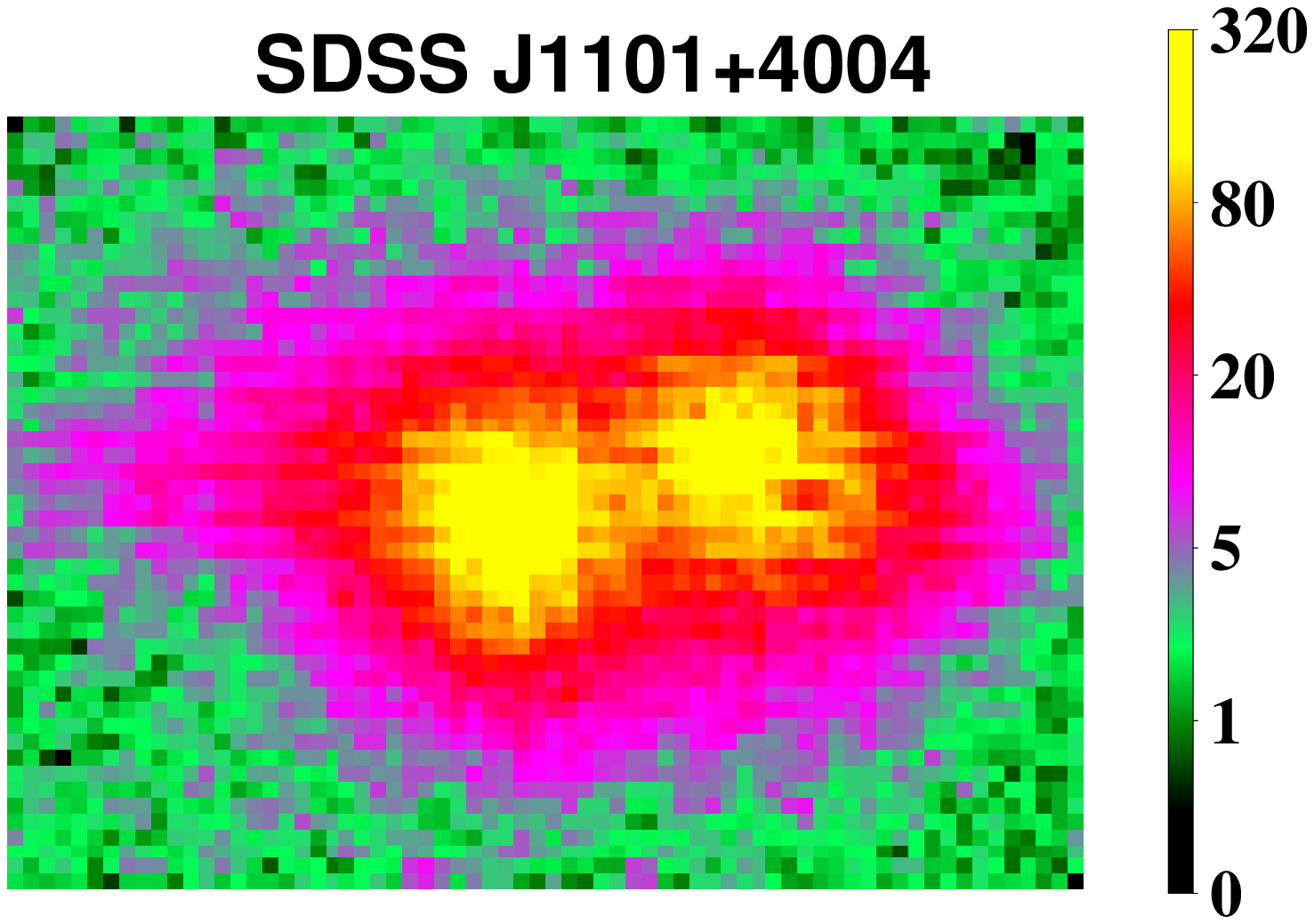}\\
\end{flushleft} 
\caption{Maps of $S/N$ ratios at the peak of the \oiii\ emission line profile which allow for seeing low surface-brightness features.}
\label{fig:lowsb}
\end{figure*}

A different type of extended morphological and kinematic features are present in SDSS J0319$-$0019 (Figure \ref{fig:bubbles}, top). In most objects, the surface brightness of [O {\sc iii}] emission is a steeply declining function of distance ($I\propto R^{-3.5}$ or so). Thus, the extent and shape of the nebulae are largely insensitive to the exact surface brightness level at which we cut off our maps (Figure \ref{fig:lowsb}). In this source, however, there is weak, very extended emission with typical peak $S/N$ of the \oiii\ emission of 1.5 (although the significance of detection is much higher since many correlated spectral pixels are used in profile fitting). The low surface brightness emission in this source has a symmetric X-shaped morphology. We hypothesize that the quasar wind in this source broke out of the high density interstellar medium and is now expanding into the intergalactic medium largely perpendicularly to the main plane of the galaxy, in ``super-bubbles'' that extend at least out to 15 kpc from the central quasar on either side. Such features (albeit with more modest extents) are seen in local starburst galaxies (e.g., NGC 3079, \citealt{veil94}) and have been proposed as the typical expected morphology of quasar-driven winds \citep{fauc12}. The symmetry of the features relative to the main source and the X-shape suggestive of limb brightening of the gas that is expanding sideways and plowing into intergalactic medium argue against the origin of this gas in a companion galaxy of the kind we likely see in SDSS~J0210$-$1001. 

Even in the high-quality IFU observations presented here it may be difficult to distinguish ionized dwarf companion galaxies and super-bubbles. Both types of features may show median velocity that is significantly distinct from the main source -- in the former case, due to the orbital velocity of the companion and in the latter case, due to the bulk velocity of the flow in the bubbles. Both types of features may show relatively low velocity dispersion -- in the former case, due to the weak gravity of the companion galaxy and in the latter case because most of the emission may be coming out not from the bulk of the bubbles but rather from the high-density material accumulated in walls of the bubbles. We previously found a quasar-driven super-bubble candidate SDSS~J1356+1026 via long-slit observations \citep{gree12} which is limb-brightened as well and in which each kinematic component has a velocity dispersion of about 100 km s$^{-1}$.

The discriminating observable is whether there is stellar continuum emission associated with the line emission in the extended feature -- we do not expect to see any in the bubbles (nor do we in the case of SDSS~J1356+1026, \citealt{gree12}), but we do expect to see it in an illuminated companion galaxy. Unfortunately, our data are not very sensitive to the continuum, and thus the non-detection of the continuum in the extended components is not constraining. This issue can be resolved with HST observations sensitive to stellar light continuum. 

To expand our search for super-bubble candidates, we now focus on the very low surface brightness emission and look for signs of elongated extended structures around our quasars (Figure \ref{fig:lowsb}). In addition to SDSS~J1356+1026 presented by \citet{gree12} we identify the following quasar-driven super-bubble candidates detectable on projected scales 10--15 kpc from the quasar:
\begin{itemize}
\item SDSS~J0319$-$0019 (Figure \ref{fig:bubbles}): strong candidate with X-shaped extended emission symmetric around the center. The maximal velocity difference between the opposing sides is $\sim$200 km s$^{-1}$ and the typical line width in the emission associated with bubbles is $W_{80}\sim200$ km s$^{-1}$. In addition to hosting a luminous obscured quasar, this object shows a classical post-starburst (A-star dominated) spectrum \citep{liu13a}. 
\item SDSS~J0321+0016 (Figure \ref{fig:VWAK}): strong candidate with two extended regions on the opposite sides of the nucleus. The maximal velocity difference between the opposing sides is $\sim$600 km s$^{-1}$ and the typical line width in the extended regions is $W_{80}\sim200$ km s$^{-1}$. The half-light size of the continuum-emitting region (4.3 kpc) is smaller than the half-light size of the narrow-line region (5.2 kpc, \citealt{liu13a}). While this is not a proof that the very extended regions are continuum-free (because of the low sensitivity of our data to continuum emission, there could be continuum light there that we cannot detect), it is an indication that the line-to-continuum ratio increases from the center to the outer parts. There is no evidence that the extended regions are limb-brightened, so we may be seeing the bulk flow within the bubbles in this object.
\item SDSS~J0842+3625 (Figure \ref{fig:VWAK}): moderate candidate with symmetric regions extended along the main illumination axis of the quasar as determined by ground-based polarimetric observations \citep{zaka05, liu13a}; this geometry is similar to that seen in SDSS~J1356+1026 \citep{gree12}. In this object the region of higher $W_{80}$ is elongated and oriented nearly perpendicular to the illumination / elongation axis which has lower $W_{80}$.
\item SDSS~J1039+4512 (Figure \ref{fig:bubbles}, bottom row): moderate candidate in which one-sided extended region with low $W_{80}$ is revealed in low $S/N$ maps. The counter-feature may also be detected with a much smaller extent. 
\item SDSS~J0210$-$1001: unlikely candidate in which the extended region of distinct kinematics is more likely associated with a quasar-illuminated companion galaxy. We include this object on the list because there is a slight hint of a counter-feature in the surface-brightness maps (Fig. \ref{fig:lowsb}). The depth of our data are not sufficient to examine this further. 
\end{itemize}

The four strong / moderate candidates demonstrate (i) high ellipticity of outer surface brightness contours; (ii) extended features located opposite one another relative to the nucleus (or in the case of SDSS~J1039+4512, a hint of a counter-feature); (iii) smaller $W_{80}$ in the outer parts consistent with the limited range of angles of propagation, as discussed in Section \ref{sec:dispmodels}. For example, in SDSS~J0321+0016 the line-of-sight velocity dispersion drops by a factor of 3--3.5 in the extended super-bubbles compared to the central parts. Using equation (\ref{eq:angle}), we find the full opening angle of the cone of $2\theta\approx 30^{\circ}$, consistent with the observed morphology of this source. The X-shape of the other strong super-bubble candidate SDSS~J0319$-$0019 suggests that in this case we are likely seeing the walls rather than the bulk of the conical outflow, and thus the approximation used in equation (\ref{eq:angle}) is not applicable. 

In addition to these major extended features, we detect a number of weaker emission-line blobs (Fig. \ref{fig:lowsb}). For example, in SDSS J0759+1339 we find a compact companion at a distance of 3.6\arcsec\ (25 kpc) from the quasar. It is an emission-line object at the redshift of the quasar, with a relative line-of-sight velocity of 250 km s$^{-1}$ and an apparent velocity dispersion of 80 km s$^{-1}$ (i.e., marginally resolved at our spectral resolution). Similarly, in SDSS J0319$-$0058 we find an emission line object 3\arcsec\ (20 kpc) away from the quasar (top right corner of the field of view of SDSS~J0319$-$0058, Figure \ref{fig:lowsb}), with a relative line-of-sight velocity of about $-40$ km s$^{-1}$ and an apparent velocity dispersion of 53 km s$^{-1}$ (unresolved at our spectral resolution). Another fainter emission line object (2.2\arcsec, or 15 kpc, below the quasar in the Figure) has a relative line-of-sight velocity of $-130$ km s$^{-1}$ and an apparent velocity dispersion of 70 km s$^{-1}$ (marginally resolved at our spectral resolution). This source was also identified by \citet{liu09} in long-slit observations with a relative line-of-sight velocity of $-198$ km s$^{-1}$. The difference in the two reported values of line-of-sight velocity of $\sim 60$ km s$^{-1}$ is due to the difference in the assumed redshift for the main quasar and is part of the uncertainty in the host galaxy redshift discussed in Section \ref{sec:parameters}. We assume that these emission-line sources are emission-line companion galaxies to the quasar host galaxy that happen to be within the small field of view of the IFU observations. In this case they are likely unrelated to quasar activity and quasar winds, although in some cases they may be photo-ionized by the quasar producing a high \oiii/H$\beta$ ratio \citep{liu09}.

The morphologies and kinematics of radio-loud objects in our sample are noticeably more complex; this situation is similar to that in the sample of radio-quasars studied by \citep{fu09}. It appears that in these sources, the rotation of the host galaxy, the outflow due to the jet, companion galaxies and illuminated tidal debris all contribute to the observed diversity of features. 

\section{Energetics of the wind}
\label{sec:energy}

\subsection{Standard ``Case B'' estimate}

A big difficulty in estimating the energy rate of the wind is presented by its multi-phase nature, which likely consists of a low-density, high-temperature outflow with clouds of varying densities embedded in it \citep{zubo12}. It is likely that only a fraction of the mass of these clouds is in the warm ionized phase that we detect using optical emission lines. For example, \citet{rupk13} directly observe a higher mass in the neutral gas phase than in the ionized gas phase in the outflows from star-bursting ultra-luminous infrared galaxies, although, granted, in these systems one may expect milder ionization conditions than in luminous quasars we present here. The lowest limit on the mass and energy of the wind can be obtained just by counting photons emitted by recombining hydrogen atoms. Under the ``Case B'' assumption and adopting an intrinsic line ratio of $\rm H\alpha/H\beta=2.9$ following \citet{nesv11} and \citet{oste06}, the total mass of the observed ionized gas can be expressed as
\begin{equation}
\frac{M_{\rm gas}}{2.82\times10^9~M_{\odot}}=\left(\frac{L_{\rm H\beta}}{10^{43}~{\rm erg~s^{-1}}}\right)\left(\frac{n_e}{100~{\rm cm^{-3}}}\right).
\end{equation}
The \oiii\ line luminosities of our eleven radio-quiet quasars span a range $L_{\rm [O~{\scriptscriptstyle III]}}=10^{42.7\mhyphen43.6}$ erg s$^{-1}$ with a median of 10$^{43.3}$ erg s$^{-1}$. Since these recombination lines are dominated by the spaxels close to the bright centers where [O {\sc iii}]/H$\beta\sim10$, $L_{\rm H\beta}$ is typically $2\times 10^{42}$ erg s$^{-1}$. Adopting an electron density $n_e=100$ cm$^{-3}$ (this is around the critical density of the lines and is what is inferred for the densities from [S {\sc ii}] in \citealt{gree11}; radio galaxies at $z$=2--3 also have $n_e$ of a few 100 cm$^{-3}$ as found by \citealt{nesv06,nesv08}), we find $M_{\rm gas}$=(0.2--1)$\times10^9~M_{\odot}$, with a median of $6\times10^8~M_{\odot}$. For an outflow with a constant velocity of 760 km s$^{-1}$, we estimate the total kinetic energy of the outflowing ionized gas to be 
\begin{equation}
E_{\rm kin}=\frac{1}{2} M_{\rm gas}v_{\rm gas}^2=10^{56.9\mhyphen57.8}~{\rm erg}.
\end{equation}
\citet{nesv08} find extinction $A_V$=1--4 mag in high redshift radio galaxies, and taking into account the dust extinction can roughly scale up this estimate by an order of magnitude. Hence, the standard calculation using Balmer lines yields a total kinetic energy of outflowing ionized gas of $10^{58\mhyphen59}$ erg. The energy injection rate required to yield this amount of energy over the life time of the nebula $\tau=1.8\times 10^7$ years is therefore $\dot{E}_{\rm kin}\sim 2\times 10^{43\mhyphen44}$ erg s$^{-1}$.

\subsection{Kinetic energy and mass flow}

We have additional information on our winds which allows us to improve on this calculation. The narrow-line--emitting gas is likely in the form of relatively dense clouds embedded in hot, low density, volume-filling wind. If the clouds are dense and big enough, they remain largely optically thick to the ionizing quasar radiation. Only a thin shell on the surface of these so-called ionization-bounded clouds would produce emission lines. As the clouds move out with the wind, the ambient pressure declines, the clouds expand and they become optically thin to ionizing radiation (i.e., they enter ``matter-bounded'' regime). Both the size-luminosity relationship we see in quasar nebulae ($R_{\rm [O~{\scriptscriptstyle III}]}\propto L_{\rm [O~{\scriptscriptstyle III}]}^{0.25\pm 0.02}$) and the [O {\sc iii}]/H$\beta$ ratios strongly declining in the outer parts of the nebulae indicate that we have likely detected this transition which occurs at the mean distance of $D=7.0\pm1.8$ kpc among the objects in our sample \citep{liu13a}.

We now estimate the rate of kinetic energy flow at distance $D$ from the quasar. To this end, we use the clouds that are transitioning from the ionization-bounded to the matter-bounded regime and are thus in the ``sweet spot'' for such calculation. On the one hand, these clouds are fully ionized and therefore we are not missing any neutral mass in this estimate. On the other hand, the number of ionizing photons in these clouds is balanced by the number of recombinations and therefore these clouds do not become over-ionized to higher ionization states and temperatures than would be observable with our data \citep{liu13a}. 

Let $\xi$ be the number density of such clouds per unit volume at distance $D$, $m_{\rm c}$ their mass and $v_{\rm c}$ their velocity. Then the kinetic energy flow due to these clouds through the spherical shell of radius $D$ from the quasar is
\begin{equation}
\dot{E}_{\rm kin}=2 \pi D^2 \xi m_{\rm c} v^3_{\rm c}. \label{eq:energy}
\end{equation}
Both $\xi$ and $m_{\rm c}$ are at the moment very poorly known, but fortunately their product is directly related to the amount of luminosity density in recombination photons that these clouds produce:
\begin{equation}
j_{\rm H\beta}=h\nu_{\rm H\beta}R_{\rm H\beta}n_{\rm H}n_e \xi m_{\rm c}\rho_{\rm c}^{-1}.
\end{equation}
Here $R_{\rm H\beta}\approx 1.6\times 10^{-14}$ cm$^3$ s$^{-1}$ is the emissivity coefficient of H$\beta$ at temperature 20,000 K \citep{oste06}, $h\nu_{\rm H\beta}$ is the energy of each photon, $n_e$ and $n_{\rm H}$ are electron density and hydrogen density related via $n_e\approx 1.2 n_{\rm H}$ for the mix of ionized hydrogen and helium, $\rho_{\rm c}=n_{\rm H} m_{\rm proton}/X$ is the mass density of the clouds and $X=0.7$ is the hydrogen fraction by mass. 

At the same time, $j_{\rm H\beta}$ can be strongly constrained from our observations. Within the matter-bounded region, the emission line surface brightness of our nebulae falls off as $I_{\rm H\beta}(R)\propto R^{-\zeta}$ with $\zeta=3.5\pm 1.0$ among the objects in our sample \citep{liu13a}. Assuming a spherically symmetric nebula and using Abel transform, we find
\begin{equation}
j_{\rm H\beta}(D)=\frac{I_{\rm H\beta}(D)}{1.44 D},
\end{equation}
where the numerical coefficient $\sqrt{\pi}\Gamma(\frac{\zeta}{2})/\Gamma(\frac{\zeta+1}{2})=1.44$ is calculated for $\zeta=3.5$. The surface brightness at the transition point $I_{\rm H\beta}(D)$ is an observable equal to $4 \pi \Sigma_{\rm H\beta}(D) \times$ (arcsec/radian)$^2$, where $\Sigma$ is the surface brightness in units of erg s$^{-1}$ cm$^{-2}$ arcsec$^{-2}$. Using values of $\Sigma_{\rm [OIII]}(D)$ and \oiii/H$\beta$ at the break radius presented in Paper I, we find that the values of $\Sigma_{\rm H\beta}(D)$ (where the observed values need to be multiplied by $(1+z)^4$ to correct for the cosmological dimming) are distributed within a narrow range from $3.0\times 10^{-15}$ to $5.6\times 10^{-15}$ (with a median of $3.4\times 10^{-15}$) erg s$^{-1}$ cm$^{-2}$ arcsec$^{-2}$.

Putting all relations into equation (\ref{eq:energy}), we find
\begin{equation}
\dot{E}_{\rm kin}=\frac{2\pi I_{\rm H\beta}(D)m_p D v_{\rm c}^3}{1.44 X h\nu_{\rm H\beta}R_{\rm H\beta}n_e}. 
\end{equation}
The electron density inside the clouds is not well-known. In the brightest parts of the narrow-line region $n_e$ is usually assumed to be a few hundred cm$^{-3}$ from diagnostic line ratios, but no good measurements are available in the outer parts of the nebulae. If the clouds are confined by the wind whose pressure declines $\propto r^{-2}$ (as required by the nearly constant \oiii/H$\beta$ ratios seen over a large range of distances), then we expect the density inside the clouds to decline $\propto r^{-2}$ as well because the temperature established in photo-ionization balance is roughly constant \citep{liu13a}. Thus if $n_e\approx 100$ cm$^{-3}$ at 1 kpc from the quasar, we expect it to fall to at most a few cm$^{-3}$ at the transition distance $D=7$ kpc. Taking the upper limit on $n_e$ to be 10 cm$^{-3}$ following \citet{rupk13}, we find a lower limit on kinetic energy:
\begin{eqnarray}
\frac{\dot{E}_{\rm kin}}{4.1\times 10^{44}~{\rm erg~s^{-1}}}=
\left(\frac{\Sigma_{\rm H\beta}(D)}{5\times 10^{-15}~{\rm erg~s^{-1}~cm^{-2}~arcsec^{-2}}}\right) \nonumber \\
\left(\frac{D}{7~{\rm kpc}}\right)\left(\frac{v_{\rm c}}{760~{\rm km~s^{-1}}}\right)^3
\left(\frac{n_e}{10~{\rm cm}^{-3}}\right)^{-1}.
\label{eq:kinenergy}
\end{eqnarray}

This calculation assumes that clouds constitute only a fraction of the volume of the wind:
\begin{eqnarray}
F=\xi m_{\rm c}\rho_{\rm c}^{-1}=0.014 \left(\frac{\Sigma_{\rm H\beta}(D)}{5\times 10^{-15}~{\rm erg~s^{-1}~cm^{-2}~arcsec^{-2}}}\right) \nonumber \\
\left(\frac{D}{7~{\rm kpc}}\right)\left(\frac{n_e}{10~{\rm cm}^{-3}}\right)^{-2}.
\end{eqnarray}
Since this filling factor is constrained to be $\le 1$, the minimal $n_e$ of the clouds at the transition distance (corresponding to clouds filling up all the volume) is 1.2 cm$^{-3}$, giving the maximal kinetic energy of $3.7\times 10^{45}$ erg s$^{-1}$.

This is an impressively high amount of kinetic energy flux in just the ionized clouds alone. To re-iterate, in deriving this estimate we only take into account the matter that is directly observed via its hydrogen recombination lines. Neither the highest density neutral or molecular phase, nor the lowest density volume-filling phase are included in this calculation. We use two observables from Paper I: the distance from the quasar at which the clouds transition from ionization-bounded to matter-bounded, and the surface brightness of the H$\beta$ line at this distance. Both of these values have very small spread among the objects in our sample, perhaps because our targets were originally selected to be within a narrow range of [O {\sc iii}] luminosity. Our third observable -- the velocity of the outflow -- is the result of measurements in this paper. The energy injection rate derived using our new method is higher than the standard Case B estimates. The Case B estimate accounts for just the mass of the nebula that is currently in the ionized form: if any clouds are ionization-bounded and retain a neutral core, only the kinetic energy of the ionized part is taken into account. In our new estimate, we use the fact that the entire mass of the clouds becomes ionized (and thus visible in emission lines) at the transition distance.

The mass outflow rate in the observed ionized gas is $\dot{M}=2\dot{E}_{\rm kin}/v_{\rm c}^2$, or
\begin{eqnarray}
\frac{\dot{M}}{2240~M_{\odot}~{\rm yr}^{-1}}=\left(\frac{\Sigma_{\rm H\beta}(D)}{5\times 10^{-15}
{\rm erg~s^{-1}~cm^{-2}~arcsec^{-2}}}\right)\nonumber \\
\left(\frac{D}{7~{\rm kpc}}\right)\left(\frac{v_{\rm c}}
{760~{\rm km s^{-1}}}\right)\left(\frac{n_e}{10~{\rm cm}^{-3}}\right)^{-1}.\label{eq:mass}
\end{eqnarray}
Furthermore, if the lower limit to the electron density is again given by the constraint on the volume filling factor of the ionized gas and is equal to $n_e=1.2$ cm$^{-3}$, then we get an upper limit on mass flux of $\dot{M}=1.9\times 10^4$ $M_{\odot}$ yr$^{-1}$. Thus if the feedback episode lasts for $10^{6\mhyphen7}$ years with the same intensity as observed at the current epoch, the wind is capable of eliminating $2\times 10^{9\mhyphen10} M_{\odot}$ worth of gas from the galaxy at the rate given by equation (\ref{eq:mass}) and an order of magnitude more if the density $n_e$ is closer to its lower limit. If the clouds originated close to the center of the host galaxy, then the time required for them to travel out to the median extent of the nebulae of $14\pm 4$ kpc at 760 km s$^{-1}$ is $1.8 \times 10^7$ years; this sets the minimal duration of the activity episode. This is a sufficient time to eliminate much of the gas available for star formation even in a massive, gas-rich galaxy at the rate given by equation (\ref{eq:mass}). We discuss the application of equations (\ref{eq:kinenergy}) and (\ref{eq:mass}) to individual objects in the next subsection. 

\subsection{Powering the outflow}

What can power an outflow of such large kinetic energies? Although star-bursting and star-forming galaxies can produce spectacular galaxy-wide outflows, this is likely an insufficient source of energy in our objects. Indeed (repeating the arguments we presented in \citealt{gree12}), for every solar mass of stars formed the stellar winds and supernovae can yield at most $7\times 10^{41}$ erg s$^{-1}$ worth of kinetic energy of the outflow \citep{leit95, veil05}. Thus in order to produce outflow energies in the range $4\times 10^{44\mhyphen45}$ erg s$^{-1}$, we need infrared luminosities due to star formation $L_{\rm FIR}>10^{46}$ erg s$^{-1}$. Although we do not know the star formation rates in the objects in our IFU sample, the median star-formation luminosity of the host galaxies of luminous obscured quasars is $2\times 10^{45}$ erg s$^{-1}$ \citep{zaka08} which would be inadequate for producing the observed outflows by at least an order of magnitude. 

A more directly observable way to compare outflows from star-forming galaxies and quasars appears to be outflow velocity. Although without a mass measurement the velocities do not directly translate into outflow kinetic energy, they are easily measurable and appear to be significantly different in our quasars compared to even the most extreme star-forming objects. \citet{rupk05} and \citet{rupk13} conducted a study of outflows from ultraluminous infrared galaxies and concluded that typical spatially averaged maximal outflow velocities in starburst-powered sources are $\langle v_{02} \rangle \sim -500$ km s$^{-1}$, whereas the quasar-dominated objects in their sample tended to show much higher values -- a median of $\langle v_{02} \rangle \sim -1000$ km s$^{-1}$ in the ionized gas outflow from the three quasars presented by \citealt{rupk13}. We find an excellent agreement between this value and the one derived from our sample, in which the sample median and dispersion is $\langle v_{02} \rangle \sim -950\pm 370$ km s$^{-1}$. The maximal values $v_{\rm 02,max}$ are also more extreme in quasar-driven outflows than they are in starburst-driven ones, by about 500 km s$^{-1}$ on average in \citet{rupk13}. These values reach $-2400$ km s$^{-1}$ in our sample and $-3350$ km s$^{-1}$ in quasars in their sample, although there is always a chance that beam smearing in our much more distant objects biases this value low.

\begin{figure*}%[h!]
\centering
    \includegraphics[scale=0.7,trim=5cm 10cm 4cm 0cm]{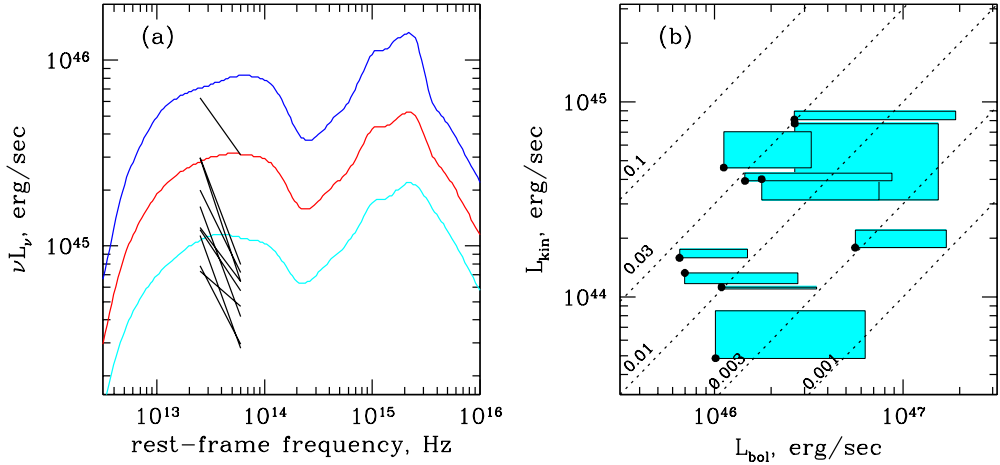}
\caption{Left: monochromatic luminosities of 10 obscured quasars (black) at 5 and 12 \micron\ in the rest-frame connected using a power-law fit to the spectral energy distribution. Also shown are composite spectra of unobscured quasars from \citet{rich06} (bright, reddened and faint quasars from top to bottom). Right: the relationship between bolometric luminosities of the quasars and the kinetic energies of their outflows as measured from the IFU data. Black points show bolometric luminosities computed using $\nu L_{\rm \nu,12\mu m}$ with a bolometric correction of 9 from \citet{rich06} and kinetic luminosities computed from the $W_{80}$ of the entire line profile fit. The blue rectangles show the range of values we obtain when we correct for obscuration and / or allow for a two-component outflow. Slanted dashed lines show different values of the fraction of the bolometric luminosity converted to the kinetic energy of the outflow. }
\label{fig:energetics} 
\end{figure*}

We therefore assume that the observed outflows are powered by the quasar activity and set out to determine the fraction of the bolometric luminosity that is converted into the kinetic energy of the ionized gas. For this calculation we use ten objects -- all our radio-quiet quasars except SDSS J0319$-$0019 in which the transition from the ionization-bounded to the matter-bounded regime was not detected in Paper I. Determining the bolometric luminosity of obscured quasars is not a straightforward enterprise since they are obscured at X-ray, optical and even mid-infrared frequencies \citep{zaka08, jia12}. Narrow emission lines themselves can be used as a tracer of the bolometric activity, but line luminosity vs. bolometric luminosity correlations have a large scatter ($\sigma\sim 0.5$ dex). The best data available to us at the moment for the determination of the bolometric luminosity is from the Wide-field Infrared Survey Explorer (WISE; \citealt{wrig10}) which provides all-sky catalogs at 3.4, 4.6, 12 and 22\micron. All the sources in this paper are detected by WISE which allows us to calculate K-corrected monochromatic luminosities $\nu L_{\nu}$ at rest-frame 5 and 12\micron. These values are shown in Figure \ref{fig:energetics}a together with spectral energy distributions of unobscured quasars \citep{rich06}. 

The simplest approach is to apply the standard photometric corrections derived from unobscured quasars (e.g., a factor of 9 at 12\micron, \citealt{rich06}). However, the spectral energy distributions of obscured quasars are strikingly redder than those of unobscured ones, indicating obscuration even at mid-infrared wavelengths \citep{zaka08}. Therefore, such a basic photometric correction would likely result in an underestimate of the bolometric luminosity, since WISE data clearly do not probe the peak of the spectral energy distributions of our sources. Another approach is to extrapolate the spectral energy distribution of our sources to 30\micron\ using the observed slopes, assume that the emission at that wavelength is not significantly affected by absorption, and again use the bolometric correction derived from unobscured quasars (factor of 13 at this wavelength, \citealt{rich06}). We use both approaches to bracket our estimate of the bolometric luminosity which is now contained to $\sim 0.7$ dex intervals (Figure \ref{fig:energetics}b) but is still rather uncertain. The black points in Figure \ref{fig:energetics}b derived from the $\nu L_{\rm \nu,12\mu m}$ data provide strict lower limits on the bolometric luminosities. 
Although our data does not allow for calculating the black hole masses, the median value of the derived bolometric luminosities of our sample ($10^{46.8}$ erg s$^{-1}$) is close to that of \citet[][$10^{46.7}$ erg s$^{-1}$]{liu09}. Adopting their median Eddington ratio ($\sim$0.7), we estimate the Eddington luminosity of our sample to be $L_{\rm Edd}\sim10^{47.0}$ erg s$^{-1}$.

As for the kinetic energy of the outflow, all values which enter equation (\ref{eq:kinenergy}) are tabulated in Paper I and repeated in Table \ref{tab2}, with the exception of $v_{\rm c}$ which we take to be $W_{80}/1.3$ with $W_{80}$ from Table \ref{tab1}. Splitting the wind into a slow component and a fast one as discussed in Section \ref{sec:modsum} results in a small change in the kinetic energy estimate. 

Taking just the black points in Figure \ref{fig:energetics}-b, we find that there may be a weak positive correlation between the kinetic energy of the ionized gas outflow and the estimated bolometric luminosity, but it is not yet established with reasonable statistical confidence. The probability of the null hypothesis (uncorrelated datasets) as measured by the Spearman rank correlation test is about 4\%. The fraction of the bolometric luminosity converted to the kinetic energy of the ionized gas is between 0.3\% and 10\% in all cases, with the median of 2.3\%. These rates of conversion of bolometric luminosity to kinetic energy of the outflow are not dissimilar from those found by \citet{harr12}, and are also consistent with the recent spectroscopic observation of a broad absorption line quasar by \citet{arav13} that finds a large-scale outflow (3 kpc from the central source) with a similar kinetic luminosity ($\dot{E}_{\rm kin}\sim10^{45}$ erg s$^{-1}$, $\sim$1\% of the bolometric luminosity).

\begin{table*}
\caption{Derived properties of ten radio-quiet quasars.}
\setlength{\tabcolsep}{1.4mm}
\label{tab2}
%\begin{small}
\begin{center}
\begin{tabular}{crcccccccc}
\hline
\noalign{\smallskip}
\multicolumn{1}{c}{Object name} &
\multicolumn{1}{c}{$R_{\rm br}$} &
\multicolumn{1}{c}{$\Sigma_{\rm [O~{\scriptscriptstyle III}],br}$} &
\multicolumn{1}{r}{[O {\sc iii}]/H$\beta$} &
\multicolumn{1}{c}{$\nu L_{\rm \nu,5\mu m}$} &
\multicolumn{1}{c}{$\nu L_{\rm \nu,12\mu m}$} &
\multicolumn{1}{c}{$L_{\rm bol,12\mu m}$} &
\multicolumn{1}{c}{$L_{\rm bol,30\mu m}$} &
\multicolumn{1}{c}{${\dot E}_{\rm kin,W_{80}}$} &
\multicolumn{1}{c}{${\dot E}_{\rm kin,2G}$} \\
\multicolumn{1}{c}{(1)} &
\multicolumn{1}{c}{(2)} &
\multicolumn{1}{c}{(3)} &
\multicolumn{1}{c}{(4)} &
\multicolumn{1}{c}{(5)} &
\multicolumn{1}{c}{(6)} &
\multicolumn{1}{c}{(7)} &
\multicolumn{1}{c}{(8)} &
\multicolumn{1}{c}{(9)} &
\multicolumn{1}{c}{(10)} \\
\hline

SDSS J014932.53$-$004803.7 &  4.1 & $-$13.6 & 31.8 & 44.62 & 45.21 & 46.16 & 46.94 & 44.59 & 44.64 \\
SDSS J021047.01$-$100152.9 &  7.5 & $-$14.0 & 10.0 & 44.67 & 44.87 & 45.82 & 46.17 & 44.20 & 44.24 \\
SDSS J031950.54$-$005850.6 &  7.5 & $-$14.3 & 11.9 & 44.47 & 44.89 & 45.84 & 46.44 & 44.12 & 44.07 \\
SDSS J032144.11$+$001638.2 & 11.0 & $-$14.1 & 16.2 & 44.81 & 45.10 & 46.05 & 46.51 & 44.66 & 44.85 \\
SDSS J075944.64$+$133945.8 &  7.5 & $-$14.0 & 16.1 & 44.81 & 45.47 & 46.42 & 47.28 & 44.91 & 44.95 \\
SDSS J084130.78$+$204220.5 &  6.4 & $-$14.1 & 15.7 & 44.76 & 45.09 & 46.04 & 46.54 & 44.05 & 44.04 \\
SDSS J084234.94$+$362503.1 &  9.0 & $-$14.2 & 10.3 & 44.45 & 45.06 & 46.01 & 46.80 & 43.69 & 43.93 \\
SDSS J085829.59$+$441734.7 &  5.6 & $-$14.0 & 11.9 & 45.49 & 45.80 & 46.75 & 47.23 & 44.25 & 44.34 \\
SDSS J103927.19$+$451215.4 &  5.8 & $-$14.1 & 12.2 & 44.86 & 45.30 & 46.25 & 46.87 & 44.60 & 44.50 \\
SDSS J104014.43$+$474554.8 &  7.6 & $-$14.2 &  8.7 & 44.90 & 45.48 & 46.43 & 47.19 & 44.89 & 44.50 \\

\hline
\end{tabular}
\tablenotes{{\bf Notes.} -- 
{\bf (1)} Object name.
{\bf (2)} Break radius (semi-major axis) of the [O {\sc iii}]/H$\beta$ radial profile \citep[in kpc, from][]{liu13a}.
{\bf (3)} [O {\sc iii}] surface brightness where the break of the [O {\sc iii}]/H$\beta$ radial profile happens \citep[logarithmic scale, in units of erg s$^{-1}$ cm$^{-2}$ arcsec$^{-2}$; cf. Figure 6 in][]{liu13a}, not corrected for cosmological dimming.
{\bf (4)} [O {\sc iii}]-to-H$\beta$ line ratio in the plateau region of its radial profile \citep[cf. Figure 6 in][]{liu13a}. 
{\bf (5)} WISE luminosity at rest-frame 5 $\mu$m (logarithmic scale, in erg s$^{-1}$).
{\bf (6)} WISE luminosity at rest-frame 12 $\mu$m (logarithmic scale, in erg s$^{-1}$).
{\bf (7)} Bolometric luminosity (logarithmic scale, in erg s$^{-1}$), derived from 12 $\mu$m WISE luminosity 
   with a bolometric correction of 9 from \citet{rich06}.
{\bf (8)} Bolometric luminosity (logarithmic scale, in erg s$^{-1}$), derived from 30 $\mu$m WISE luminosity 
   (extrapolated from the measurements) with a bolometric correction of 13 from \citet{rich06}.
{\bf (9)} Kinetic energy (logarithmic scale, in erg s$^{-1}$), derived from $W_{80}$ values in Table \ref{tab1} using 
   Equation \ref{eq:kinenergy}.
{\bf (10)} Kinetic energy (logarithmic scale, in erg s$^{-1}$), derived from two-component Gaussian fits using 
   Equation \ref{eq:mass}.
}
\end{center}
%\end{small}
\end{table*}

\section{Conclusions}
\label{sec:summary}

In this paper, we present kinematic measurements of the warm ($\sim10^4$ K) ionized gas surrounding eleven luminous obscured quasars obtained using Gemini-GMOS IFU observations. Our key observational findings are as follows:
\begin{itemize}
\item The radial velocity differences of the emission-line gas across the nebulae are modest, ranging between 90 and 520 km s$^{-1}$ for the objects in our sample.
\item The range of velocities comprising 80\% of the total line power ($W_{80}$; a measure that for a Gaussian profile is very similar to FWHM) is between 500 and 1800 km s$^{-1}$ for the spatially integrated emission from our sources, with some parts of some objects displaying $W_{80}$ up to 2300 km s$^{-1}$. The high values of the line-of-sight velocity dispersion rule out the possibility that the gas clouds are gravitationally bound to the host galaxy or are in an equilibrium configuration in a galactic disk. 
\item The combination of modest velocity differences and high velocity dispersions, as well as nearly flat profiles $W_{80}(R)$ lead us to propose a model of quasi-spherical outflows with a median velocity of 760 km s$^{-1}$ among the objects in our sample which qualitatively explains our kinematic data.  Escape velocities from galaxies are poorly known, but in the massive galaxies characteristic of our sample the gas needs to move at 500--1000 km s$^{-1}$ to reach 100 kpc \citep{gree11}; the inferred outflow velocities are comparable to these values. 
\item A spherically symmetric outflow would show a zero velocity gradient on the sky and a line-of-sight velocity dispersion closely tied to the outflow velocity. Partial reddening of parts of the outflow due to dust within the host galaxy or a slight collimation of the wind into a wide-angle bi-conical configuration is able to simultaneously explain the relatively small velocity differences $\Delta v_{\rm max}$ seen in our data and the high velocity dispersions.
\item Line profiles are highly non-Gaussian, typically displaying a relatively narrow core and a broad base. This likely reflects the range of cloud velocities at every volume element of the outflow. In the central parts of the nebulae, all objects except one show line profiles with excess blueshifted gas, a classical characteristic of an outflow whose far (redshifted) side is partially extincted by the host galaxy. 
\item The emission line widths $W_{80}$ are nearly constant as a function of projected radius from the center, with indications of a slow decline at large distances. This may be due to the cloud velocities becoming less random and more radial in the outer parts of the nebula. 
\item In Section \ref{sec:bubbles}, we present four new candidates for quasar-driven ``super-bubbles'', which are in addition to the candidate we presented in an earlier paper (SDSS J1355+1026; \citealt{gree12}). In such objects, the quasar-driven wind breaks out of the high-density host galaxy and expands into the lower-density inter-galactic medium, preferentially in the direction perpendicular to the disk of the host galaxy or the disk of circumnuclear material. The sharp decline in the observed line width within the bubbles compared to the central parts of these objects is roughly consistent with a purely geometric model in which the outflow switches from a spherically symmetric to a conical one. 
\item Ionized gas nebulae around radio-quiet obscured quasars are rounder and better kinematically organized than those around radio-loud objects which tend to be elongated, lumpy and with multiple kinematic components. 
\end{itemize}

We previously demonstrated (Paper I; \citealt{liu13a}) that the ionized gas extends to large ($>10$ kpc) distances from the active nucleus and that the emission line nebulae around radio-quiet quasars are significantly rounder and less structured than those around radio-loud quasars. These observations indicate that large amounts of gas are present throughout the host galaxy of radio-quiet quasars with large covering factors. In combination with the kinematic data presented in this paper, we find strong evidence that the ionized gas is in fact embedded in a high-velocity wind ($\sim 760$ km s$^{-1}$, our median value from the spherical model) which engulfs the entire galaxy and has a wide opening angle.

Furthermore, in Paper I we showed that ionized gas clouds are likely to be present beyond the visible nebulae, but having followed the declining pressure of the wind they are at low densities and are thus not visible in the optical transitions we can study with our data. We observed a strong decline in the [O {\sc iii}]/H$\beta$ ratio beyond about 7 kpc from the quasar and interpreted this observation as the transition of the clouds from the ionization-bounded to matter-bounded regime. This interpretation is further supported by the slow increase of the narrow-line region size with the [O {\sc iii}] luminosity ($R_{\rm [O~{\scriptscriptstyle III}]}\propto L_{\rm [O~{\scriptscriptstyle III}]}^{0.25\pm 0.02}$). 

In this paper, using kinematic information in combination with the measurements of the transition distance and the surface brightness of the nebulae at this distance we robustly determine lower and upper bounds on the kinetic energy carried by ionized gas clouds to be from $4.1 \times 10^{44}$ to $3.4 \times 10^{45}$ erg s$^{-1}$. Both these values scale as $(v/760\mbox{ km s}^{-1})^3$ with the velocity of the outflow, but otherwise are fairly insensitive to the largely unknown physical conditions in the nebulae and are determined largely based on the previous observations at the transition regime. Furthermore, the transition distances and especially the surface brightness values at the transition distance that underlie the energy calculation have very small variations among the objects in our sample. The corresponding mass outflow rates are between $2.2\times 10^3$ and $1.9 \times 10^4 M_{\odot}$ yr$^{-1}$ and scale as $(v/760\mbox{ km s}^{-1})$. The life-time of the episode of quasar wind activity may be estimated at $\sim$1.8$\times$10$^7$ years as the travel time of clouds to reach the observed distances from the center. Our derived values of outflow velocities and mass outflow rates are not dissimilar from those predicted by models of feedback necessary to regulate the black hole / galaxy correlations \citep{king10, zubo12}.

Bolometric luminosity measurements in obscured quasars are rather uncertain. Using various methods, we estimate that within our sample the median conversion factor from bolometric luminosity to kinetic energy of ionized gas is 2.3\%; there could be additional outflow components whose energy is not captured by our calculations. These large values of kinetic energy flux and mass flux, combined with outflow velocities similar to or exceeding the escape speed from the host galaxy, suggest that winds observed in obscured (type 2) quasars can be the long-sought signature of powerful and ubiquitous quasar feedback. At these energies and mass outflow rates, the winds are bound to make a strong impact on the formation of massive galaxies. 

\section*{Acknowledgments}

N.L.Z. and J.E.G. would like to thank R.D. Blandford, E.C. Ostriker, J.P. Ostriker, G.S. Novak, and S.D. Tremaine for the useful discussions. N.L.Z. is grateful to the Institute for Advanced Study in Princeton, NJ (where part of this work was performed) for hospitality. Part of this work was performed while N.L.Z. was supported by a Kavli Fellowship at SLAC / Stanford University. N.L.Z. and J.E.G. are supported in part by the Alfred P. Sloan fellowship. G.L. and N.L.Z. acknowledge support from the Theodore Dunham, Jr. Grant of the Fund for Astrophysical Research. Support for the work of X.L. was provided by NASA through Hubble Fellowship grant number HST-HF-51307.01 awarded by the Space Telescope Science Institute, which is operated by the Association of Universities for Research in Astronomy, Inc., for NASA, under contract NAS 5-26555.

Funding for SDSS-III has been provided by the Alfred P. Sloan Foundation, the Participating Institutions, the National Science Foundation, and the U.S. Department of Energy Office of Science. The SDSS-III web site is http://www.sdss3.org/.

SDSS-III is managed by the Astrophysical Research Consortium for the Participating Institutions of the SDSS-III Collaboration including the University of Arizona, the Brazilian Participation Group, Brookhaven National Laboratory, Carnegie Mellon University, University of Florida, the French Participation Group, the German Participation Group, Harvard University, the Instituto de Astrofisica de Canarias, the Michigan State/Notre Dame/JINA Participation Group, Johns Hopkins University, Lawrence Berkeley National Laboratory, Max Planck Institute for Astrophysics, Max Planck Institute for Extraterrestrial Physics, New Mexico State University, New York University, Ohio State University, Pennsylvania State University, University of Portsmouth, Princeton University, the Spanish Participation Group, University of Tokyo, University of Utah, Vanderbilt University, University of Virginia, University of Washington, and Yale University.

\bibliographystyle{mn2e}
\bibliography{master}

%\begin{thebibliography}{}

%\bibitem[Zakamska et al.(2006)]{Zakamska06} Zakamska,.................

%\end{thebibliography}

\label{lastpage}

\end{document}